\documentclass[]{spie}  

\usepackage{amsmath,amsfonts,amssymb}
\usepackage{graphicx}
\usepackage{multirow}
\usepackage{float}
\usepackage[dvipsnames]{xcolor}
\usepackage[colorlinks=true, allcolors=blue]{hyperref}

\newcommand{\tpr}[1]{\textcolor{PineGreen}{#1}}
\newcommand{\fdr}[1]{\textcolor{BrickRed}{#1}}

\title{Exoplanet Imaging Data Challenge: benchmarking the various image processing methods for exoplanet detection}

\author[a]{Cantalloube~F.}
\author[b]{Gomez-Gonzalez~C.}
\author[c]{Absil~O.}
\author[c,d]{Cantero~C.}
\author[e]{Bacher~R.}
\author[f]{Bonse~J.~M.}
\author[g]{Bottom~M.}
\author[c]{Dahlqvist~C.-H.}
\author[a,h]{Desgrange~C.}
\author[i]{Flasseur~O.}
\author[j]{Fuhrmann~T.}
\author[a]{Henning~Th.}
\author[k]{Jensen-Clem~R.}
\author[l]{Kenworthy~M.}
\author[m]{Mawet~D.}
\author[n]{Mesa~D.}
\author[m]{Meshkat~T.}
\author[e]{Mouillet~D.}
\author[a]{M\"uller~A.}
\author[a]{Nasedkin~E.}
\author[o]{Pairet~B.}
\author[d]{Pierard~S.}
\author[m]{Ruffio~J.-B.}
\author[p]{Samland~M.}
\author[q]{Stone~J.}
\author[d]{Van~Droogenbroeck~M.}

\affil[a]{Max-Planck-Institut f\"{u}r Astronomie, K\"{o}nigstuhl 17, Heidelberg 69117, Germany}
\affil[b]{Barcelona Supercomputing Center, Spain}
\affil[c]{Space Sciences, Technologies \& Astrophysics Research (STAR) Institute, Universit\'e de Li\`ege, All\'ee du Six Ao\^{u}t 19c, B-4000 Li\`ege, Belgium}
\affil[d]{Montefiore Institute, Universit\'e de Li\`ege, 4000 Li\`ege, Belgium}
\affil[e]{Université Grenoble Alpes, CNRS, IPAG, 38000, Grenoble, France}
\affil[f]{Institute for Particle Physics and Astrophysics, ETH Zurich, Wolfgang-Pauli-Strasse 27, 8093, Zurich, Switzerland}
\affil[g]{Jet Propulsion Laboratory, California Institute of Technology, Pasadena, California, United States}
\affil[h]{Department of Physics, Ecole Normale Supérieure de Lyon, 69364, Lyon, France}
\affil[i]{Université de Lyon, Université Lyon1, ENS de Lyon, CNRS, Centre de Recherche
Astrophysique de Lyon UMR 5574, F-69230, Saint-Genis-Laval, France}
\affil[j]{Ostbayerische Technische Hochschule, Regensburg, Germany}
\affil[k]{Astronomy Department, University of California Santa Cruz, 1156 High St., Santa Cruz, CA 95064, USA}
\affil[l]{Leiden Observatory, Leiden University, P.O. Box 9513, 2300 RA Leiden, The Netherlands}
\affil[m]{Department of Astronomy, California Institute of Technology, 1200 E. California Blvd., Pasadena, CA 91125, USA}
\affil[n]{INAF-Osservatorio Astronomico di Padova, Vicolo dell'Osservatorio 5, I-35122 Padova, Italy}
\affil[o]{ISPGroup, ELEN/ICTEAM, UCLouvain, B-1348 Louvain-la-Neuve, Belgium}
\affil[p]{Department of Astronomy, Stockholm University, Stockholm, Sweden}
\affil[q]{Steward Observatory, University of Arizona, 933 N. Cherry Ave., Tucson, AZ 85721-0065, USA}

\authorinfo{Further author information: (Send correspondence to F.C)\\
F.C.: E-mail: cantalloube@mpia.de}

\begin{document} 
\maketitle

\begin{abstract}
The \emph{Exoplanet Imaging Data Challenge} is a community-wide effort meant to offer a platform for a fair and common comparison of image processing methods designed for exoplanet direct detection. 
For this purpose, it gathers on a dedicated repository (Zenodo), data from several high-contrast ground-based instruments worldwide in which we injected synthetic planetary signals. The data challenge is hosted on the CodaLab competition platform, where participants can upload their results. The specifications of the data challenge are published on our website \url{https://exoplanet-imaging-challenge.github.io/}. 
The first phase, launched on the 1st of September 2019 and closed on the 1st of October 2020, consisted in detecting point sources in two types of common data-set in the field of high-contrast imaging: data taken in pupil-tracking mode at one wavelength (subchallenge 1, also referred to as ADI) and multispectral data taken in pupil-tracking mode (subchallenge 2, also referred to as ADI+mSDI). 
In this paper, we describe the approach, organisational lessons-learnt and current limitations of the data challenge, as well as preliminary results of the participants' submissions for this first phase. 
In the future, we plan to provide permanent access to the standard library of data sets and metrics, in order to guide the validation and support the publications of innovative image processing algorithms dedicated to high-contrast imaging of planetary systems. 
\end{abstract}


\section{INTRODUCTION}
\label{sec:intro}  
As of today, there exist a large number of algorithms to detect exoplanets within high-contrast images. During the last few years, this field of research was very active in order to reach fainter signals and accurately recover their properties, leading to a few publications per year. However, in each publication of a new algorithm, the authors choose to demonstrate their method on a different data set and using different metrics compared to previous publications. In order to avoid confusion in the community, and to offer a catalog of existing methods adapted to various science cases (e.g. large surveys or in-depth characterisation), we propose to homogenize the comparison between methods through the \emph{exoplanet imaging data challenge} (EIDC). The EIDC offers a common ensemble of representative data sets from different high-contrast instruments, as well as a common set of metrics to compare and validate current and new methods.

This data challenge is the outcome of a collaboration between several institutes worldwide and is an open-source community effort, allowing anyone to contribute. The underlying idea is to share knowledge and spark new collaborations to make the best use of the data we acquired from various telescopes. The ensemble of data sets are hosted on a \emph{Zenodo}\footnote{\url{https://zenodo.org/record/3361544\#.X8pxK9so9GE}} open-access repository. The participants are invited to submit their results via the \emph{CodaLab}\footnote{\url{https://competitions.codalab.org/competitions/20693\#participate}} competition platform. Upon reception, CodaLab computes various metrics that we have defined and calculates a score publicly displayed on a leader-board. At last, relevant information about the exoplanet imaging data challenge is gathered on a dedicated website \url{https://exoplanet-imaging-challenge.github.io/}. 

We communicated about this data challenge via various media (mailing-lists, social networks, instrument consortia and advertisement during conferences). The first phase was officially launched on the 1st of September 2019, with a preliminary deadline on the 30th of January 2020. We discussed the outcome of this preliminary phase during a dedicated workshop, called \emph{Post-processing for high-contrast imaging of exoplanets and circumstellar disks}, which took place at the end of January 2020 at the Max Planck society Harnack-haus (Berlin, Germany). During this workshop we discussed the feedback from participants and the relevance of the metrics used. At the end of the workshop we decided to form a working group (contact: \url{exoimg.datachallenge@gmail.com}), to add more information about the data sets on the website and to reveal the results from one single data set (\url{https://exoplanet-imaging-challenge.github.io/preresults/}). The latter proved useful for several reasons, in particular: (1) verifying the convention used when running the algorithms on the data, (2) making sure that the intention of the EIDC is not to reduce the methods to a single grade, but to broadly discuss the variety of existing algorithms and (3) encouraging other teams to participate to the data challenge. Following this, we extended the deadline for the first phase to the 1st of October 2020.

In the following, we first describe the data sets provided to the participants via the Zenodo repository (Sect.~\ref{sec:data}) and then describe the evaluation procedure used on the results collected during the first phase of the EIDC (Sect.~\ref{sec:metrics}). We then discuss the submissions we received and the rankings obtained when applying our evaluation procedure (Sect.~\ref{sec:results}). At last, we propose a path-forward for the continuation of the EIDC and for leveraging the limitations noted during this first phase.

\section{Data sets}
\label{sec:data}  
Several typical high-contrast data sets are available on the EIDC \emph{Zenodo} repository. Participants must submit the results of a single algorithm applied to the complete ensemble of data sets described in Sect.~\ref{ssec:predata}. It is possible for participants to provide several submissions, each resulting from one single algorithm applied to the ensemble of data sets. However, the \emph{CodaLab} competition platform only displays one result per participant (selected by the participant) on the public leader-board. We therefore warn users to not rely on the results shown on the \emph{CodaLab} EIDC competition web-page.
In each image cube, we injected synthetic planetary signals following the procedure described in Sect.~\ref{ssec:inj}.

\subsection{Pre-reduced data provided to the participants}
\label{ssec:predata}
The performance of an image processing algorithm can be highly dependent upon the type of data through the various temporal and spatial distribution of the starlight residuals (generally referred to as speckles) they can exhibit. To mitigate this effect and to discuss which method is best suited to which type of data, we used data from various high-contrast instruments installed on 8-m class telescopes around the world. In addition, the observing conditions shape the type of starlight residuals prevailing in the images. Again, to mitigate this effect, we selected several data sets from the same instrument, but taken under various observing conditions. 
In this section, we describe the data selected for both the ADI subchallenge (Sect.~\ref{ssec:sub1}) and the ADI+mSDI subchallenge (Sect.~\ref{ssec:sub2}). The data have been kindly provided by collaborators from the instruments consortia. 

\subsubsection{Subchallenge 1: angular diversity}
\label{ssec:sub1}
The most common observing mode to search for exoplanet candidates is to use the pupil tracking mode of the telescope (providing the telescope has an alt-az mount). In this mode the circumstellar signals rotate at a deterministic speed (following the parallactic angles), while the optical aberrations, responsible for the starlight residuals observed in the field of view, are stabilized. This angular diversity makes it possible to use image processing techniques based on Angular Differential Imaging\cite{Marois2006} (ADI).

In this first subchallenge, we provided temporal image cubes from three different instruments: VLT/SPHERE-IRDIS\cite{Beuzit2019,Dohlen2008irdis}, Keck/NIRC2\cite{serabyn2017nirc2}, and LBT/LMIRCam\cite{skrutskie2010lmircam}. For each instrument, three representative data sets are provided, making a total of nine data sets. 
The VLT/SPHERE instrument is a second generation instrument equipped with an extreme adaptive optics\cite{Fusco2006saxo} (AO) and an apodized Lyot coronagraph\cite{soummer2005aplc} (APLC), feeding light into three subsystems, including the IRDIS dual band imager\cite{Vigan2010dbi}, working from the Y2-band ($1.02~\mathrm{\mu m}$) to the K2-band ($2.25~\mathrm{\mu m}$). 
The Keck/NIRC2 instrument is equipped with an adaptive optics system\cite{wizinowich2000aokeck} and an Annular Groove Phase Mask (AGPM) vector vortex coronagraph\cite{Mawet2005agpm}, optimized for observations in the Lp-band\cite{nirc2agpm} ($3.78~\mathrm{\mu m}$).  
The LBT/LMIRCam instrument is equipped with an extreme adaptive optics system\cite{esposito2010flao} and the data for the EIDC are taken without coronagraph, in the Lp-band ($3.78~\mathrm{\mu m}$).

Each data set is constituted of four files (all given in \emph{.fits} format\cite{Wells1981}): (1) the image cube, (2) the corresponding parallactic angles variation corrected from true North, (3) the pixel scale of the detector and (4) a non-coronagraphic or non-saturated point spread function (PSF) of the instrument, taken either before or after the observing sequence, used to calibrate the detection in terms of contrast and which can be used as a model of a planetary signal (e.g. for forward modelling or supervised machine learning methods). 
The images provided are pre-reduced in terms of bad pixels removal, dark current and background subtraction, flat fielding and normalization of the flux for the PSF. We assume that the users do not perform further frame selection and that all the images are exploited. 
The VLT/SPHERE-IRDIS data were pre-reduced with the SPHERE Data Center\cite{Delorme2017sphereDC}, using the SPHERE Data Reduction and Handling pipeline\cite{Pavlov2008drh}. 
The Keck/NIRC2 data were pre-reduced by the dedicated pre-processing pipeline for AGPM images\cite{Xuan2018nirc2drh}\footnote{\url{https://github.com/vortex-exoplanet/NIRC2_Preprocessing}}. 
The LBT/LMIRCam data were pre-reduced by the pipeline developed for the LEECH exoplanet survey\cite{Stone2018leech}. 
The properties of the nine image cubes are summarized in Tab.~\ref{tab:data1}.

\begin{table}[!h]
\caption[example] 
   {\label{tab:data1} 
Description of the nine data sets provided for the EIDC ADI subchallenge. 
$D_{tel}$ is the effective telescope diameter (downstream the coronagraph when any); 
$\lambda_{obs}$ is the observation wavelength; 
Plate-scale is the pixel scale of the detector used;
$N_{img}$ is the size of each image frame; 
$N_{t}$ is the number of frames along the temporal axis;
$\Delta_{field}$ is the total field rotation of circumstellar objects (or parallactic angle variation). 
}
\begin{center}
\begin{tabular}{|l| c | c | c | c | c | c | c |  } 
\hline
Instrument & ID & $D_{tel}$ & $\lambda_{obs}$ & Plate-scale & $N_{img}$ & $N_{t}$ & $\Delta_{field}$ \\
           &    & $[\mathrm{m}]$ & $[\mathrm{\mu m}]$ & $[\mathrm{mas/pixel}]$ & $[\mathrm{px}] \times [\mathrm{px}]$ &  & $[\mathrm{^\circ}]$ \\
\hline \hline
                          & sph1 & $7.87$ & $1.625\pm 0.29$ & $12.255$ & $160 \times 160$ & $252$ & $40.3$ \\
\textbf{VLT/SPHERE-IRDIS} & sph2 & $7.87$ & $1.593\pm 0.052$ & $12.255$ & $160 \times 160$ & $80$  & $31.5$ \\
                          & sph3 & $7.87$ & $1.593\pm 0.052$ & $12.255$ & $160 \times 160$ & $228$ & $80.5$ \\
\hline
                    & nrc1 & $8.72$ & $3.776\pm 0.70$ & $20.00$ & $321 \times 321$ & $29$ & $53.0$ \\
\textbf{Keck/NIRC2} & nrc2 & $8.72$ & $3.776\pm 0.70$ & $20.00$ & $321 \times 321$ & $40$ & $37.3$ \\
                    & nrc3 & $8.72$ & $3.776\pm 0.70$ & $20.20$ & $321 \times 321$ & $50$ & $166.9$ \\
 \hline
                     & lmr1 & $8.40$ & $3.780\pm 0.10$ & $10.70$ & $200 \times 200$ & $4838$ & $153.4$ \\
\textbf{LBT/LMIRCam} & lmr2 & $8.40$ & $3.780\pm 0.10$ & $10.70$ & $200 \times 200$ & $3219$ & $60.6$ \\
                     & lmr3 & $8.40$ & $3.780\pm 0.10$ & $10.70$ & $200 \times 200$ & $4620$ & $91.0$ \\
 \hline
\end{tabular}
\end{center}
\end{table} 

\subsubsection{Subchallenge 2: angular and spectral diversity}
\label{ssec:sub2}
The second generation of instrument dedicated to exoplanet imaging, such as VLT/SPHERE and Gemini-S/GPI, are equipped with integral field units (IFU), with relatively low spectral resolving power ($R < 100$), producing images taken simultaneously at different wavelengths. In the images, the starlight residuals move radially with the wavelength, while the circumstellar signals remain fix in the field of view whatever the wavelength. This additional spectral diversity makes it possible to use image processing techniques based on simultaneous multiple Spectral Differential Imaging\cite{Racine1999,Sparks2002} (mSDI). In practice the observations are carried out in pupil tracking mode to combine ADI and mSDI techniques.

In this second subchallenge, we provided 4D image cubes from two different instruments: VLT/SPHERE-IFS\cite{Beuzit2019,Antichi2008}, and Gemini-S/GPI\cite{Macintosh2008}. For each instrument, five representative data sets are provided, making a total of ten data sets. 
The VLT/SPHERE-IFS has two wavelength ranges, either covering the YJH bands ($0.95$ to $1.64~\mathrm{\mu m}$) or the YJ bands ($0.95$ to $1.33~\mathrm{\mu m}$), with spectral resolving power of 30 and 50 respectively. Gemini-S/GPI is similarly equipped with an extreme AO system and an APLC, and works from the J-band ($0.95~\mathrm{\mu m}$) to the K-band ($2.12~\mathrm{\mu m}$) with a resolving power of up to 80 in the latter case. 

Each multispectral data set is constituted of five files (again given in \emph{.fits} format): The same four files as for the ADI data sets, plus a vector containing the wavelengths corresponding to each spectral channel of the IFU. Again, the provided images are pre-reduced. 
The VLT/SPHERE-IFS data were pre-reduced with the SPHERE Data Center\cite{Delorme2017sphereDC}, using the SPHERE Data Reduction and Handling pipeline\cite{Pavlov2008drh}. 
The Gemini-S/GPI data were pre-reduced with the GPI Data Cruncher\cite{Wang2018DataCruncher}, making use of the GPI reduction pipeline\cite{Perrin2016GPIRP}. 
The properties of the ten image cubes are summarized in Tab.~\ref{tab:data2}.

\begin{table}[!h]
\caption[example] 
   {\label{tab:data2} 
Description of the ten data sets provided for the EIDC ADI+mSDI subchallenge. 
$D_{tel}$ is the effective telescope diameter (downstream the coronagraph); $\lambda_{range}$ is the range of observation wavelength; 
Plate-scale is the pixel scale of the detector used; 
$N_{img}$ is the size of each image frame; 
$N_{t}$ is the number of frames along the temporal axis; 
$N_{\lambda}$ is the number of channels along the spectral axis; 
$\Delta_{field}$ is the total field rotation of circumstellar objects. 
}
\begin{center}
\begin{tabular}{|l| c | c | c | c | c | c | c | c | } 
\hline
Instrument & ID & $D_{tel}$ & $\lambda_{range}$ & Plate-scale & $N_{img}$ & $N_{t}$ & $N_{\lambda}$ & $\Delta_{field}$ \\
           &    & $[\mathrm{m}]$ & $[\mathrm{\mu m}]$ & $[\mathrm{mas/pixel}]$ & $[\mathrm{px}] \times [\mathrm{px}]$ & & & $[\mathrm{^\circ}]$\\
\hline \hline
                        & ifs1 & $7.87$ & $0.957 - 1.329$ & $7.46$ & $200 \times 200$ & $128$ & $39$ & $25.6$  \\
                        & ifs2 & $7.87$ & $0.957 - 1.636$ & $7.46$ & $200 \times 200$ & $112$ & $39$ & $131.9$ \\
\textbf{VLT/SPHERE-IFS} & ifs3 & $7.87$ & $0.957 - 1.636$ & $7.46$ & $200 \times 200$ & $109$ & $39$ & $43.5$  \\
                        & ifs4 & $7.87$ & $0.957 - 1.636$ & $7.46$ & $200 \times 200$ & $112$ & $39$ & $46.2$   \\
                        & ifs5 & $7.87$ & $0.957 - 1.329$ & $7.46$ & $200 \times 200$ & $80$  & $39$ & $31.8$   \\
\hline
                    & gpi1 & $7.57$ & $1.495 - 1.797$ & $14.17$ & $161 \times 161$ & $35$ & $37$ & $54.7$   \\
                    & gpi2 & $7.57$ & $1.495 - 1.797$ & $14.17$ & $161 \times 161$ & $41$ & $37$ & $22.5$   \\
\textbf{Gemini-S/GPI} & gpi3 & $7.57$ & $1.495 - 1.797$ & $14.17$ & $161 \times 161$ & $38$ & $37$ & $12.6$   \\
                    & gpi4 & $7.57$ & $1.495 - 1.797$ & $14.17$ & $161 \times 161$ & $37$ & $37$ & $13.8$   \\
                    & gpi5 & $7.57$ & $1.495 - 1.797$ & $14.17$ & $161 \times 161$ & $44$ & $37$ & $24.4$   \\
\hline
\end{tabular}
\end{center}
\end{table}

\subsection{Injected planetary signals}
\label{ssec:inj}
\paragraph{Injection procedure}
Within each data cube we injected from 0 to 5 point sources using the VIP package\cite{gonzalez2017vip}. To avoid interfering with potential real planetary signals or extended sources in the images, we injected the point sources using the opposite parallactic angles. The latter procedure keeps the temporal statistical behavior of the starlight residuals, while smearing out the potential circumstellar signals. We randomly picked separations from the inner working angle of the coronagraph to the end of the AO correction zone, at a random position angle, making sure that the injected signals will not overlap. For a given position, we randomly picked a contrast value in a range of $\pm 3~\sigma$ from the $5~\sigma$ contrast curve provided when running the chosen baseline algorithm, which is a regular annular Principal Component Analysis\cite{Amara2012,Soummer2012} (PCA), using the VIP package implementation. In the 9 ADI subchallenge data sets, we injected a total of 20 planetary signals, including two data sets without injection. In the 10 ADI+mSDI subchallenge data sets, we injected a total of 23 planetary signals, including two data sets without injection.    

\paragraph{Limitations of the injection procedure}
The injected synthetic planetary signals are modeled with the non-coronagraphic (or unsaturated) PSF used for calibration, which is included in each data set. This injection procedure is therefore limited: (i) the flux of the synthetic planetary signal is constant throughout the observing sequence, (ii) the center of the image (position of the host star) is fixed when injecting the signals, that is to say potential mis-centering of the star throughout the sequence is not taken into account, and (iii) we do not take into account the azimuthal smearing of the signal that could appear for long exposure times and/or at large angular separation from the star. For the injection in the multispectral data (ADI+mSDI subchallenge), we used the trend from typical L-type and T-type brown dwarf spectra obtained from the SpexPrism\cite{Burgasser2014} data base.

\section{Evaluation procedure}
\label{sec:metrics}  
The first phase of the exoplanet imaging data challenge is primarily focused on the point source detection capabilities of the classical and latest image processing methods dedicated to high-contrast imaging. We therefore decided to analyze the submitted material as a classification problem, that is to say in terms of counting detections and non-detections. The EIDC participants were requested to submit their results by including three types of files (in \emph{.fits} format): (1) the detection maps for each data set, (2) one threshold common to all the data set, and optionally (3), the full width at half maximum (FWHM) of the expected planetary signal in the detection map. 
The results of the first phase of the EIDC are available upon request if one wants to apply different evaluation tools to the detection maps. In the future, we plan to apply more advanced metrics that take into consideration aspects such as the working nature of each algorithm, the difficulty of a given detection (e.g. in terms of contrast and separation to the star) and a proper ranking procedure (Pi\'erard et al., in prep.).

With this data challenge, we do not intend to reduce the performance of a given algorithm to a single number, which would be pointless since the efficiency of each method is highly dependent on the type of instrument and observing conditions. Instead, we aim at providing the community with homogeneous information about the different algorithm capabilities so that they can build their own opinion based on what type of application they need. In this section, we propose to evaluate the results by using a set of metrics, with known short-comings, in order to guide the comparison, from both a qualitative and quantitative standpoint. 

\subsection{Detection considerations}
Based on the planetary signal injections in the nine (resp.~ten) data sets available for the ADI subchallenge (resp.~ADI+mSDI subchallenge), we counted the number of true positives (TP), false positives (FP), true negatives (TN) and false negatives (FN) at the detection threshold provided by the participants. We then repeated this procedure for a range of thresholds discussed below. A TP is defined as a detection within one resolution element from the position of an injected companion, while a FP is a detection at any other location. The FN and TN respectively gather the non-detections at the position of injected companions and the non-detections at any other location considering the FWHM aperture grid applied to the image.

Any signal above the chosen threshold is considered as a detection: we do not take into account neither visual detections (detections that are obvious visually but below the threshold) nor any type of post-sorting (signals that are above threshold but obviously artefacts from a visual inspection or automatic rejection as in Ref.\cite{Cantalloube2015}). In the future, we will alleviate this limitation by asking the participants for their detection parameters (e.g. position, signal-to-noise ratio or probability of presence). 

For the non-detections, we counted the number of possible detections in the image and subtract the number of detections. 
We decided not to use the FWHM submitted by participants since it could (dis)favor some algorithms. Instead, we used the instrumental PSF FWHM (or resolution element) computed from the observation wavelength, the telescope effective diameter and the detector pixel scale. In practice, a planetary signal should always be of about this size, as none of the submitted algorithm perform any type of deconvolution. 

In addition, each algorithm has its own inner and outer working angle (smallest and largest angular separation processed). To allow a fair comparison, we applied to all the detection maps received for a given data set, a unique binary mask to select the smallest inner working angle and largest outer working angle submitted for this data set (see the images galleries in Appendix, bottom-right).

In order to guarantee that the results are not biased by the person analyzing the data, and to minimize the errors, we analyzed the data in double blind. Two authors evaluated the data without knowing their reference, and analyzed the results with their own tools to make sure that the results are consistent.

\subsection{Metrics}
From these counts, we compute different metrics:

\vspace{-0.4cm}
\begin{enumerate}
    \item the true positive rate (also called sensitivity or recall): $TPR = TP / (TP+FN)$, 
    \item the false positive rate (also called fall-out): $FPR = FP/(FP+TN)$, 
    \item the false discovery rate (also called precision): $FDR = FP/(FP+TP)$,
    \item the F1-score (harmonic mean of precision and sensitivity): F1-score $ = 2 \cdot TP/(2 \cdot TP+FP+FN)$.
\end{enumerate}

\vspace{-0.6cm}
\paragraph{True positive rate, TPR:} If the number of injection is non-null, the TPR as a function of the threshold is a decreasing step function. An ideal algorithm should provide a TPR of 1 whatever the threshold (up to the maximum value of the detection map). We additionally compute the area under the curve (AUC) of the TPR as a function of various threshold values, which must be as close as possible to one. 

\vspace{-0.5cm}
\paragraph{False positive rate, FPR:} On the contrary, the FPR as a function of the threshold is decreasing monotonically. An ideal algorithm should provide a FPR of 0 whatever the threshold. This quantity characterizes the residual noise in the detection map. 
At a threshold of zero, this FPR indicates the fraction of resolution elements above zero in the detection map. 
For a threshold value corresponding to the minimum of the detection map, the FPR is equal to one.

\vspace{-0.5cm}
\paragraph{False discovery rate, FDR:}
The FDR as a function of the threshold is a decreasing step function. An ideal algorithm should also provide a FDR of 0 whatever the threshold. This figure of merit gives information about the detections, regardless of the true negatives, which is more important for our science case as we want to detect planetary signals (and we are not interested in knowing how we cannot detect them). We also compute the AUC of the FDR as a function of various threshold values, which must be as close as possible to zero. 

As a result, we chose to use three figures of merit: (1) the F1-score computed at the submitted threshold, (2) the AUC of the TPR (must be close to 1), and (3) the AUC of the FDR (must be close to 0). The three scores have values between 0 and 1. If no synthetic planetary signals have been injected into the data set, these three scores are undefined. We computed these scores for each data set separately, and averaged them for each instrument, then finally averaged them for all the data sets of a given subchallenge, as gathered in Tab.~\ref{tab:resadiss}, Tab.~\ref{tab:resadimap}, Tab.~\ref{tab:res_adi_sml} and Tab.~\ref{tab:res_sadi}. We ranked each submission according to these three scores, not only to verify the relevance of the F1-score, but also to unveil more information about the performance of each algorithm (Fig.~\ref{fig:rnk_adi} and Fig.~\ref{fig:rnk_asdi}).

\subsection{Graphical comparison}
To support the scores mentioned above, we also provide with qualitative and graphical comparisons. 
First, the detection map shows visually the type of residuals (not quantified in the current paper), as well as the inner and outer working angles of each submitted algorithm (see Fig.~\ref{fig:img_sph3_baseline}, left). 
As mentioned previously, we apply a common binary mask to all the detection maps for counting the detections and non-detections within the same region for all the algorithms (see Fig.~\ref{fig:img_sph3_baseline}, right). 
In addition, we plot the TPR, FDR and FPR as a function of a given range of thresholds to visualize the trend of these three quantities (see Fig.~\ref{fig:img_sph3_baseline}, middle). 
We arbitrarily chose a range of threshold from 0 to twice the submitted threshold. However for some type of algorithms (e.g. the ones providing close to a binary detection map), this range of threshold can (dis)favor the AUC of the TPR. We therefore separated the algorithms based on their underlying concept and compared them with each other.  


\begin{figure}
    \centering
    \resizebox{\hsize}{!}{\includegraphics{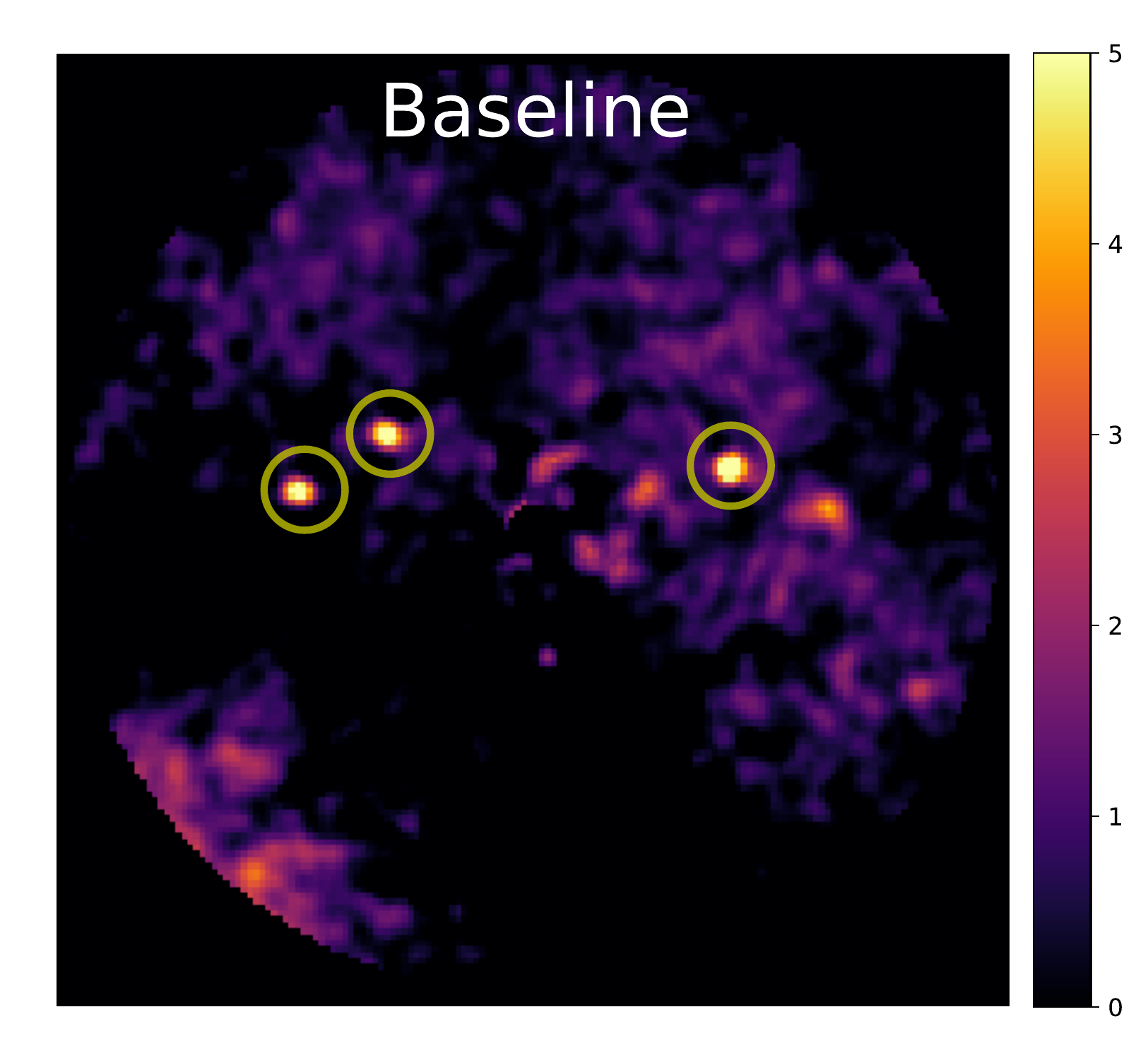}\includegraphics{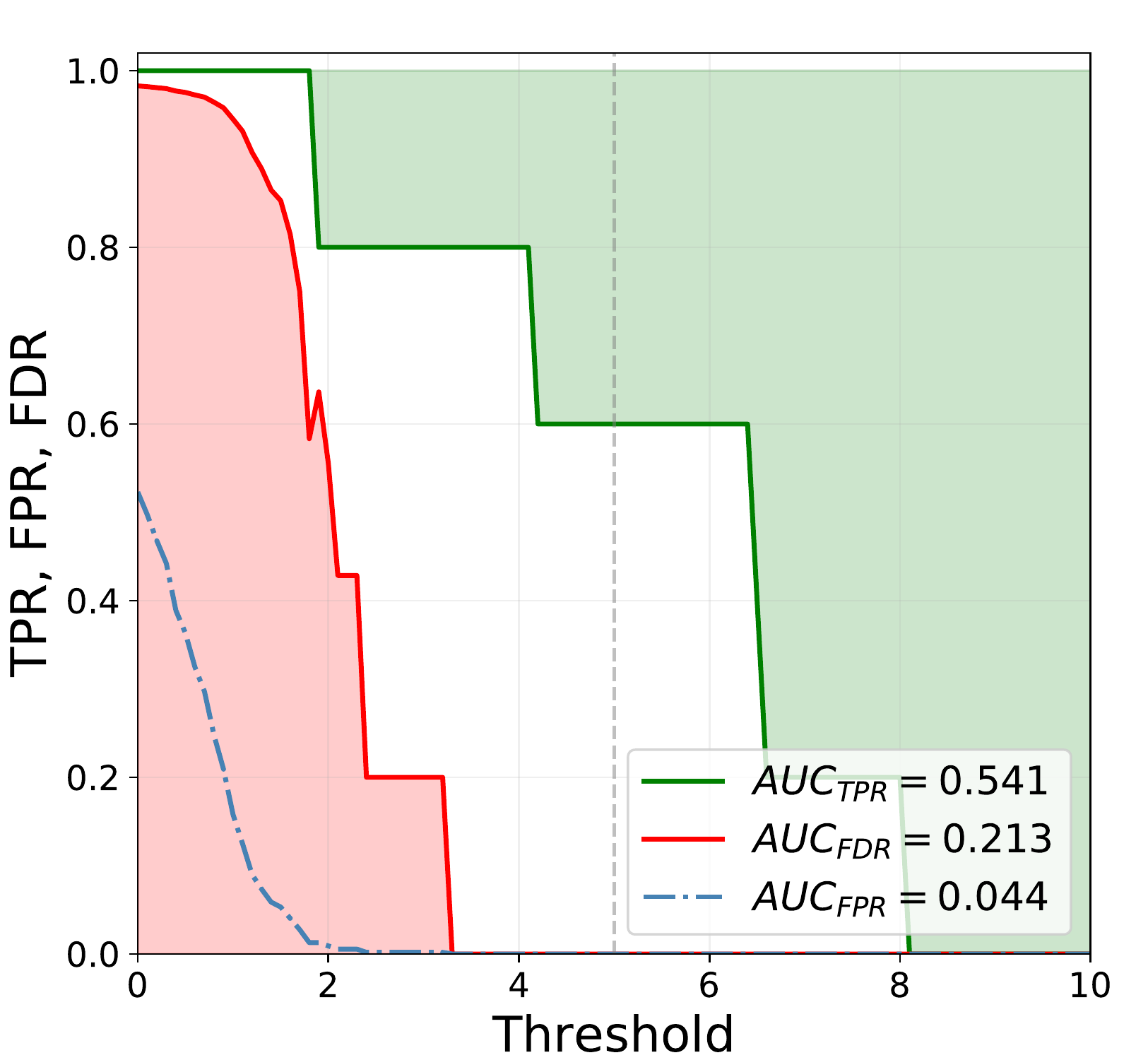}\includegraphics{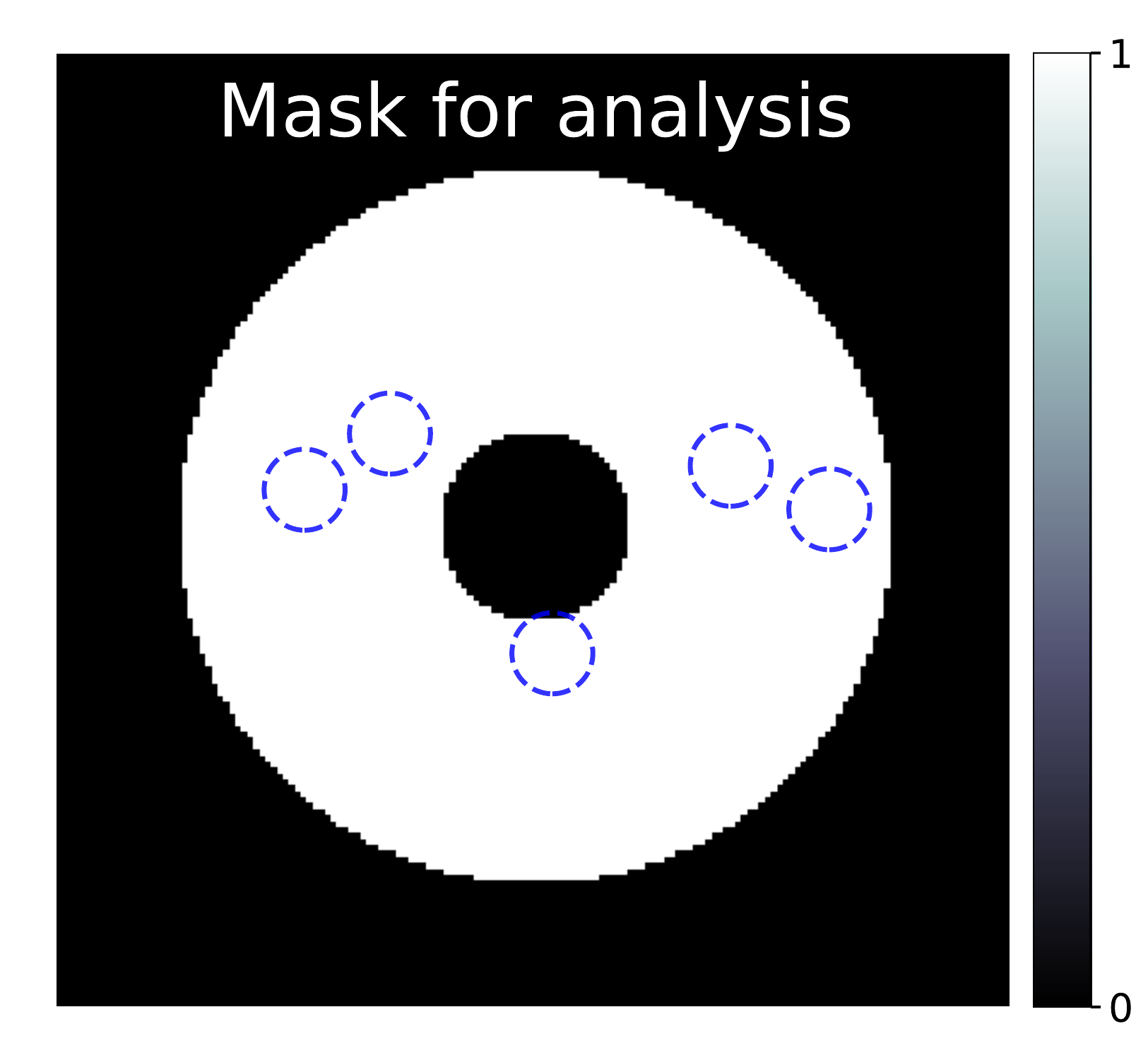}}
    \caption{Example of the results displayed for the chosen baseline algorithm applied on the third VLT/SPHERE-IRDIS data set (sph3). 
    Left: Detection map with a color-bar ranging from 0 to the submitted threshold. True positives at the submitted threshold are indicated with yellow circles (the diameter of the circles correspond to the instrumental PSF FWHM). False positives at the submitted threshold are indicated with red squares. 
    Middle: True Positive Rate (TPR, in green), False Discovery Rate (FDR, in red) and False Positive Rate (FPR, dash-dotted blue line), for a range of threshold varying from 0 to twice the submitted threshold. 
    Right: Common mask applied to all the detection maps of this specific sph3 data set. The dashed blue circles indicate the positions of the 5 synthetic planetary signals injected in this data set.}
    \label{fig:img_sph3_baseline}
\end{figure}

\section{Submissions results}
\label{sec:results}  

\subsection{Subchallenge 1: ADI}
For this challenge, we received $22$ valid submissions from $12$ participants. We splitted these submissions into three main families on which they are based: (1) speckle subtraction techniques, (2) inverse problem approaches, and (3) supervised machine learning techniques. In the following, we show the detailed plots for the third VLT/SPHERE-IRDIS data set (sph3), which contains 5 injected planetary signals (indicated with blue circles in Fig.~\ref{fig:img_sph3_baseline}, right). The 22 submitted detection maps obtained for the nine data sets of this subchallenge can be found in App.~\ref{app_adi}. 

\subsubsection{Speckle subtraction techniques}
We received $12$ submissions based on speckle subtraction techniques, which is the most commonly used technique, consisting of three steps: (i) estimating the stellar light residuals (or speckle field) directly from the image cube, (ii) subtracting the model(s) of the stellar light residuals to each frame of the cube and (iii) de-rotating the images to align the circumstellar signals with the true North and combine all these images to produce a so-called residual map. In addition to or instead of this last step, some techniques build detection maps where the pixel values correspond to a probability of presence of a planetary signal.

$\mathrm{cADI_{SpeCal}}$ stands for classical Angular Differential Imaging, as implemented in the SpeCal pipeline\cite{Galicher2018} dedicated to process VLT/SPHERE images. 
$\mathrm{LOCI}$ stands for Locally Optimized Combination of Images\cite{Lafreniere2007}, and is the version implemented in the VIP package\cite{gonzalez2017vip}.
$\mathrm{TLOCI_{SpeCal}}$ is an improved version of LOCI called Template LOCI\cite{Marois2014}, as implemented in the SpeCal pipeline. 
$\mathrm{PCA_{SpeCal}}$ is a Principal Component Analysis (PCA\cite{Amara2012,Soummer2012}), as implemented in the SpeCal pipeline, using a default value of 5 principal components. 
$\mathrm{PCA_{MPIA}}$ is a full-frame PCA used for the VLT/NACO ISPY\cite{Launhardt2020} survey. 
$\mathrm{PCA_{Padova}}$\cite{Mesa2015} is an improved PCA used for the VLT/SPHERE SHINE\cite{Chauvin2017} survey and also included in the SpeCal pipeline. $\mathrm{Unknown}$ stands for a reference-less submission we received. 
$\mathrm{STIM_{fullframe}}$ is the Standardized Trajectory Intensity Mean (STIM) detection map\cite{Pairet2019} computed from a fullframe PCA subtraction. 
$\mathrm{STIM_{annuli}}$ is the STIM detection map computed from an annular PCA subtraction. 
$\mathrm{STIM_{hpf}}$ is the STIM detection map obtained from a fullframe PCA subtraction applied on previously high-pass filtered  images (using a filtering fraction of 0.25). 
$\mathrm{SLIMask}$\cite{ThesePairet} for STIM Largest Intensity Mask, is an improvement of the STIM detection map where several PCA with different number of PCs are run to produce a mask and discard unlikely signals.
$\mathrm{RSM}$\cite{Dahlqvist2020}\footnote{\url{https://github.com/chdahlqvist/RSMmap}}, standing for Regime-Switching Model detection map, traces the deviation form a noise to a noise+planet regime in previously speckle-subtracted images with various speckle subtraction techniques, using in this case annular PCA, Non-negative Matrix Factorization (NMF\cite{ren2018nmf}), and low rank plus sparse decomposition (LLSG\cite{gonzalez2016llsg}). 

\begin{figure}
    \centering
    \resizebox{15.5cm}{!}{\includegraphics{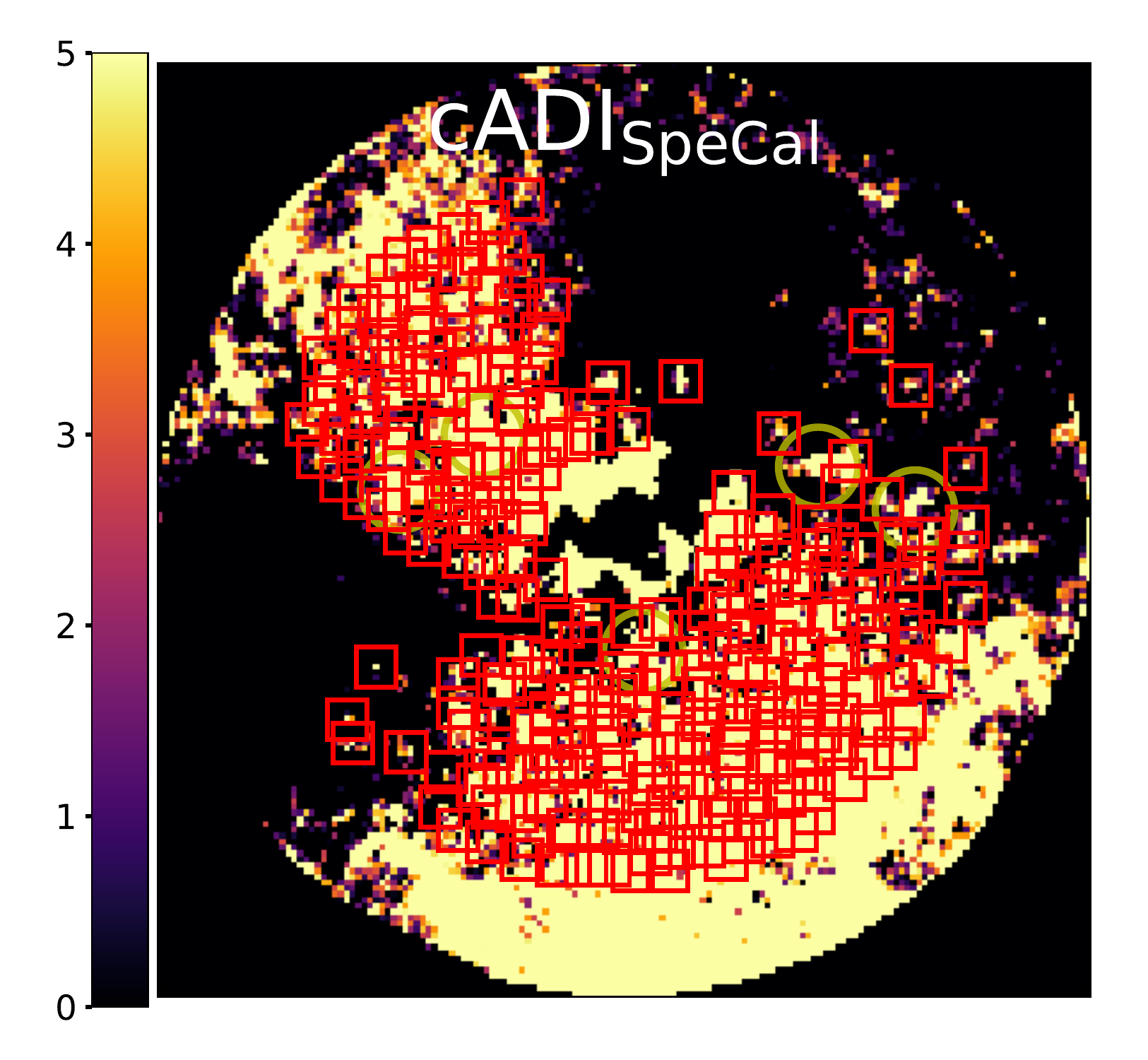}\includegraphics{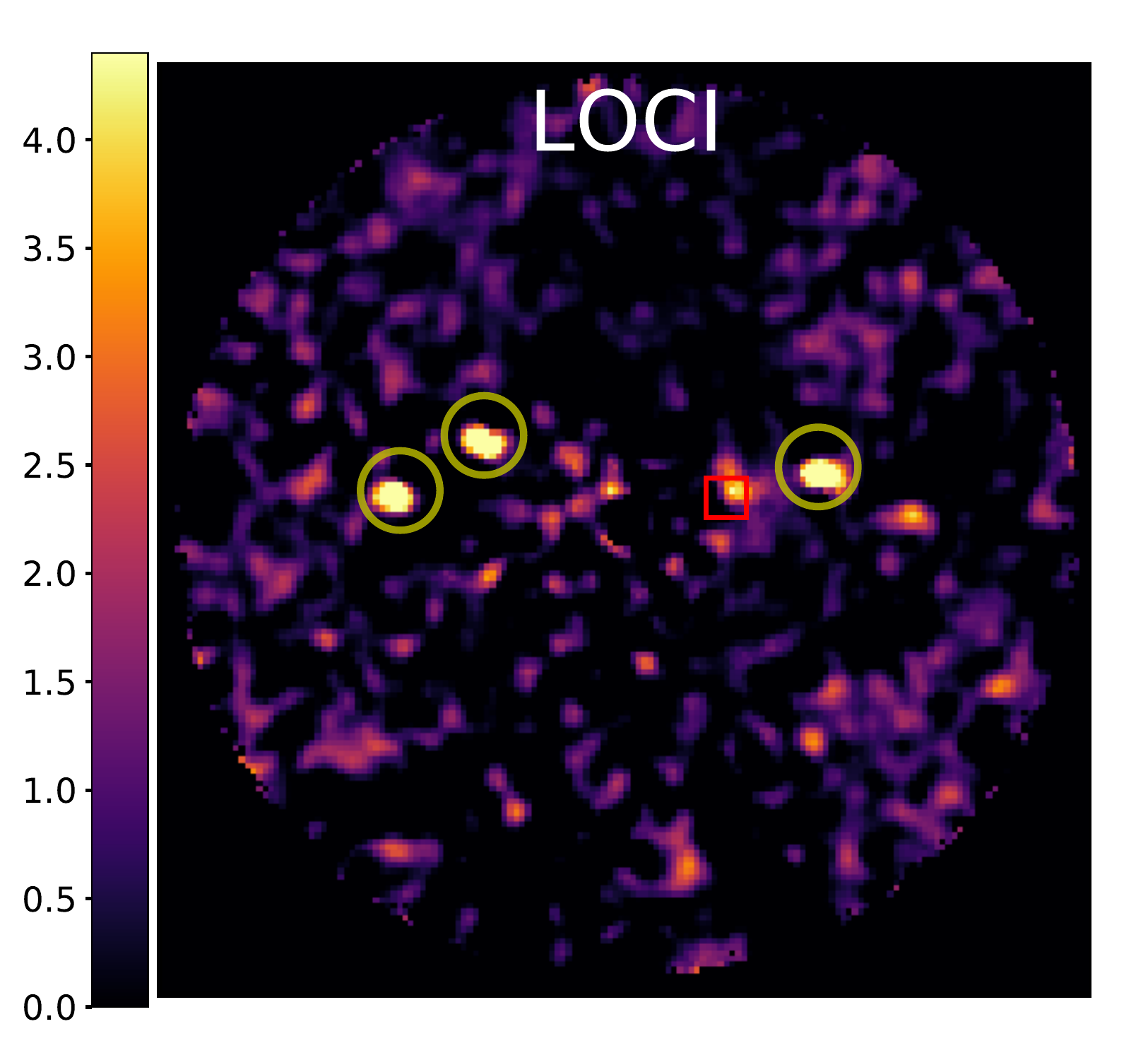}\includegraphics{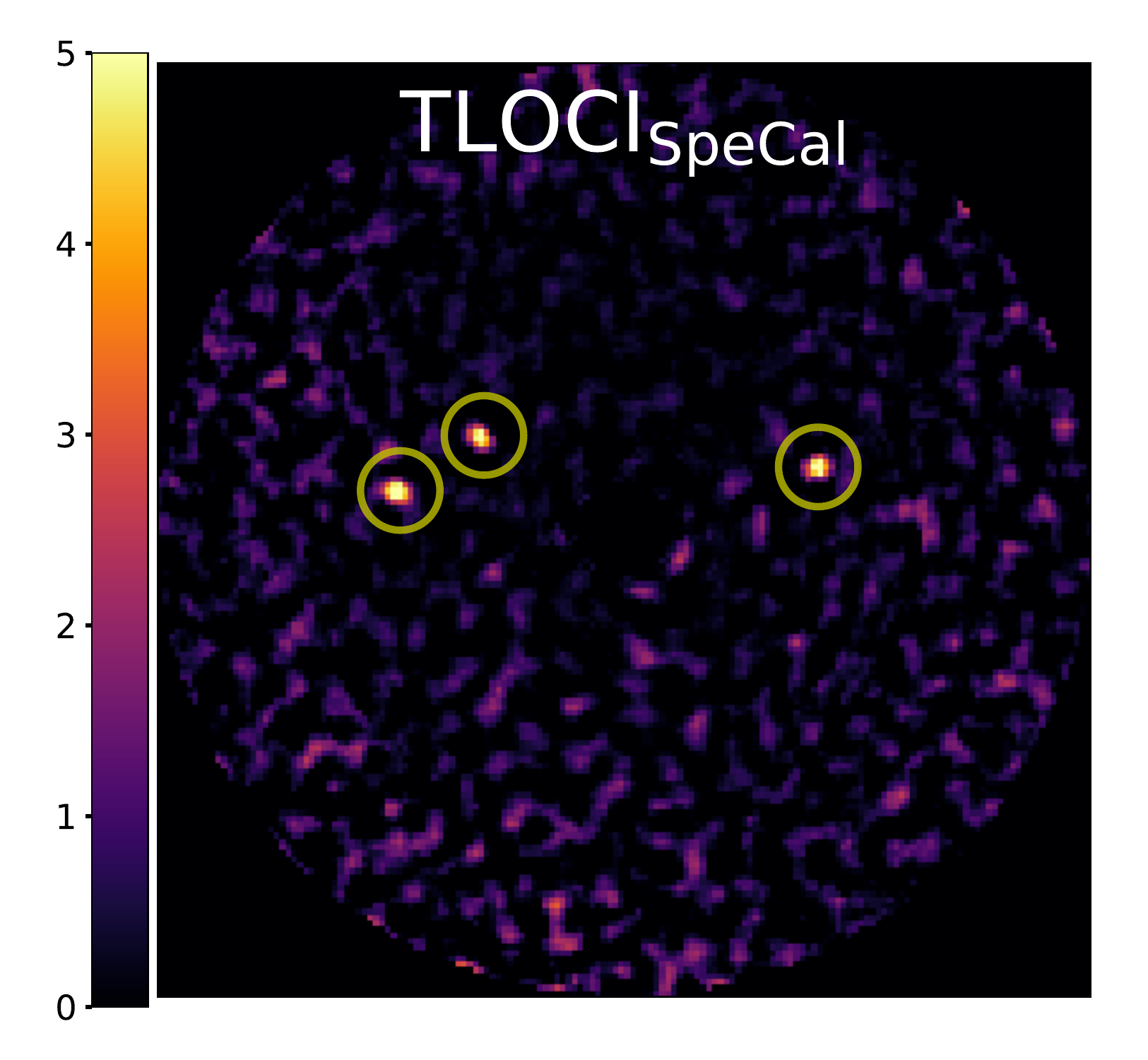}\includegraphics{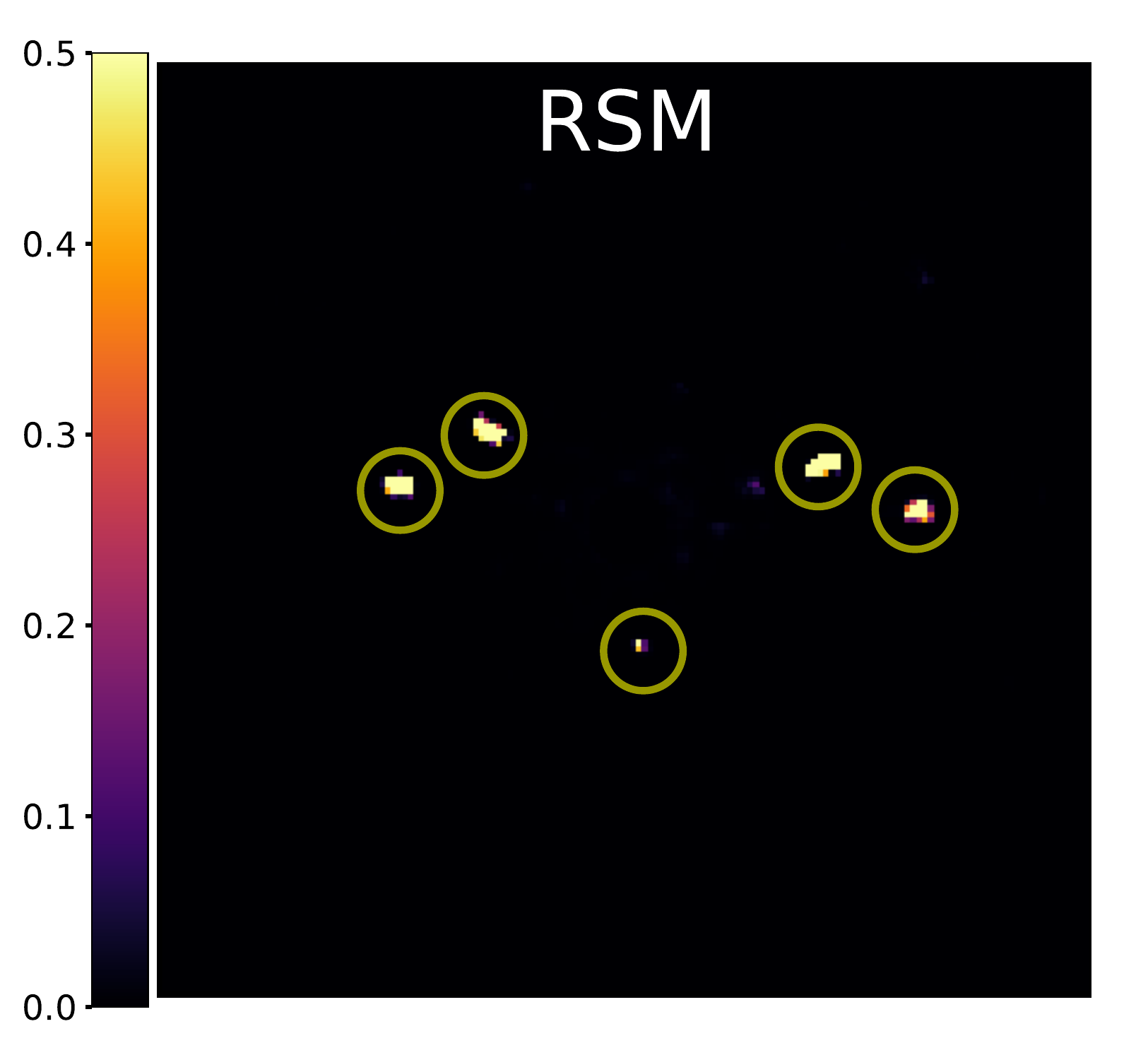}}
   \resizebox{15.5cm}{!}{\includegraphics{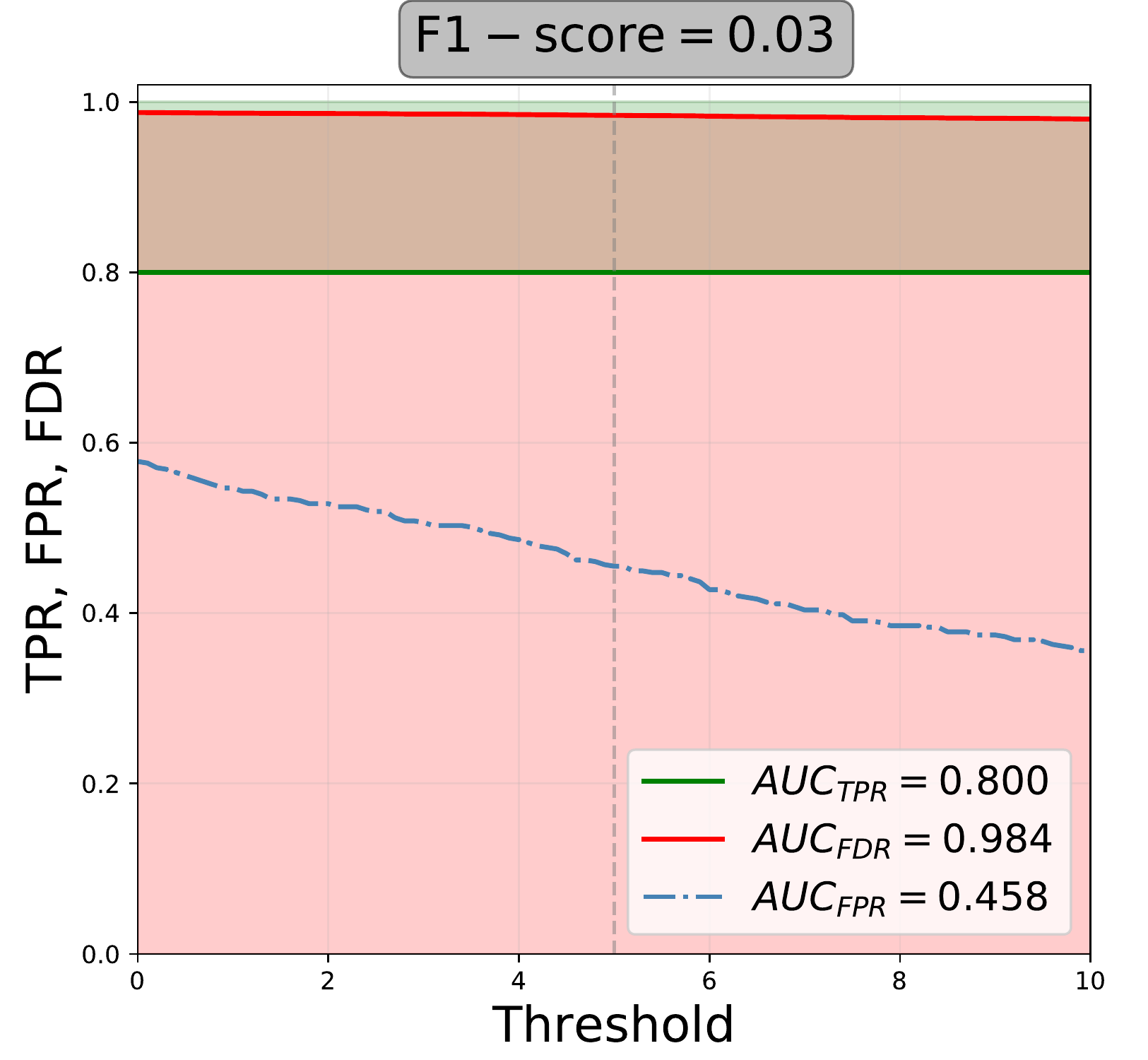}\includegraphics{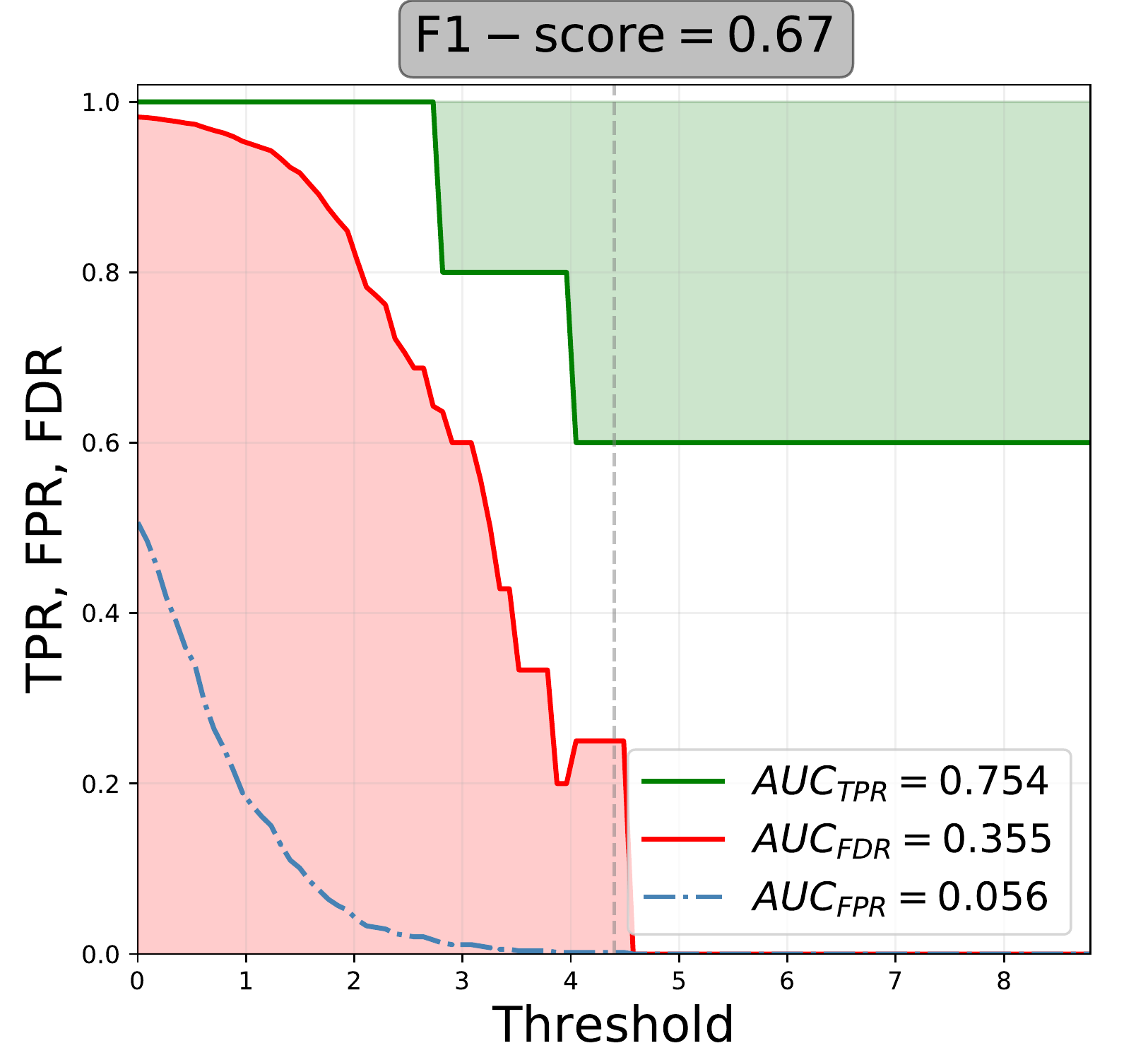}\includegraphics{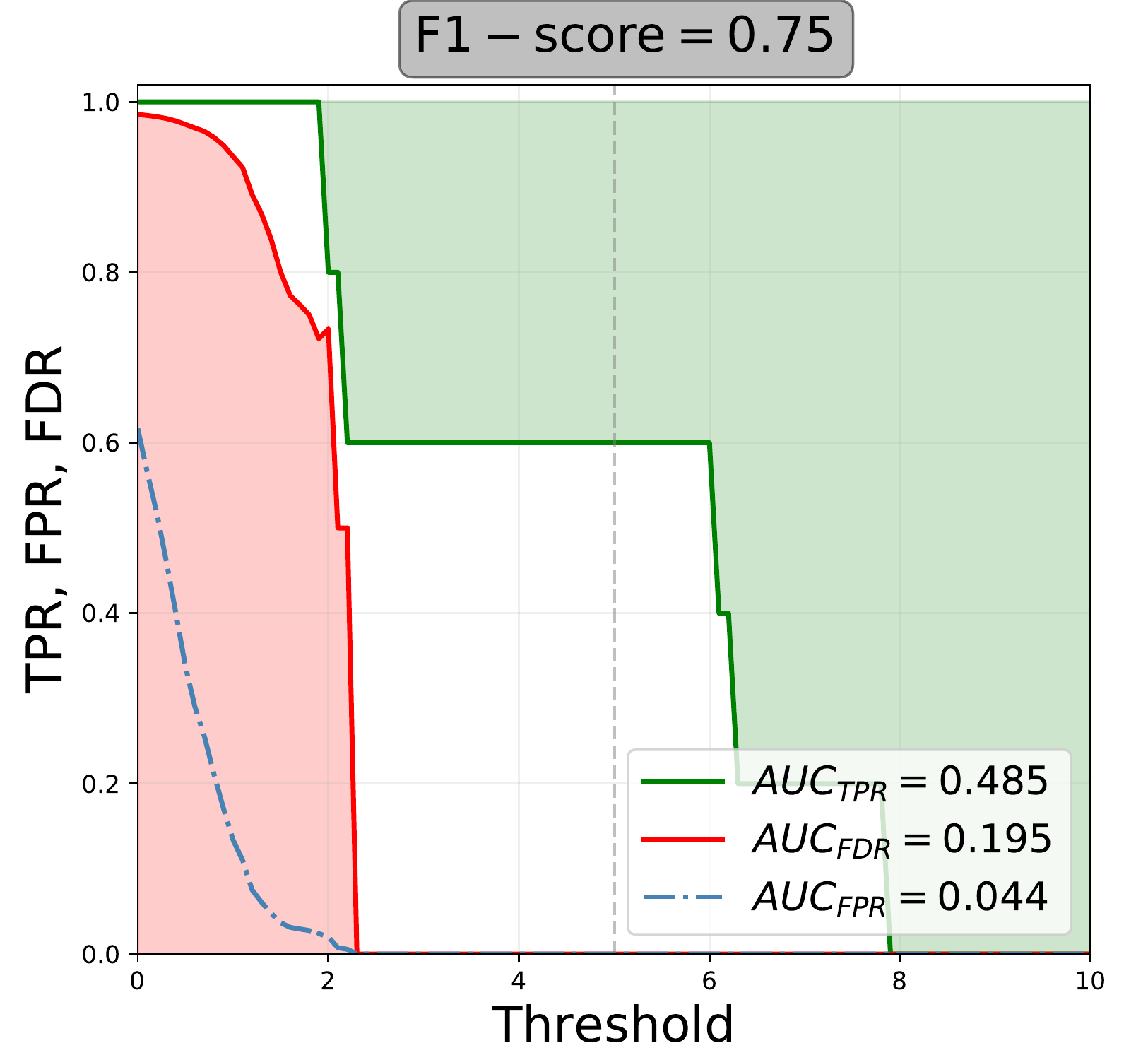}\includegraphics{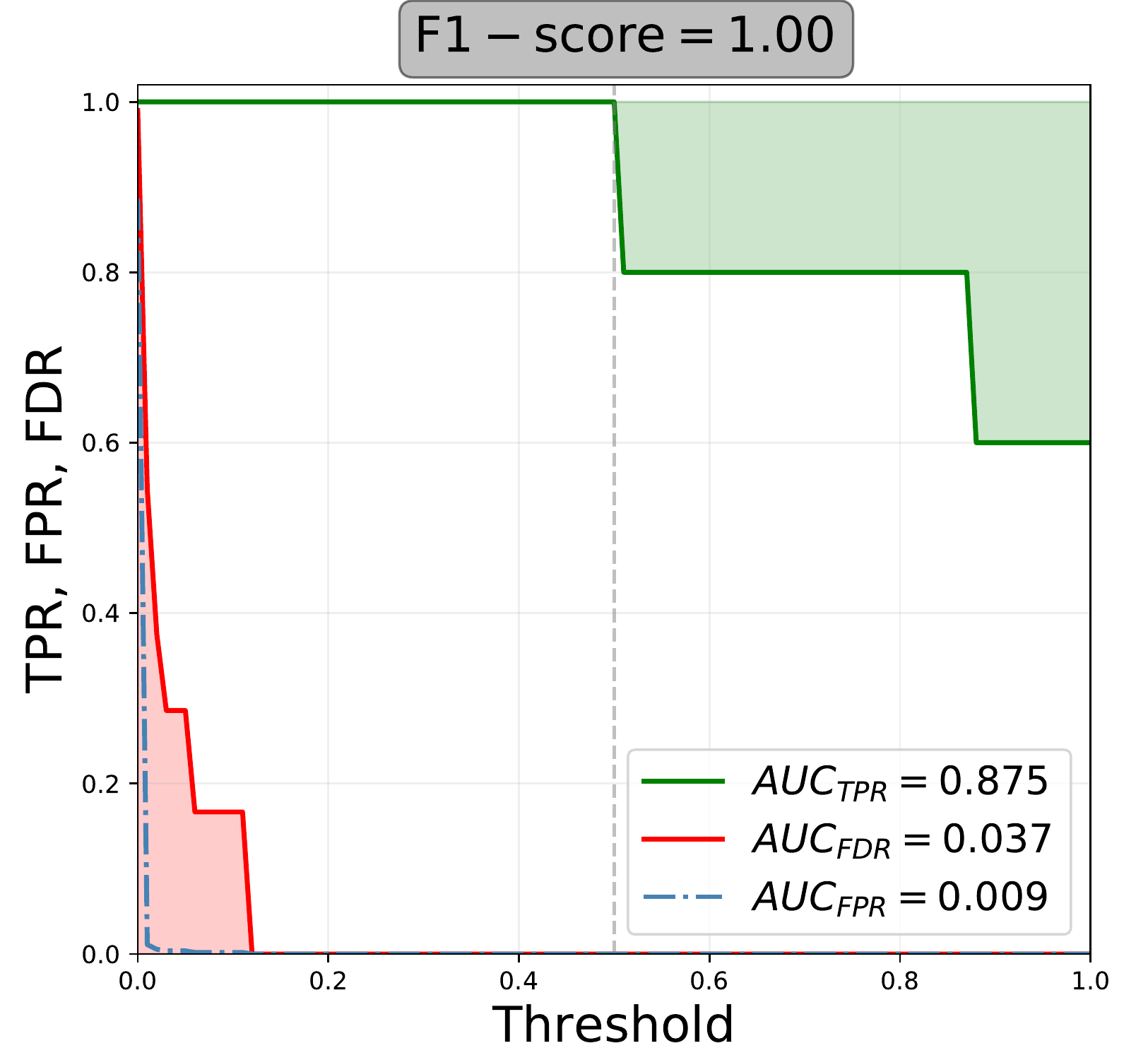}}

    \resizebox{15.5cm}{!}{\includegraphics{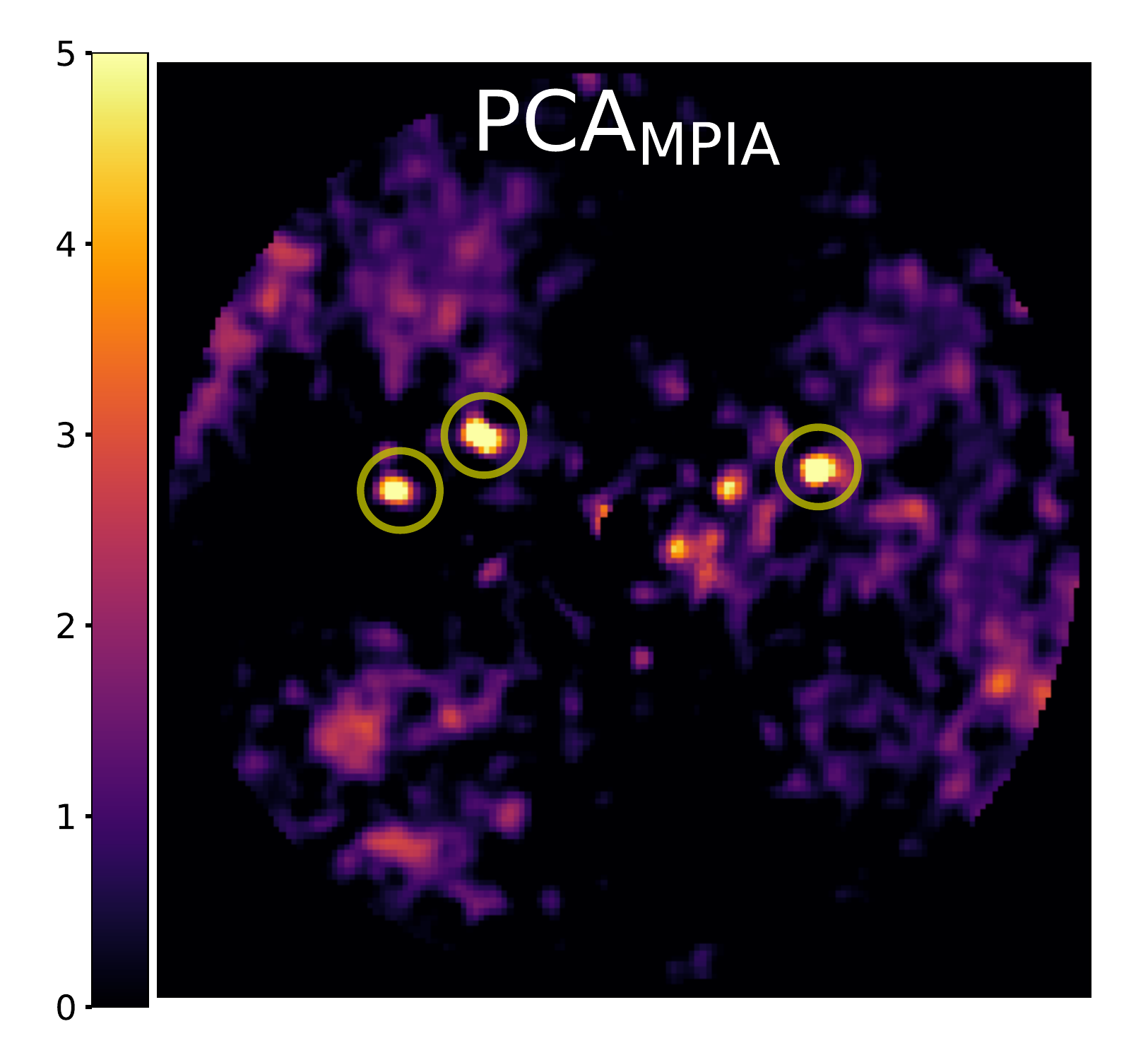}\includegraphics{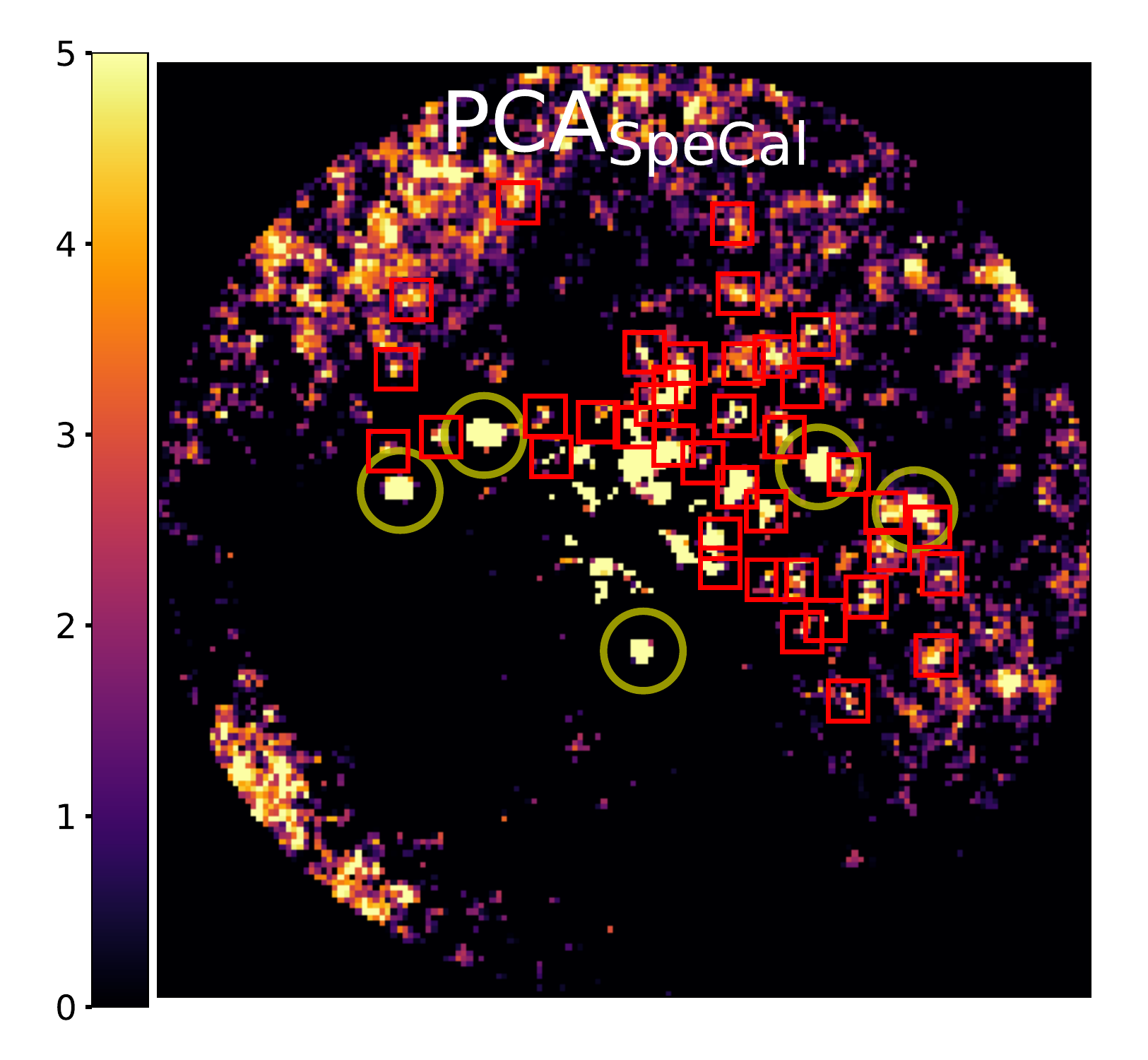}\includegraphics{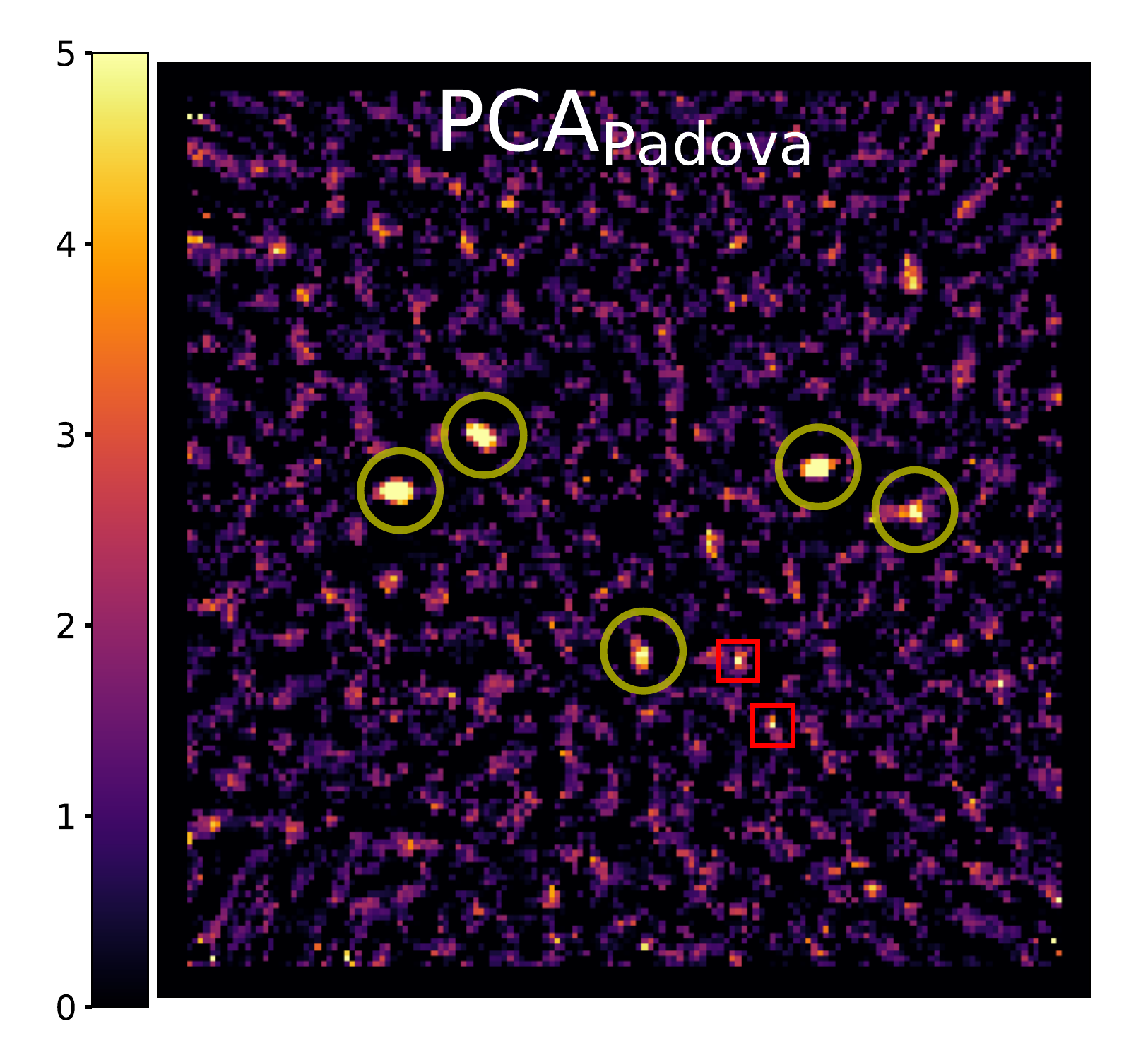}\includegraphics{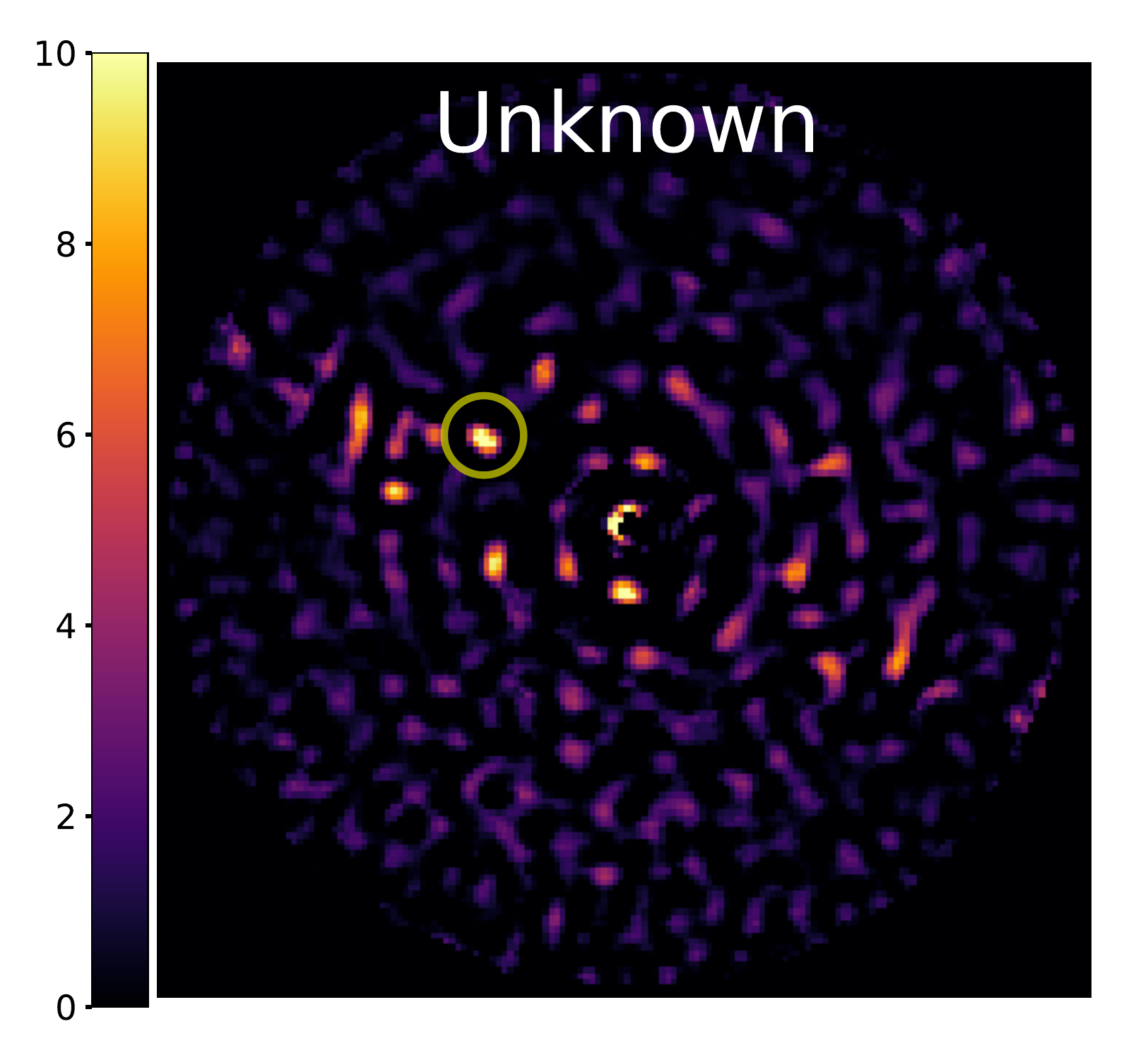}}
   \resizebox{15.5cm}{!}{\includegraphics{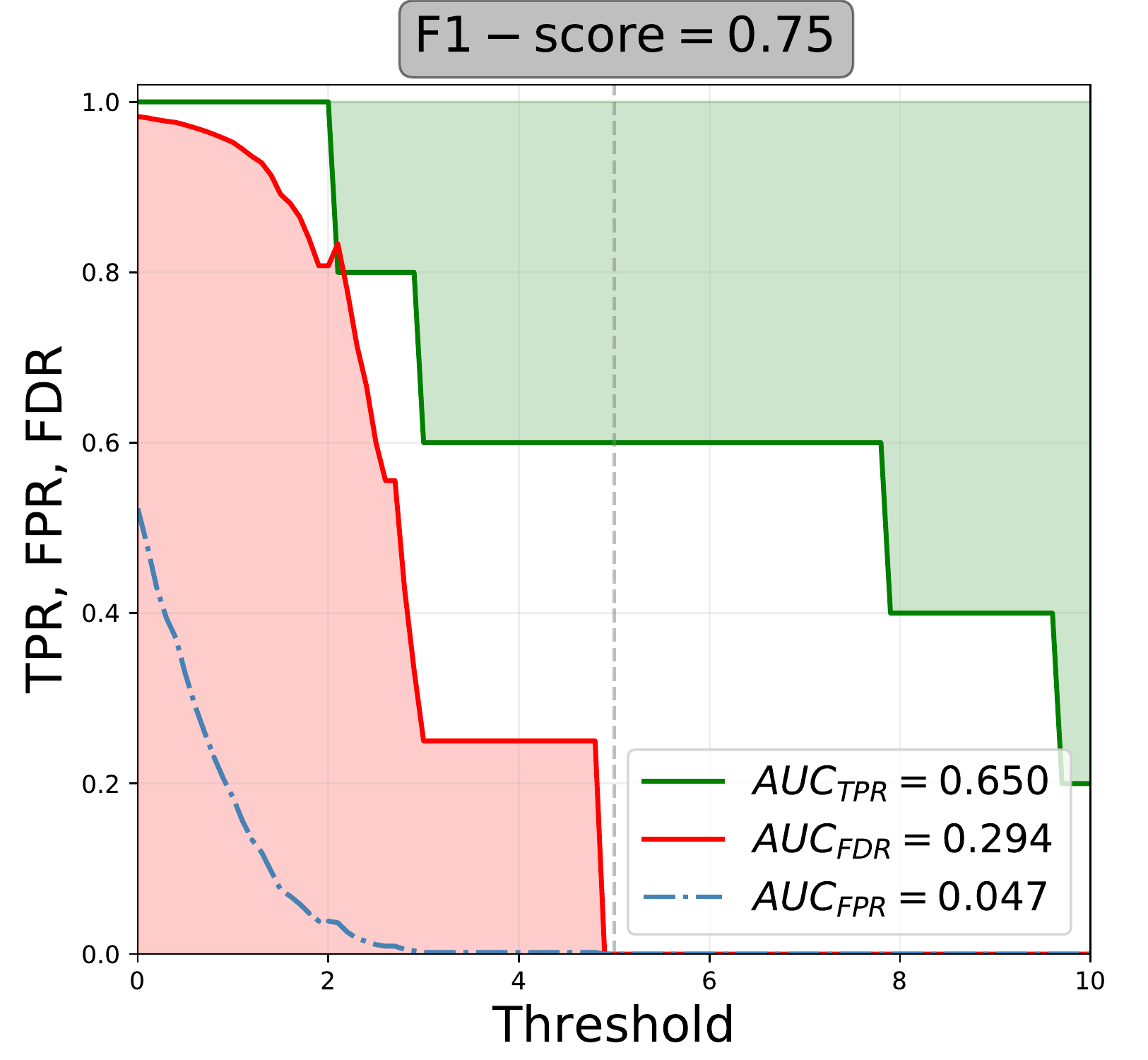}\includegraphics{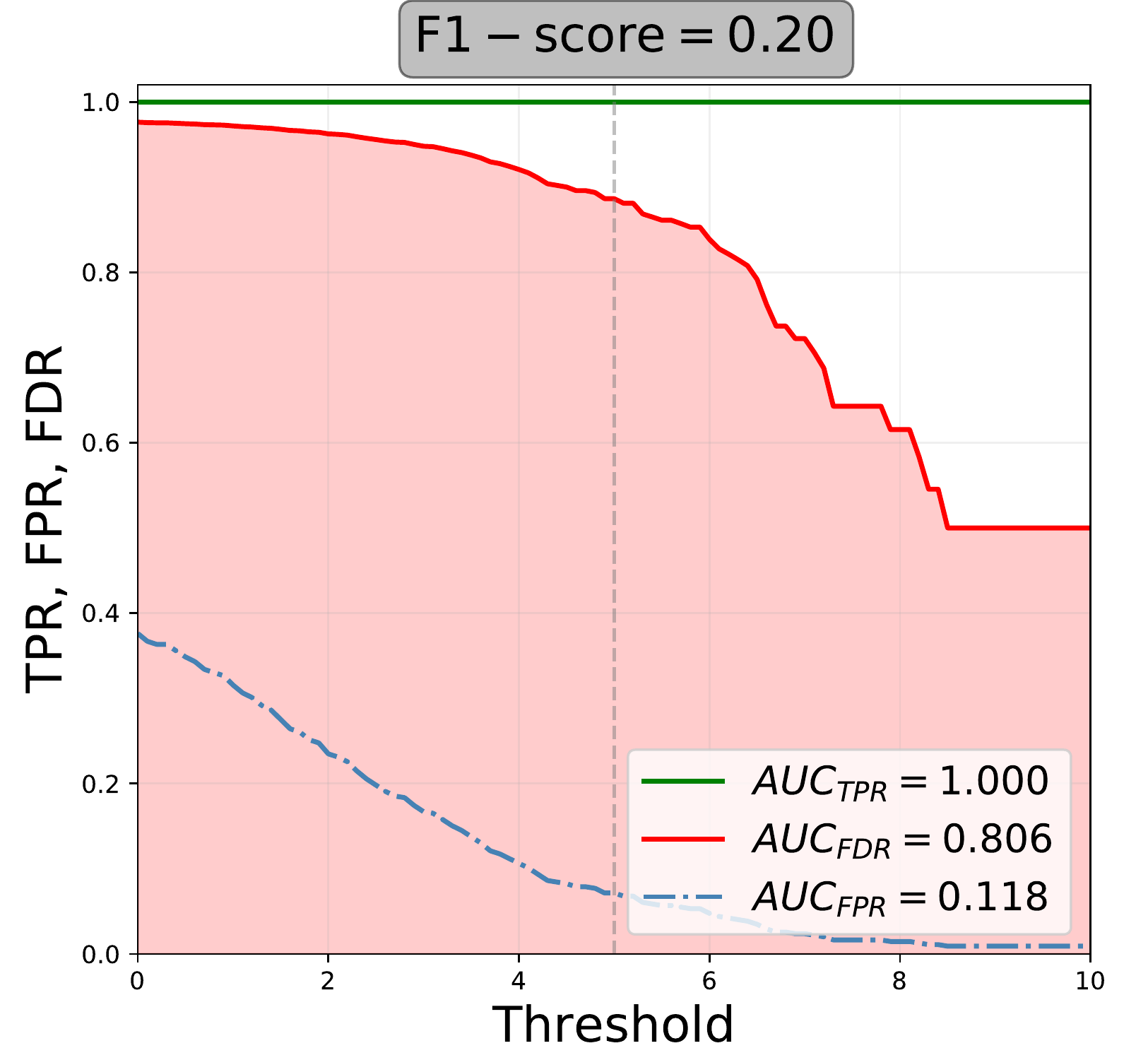}\includegraphics{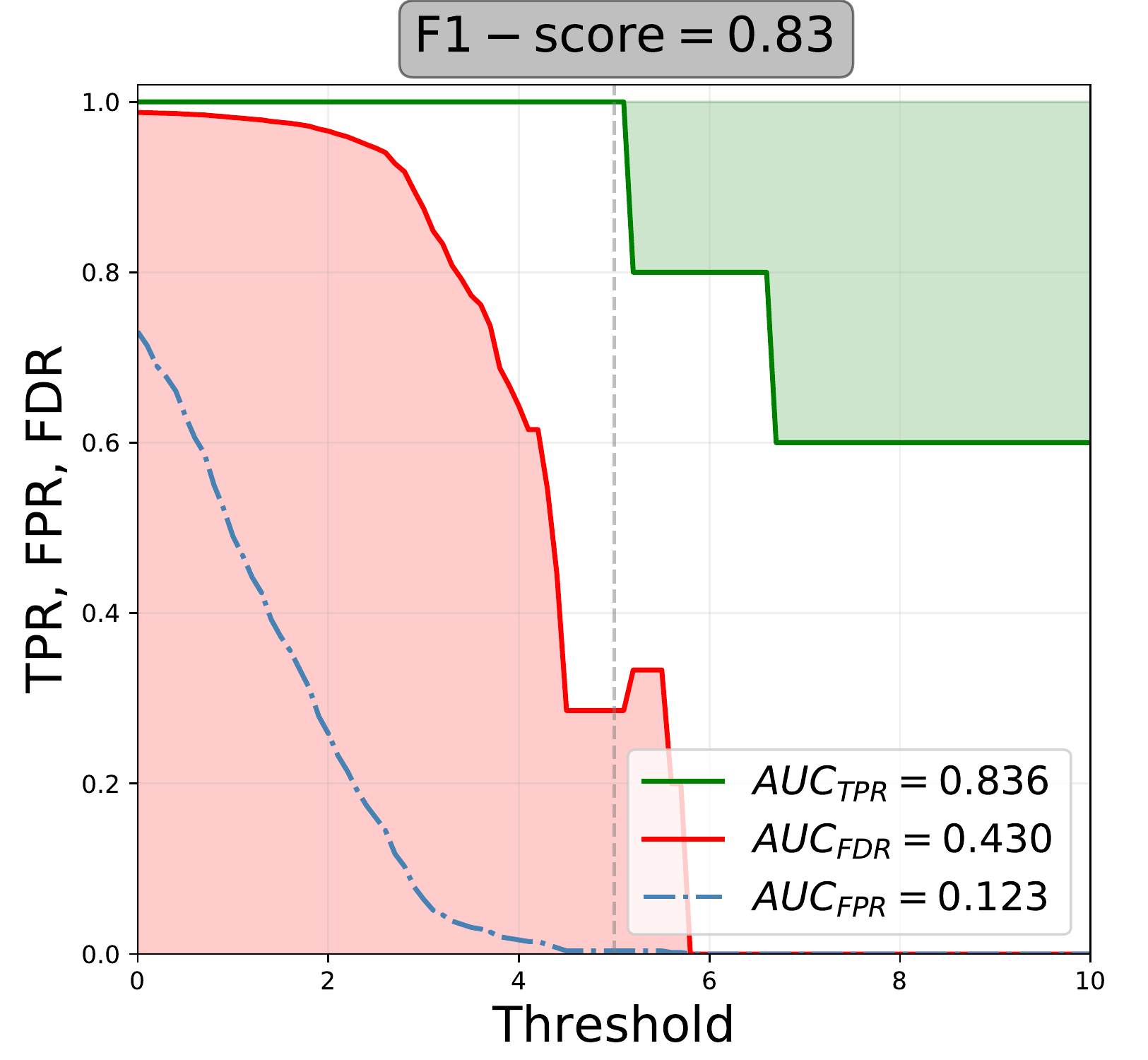}\includegraphics{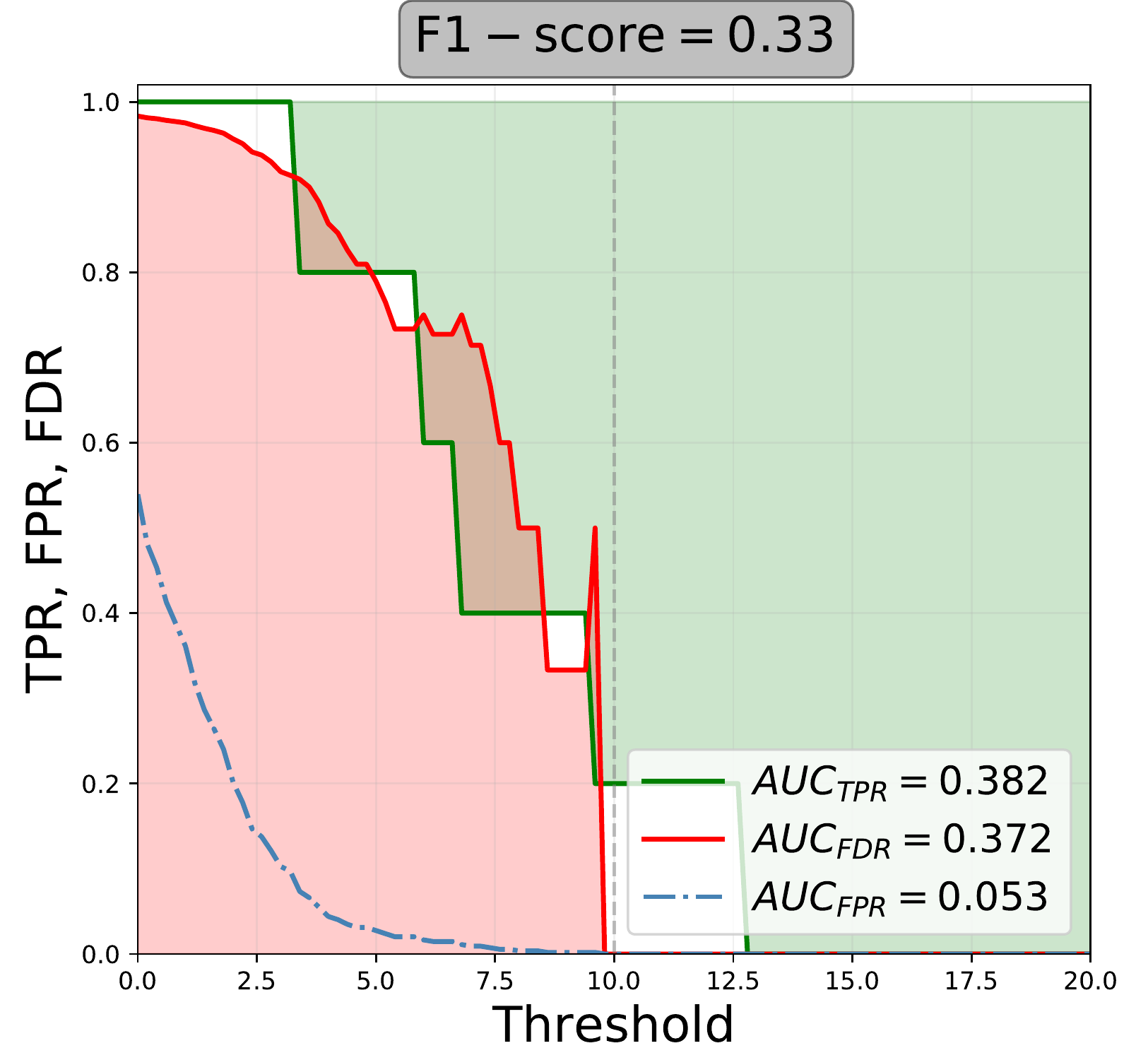}}
   
     \resizebox{15.5cm}{!}{\includegraphics{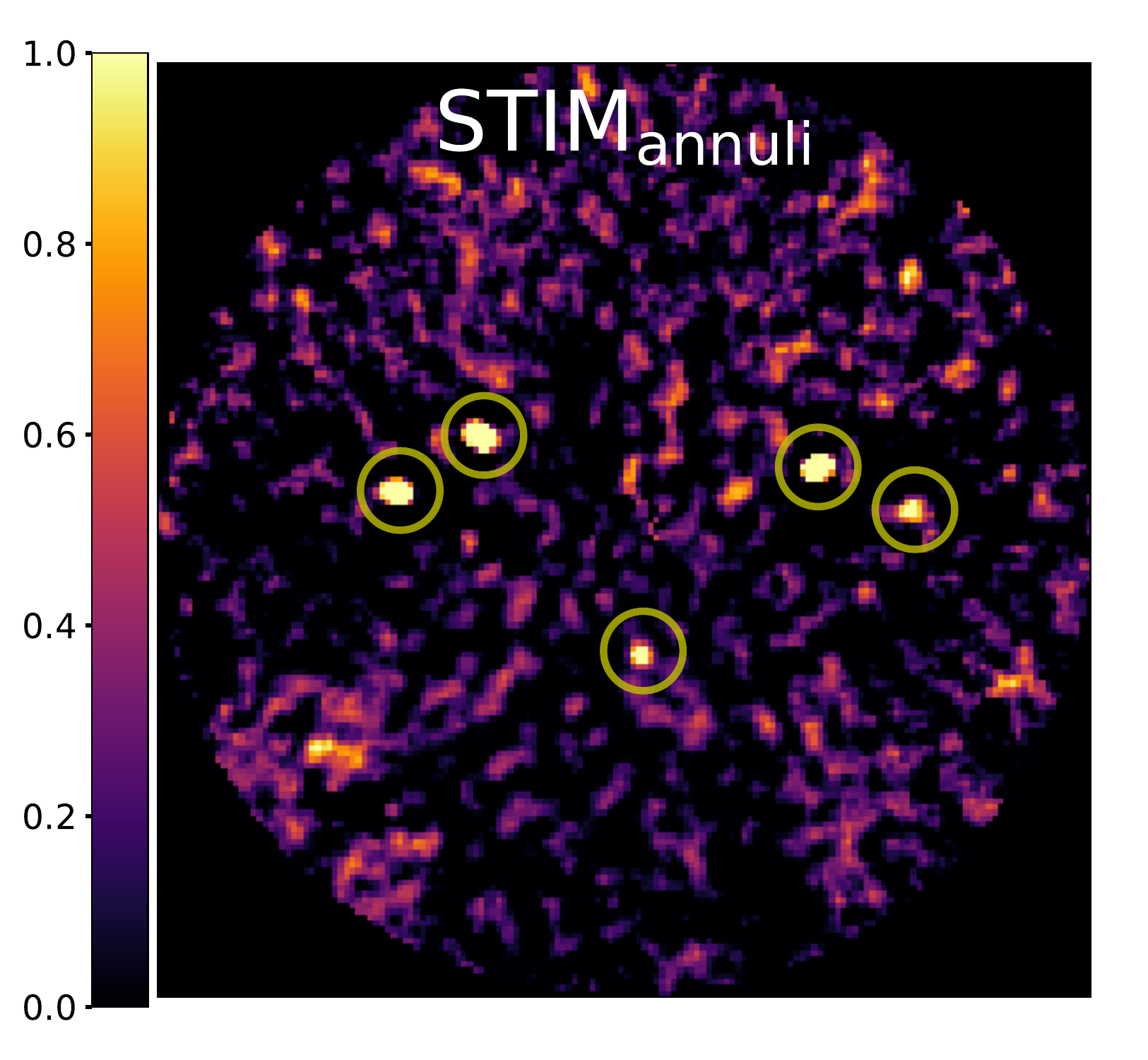}\includegraphics{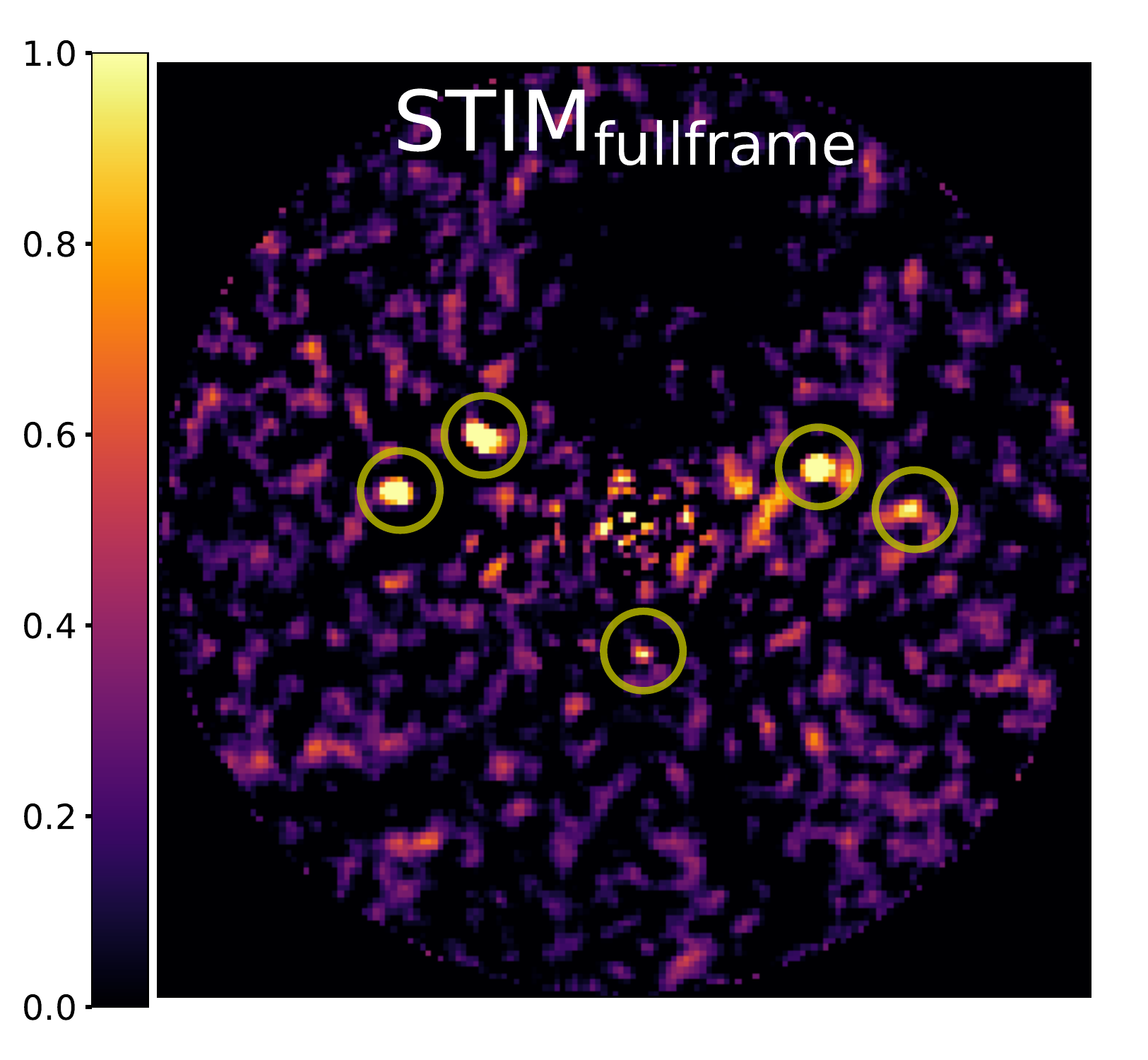}\includegraphics{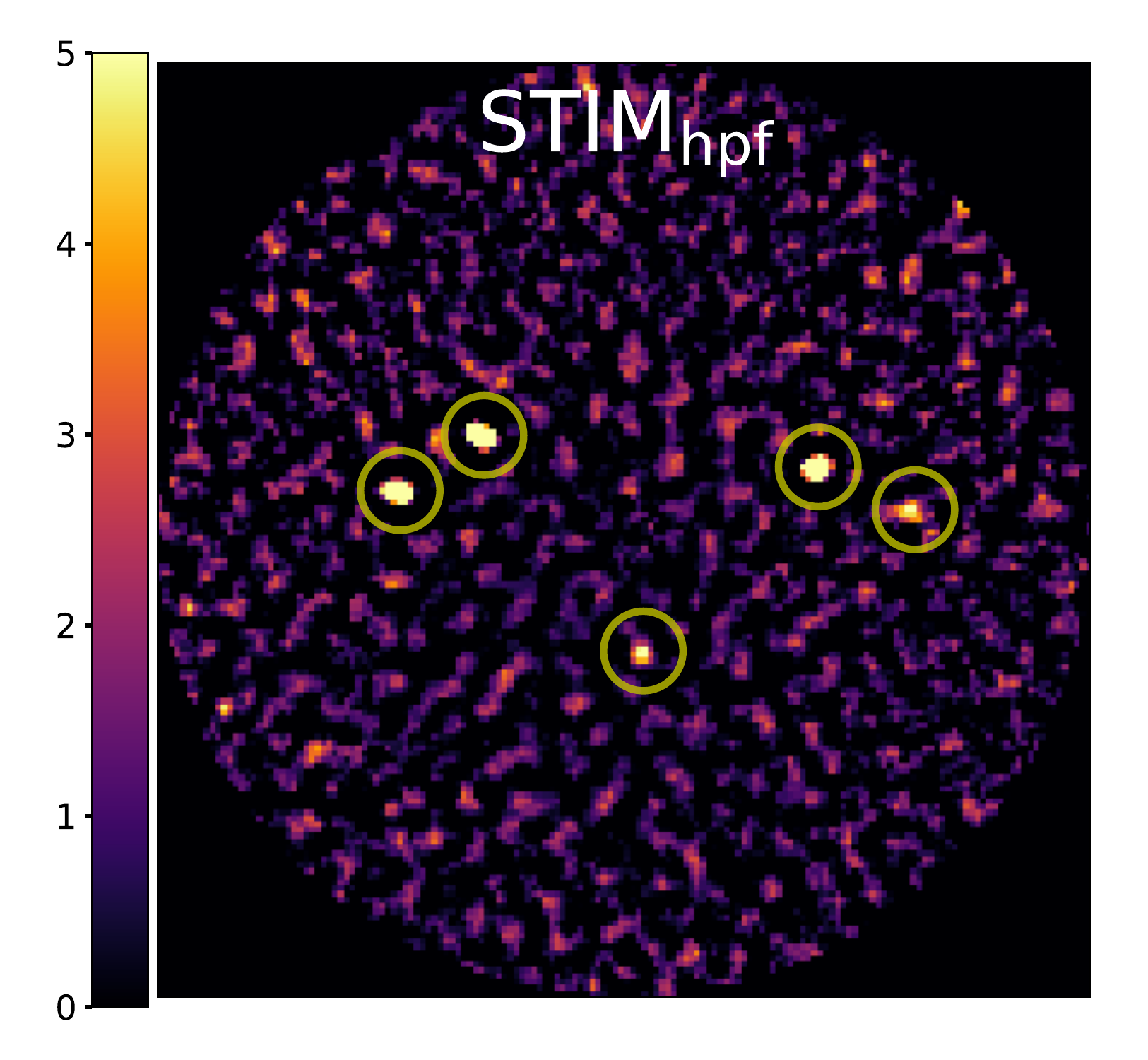}\includegraphics{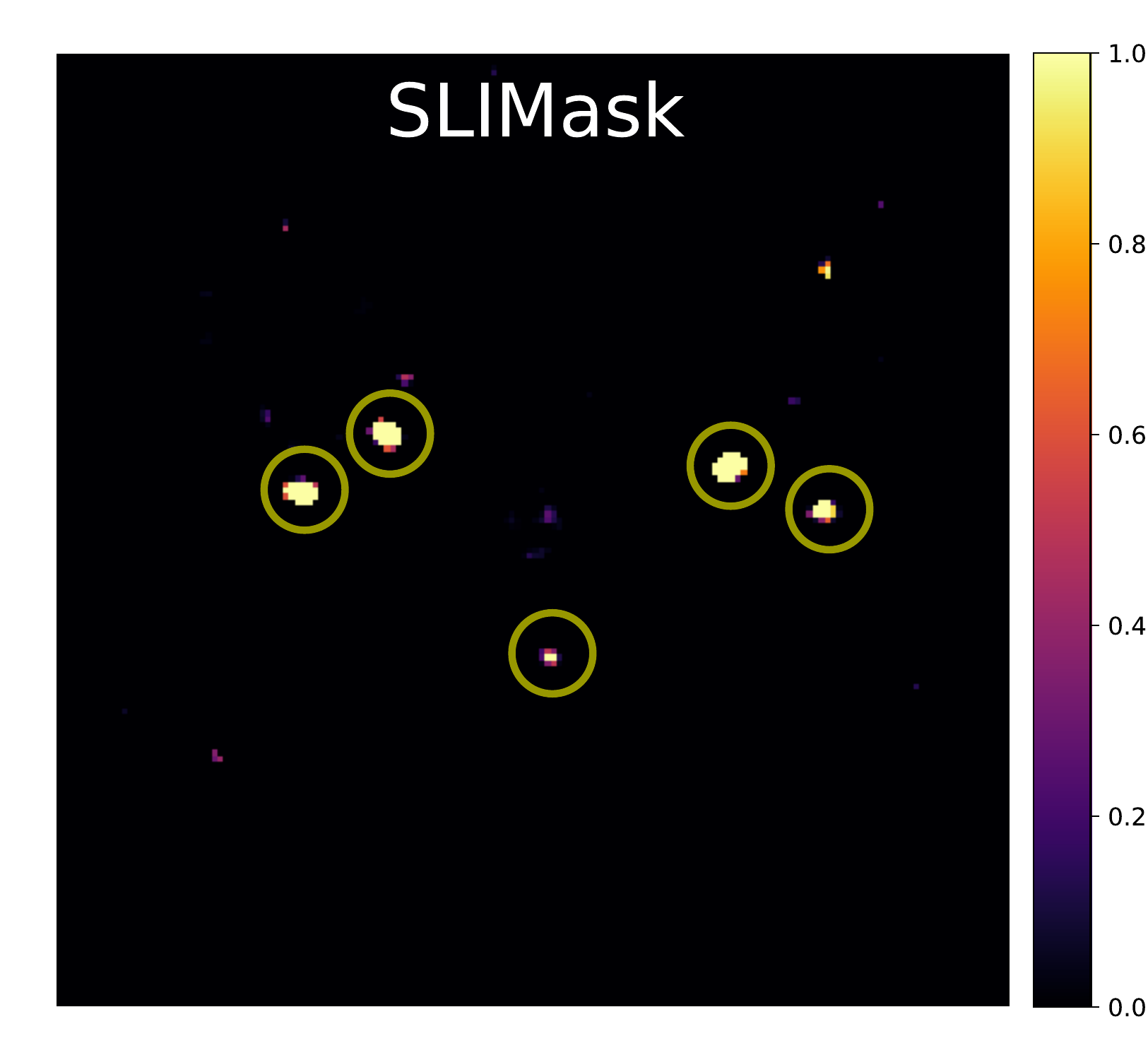}}
   \resizebox{15.5cm}{!}{\includegraphics{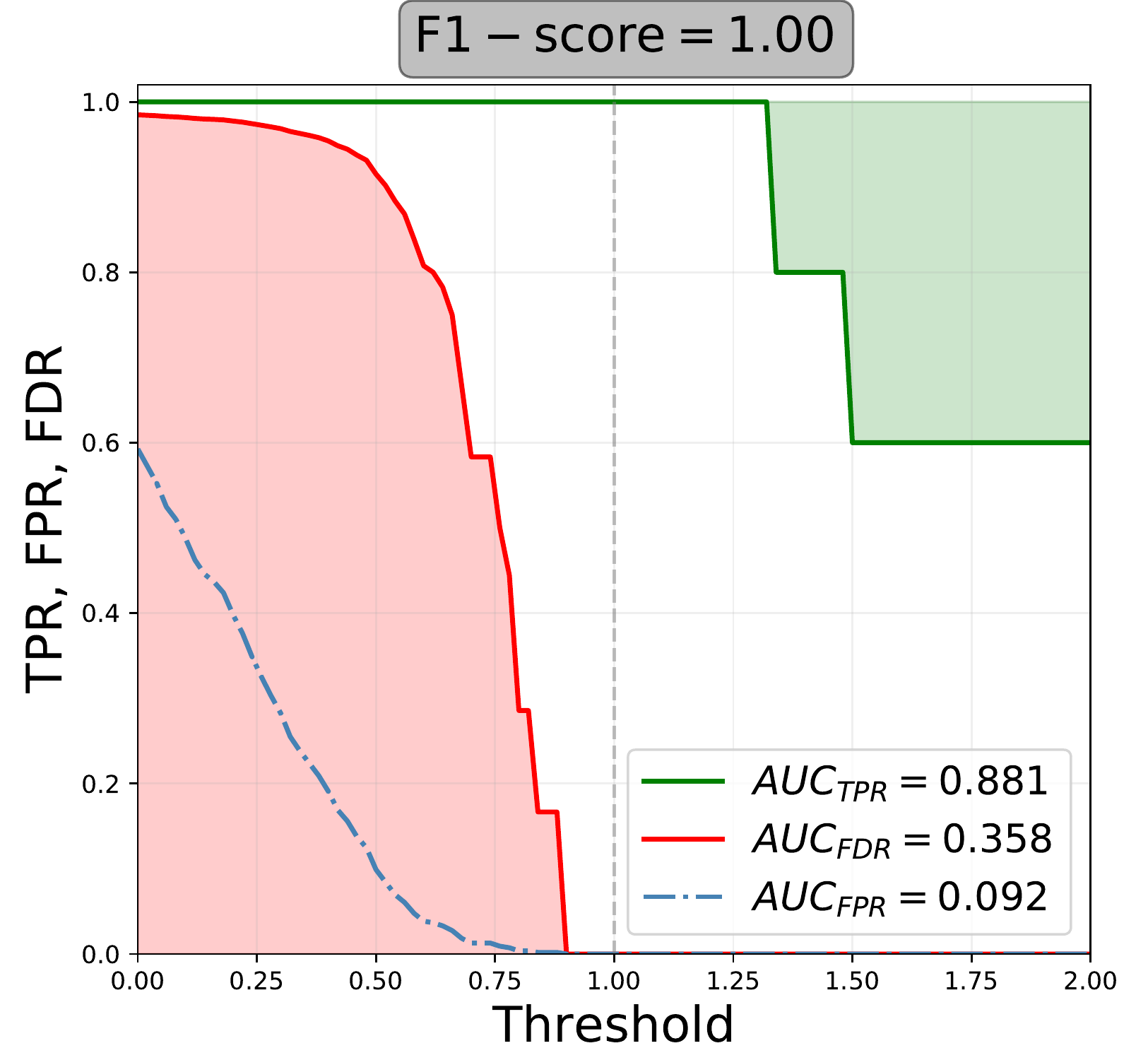}\includegraphics{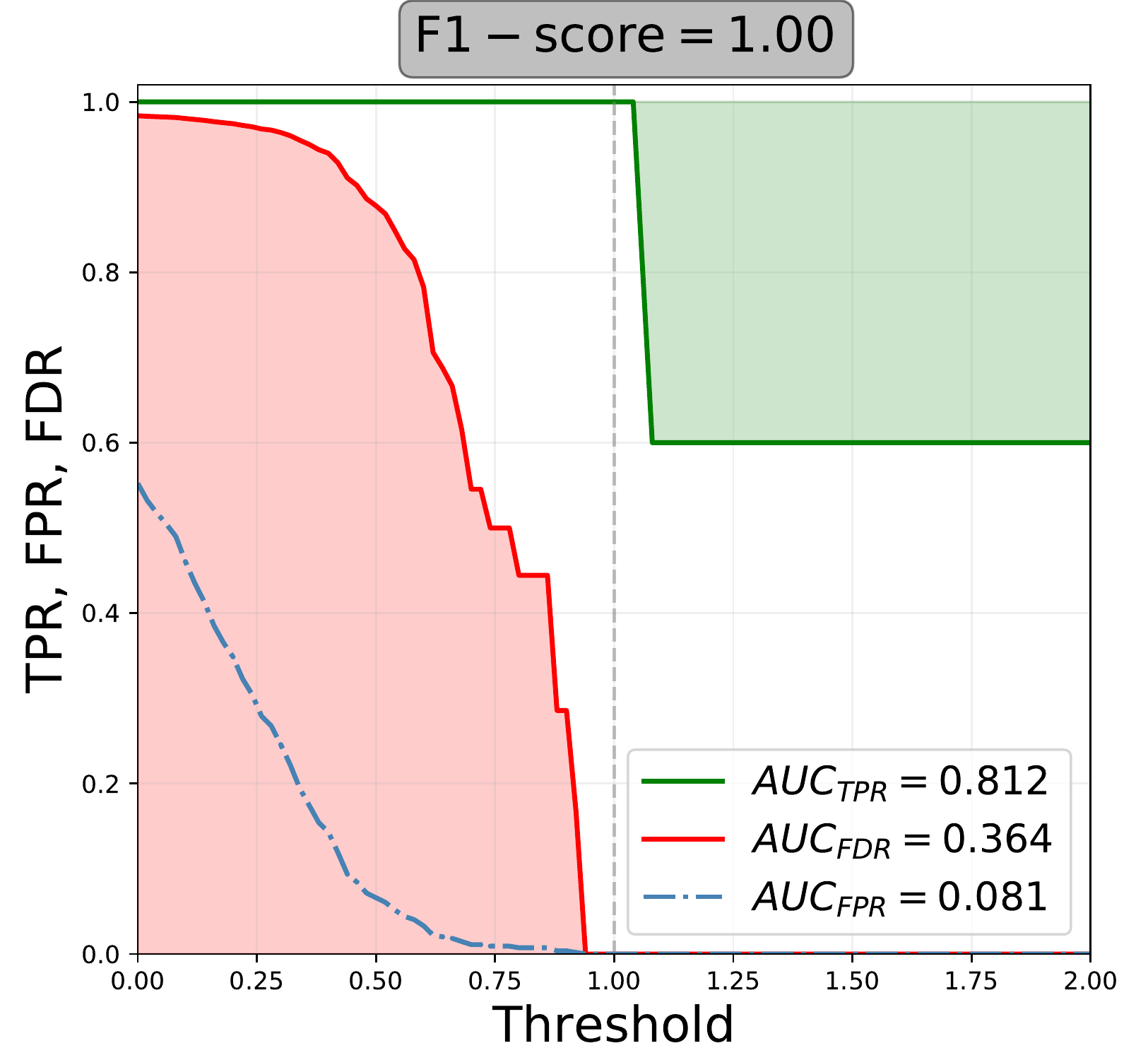}\includegraphics{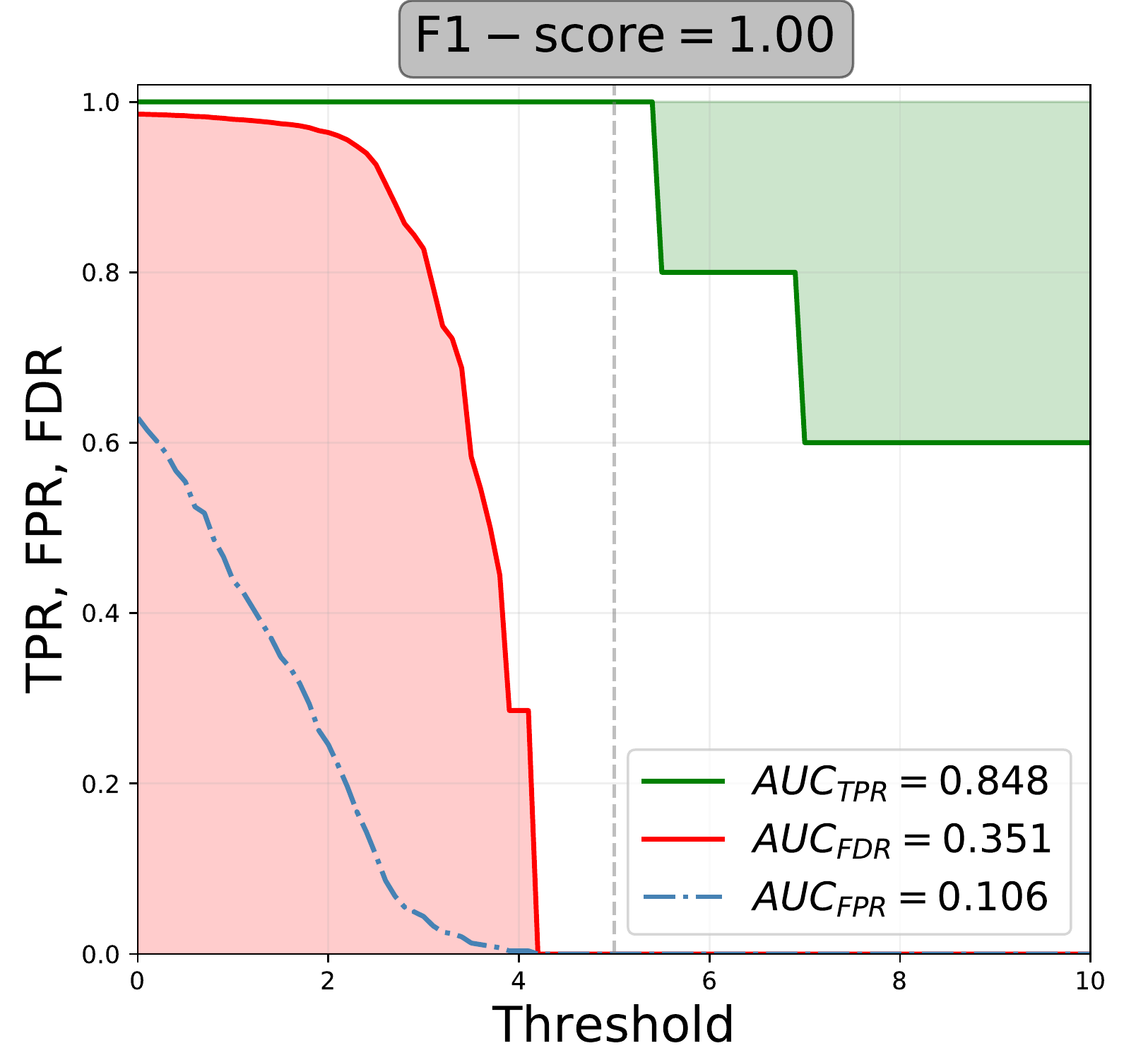}\includegraphics{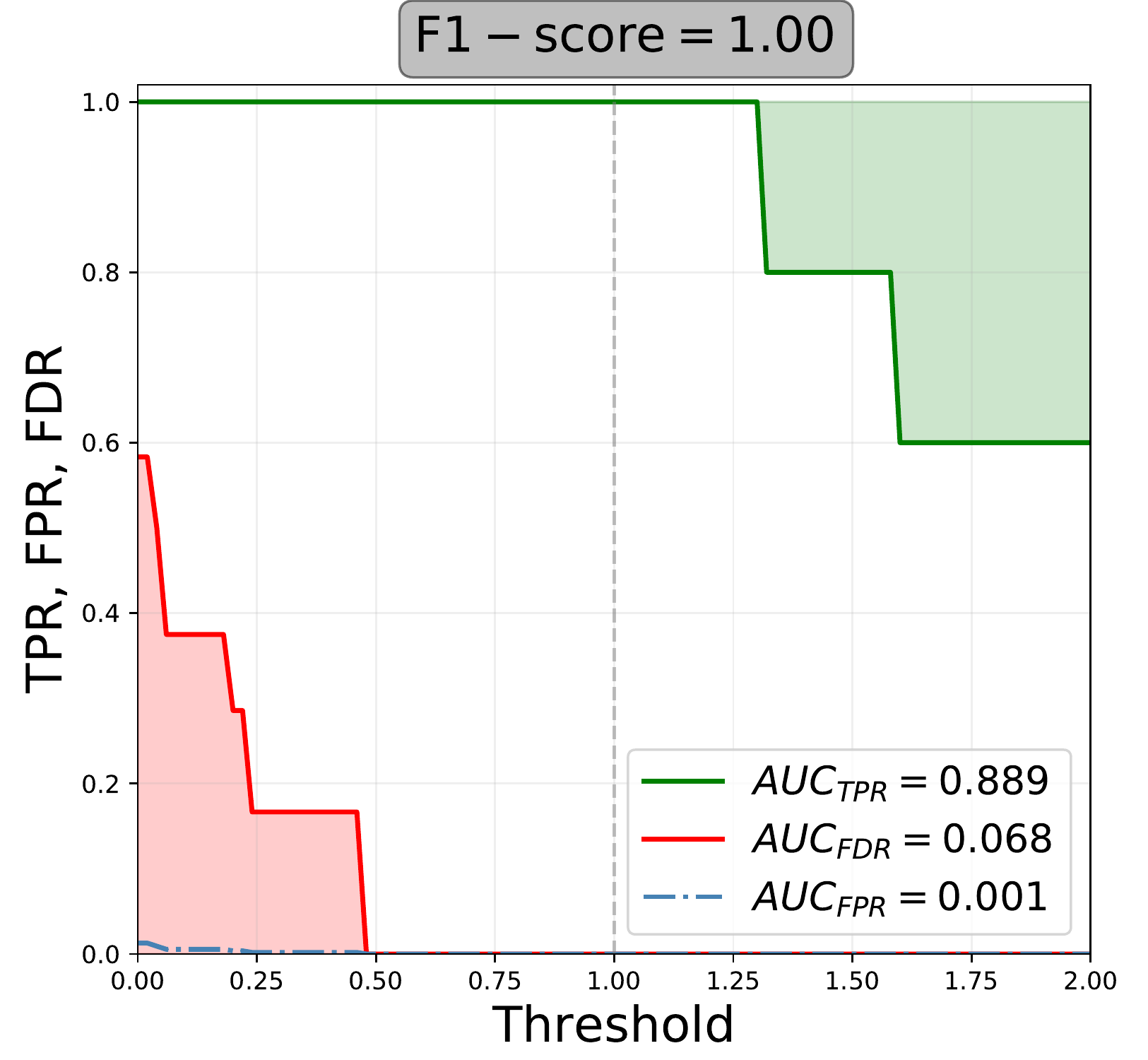}}
   
    \caption{Results of the ADI subchallenge for the sph3 VLT/SPHERE-IRDIS dataset: speckle subtraction techniques.}
    \label{fig:img_sph3_sst}
\end{figure}

For these post-processing methods, the images and corresponding counts of detections and non-detections as a function of the threshold are displayed in Fig.~\ref{fig:img_sph3_sst} for the case of the third VLT/SPHERE-IRDIS data set (sph3). The scores proposed in Sect.~\ref{sec:metrics} computed on the $9$ data sets from this ADI subchallenge are gathered in Tab.~\ref{tab:resadiss}. The final ranking is shown in Fig.~\ref{fig:rnk_adi}, in red for the residual maps and in orange for the detection maps. 
For the specific case of the three SpeCal algorithms, we would like to emphasize that the SpeCal detections are usually done by visual inspection and are not optimized for the use of a preset detection threshold. Since the images shown in the galleries displayed in App.~\ref{app_adi} scale from $0$ to the threshold set, the SpeCal images sometimes appear very dark and sometimes very bright, confirming that the SpeCal output is poorly adapted to $5\sigma$ threshold we used for the systematic comparison. 

\subsubsection{Inverse  problem techniques:} 
We received $5$ submissions based on inverse problem approaches, which consist in modeling the expected planetary signal and tracking it in the image cube. By construction, this method estimates the contrast of the potential planetary signal, at any position in the field of view, via a maximum likelihood estimation, itself relying upon a specific noise distribution. The provided detection map is the signal-to-noise ratio map computed as the estimated contrast divided by the uncertainty on this estimated contrast. 

$\mathrm{ANDROMEDA}$\cite{Cantalloube2015} first performs a pair-wise image subtraction before tracking the planetary signal under the assumption that the residual noise (after the subtraction) is white and Gaussian. $\mathrm{ANDROMEDA}$ is included in the SpeCal pipeline and implemented in the VIP package. 
$\mathrm{FMMF}$\cite{Ruffio2017}\footnote{\url{https://bitbucket.org/pyKLIP/pyklip/}} performs a PCA subtraction before matching a forward-modeled planetary template to the data, assuming that the residual noise is white and Gaussian. $\mathrm{FMMF}$ is included in the pyKLIP pipeline\cite{Wang2015pyklip} and has been used to process the GPI survey demographic analysis data (GPIES\cite{Nielsen2019}). 
$\mathrm{PACO}$\cite{Flasseur2018} does not perform any subtraction prior to tracking the planetary signal, and assumes that the noise is a multivariate Gaussian noise, self-calibrated on the data, accounting for the spatial covariances of the speckles at the scale of small patches. 
$\mathrm{pyPACO}$ is the python version of PACO, currently under implementation in the VIP package. 
At last, $\mathrm{TRAP}$\cite{Samland2020}\footnote{\url{https://github.com/m-samland/trap}} estimates both the starlight residuals and the planetary signal along the temporal axis (contrary to other methods modeling the signal for each frame) and models the starlight residuals as principal components derived from specific areas that do not include the planetary signal but share similar noise statistical properties.

\begin{figure}
    \centering
    \resizebox{\hsize}{!}{\includegraphics{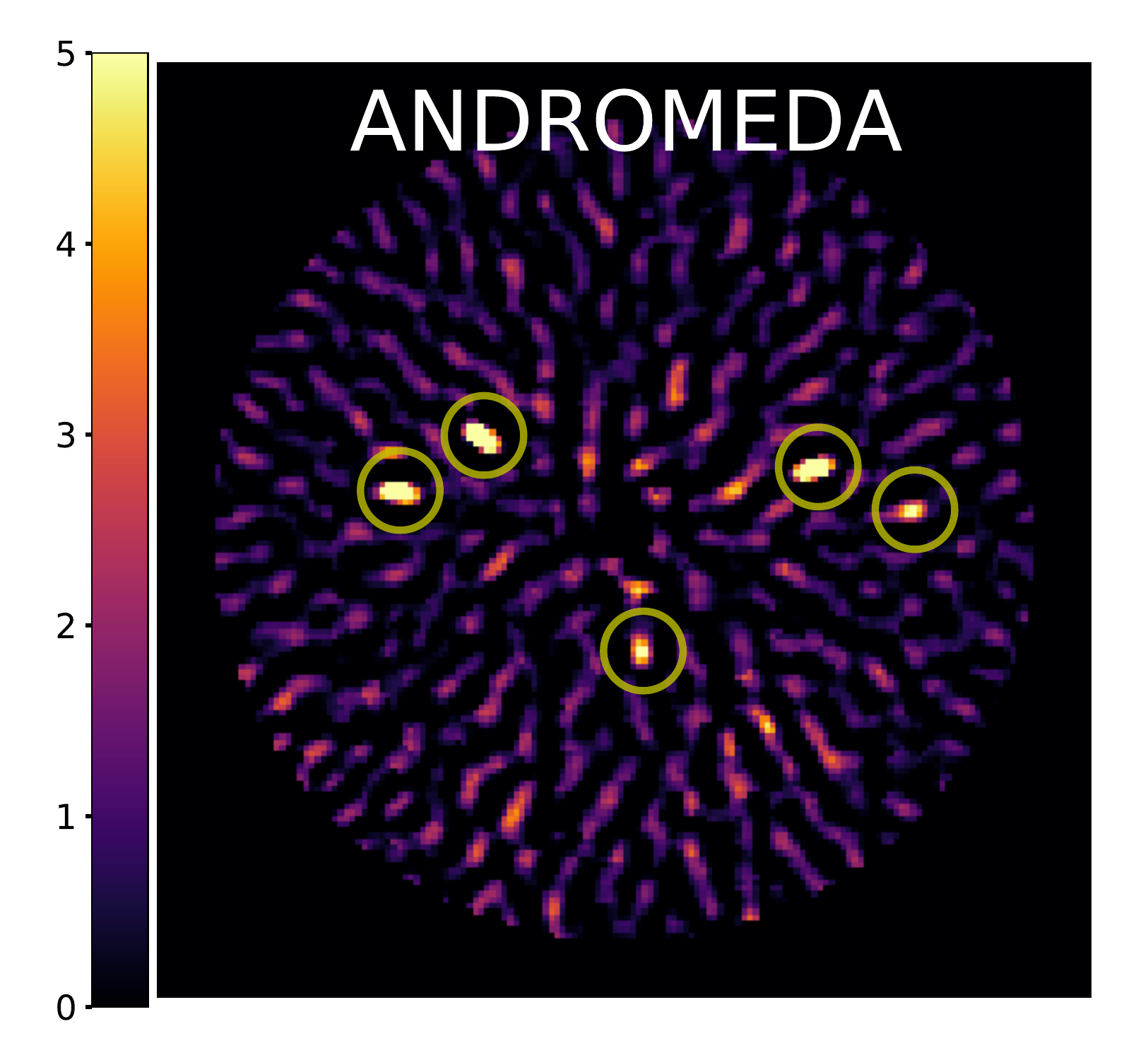}\includegraphics{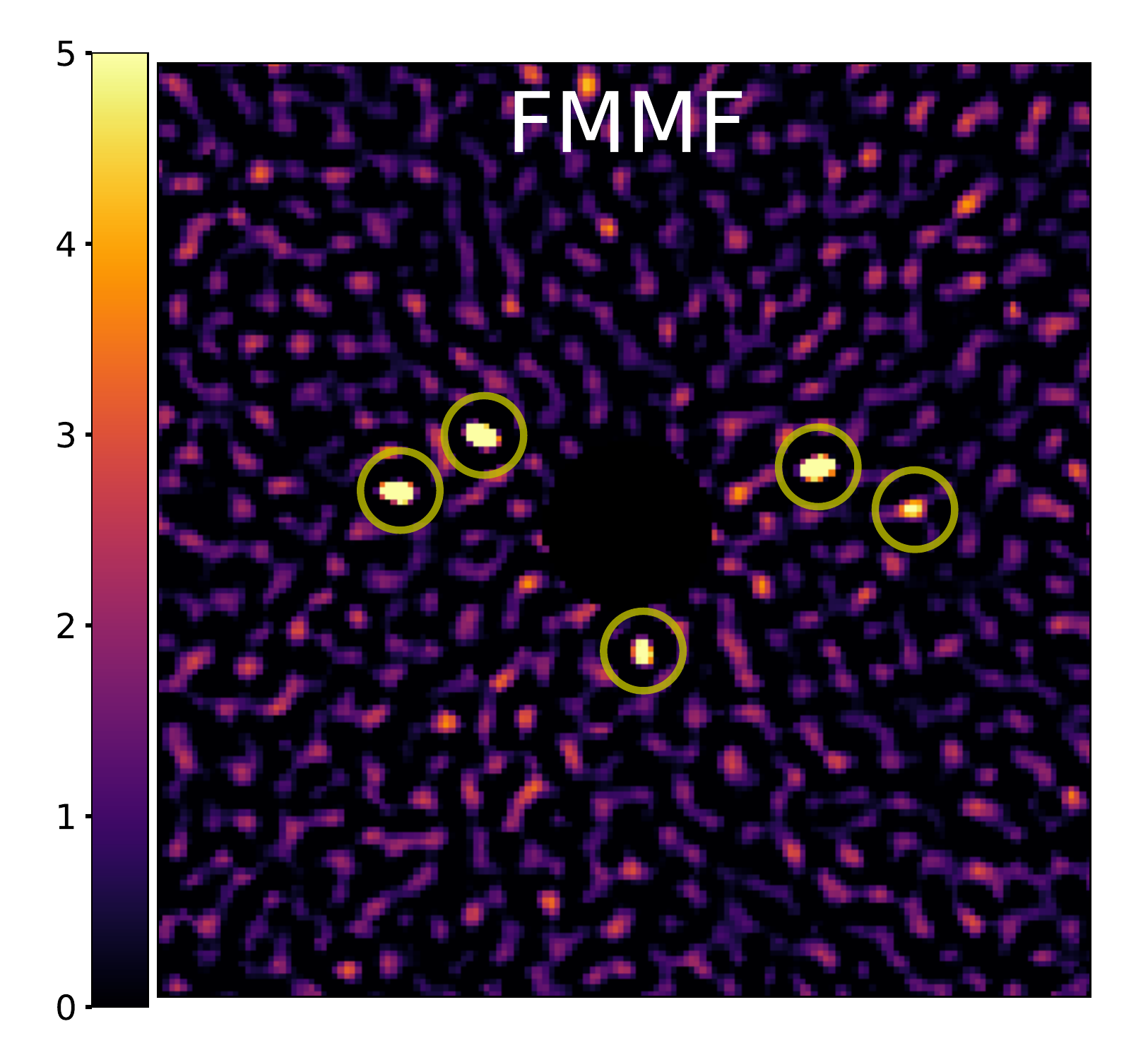}\includegraphics{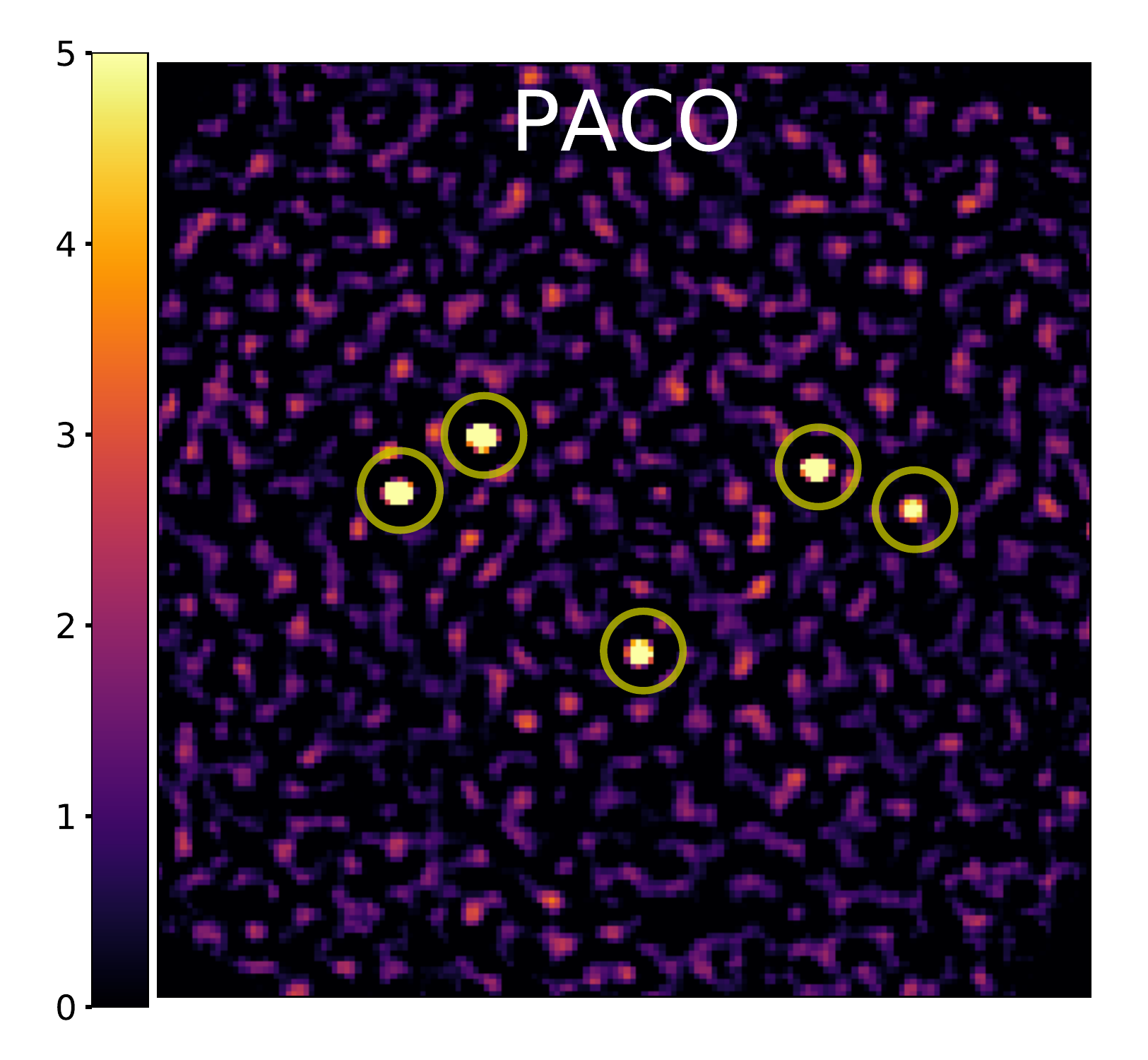}\includegraphics{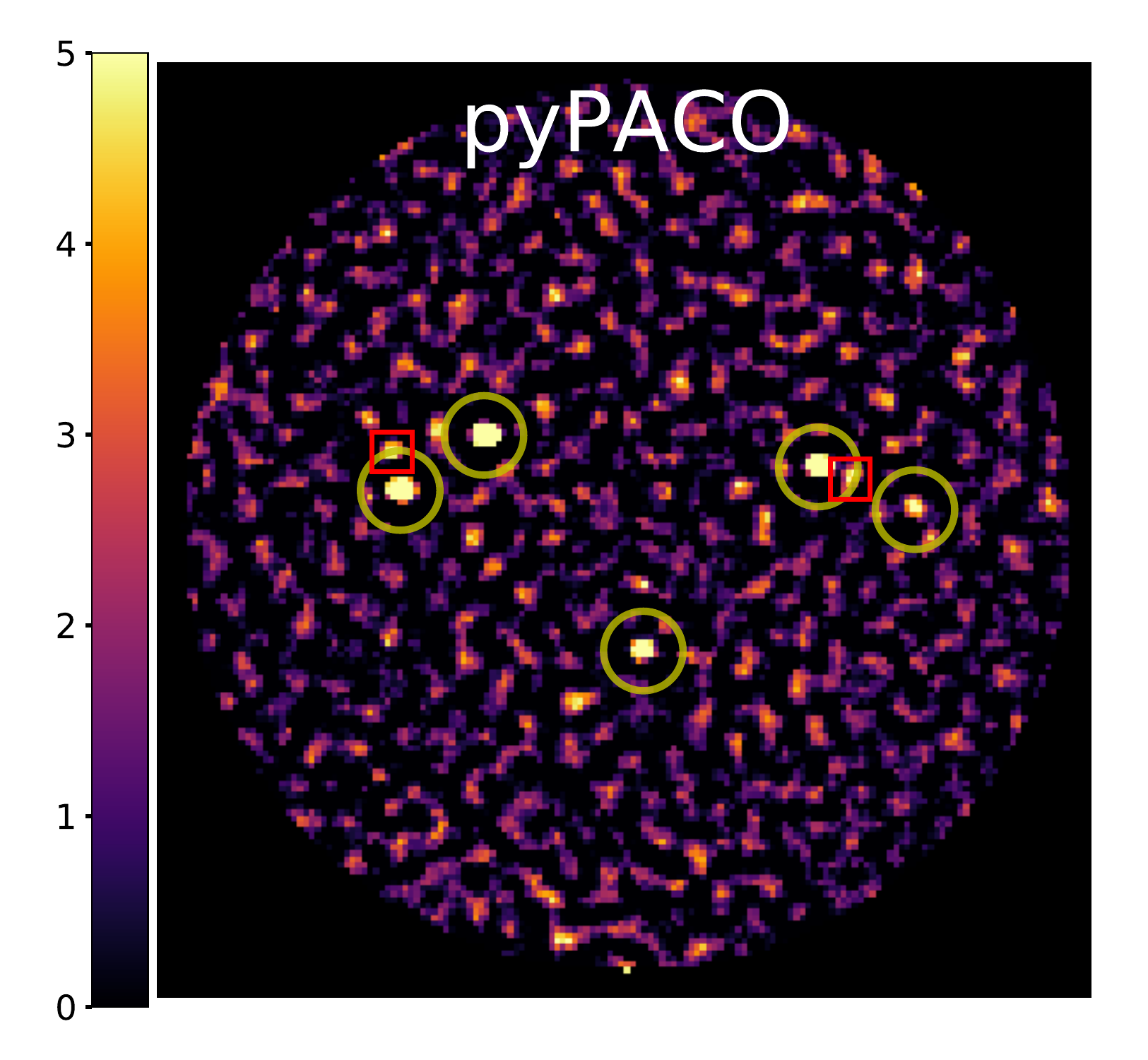}\includegraphics{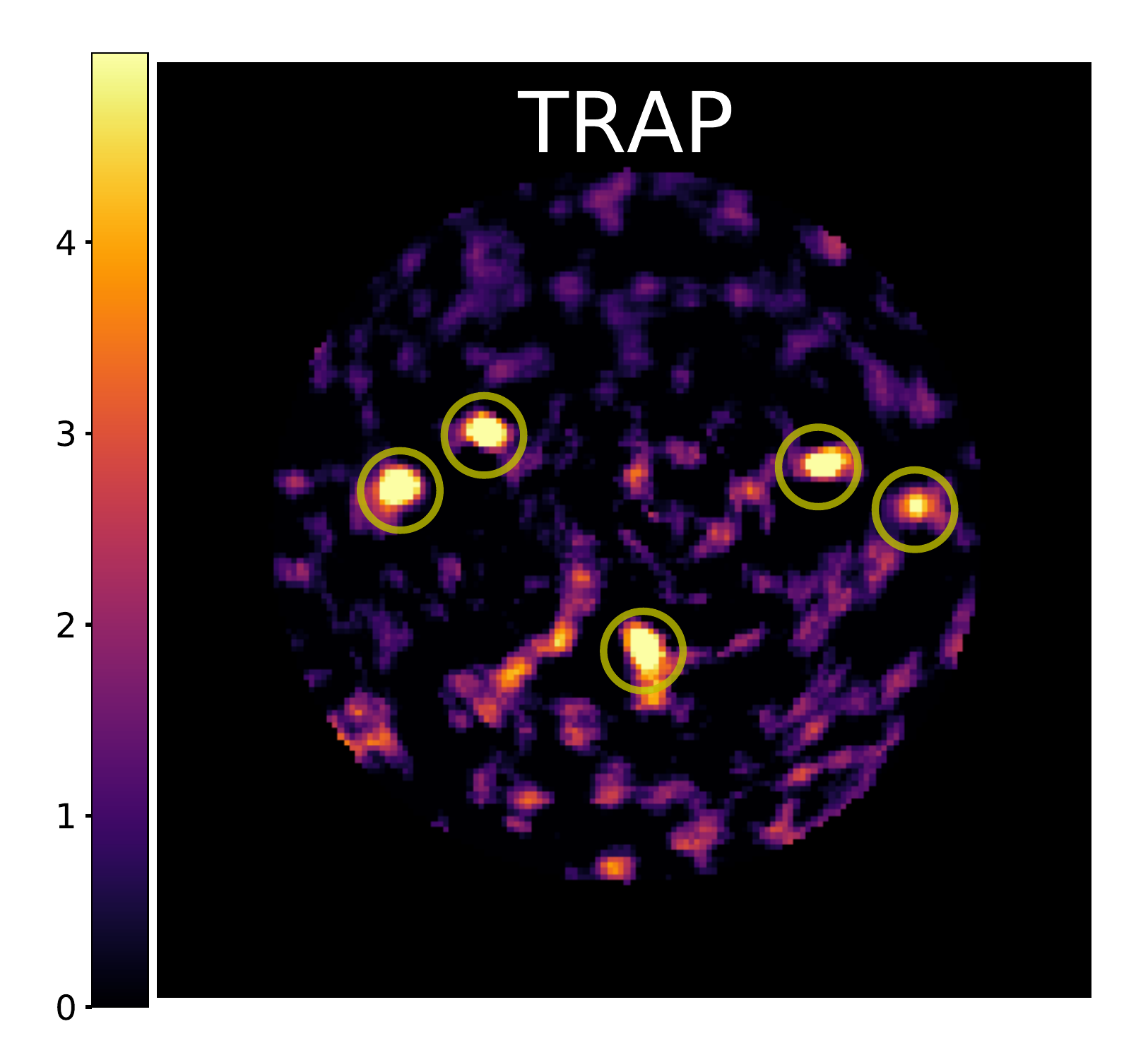}}
   \resizebox{\hsize}{!}{\includegraphics{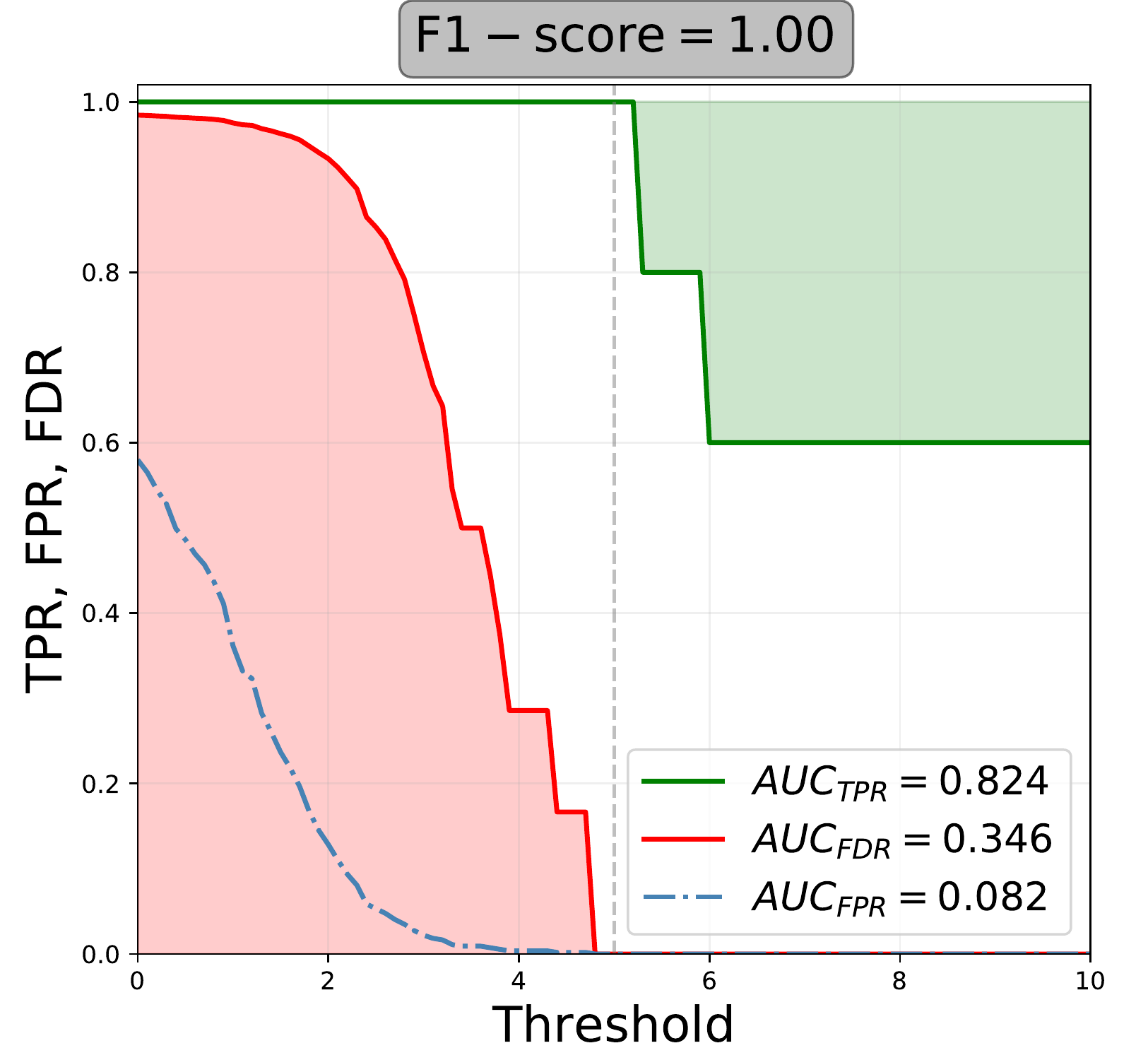}\includegraphics{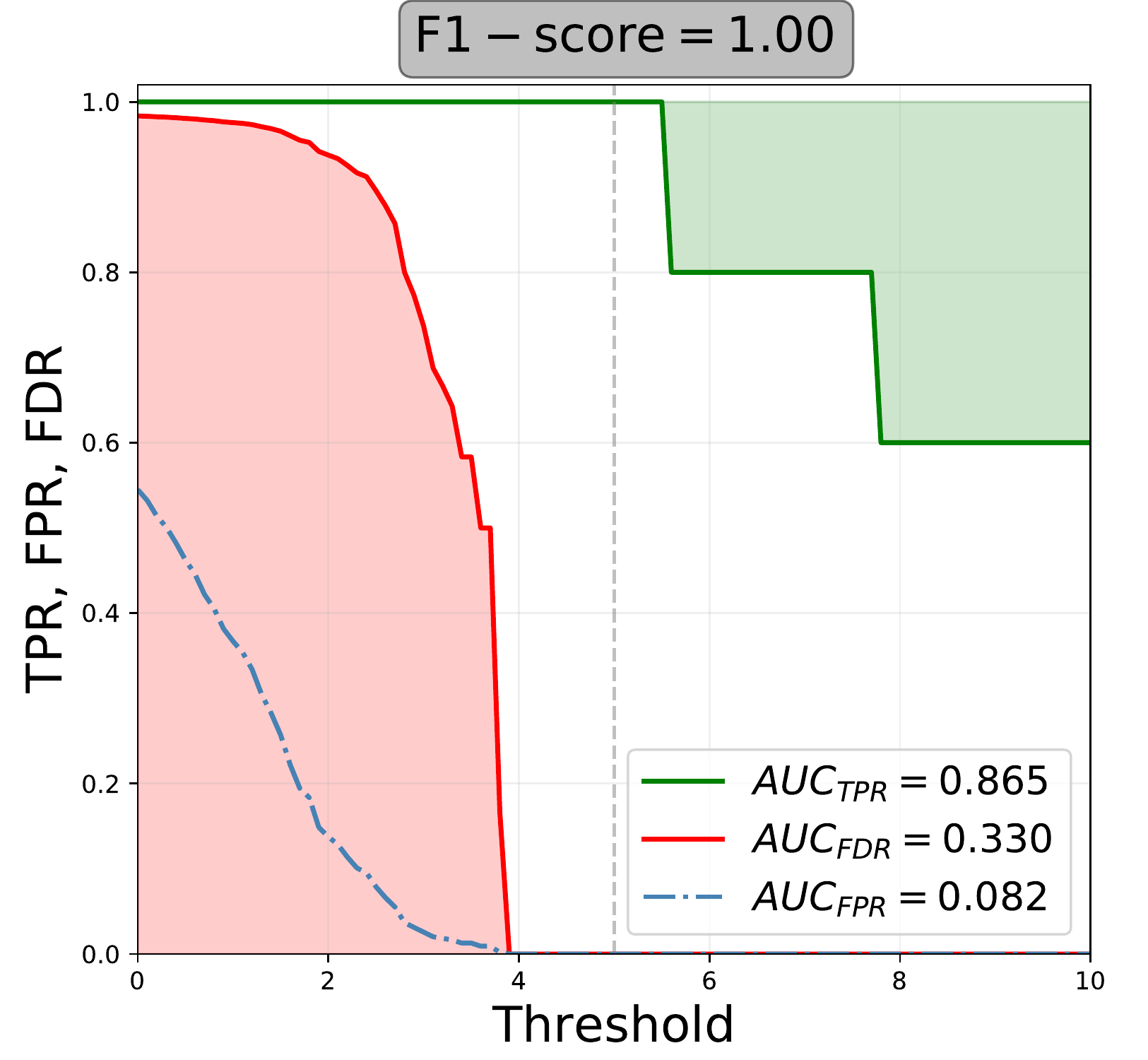}\includegraphics{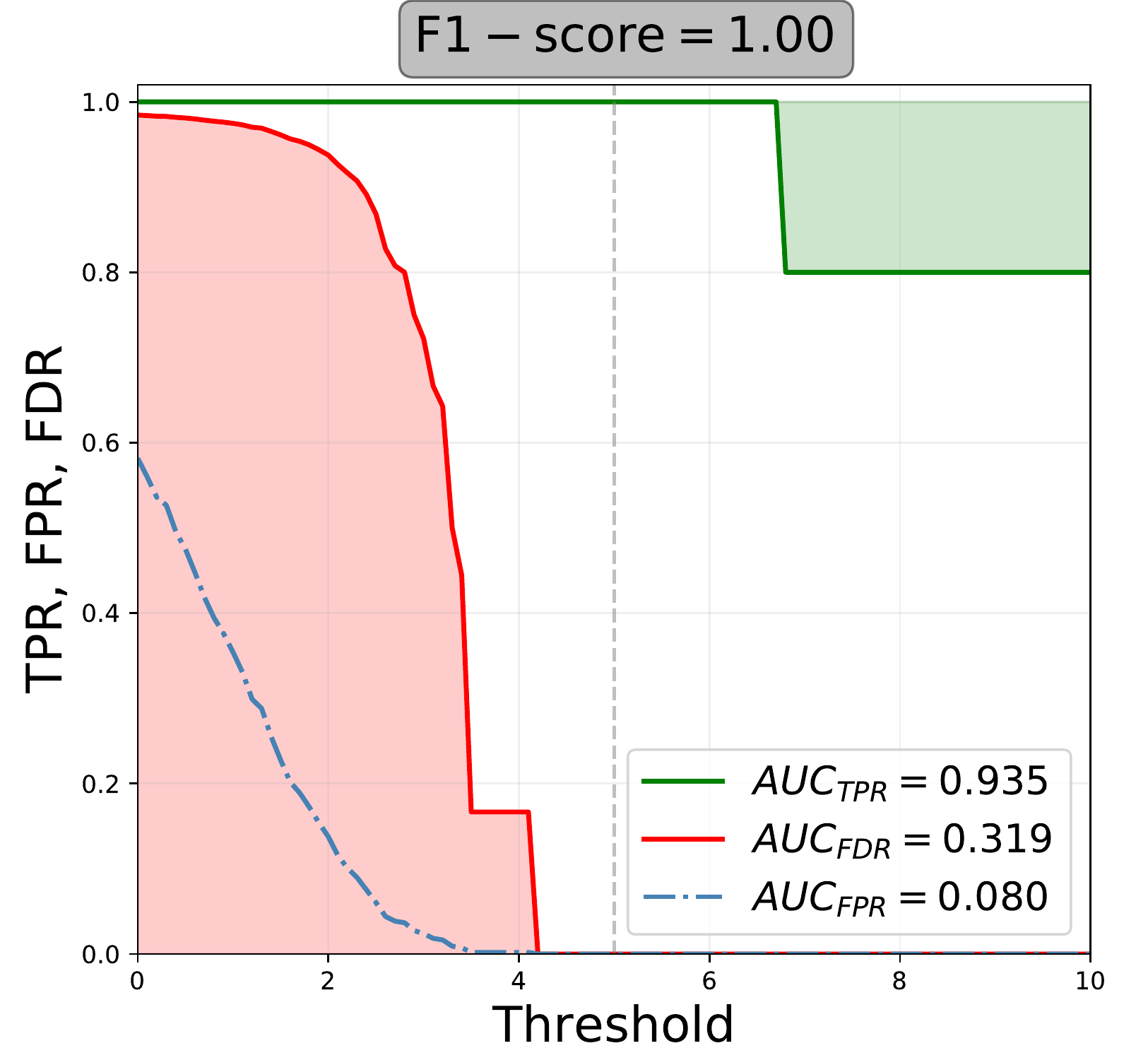}\includegraphics{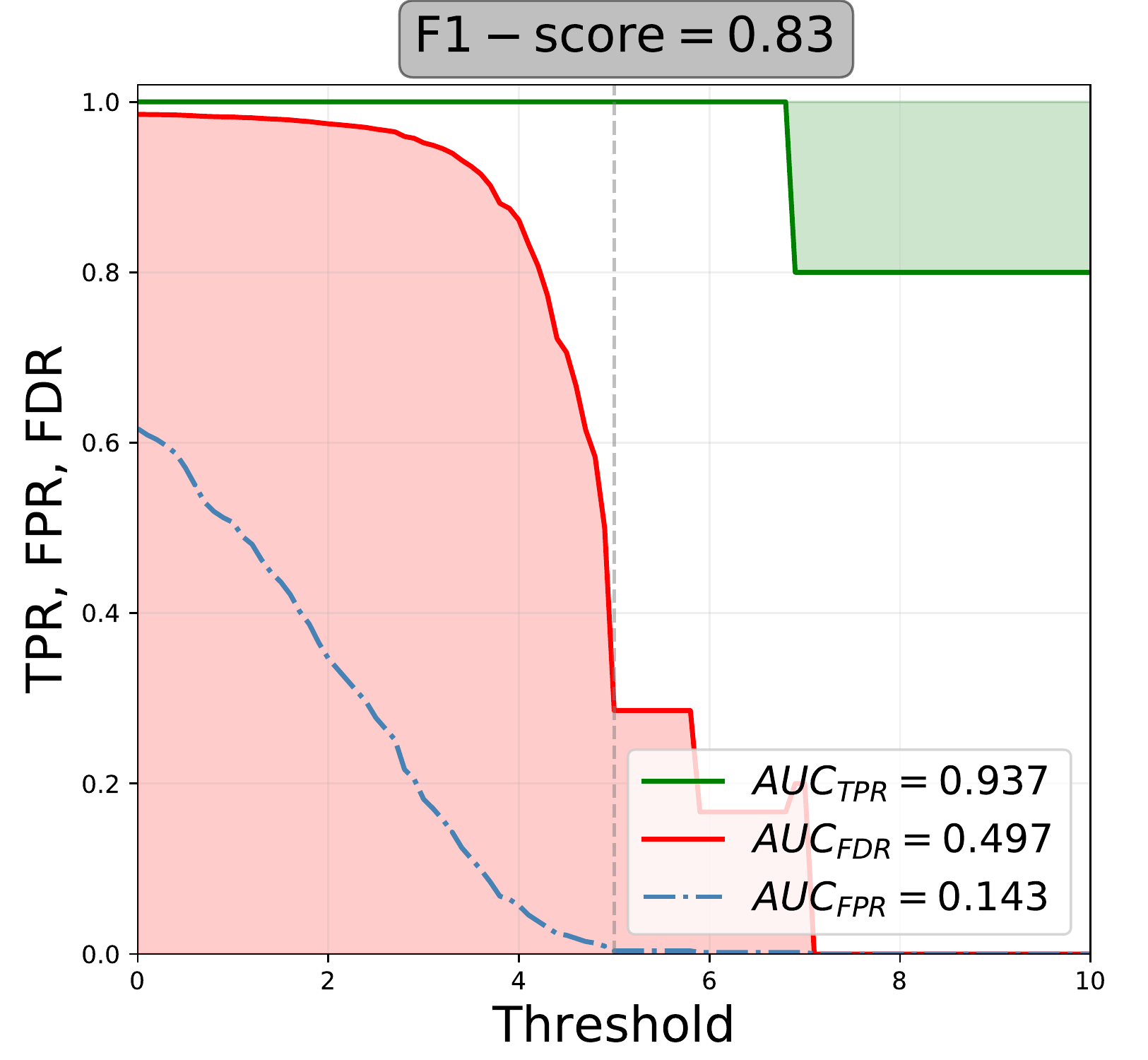}\includegraphics{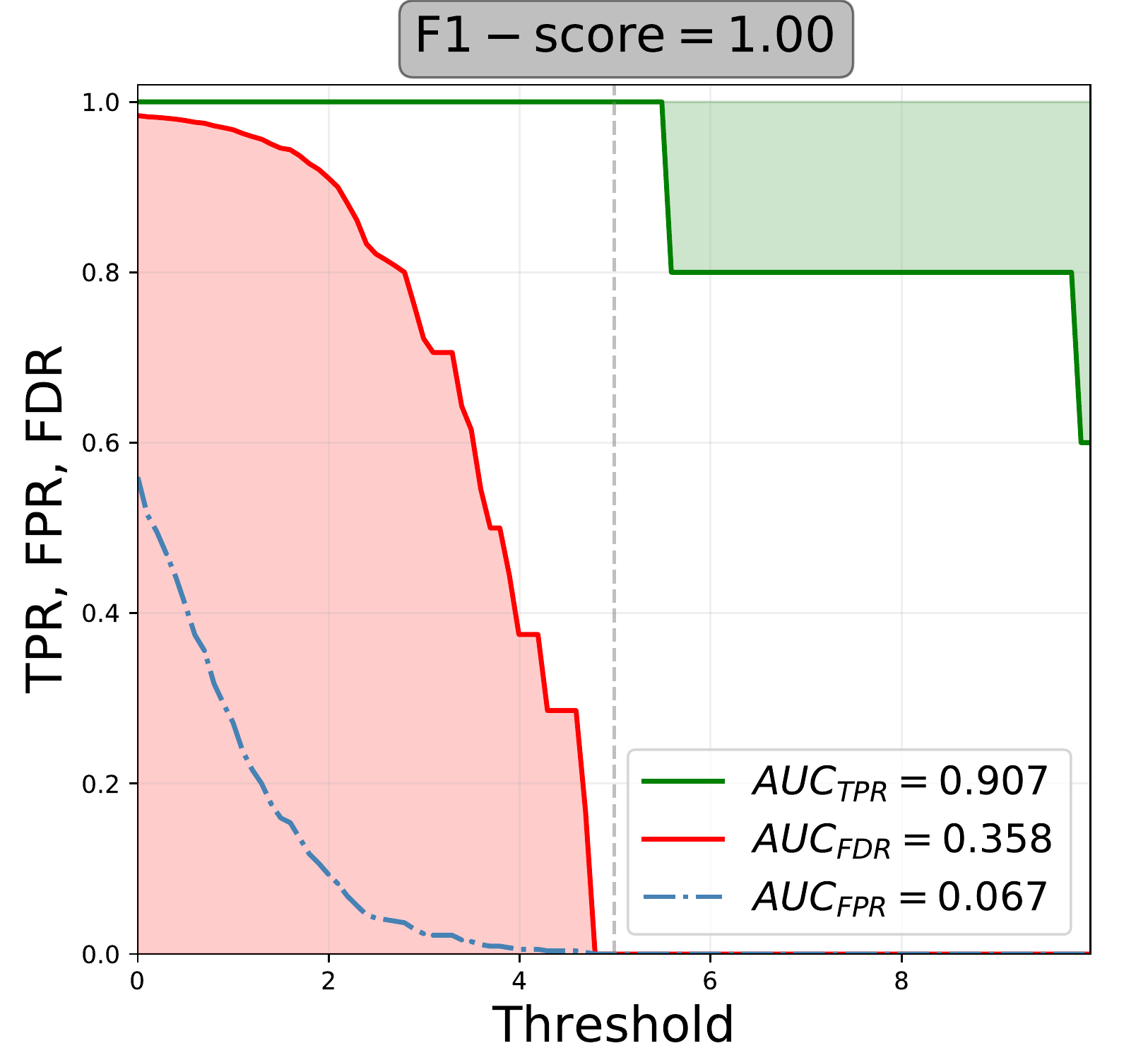}}
    \caption{Results of the ADI subchallenge for the sph3 VLT/SPHERE-IRDIS dataset: inverse problem approaches.}
    \label{fig:img_sph3_ipt}
\end{figure}

The images and corresponding counts of detections and non-detections as a function of the threshold are displayed in Fig.~\ref{fig:img_sph3_ipt} for the case of the third VLT/SPHERE-IRDIS data set (sph3). The scores proposed in Sect.~\ref{sec:metrics}, computed on the 9 data sets from this ADI subchallenge, are gathered in Tab.~\ref{tab:resadimap}. The ranking of the different methods based on inverse problem is shown in Fig.~\ref{fig:rnk_adi} (blue).

\subsubsection{Supervised machine learning techniques:} 
We received $5$ submissions based on supervised machine learning techniques. 
SODIRF and SODINN\cite{gonzalez2018sml} are based on the random forest technique and neural networks, respectively. Both algorithms are trained on  hundreds of thousands of two class Multi-level Low-rank Approximation Residual (MLAR) samples. MLAR samples are square patch sequences of twice the instrument PSF FWHM size, centered on each pixel of the ADI sequence. Each of these patches on a MLAR sequence corresponds to a low-rank level (PCA-principal components). The full MLAR sequence shares the temporal dimension of the ADI sequence. Therefore, one MLAR class represents the evolution of an exoplanet light source along the principal component space and the other class the evolution of residual noise. Once trained, SODIRF and SODINN generate a normalized detection map in which each pixel has a probability to belong to one of these classes. 

$SODIRF_{original}$\footnote{\url{https://github.com/carlgogo/sodinn}} is the original random forest algorithm published in Ref.\cite{gonzalez2018sml}, using the same set of hyper-parameters. 
$SODIRF_{adapted}$ is a modified version in which the optimal hyper-parameter values have been found for each particular dataset by means of applying a grid search and cross-validation with three folds. 
For the case of neural networks, $SODINN_{3D}$ and $SODINN_{LSTM}$ are the original versions presented in Ref.\cite{gonzalez2018sml}, and use respectively a 3D convolutional layers and long short term memory (LSTM) layers. 
$SODINN_{BiLSTM}$ preserves the full architecture of $SODINN_{LSTM}$ except that it uses bidirectional-LSTM layers instead. 

\begin{figure}
    \centering
    \resizebox{\hsize}{!}{\includegraphics{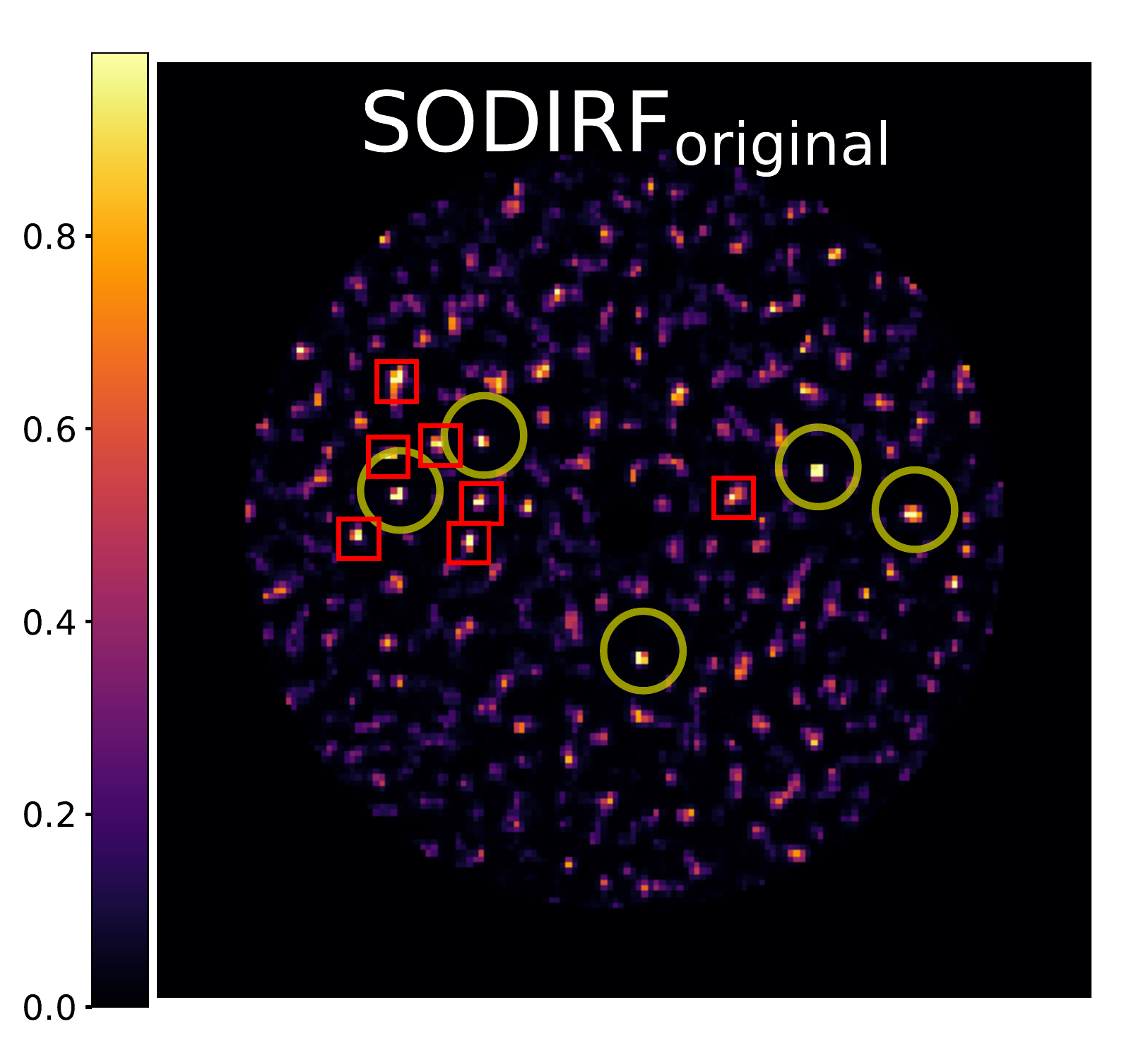}\includegraphics{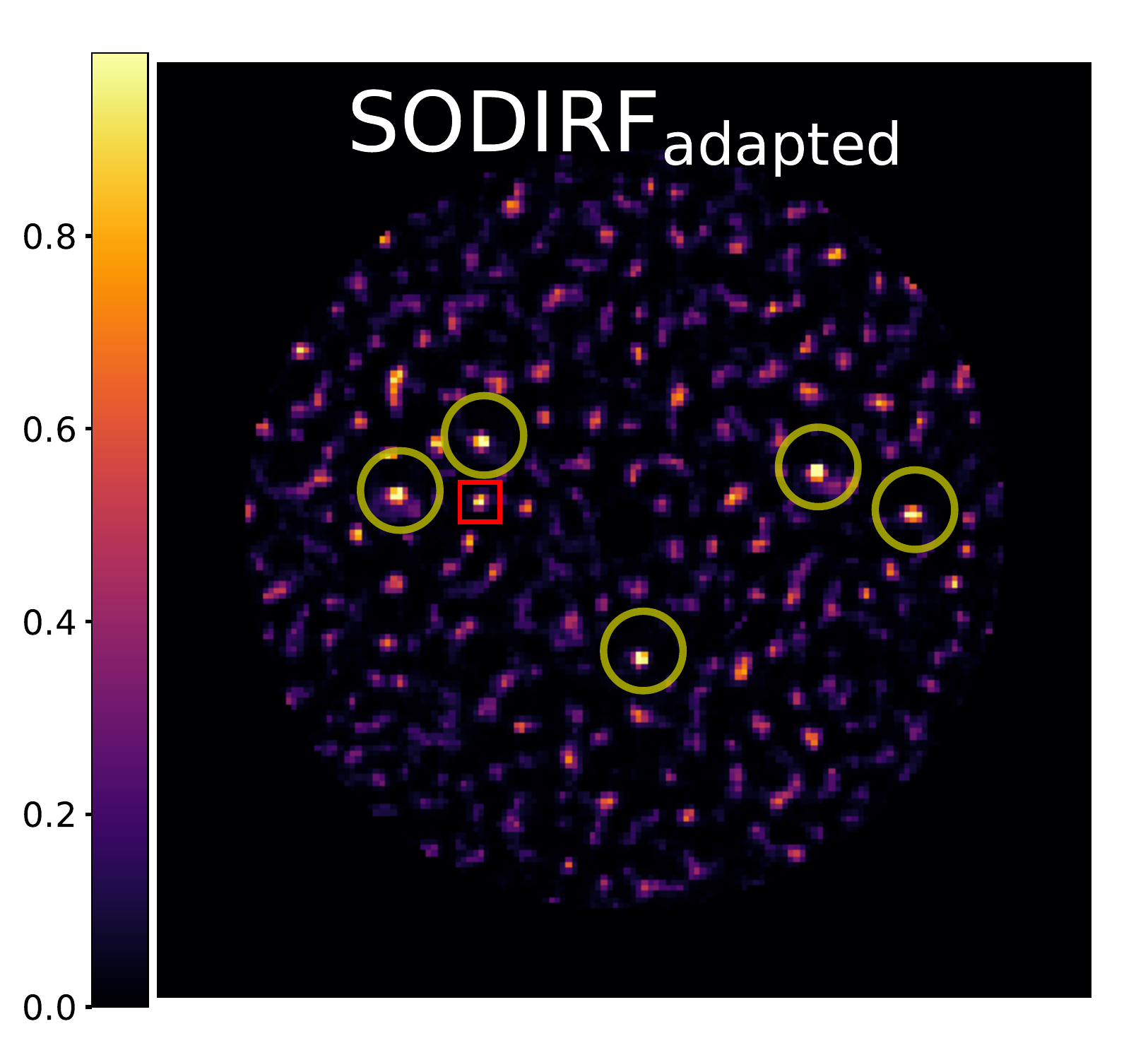}\includegraphics{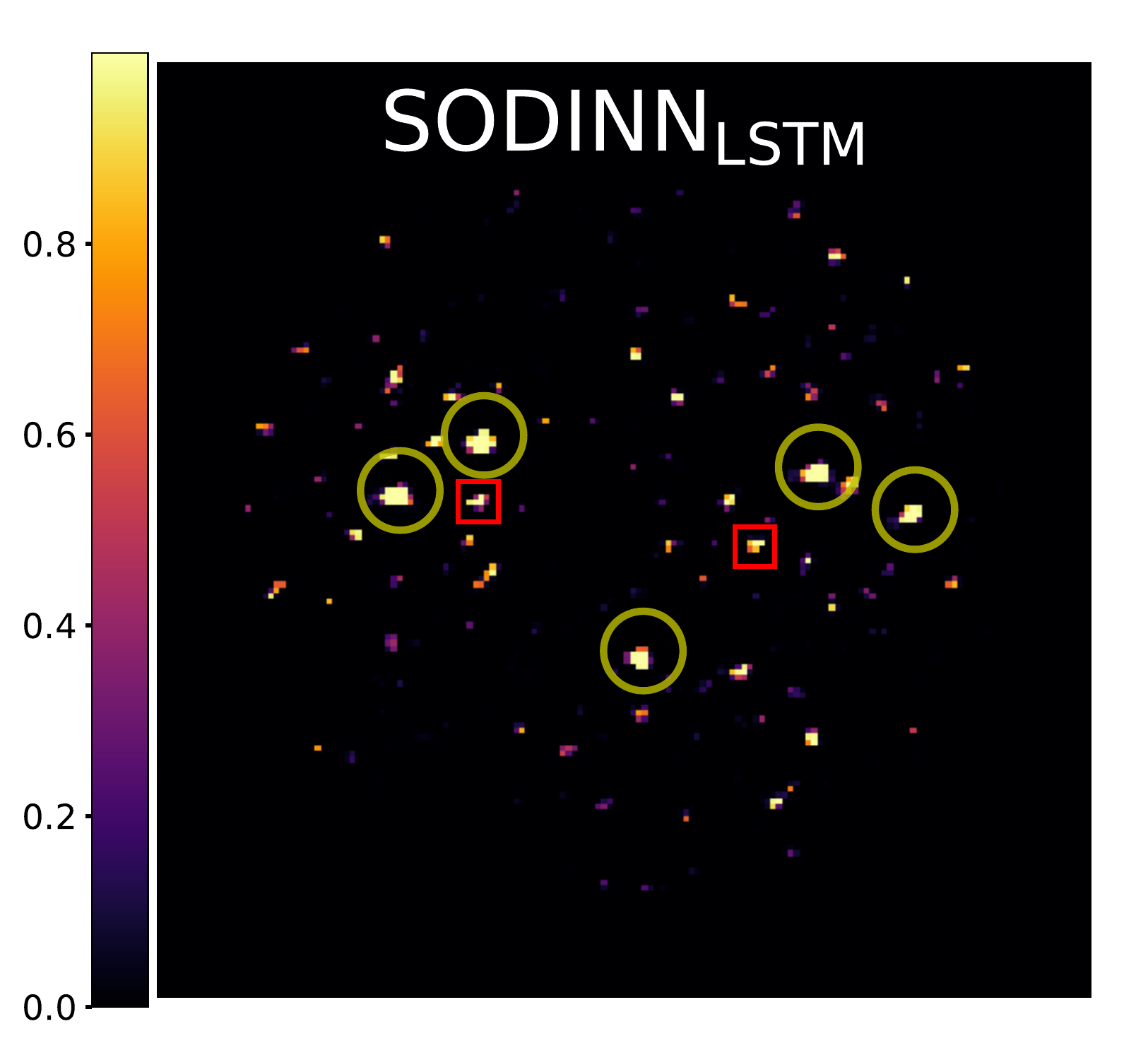}\includegraphics{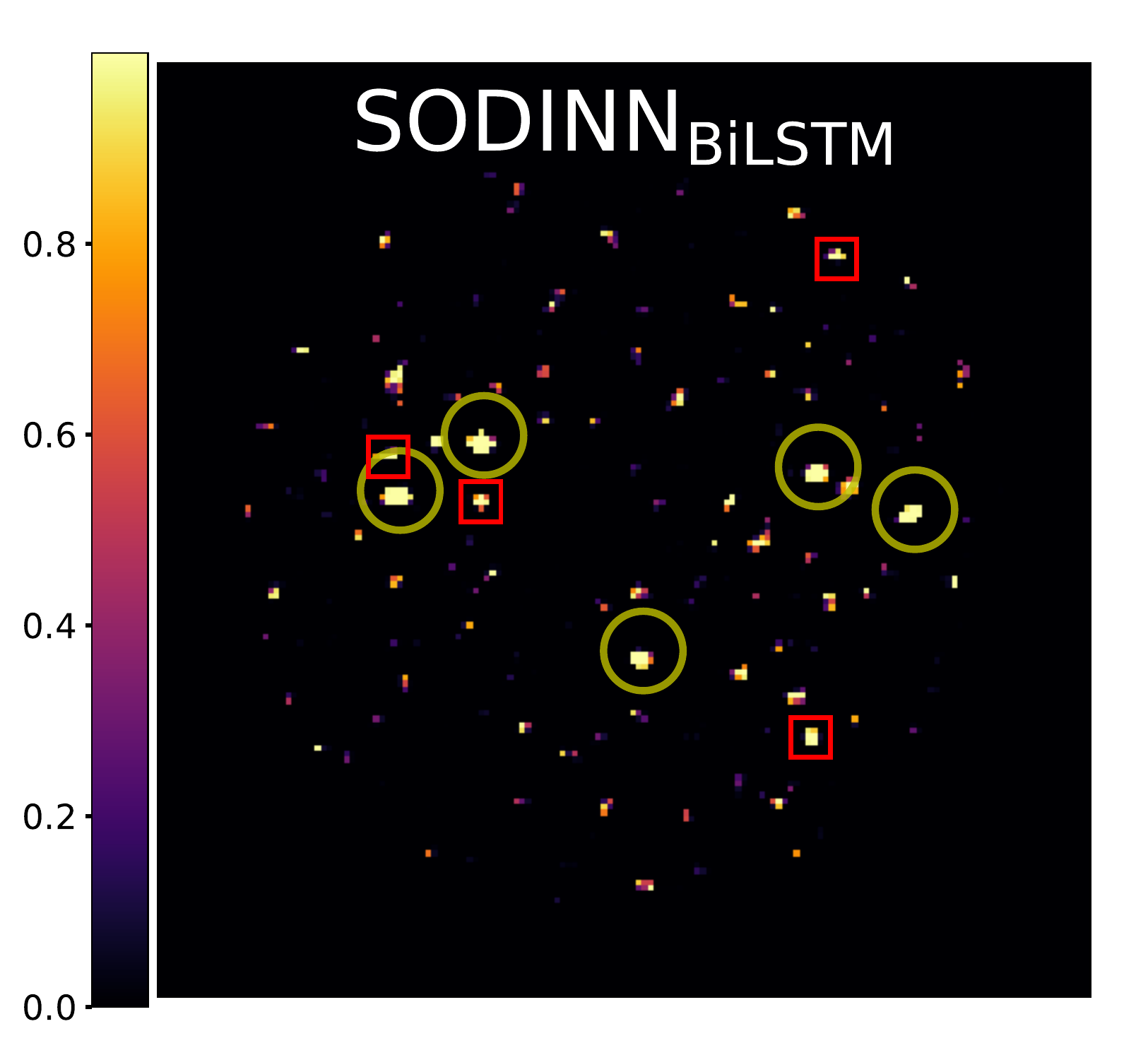}\includegraphics{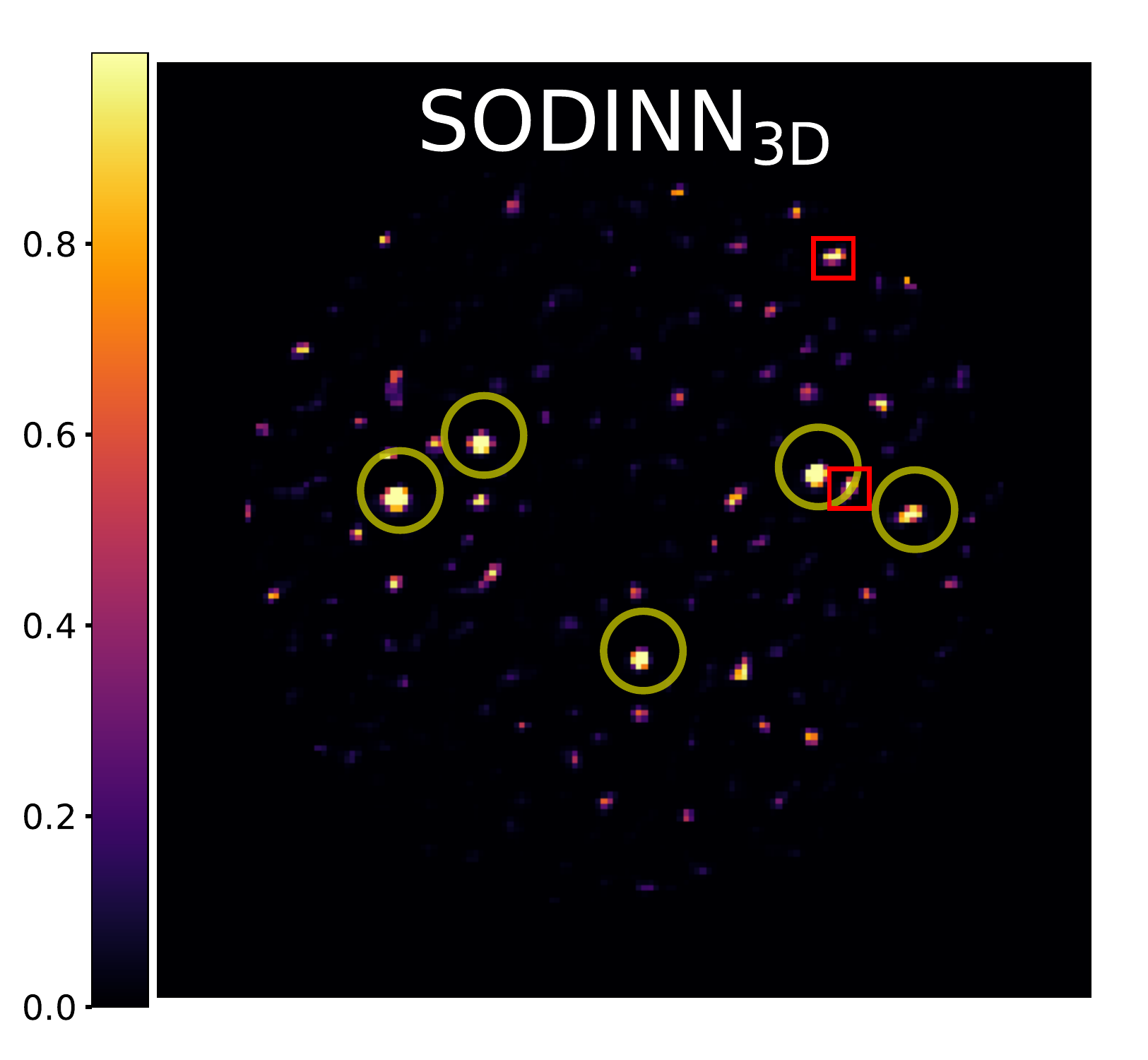}}
   \resizebox{\hsize}{!}{\includegraphics{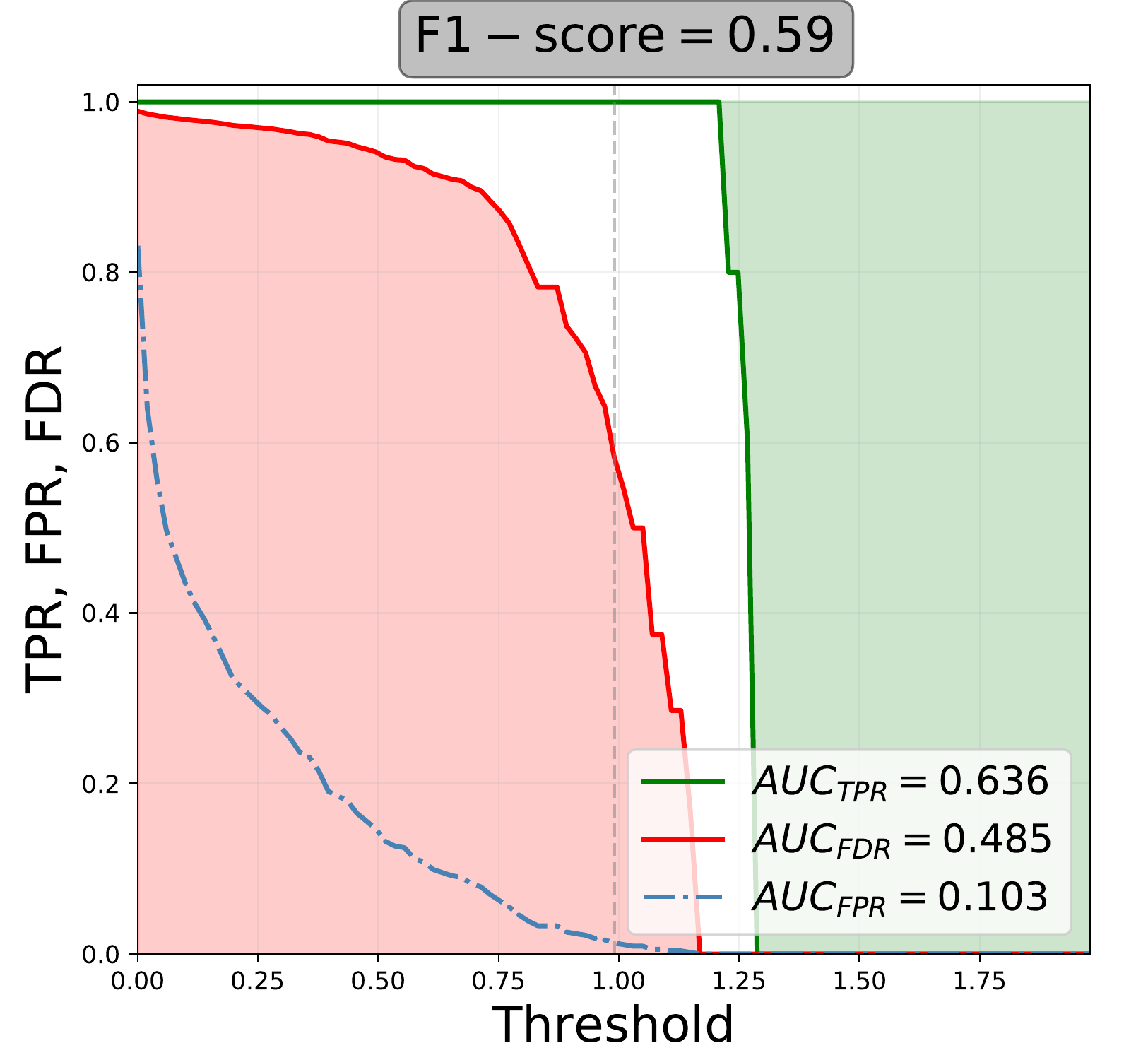}\includegraphics{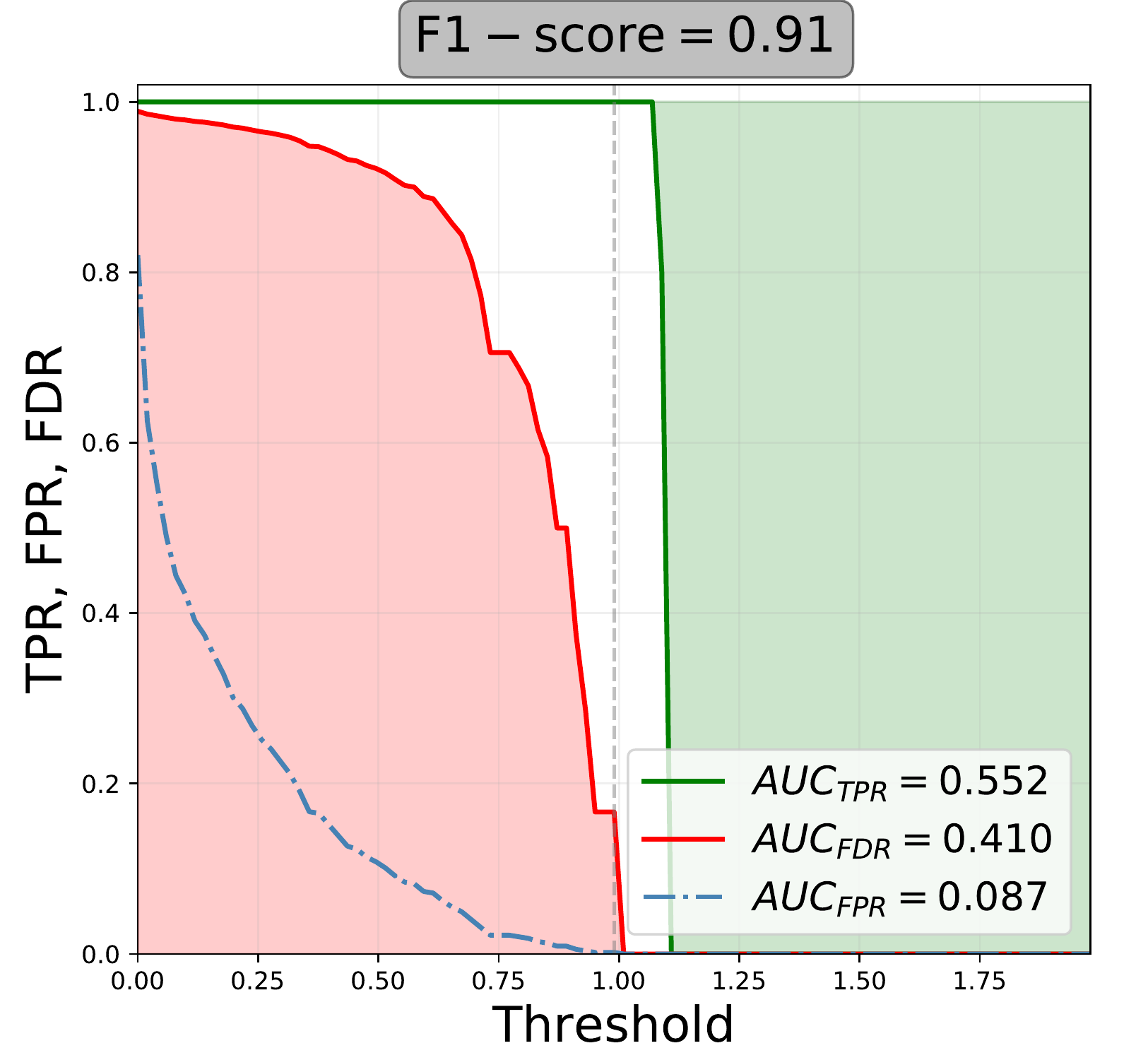}\includegraphics{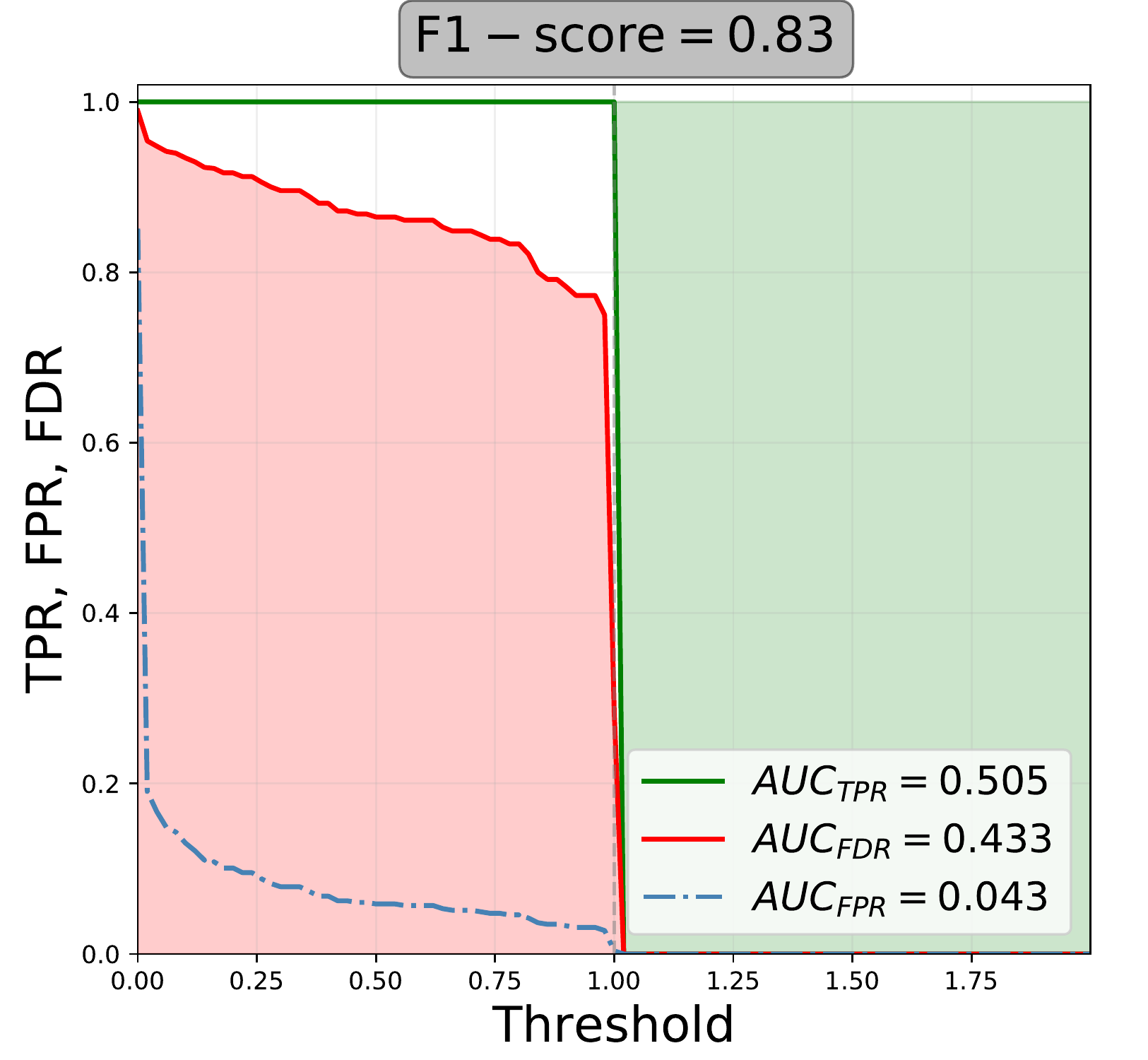}\includegraphics{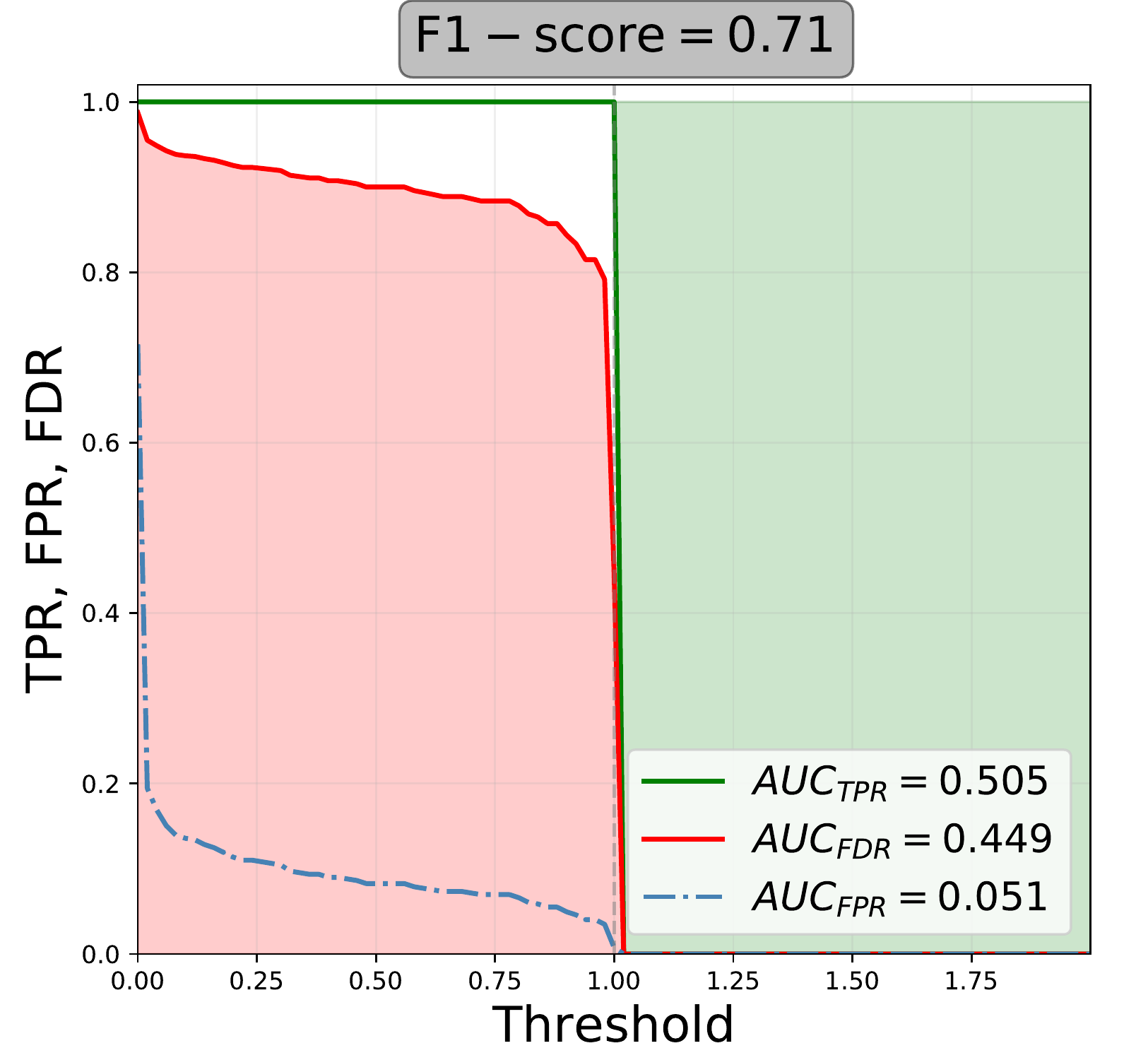}\includegraphics{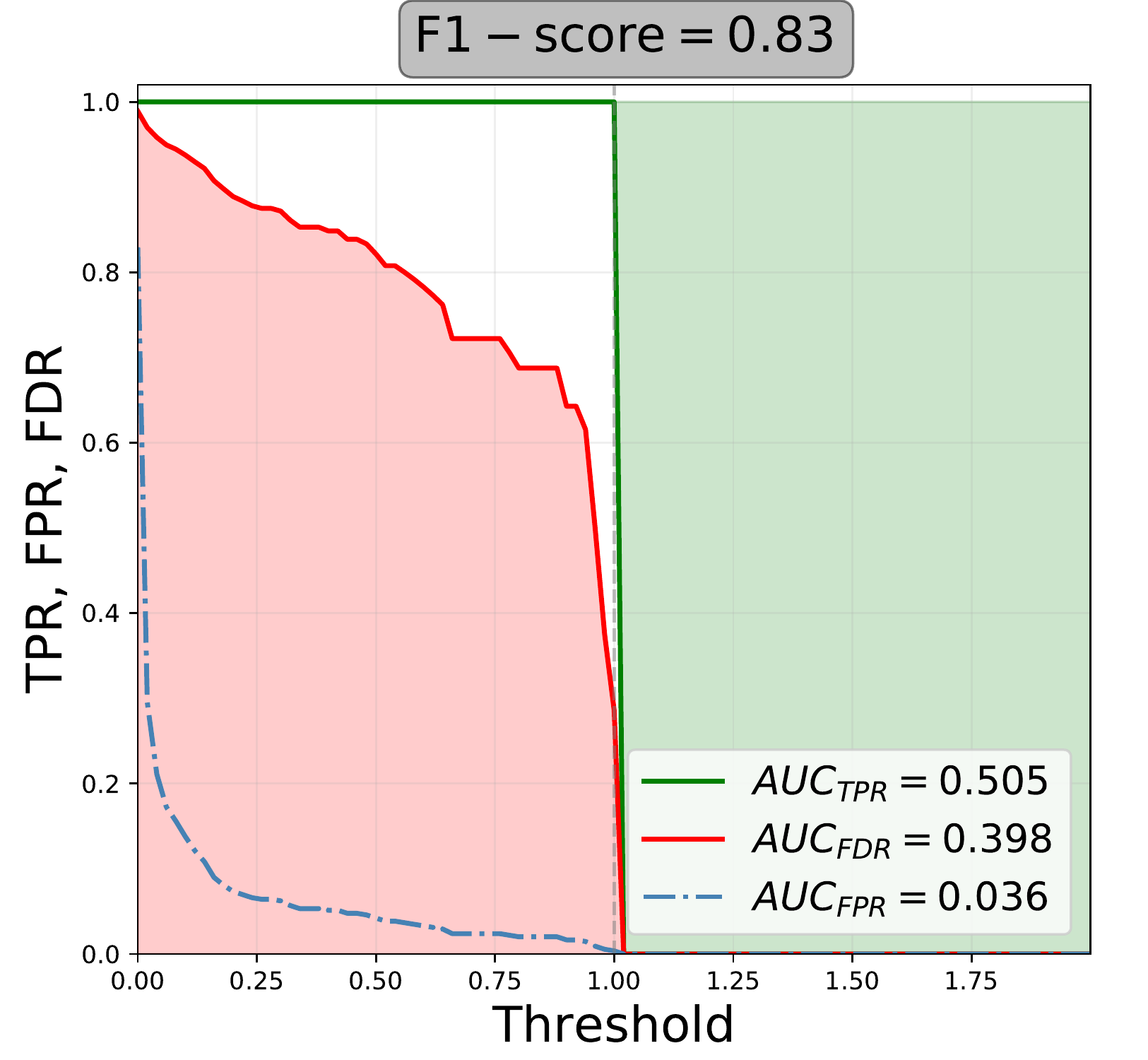}}
    \caption{Results of the ADI subchallenge for the sph3 VLT/SPHERE-IRDIS dataset: supervised machine learning.}
    \label{fig:img_sph3_sml}
\end{figure}

\begin{figure}
    \centering
    \resizebox{\hsize}{!}{\includegraphics[trim={0 1.1cm 0 0},clip]{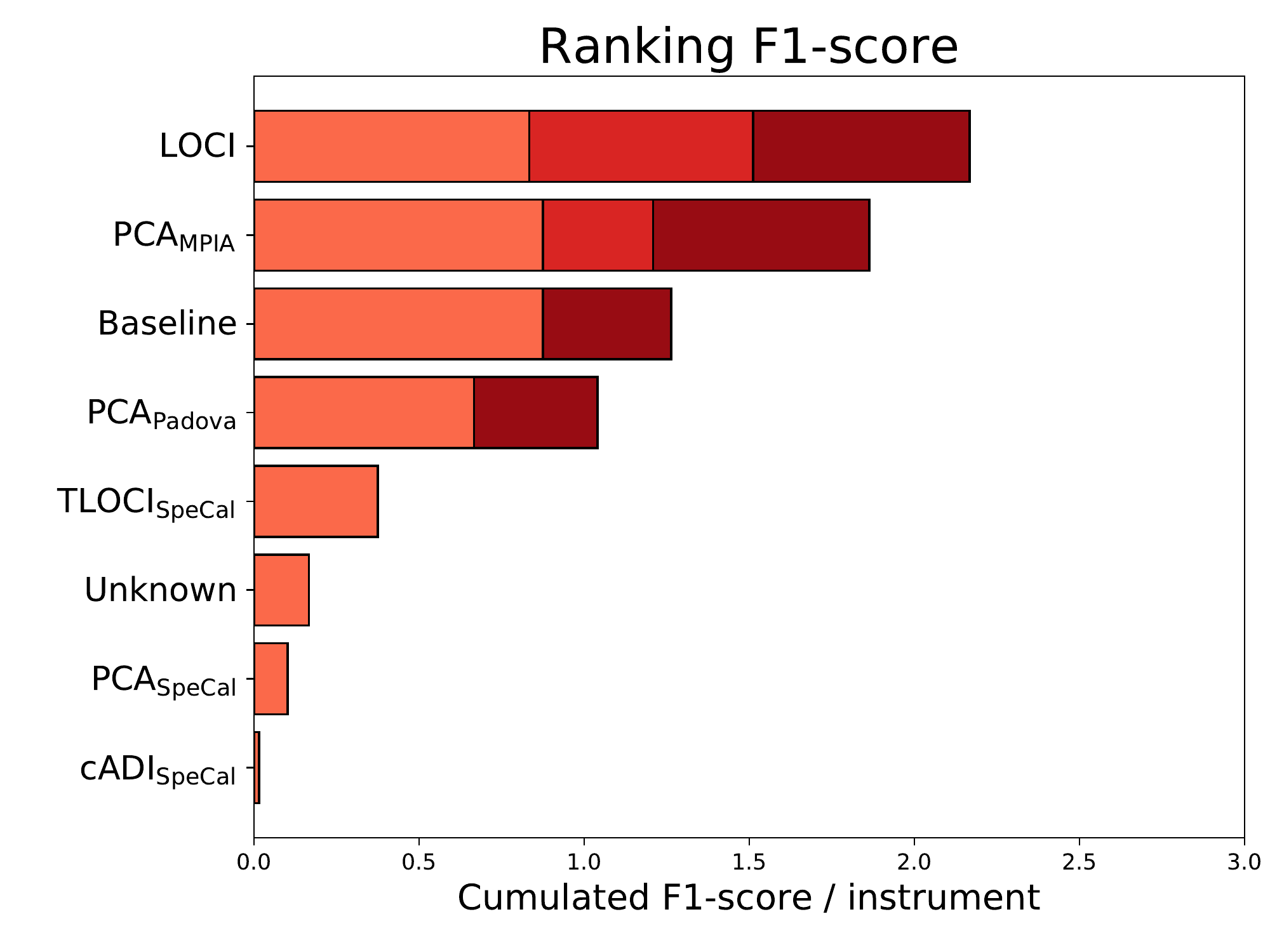}
                          \includegraphics[trim={0 1.1cm 0 0},clip]{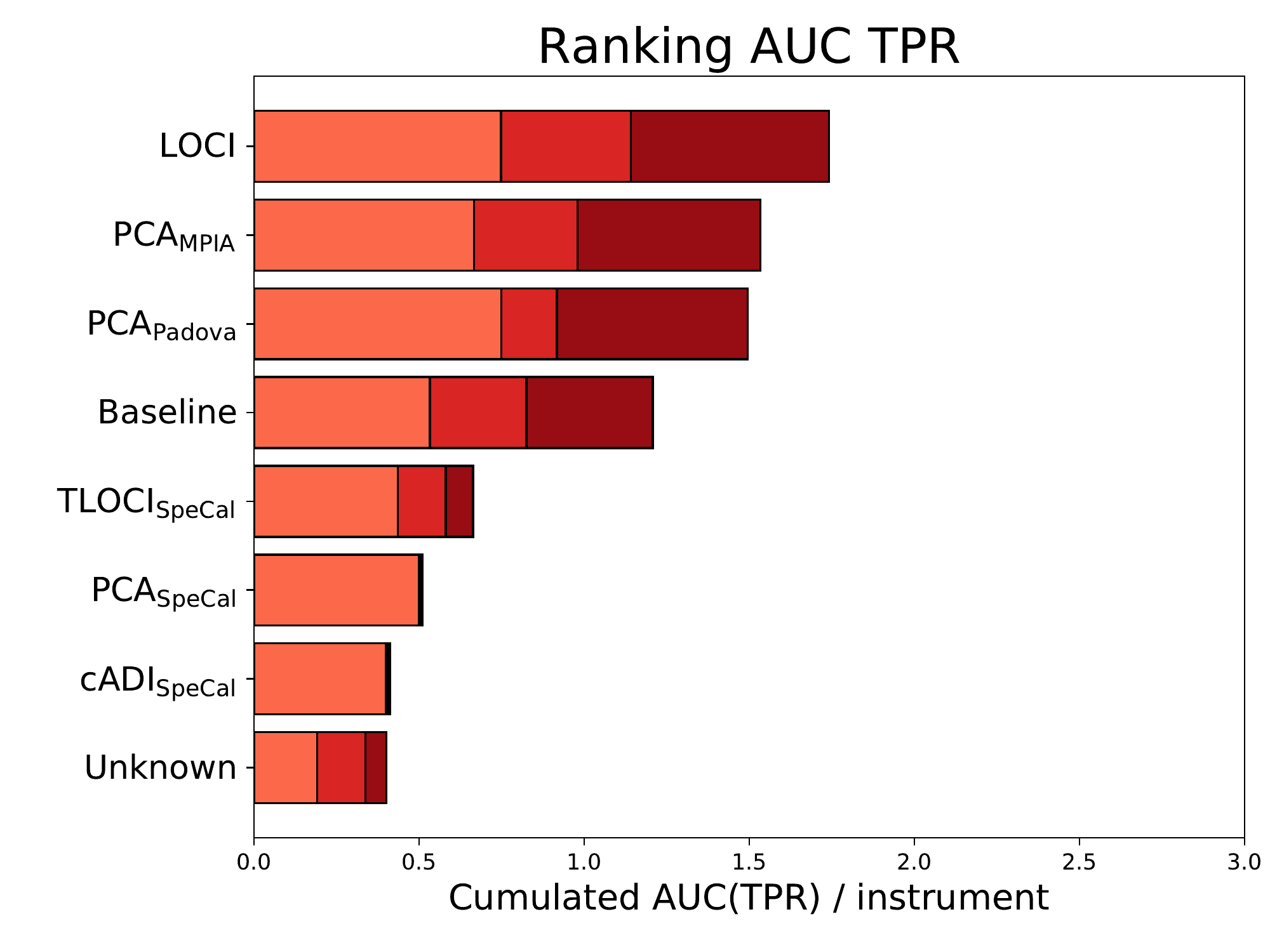}
                          \includegraphics[trim={0 1.1cm 0 0},clip]{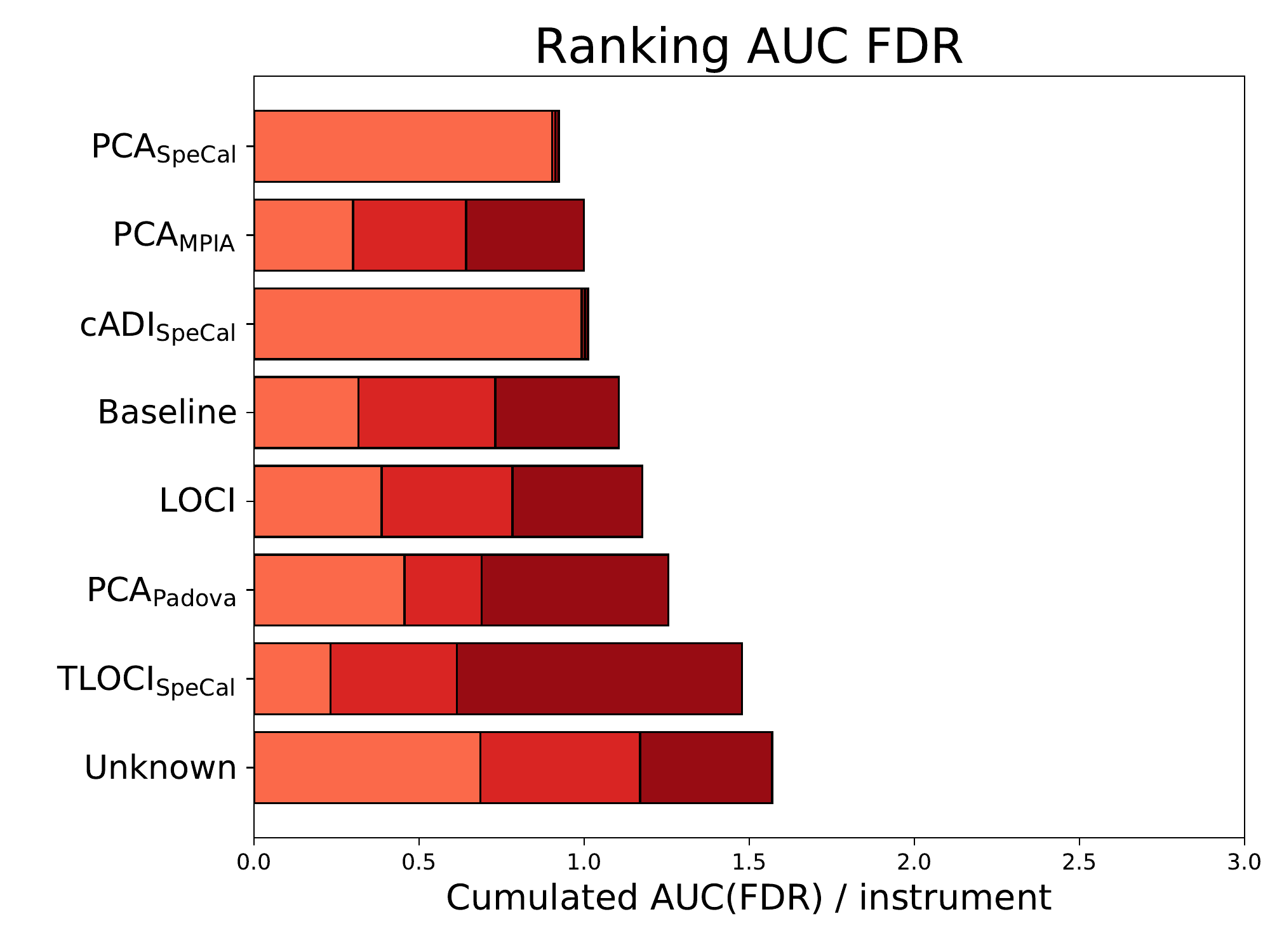}}
    \resizebox{\hsize}{!}{\includegraphics[trim={0 0.5cm 0 0.78cm},clip]{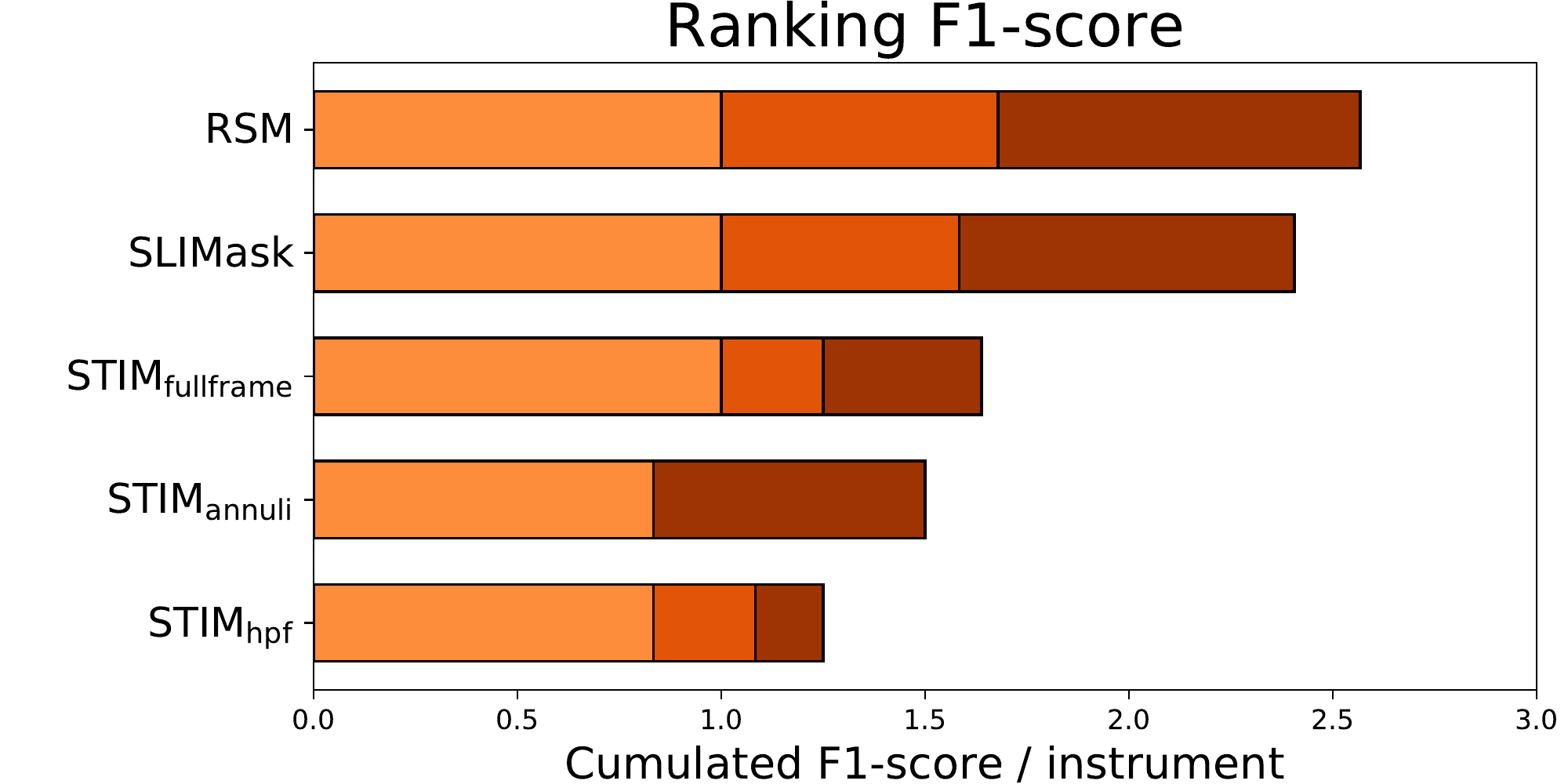}
                          \includegraphics[trim={0 0.5cm 0 0.78cm},clip]{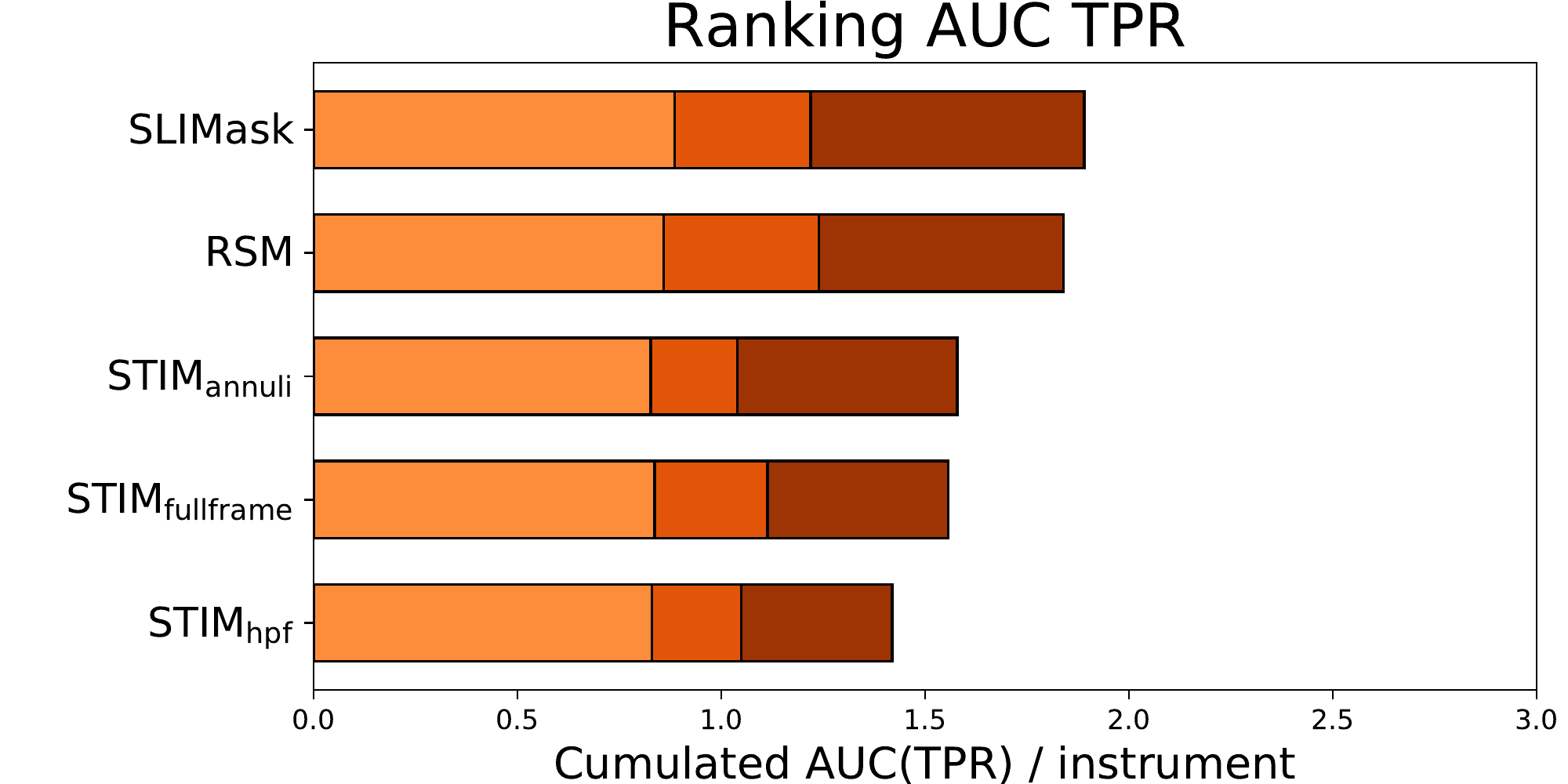}
                          \includegraphics[trim={0 0.5cm 0 0.78cm},clip]{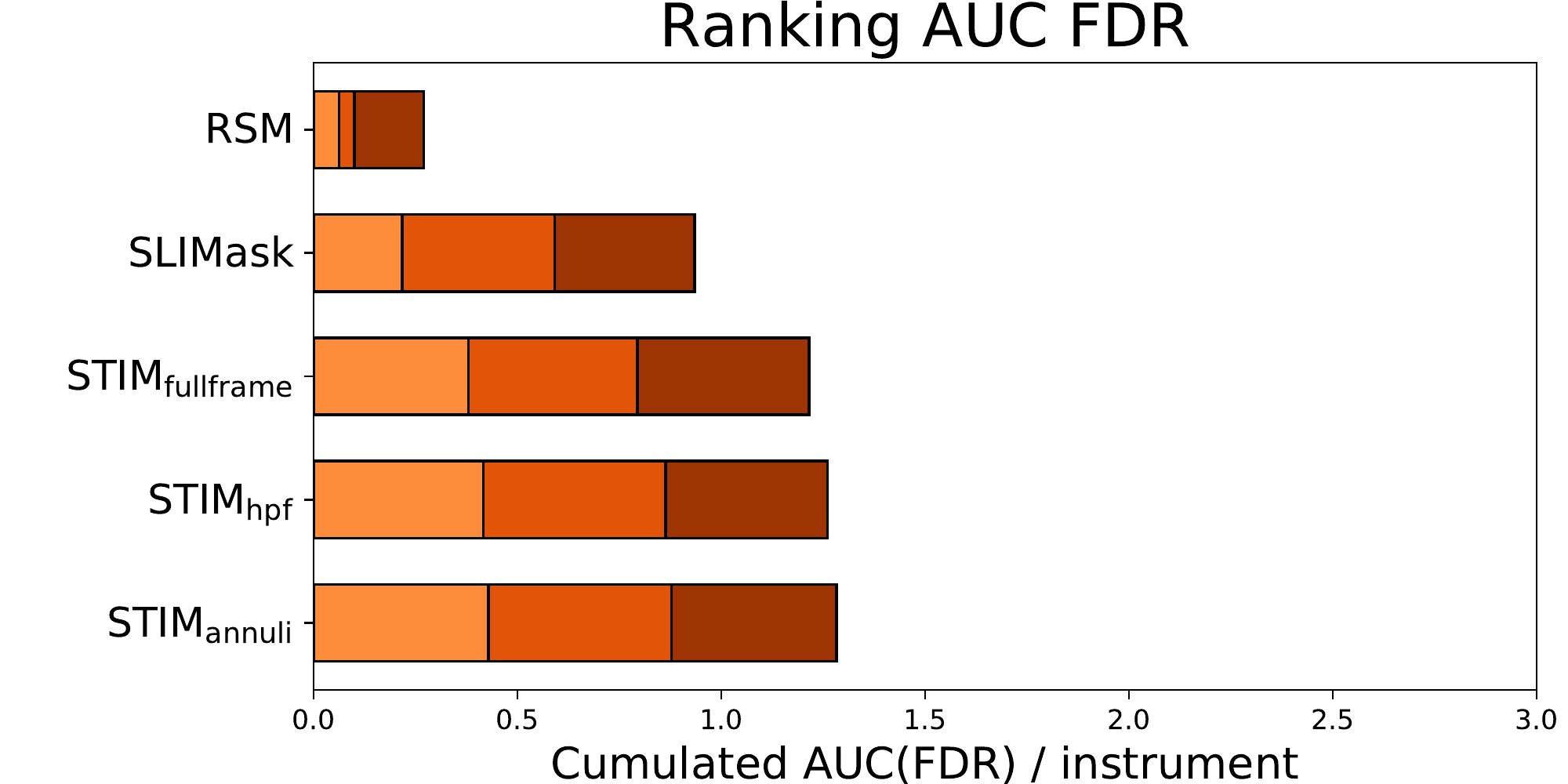}}
    \resizebox{\hsize}{!}{\includegraphics[trim={0 0.5cm 0 0.78cm},clip]{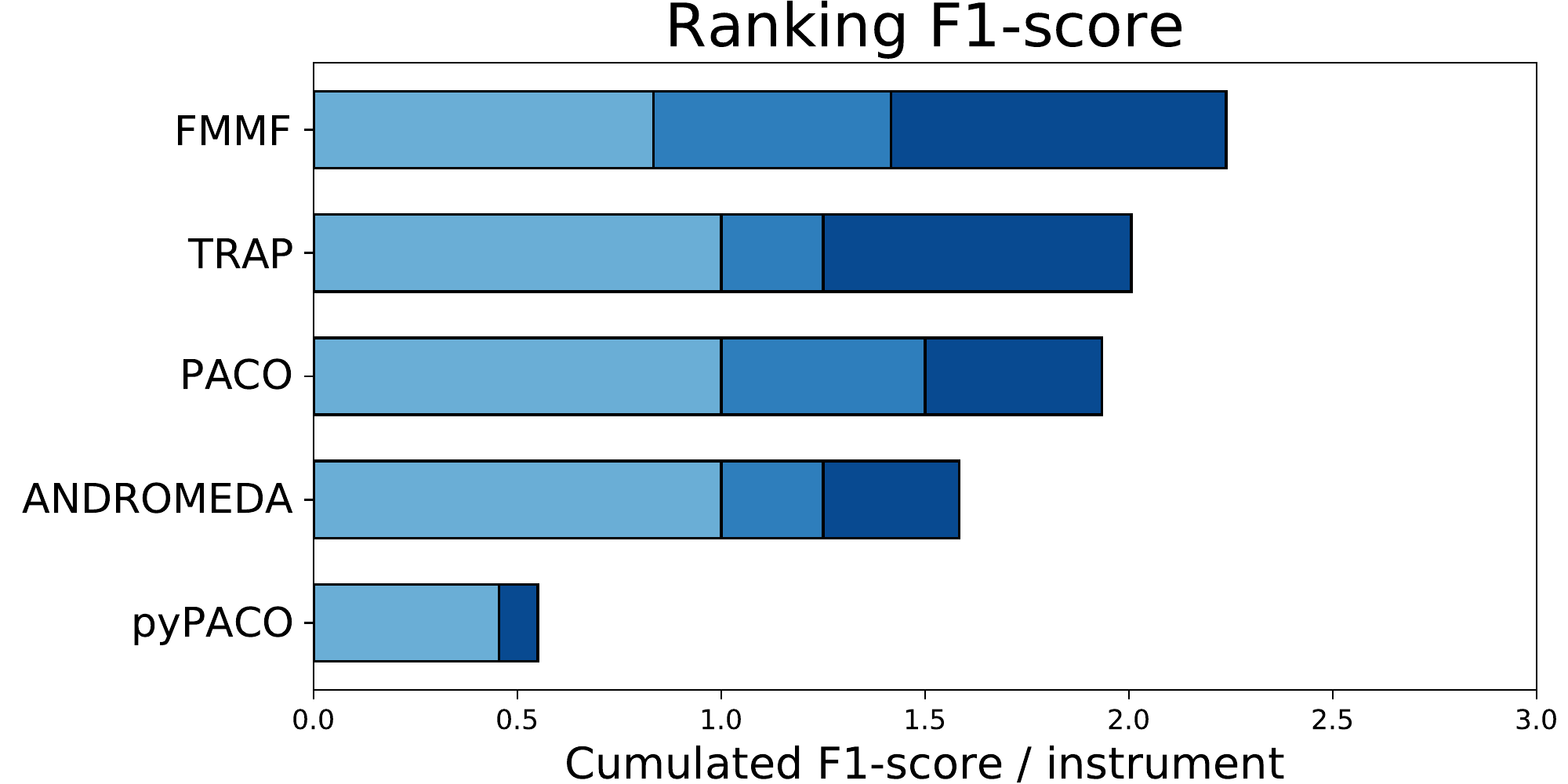}
                          \includegraphics[trim={0 0.5cm 0 0.78cm},clip]{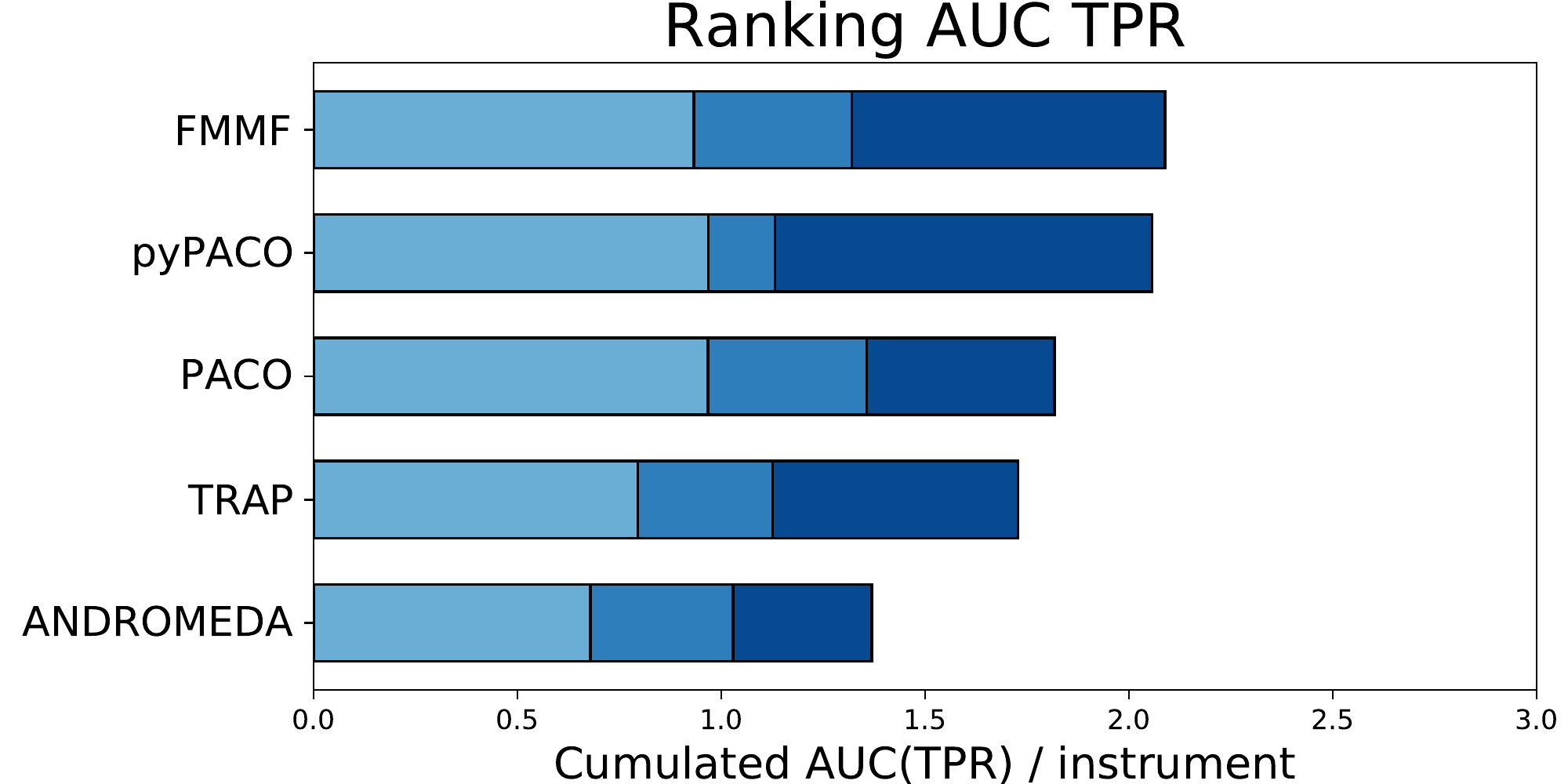}
                          \includegraphics[trim={0 0.5cm 0 0.78cm},clip]{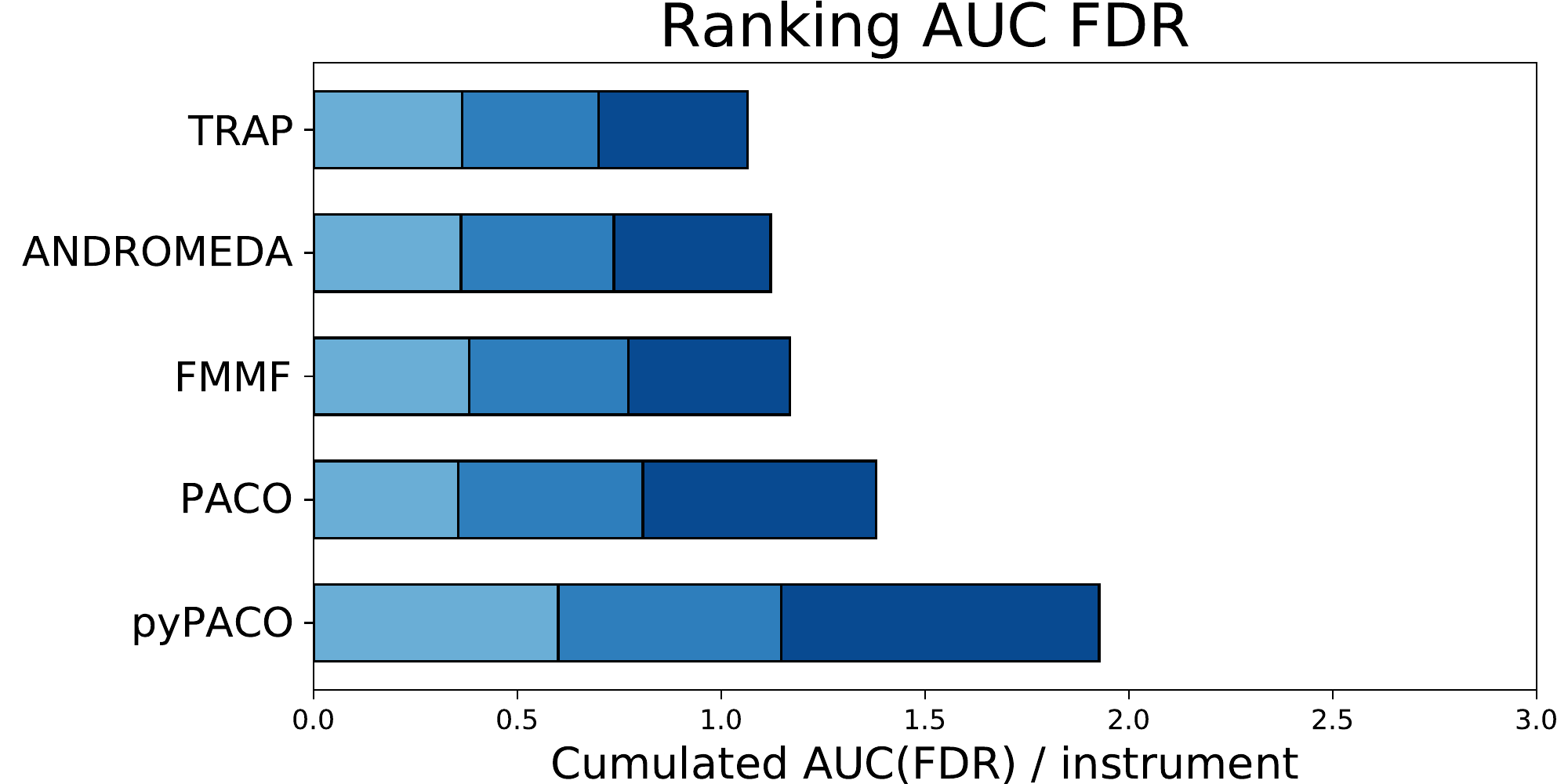}}
    \resizebox{\hsize}{!}{\includegraphics[trim={0 0 0 0.76cm},clip]{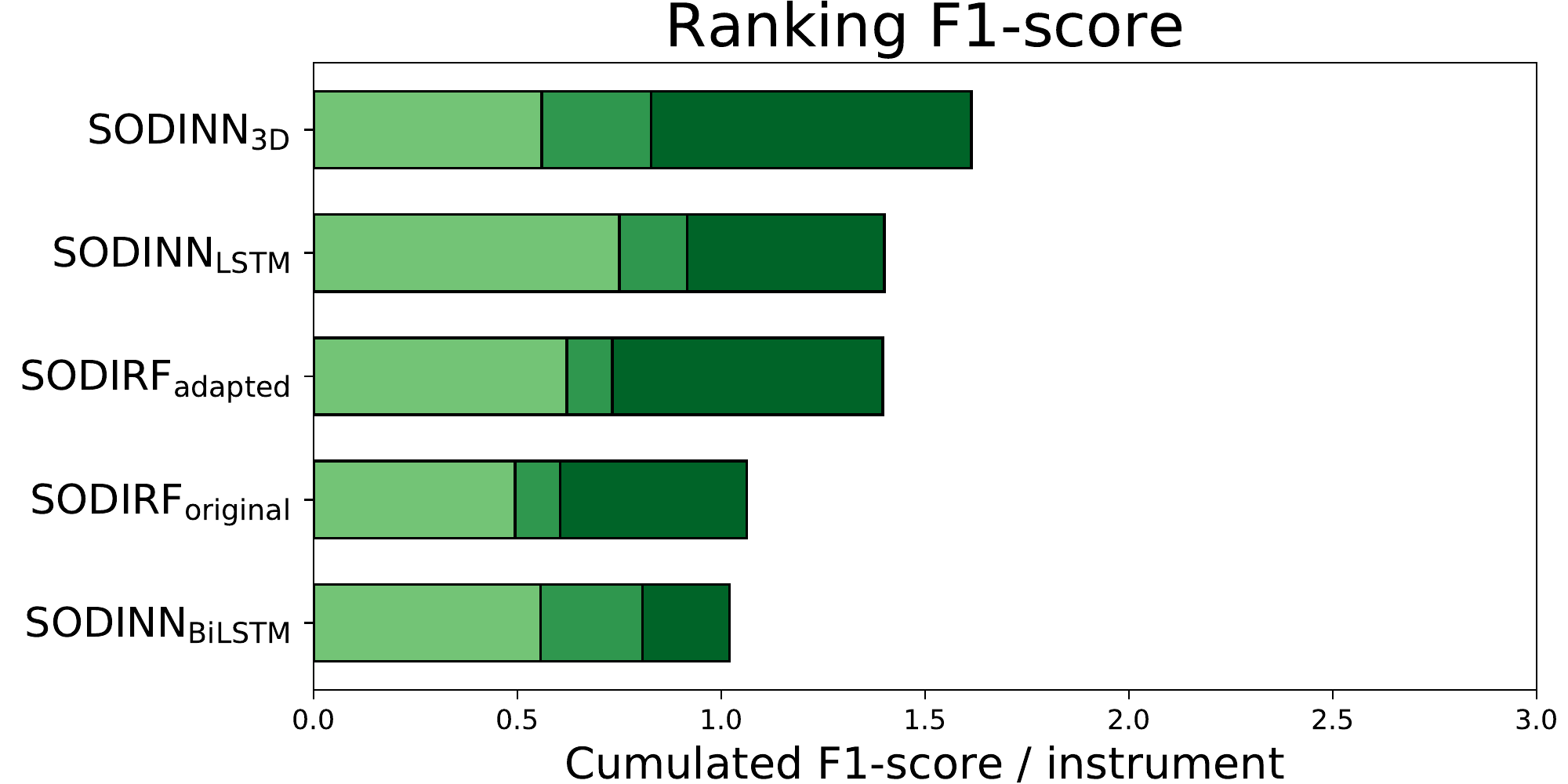}
                          \includegraphics[trim={0 0 0 0.76cm},clip]{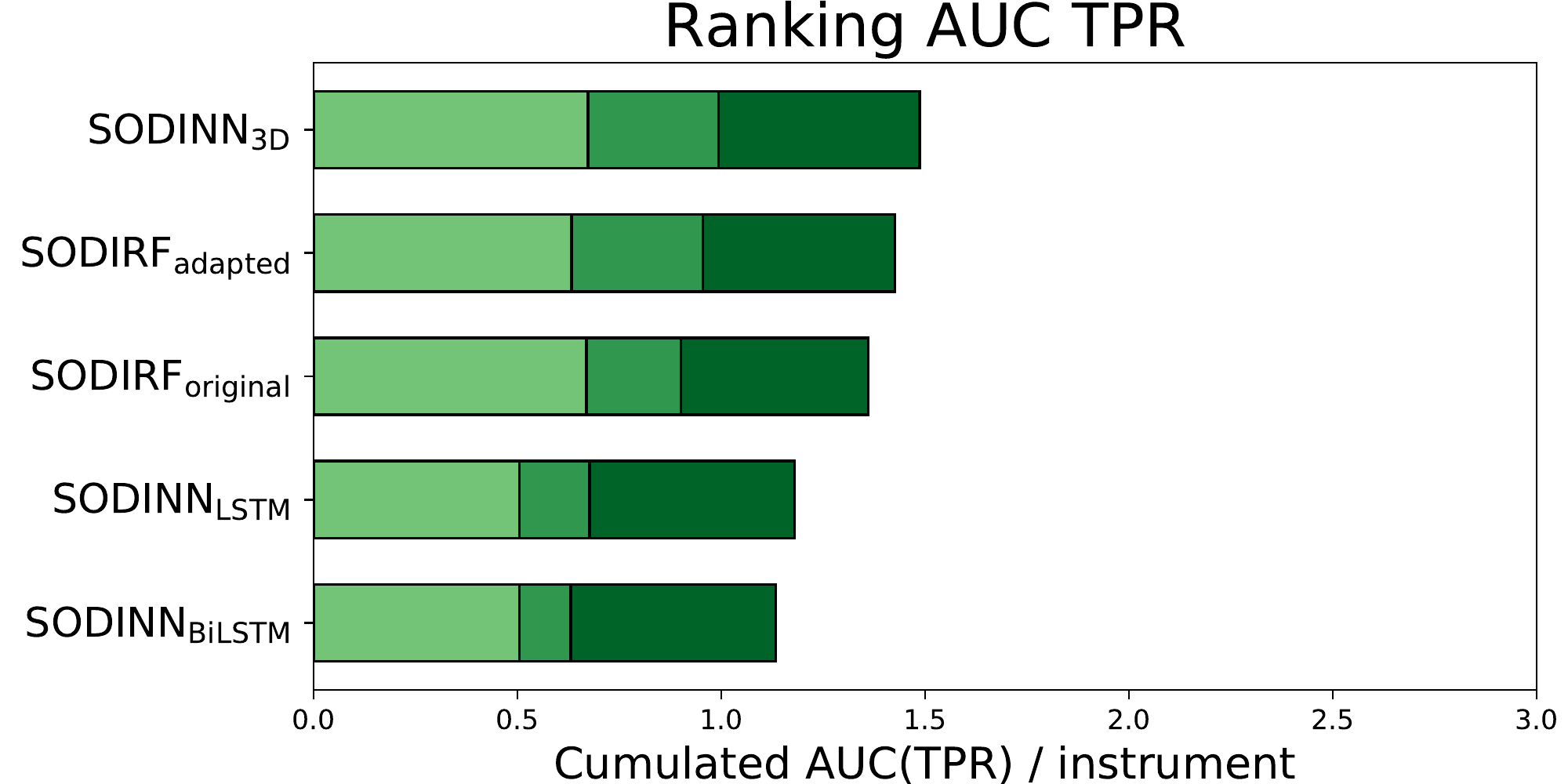}
                          \includegraphics[trim={0 0 0 0.76cm},clip]{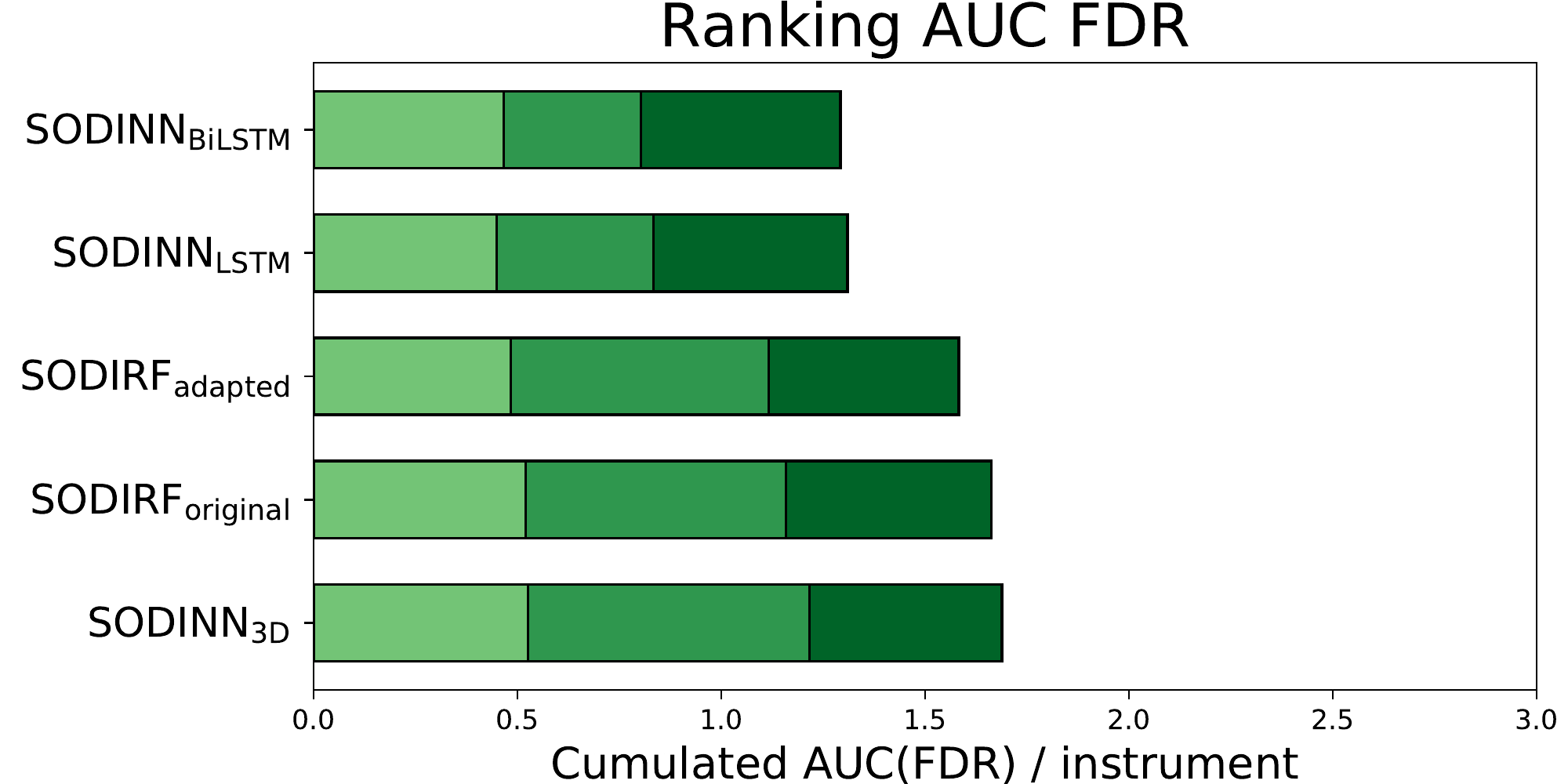}}                          
    \caption{Ranking of the different families of submitted algorithms for the EIDC ADI subchallenge. Left column: Ranking based on the F1-score. Middle column: Ranking based on the AUC of the TPR. Right column: Ranking based on the AUC of the FDR. From top to bottom, the 5 different families of algorithms: classical speckle subtraction providing residual maps (red), advanced speckle subtraction building detection maps (orange), inverse problems (blue) and supervised machine learning (green). The light, medium and dark colors correspond to the three VLT/SPHERE-IRDIS, Keck/NIRC2, and LBT/LMIRCam data sets respectively.}
    \label{fig:rnk_adi}
\end{figure}

The images and corresponding counts of detections and non-detections as a function of the threshold are displayed in Fig.~\ref{fig:img_sph3_sml} for the case of the third VLT/SPHERE-IRDIS data set (sph3). The scores proposed in Sect.~\ref{sec:metrics}, computed on the 9 data sets from this ADI subchallenge, are gathered in Tab.~\ref{tab:res_adi_sml}. The ranking of the different methods based on supervised machine learning is shown in Fig.~\ref{fig:rnk_adi} (green).

\begin{table}
\caption[example] 
   {\label{tab:resadiss} 
Results from the first subchallenge (ADI) of the EIDC: speckle subtraction based techniques. 
For each algorithm and each data set, the F1-score computed at the submitted threshold is in black (1st line), the AUC of the TPR is in green (2nd line) and the AUC of the FDR is in red (3rd line). When there are no injected planetary signals, the scores cannot be computed and are therefore undefined (indicated with the letter \emph{u}). The last column is the mean of the scores for the seven data sets that include injected planetary signal(s).
}
\begin{center}
\begin{tabular}{|l|| c | c | c | c || c | c | c | c || c | c | c | c || c |} 
\hline
\multirow{2}{*}{\textbf{Algorithm}} & \multicolumn{4}{c||}{VLT/SPHERE-IRDIS} & \multicolumn{4}{c||}{Keck/NIRC2} & \multicolumn{4}{c||}{LBT/LMIRCam} & \multirow{2}{*}{\textbf{All}} \\ 
\cline{2-13}
& (1) & (2) & (3) & Mean & (1) & (2) & (3) & Mean & (1) & (2) & (3) & Mean & \\
\hline\hline
\multirow{3}{*}{$\mathrm{Baseline}$} & 1.00 & \emph{u} & 0.75 & 0.88 & 0.00 & 0.00 & \emph{u} & 0.00 & 0.67 & 0.50 & 0.00 & 0.39 & 0.42\\
 & \tpr{0.52} & \tpr{ \emph{u}} & \tpr{ 0.54} & \tpr{ 0.53} & \tpr{ 0.35} & \tpr{ 0.24} & \tpr{ \emph{u}} & \tpr{ 0.29} & \tpr{ 0.50} & \tpr{ 0.49} & \tpr{ 0.17} & \tpr{ 0.38} & \tpr{ 0.40}\\
 & \fdr{0.42} & \fdr{ \emph{u}} & \fdr{ 0.21} & \fdr{ 0.32} & \fdr{ 0.29} & \fdr{ 0.54} & \fdr{ \emph{u}} & \fdr{ 0.41} & \fdr{ 0.32} & \fdr{ 0.49} & \fdr{ 0.31} & \fdr{ 0.37} & \fdr{ 0.37}\\
\hline\hline
\multirow{3}{*}{$\mathrm{cADI_{SpeCal}}$} & 0.00 & \emph{u} & 0.03 & 0.02 & 0.00 & 0.00 & \emph{u} & 0.00 & 0.00 & 0.00 & 0.00 & 0.00 & 0.01\\
 & \tpr{0.00} & \tpr{ \emph{u}} & \tpr{ 0.80} & \tpr{ 0.40} & \tpr{ 0.01} & \tpr{ 0.01} & \tpr{ \emph{u}} & \tpr{ 0.01} & \tpr{ 0.00} & \tpr{ 0.00} & \tpr{ 0.01} & \tpr{ 0.01} & \tpr{ 0.14}\\
 & \fdr{1.00} & \fdr{ \emph{u}} & \fdr{ 0.98} & \fdr{ 0.99} & \fdr{ 0.01} & \fdr{ 0.01} & \fdr{ \emph{u}} & \fdr{ 0.01} & \fdr{ 0.01} & \fdr{ 0.01} & \fdr{ 0.01} & \fdr{ 0.01} & \fdr{ 0.34}\\
\hline
\multirow{3}{*}{$\mathrm{LOCI}$} & 1.00 & \emph{u} & 0.67 & 0.83 & 0.86 & 0.50 & \emph{u} & 0.68 & 0.67 & 0.50 & 0.80 & 0.66 & 0.72\\
 & \tpr{0.74} & \tpr{ \emph{u}} & \tpr{ 0.75} & \tpr{ 0.75} & \tpr{ 0.48} & \tpr{ 0.30} & \tpr{ \emph{u}} & \tpr{ 0.39} & \tpr{ 0.65} & \tpr{ 0.55} & \tpr{ 0.60} & \tpr{ 0.60} & \tpr{ 0.58}\\
 & \fdr{0.42} & \fdr{ \emph{u}} & \fdr{ 0.35} & \fdr{ 0.39} & \fdr{ 0.36} & \fdr{ 0.43} & \fdr{ \emph{u}} & \fdr{ 0.40} & \fdr{ 0.39} & \fdr{ 0.43} & \fdr{ 0.36} & \fdr{ 0.39} & \fdr{ 0.39}\\
\hline
\multirow{3}{*}{$\mathrm{TLOCI_{SpeCal}}$} & 0.00 & \emph{u} & 0.75 & 0.38 & 0.00 & 0.00 & \emph{u} & 0.00 & 0.00 & 0.00 & 0.00 & 0.00 & 0.12\\
 & \tpr{0.39} & \tpr{ \emph{u}} & \tpr{ 0.49} & \tpr{ 0.44} & \tpr{ 0.21} & \tpr{ 0.08} & \tpr{ \emph{u}} & \tpr{ 0.15} & \tpr{ 0.00} & \tpr{ 0.25} & \tpr{ 0.00} & \tpr{ 0.08} & \tpr{ 0.22}\\
 & \fdr{0.27} & \fdr{ \emph{u}} & \fdr{ 0.20} & \fdr{ 0.23} & \fdr{ 0.37} & \fdr{ 0.39} & \fdr{ \emph{u}} & \fdr{ 0.38} & \fdr{ 0.82} & \fdr{ 0.98} & \fdr{ 0.79} & \fdr{ 0.86} & \fdr{ 0.49}\\
\hline
\multirow{3}{*}{$\mathrm{PCA_{SpeCal}}$} & 0.00 & \emph{u} & 0.20 & 0.10 & 0.00 & 0.00 & \emph{u} & 0.00 & 0.00 & 0.00 & 0.00 & 0.00 & 0.03\\
 & \tpr{0.00} & \tpr{ \emph{u}} & \tpr{ 1.00} & \tpr{ 0.50} & \tpr{ 0.01} & \tpr{ 0.01} & \tpr{ \emph{u}} & \tpr{ 0.01} & \tpr{ 0.00} & \tpr{ 0.00} & \tpr{ 0.01} & \tpr{ 0.00} & \tpr{ 0.17}\\
 & \fdr{1.00} & \fdr{ \emph{u}} & \fdr{ 0.81} & \fdr{ 0.90} & \fdr{ 0.01} & \fdr{ 0.01} & \fdr{ \emph{u}} & \fdr{ 0.01} & \fdr{ 0.01} & \fdr{ 0.01} & \fdr{ 0.01} & \fdr{ 0.01} & \fdr{ 0.31}\\
\hline
\multirow{3}{*}{$\mathrm{PCA_{MPIA}}$} & 1.00 & \emph{u} & 0.75 & 0.88 & 0.67 & 0.00 & \emph{u} & 0.33 & 0.67 & 0.80 & 0.50 & 0.66 & 0.62\\
 & \tpr{0.68} & \tpr{ \emph{u}} & \tpr{ 0.65} & \tpr{ 0.67} & \tpr{ 0.41} & \tpr{ 0.22} & \tpr{ \emph{u}} & \tpr{ 0.31} & \tpr{ 0.50} & \tpr{ 0.71} & \tpr{ 0.45} & \tpr{ 0.55} & \tpr{ 0.51}\\
 & \fdr{0.31} & \fdr{ \emph{u}} & \fdr{ 0.29} & \fdr{ 0.30} & \fdr{ 0.32} & \fdr{ 0.37} & \fdr{ \emph{u}} & \fdr{ 0.34} & \fdr{ 0.33} & \fdr{ 0.42} & \fdr{ 0.32} & \fdr{ 0.36} & \fdr{ 0.33}\\
\hline
\multirow{3}{*}{$\mathrm{PCA_{Padova}}$} & 0.50 & \emph{u} & 0.83 & 0.67 & 0.00 & 0.00 & \emph{u} & 0.00 & 0.40 & 0.50 & 0.22 & 0.37 & 0.35\\
 & \tpr{0.66} & \tpr{ \emph{u}} & \tpr{ 0.84} & \tpr{ 0.75} & \tpr{ 0.16} & \tpr{ 0.17} & \tpr{ \emph{u}} & \tpr{ 0.17} & \tpr{ 0.66} & \tpr{ 0.53} & \tpr{ 0.54} & \tpr{ 0.58} & \tpr{ 0.50}\\
 & \fdr{0.48} & \fdr{ \emph{u}} & \fdr{ 0.43} & \fdr{ 0.46} & \fdr{ 0.22} & \fdr{ 0.25} & \fdr{ \emph{u}} & \fdr{ 0.23} & \fdr{ 0.64} & \fdr{ 0.50} & \fdr{ 0.56} & \fdr{ 0.57} & \fdr{ 0.42}\\
\hline
\multirow{3}{*}{$\mathrm{Unknown}$} & 0.00 & \emph{u} & 0.33 & 0.17 & 0.00 & 0.00 & \emph{u} & 0.00 & 0.00 & 0.00 & 0.00 & 0.00 & 0.06\\
 & \tpr{0.00} & \tpr{ \emph{u}} & \tpr{ 0.38} & \tpr{ 0.19} & \tpr{ 0.15} & \tpr{ 0.14} & \tpr{ \emph{u}} & \tpr{ 0.15} & \tpr{ 0.00} & \tpr{ 0.12} & \tpr{ 0.07} & \tpr{ 0.06} & \tpr{ 0.13}\\
 & \fdr{1.00} & \fdr{ \emph{u}} & \fdr{ 0.37} & \fdr{ 0.69} & \fdr{ 0.60} & \fdr{ 0.37} & \fdr{ \emph{u}} & \fdr{ 0.48} & \fdr{ 0.39} & \fdr{ 0.38} & \fdr{ 0.43} & \fdr{ 0.40} & \fdr{ 0.52}\\
\hline\hline
\multirow{3}{*}{$\mathrm{STIM_{fullframe}}$} & 1.00 & \emph{u} & 1.00 & 1.00 & 0.00 & 0.50 & \emph{u} & 0.25 & 0.00 & 0.67 & 0.50 & 0.39 & 0.55\\
 & \tpr{0.79} & \tpr{ \emph{u}} & \tpr{ 0.88} & \tpr{ 0.84} & \tpr{ 0.36} & \tpr{ 0.19} & \tpr{ \emph{u}} & \tpr{ 0.28} & \tpr{ 0.23} & \tpr{ 0.48} & \tpr{ 0.62} & \tpr{ 0.44} & \tpr{ 0.52}\\
 & \fdr{0.40} & \fdr{ \emph{u}} & \fdr{ 0.36} & \fdr{ 0.38} & \fdr{ 0.44} & \fdr{ 0.39} & \fdr{ \emph{u}} & \fdr{ 0.41} & \fdr{ 0.45} & \fdr{ 0.42} & \fdr{ 0.40} & \fdr{ 0.42} & \fdr{ 0.41}\\
\hline
\multirow{3}{*}{$\mathrm{STIM_{annuli}}$} & 0.67 & \emph{u} & 1.00 & 0.83 & 0.00 & 0.00 & \emph{u} & 0.00 & 0.67 & 0.67 & 0.67 & 0.67 & 0.50\\
 & \tpr{0.84} & \tpr{ \emph{u}} & \tpr{ 0.81} & \tpr{ 0.83} & \tpr{ 0.21} & \tpr{ 0.22} & \tpr{ \emph{u}} & \tpr{ 0.21} & \tpr{ 0.26} & \tpr{ 0.68} & \tpr{ 0.67} & \tpr{ 0.54} & \tpr{ 0.53}\\
 & \fdr{0.49} & \fdr{ \emph{u}} & \fdr{ 0.36} & \fdr{ 0.43} & \fdr{ 0.52} & \fdr{ 0.38} & \fdr{ \emph{u}} & \fdr{ 0.45} & \fdr{ 0.37} & \fdr{ 0.42} & \fdr{ 0.43} & \fdr{ 0.40} & \fdr{ 0.43}\\
\hline
\multirow{3}{*}{$\mathrm{STIM_{hpf}}$} & 0.67 & \emph{u} & 1.00 & 0.83 & 0.00 & 0.50 & \emph{u} & 0.25 & 0.00 & 0.00 & 0.50 & 0.17 & 0.42\\
 & \tpr{0.81} & \tpr{ \emph{u}} & \tpr{ 0.85} & \tpr{ 0.83} & \tpr{ 0.23} & \tpr{ 0.21} & \tpr{ \emph{u}} & \tpr{ 0.22} & \tpr{ 0.33} & \tpr{ 0.35} & \tpr{ 0.43} & \tpr{ 0.37} & \tpr{ 0.47}\\
 & \fdr{0.48} & \fdr{ \emph{u}} & \fdr{ 0.35} & \fdr{ 0.42} & \fdr{ 0.49} & \fdr{ 0.41} & \fdr{ \emph{u}} & \fdr{ 0.45} & \fdr{ 0.39} & \fdr{ 0.50} & \fdr{ 0.30} & \fdr{ 0.40} & \fdr{ 0.42}\\
\hline
\multirow{3}{*}{$\mathrm{SLIMask}$} & 1.00 & \emph{u} & 1.00 & 1.00 & 0.67 & 0.50 & \emph{u} & 0.58 & 0.67 & 0.80 & 1.00 & 0.82 & 0.80\\
 & \tpr{0.88} & \tpr{ \emph{u}} & \tpr{ 0.89} & \tpr{ 0.89} & \tpr{ 0.39} & \tpr{ 0.27} & \tpr{ \emph{u}} & \tpr{ 0.33} & \tpr{ 0.67} & \tpr{ 0.64} & \tpr{ 0.71} & \tpr{ 0.67} & \tpr{ 0.63}\\
 & \fdr{0.37} & \fdr{ \emph{u}} & \fdr{ 0.07} & \fdr{ 0.22} & \fdr{ 0.34} & \fdr{ 0.40} & \fdr{ \emph{u}} & \fdr{ 0.37} & \fdr{ 0.36} & \fdr{ 0.36} & \fdr{ 0.31} & \fdr{ 0.34} & \fdr{ 0.31}\\
\hline
\multirow{3}{*}{$\mathrm{RSM}$} & 1.00 & \emph{u} & 1.00 & 1.00 & 0.86 & 0.50 & \emph{u} & 0.68 & 0.67 & 1.00 & 1.00 & 0.89 & 0.86\\
 & \tpr{0.84} & \tpr{ \emph{u}} & \tpr{ 0.88} & \tpr{ 0.86} & \tpr{ 0.55} & \tpr{ 0.21} & \tpr{ \emph{u}} & \tpr{ 0.38} & \tpr{ 0.45} & \tpr{ 0.62} & \tpr{ 0.73} & \tpr{ 0.60} & \tpr{ 0.61}\\
 & \fdr{0.09} & \fdr{ \emph{u}} & \fdr{ 0.04} & \fdr{ 0.06} & \fdr{ 0.06} & \fdr{ 0.02} & \fdr{ \emph{u}} & \fdr{ 0.04} & \fdr{ 0.14} & \fdr{ 0.29} & \fdr{ 0.08} & \fdr{ 0.17} & \fdr{ 0.09}\\
\hline
 \end{tabular}
\end{center}
\end{table}

\begin{table}
\caption[example] 
   {\label{tab:resadimap} 
Results from the first subchallenge (ADI) of the EIDC: inverse problem based techniques.
}
\begin{center}
\begin{tabular}{|l|| c | c | c | c || c | c | c | c || c | c | c | c || c |} 
\hline
\multirow{2}{*}{\textbf{Algorithm}} & \multicolumn{4}{c||}{VLT/SPHERE-IRDIS} & \multicolumn{4}{c||}{Keck/NIRC2} & \multicolumn{4}{c||}{LBT/LMIRCam} & \multirow{2}{*}{\textbf{All}} \\ 
\cline{2-13}
& (1) & (2) & (3) & Mean & (1) & (2) & (3) & Mean & (1) & (2) & (3) & Mean & \\
\hline\hline
\multirow{3}{*}{$\mathrm{ANDROMEDA}$} & 1.00 & \emph{u} & 1.00 & 1.00 & 0.00 & 0.50 & \emph{u} & 0.25 & 0.00 & 1.00 & 0.00 & 0.33 & 0.53\\
 & \tpr{0.53} & \tpr{ \emph{u}} & \tpr{ 0.82} & \tpr{ 0.68} & \tpr{ 0.30} & \tpr{ 0.40} & \tpr{ \emph{u}} & \tpr{ 0.35} & \tpr{ 0.18} & \tpr{ 0.63} & \tpr{ 0.21} & \tpr{ 0.34} & \tpr{ 0.46}\\
 & \fdr{0.38} & \fdr{ \emph{u}} & \fdr{ 0.35} & \fdr{ 0.36} & \fdr{ 0.36} & \fdr{ 0.39} & \fdr{ \emph{u}} & \fdr{ 0.38} & \fdr{ 0.38} & \fdr{ 0.39} & \fdr{ 0.38} & \fdr{ 0.38} & \fdr{ 0.37}\\
\hline
\multirow{3}{*}{$\mathrm{FMMF}$} & 0.67 & \emph{u} & 1.00 & 0.83 & 0.67 & 0.50 & \emph{u} & 0.58 & 0.67 & 0.80 & 1.00 & 0.82 & 0.75\\
 & \tpr{1.00} & \tpr{ \emph{u}} & \tpr{ 0.87} & \tpr{ 0.93} & \tpr{ 0.51} & \tpr{ 0.27} & \tpr{ \emph{u}} & \tpr{ 0.39} & \tpr{ 0.65} & \tpr{ 0.80} & \tpr{ 0.85} & \tpr{ 0.77} & \tpr{ 0.70}\\
 & \fdr{0.43} & \fdr{ \emph{u}} & \fdr{ 0.33} & \fdr{ 0.38} & \fdr{ 0.39} & \fdr{ 0.39} & \fdr{ \emph{u}} & \fdr{ 0.39} & \fdr{ 0.43} & \fdr{ 0.43} & \fdr{ 0.33} & \fdr{ 0.40} & \fdr{ 0.39}\\
\hline
\multirow{3}{*}{$\mathrm{PACO}$} & 1.00 & \emph{u} & 1.00 & 1.00 & 0.60 & 0.40 & \emph{u} & 0.50 & 0.67 & 0.57 & 0.06 & 0.43 & 0.64\\
 & \tpr{1.00} & \tpr{ \emph{u}} & \tpr{ 0.93} & \tpr{ 0.97} & \tpr{ 0.44} & \tpr{ 0.34} & \tpr{ \emph{u}} & \tpr{ 0.39} & \tpr{ 0.29} & \tpr{ 0.58} & \tpr{ 0.51} & \tpr{ 0.46} & \tpr{ 0.61}\\
 & \fdr{0.39} & \fdr{ \emph{u}} & \fdr{ 0.32} & \fdr{ 0.36} & \fdr{ 0.43} & \fdr{ 0.48} & \fdr{ \emph{u}} & \fdr{ 0.45} & \fdr{ 0.30} & \fdr{ 0.51} & \fdr{ 0.90} & \fdr{ 0.57} & \fdr{ 0.46}\\
\hline
\multirow{3}{*}{$\mathrm{pyPACO}$} & 0.08 & \emph{u} & 0.83 & 0.46 & 0.00 & 0.00 & \emph{u} & 0.00 & 0.07 & 0.09 & 0.12 & 0.09 & 0.18\\
 & \tpr{1.00} & \tpr{ \emph{u}} & \tpr{ 0.94} & \tpr{ 0.97} & \tpr{ 0.23} & \tpr{ 0.10} & \tpr{ \emph{u}} & \tpr{ 0.16} & \tpr{ 0.78} & \tpr{ 1.00} & \tpr{ 1.00} & \tpr{ 0.93} & \tpr{ 0.69}\\
 & \fdr{0.70} & \fdr{ \emph{u}} & \fdr{ 0.50} & \fdr{ 0.60} & \fdr{ 0.40} & \fdr{ 0.69} & \fdr{ \emph{u}} & \fdr{ 0.55} & \fdr{ 0.91} & \fdr{ 0.75} & \fdr{ 0.68} & \fdr{ 0.78} & \fdr{ 0.64}\\
\hline
\multirow{3}{*}{$\mathrm{TRAP}$} & 1.00 & \emph{u} & 1.00 & 1.00 & 0.00 & 0.50 & \emph{u} & 0.25 & 0.67 & 0.80 & 0.80 & 0.76 & 0.67\\
 & \tpr{0.68} & \tpr{ \emph{u}} & \tpr{ 0.91} & \tpr{ 0.80} & \tpr{ 0.33} & \tpr{ 0.33} & \tpr{ \emph{u}} & \tpr{ 0.33} & \tpr{ 0.50} & \tpr{ 0.70} & \tpr{ 0.61} & \tpr{ 0.60} & \tpr{ 0.58}\\
 & \fdr{0.37} & \fdr{ \emph{u}} & \fdr{ 0.36} & \fdr{ 0.36} & \fdr{ 0.32} & \fdr{ 0.35} & \fdr{ \emph{u}} & \fdr{ 0.33} & \fdr{ 0.32} & \fdr{ 0.45} & \fdr{ 0.33} & \fdr{ 0.37} & \fdr{ 0.35}\\
\hline
\end{tabular}
\end{center}
\end{table}

\begin{table}
\caption[example] 
   {\label{tab:res_adi_sml} 
Results from the first subchallenge (ADI) of the EIDC: supervised machine learning based techniques.}
\begin{center}
\begin{tabular}{|l|| c | c | c | c || c | c | c | c || c | c | c | c || c |} 
\hline
\multirow{2}{*}{\textbf{Algorithm}} & \multicolumn{4}{c||}{VLT/SPHERE-IRDIS} & \multicolumn{4}{c||}{Keck/NIRC2} & \multicolumn{4}{c||}{LBT/LMIRCam} & \multirow{2}{*}{\textbf{All}} \\ 
\cline{2-13}
& (1) & (2) & (3) & Mean & (1) & (2) & (3) & Mean & (1) & (2) & (3) & Mean & \\
\hline\hline
\multirow{3}{*}{$\mathrm{sodirf_{original}}$} & 0.40 & \emph{u} & 0.59 & 0.49 & 0.22 & 0.00 & \emph{u} & 0.11 & 0.18 & 0.33 & 0.86 & 0.46 & 0.35\\
 & \tpr{0.70} & \tpr{ \emph{u}} & \tpr{ 0.64} & \tpr{ 0.67} & \tpr{ 0.36} & \tpr{ 0.10} & \tpr{ \emph{u}} & \tpr{ 0.23} & \tpr{ 0.28} & \tpr{ 0.58} & \tpr{ 0.52} & \tpr{ 0.46} & \tpr{ 0.45}\\
 & \fdr{0.55} & \fdr{ \emph{u}} & \fdr{ 0.48} & \fdr{ 0.52} & \fdr{ 0.66} & \fdr{ 0.62} & \fdr{ \emph{u}} & \fdr{ 0.64} & \fdr{ 0.52} & \fdr{ 0.53} & \fdr{ 0.47} & \fdr{ 0.50} & \fdr{ 0.55}\\
\hline
\multirow{3}{*}{$\mathrm{sodirf_{adapted}}$} & 0.33 & \emph{u} & 0.91 & 0.62 & 0.22 & 0.00 & \emph{u} & 0.11 & 0.33 & 0.80 & 0.86 & 0.66 & 0.47\\
 & \tpr{0.71} & \tpr{ \emph{u}} & \tpr{ 0.55} & \tpr{ 0.63} & \tpr{ 0.42} & \tpr{ 0.22} & \tpr{ \emph{u}} & \tpr{ 0.32} & \tpr{ 0.30} & \tpr{ 0.57} & \tpr{ 0.54} & \tpr{ 0.47} & \tpr{ 0.48}\\
 & \fdr{0.56} & \fdr{ \emph{u}} & \fdr{ 0.41} & \fdr{ 0.48} & \fdr{ 0.64} & \fdr{ 0.62} & \fdr{ \emph{u}} & \fdr{ 0.63} & \fdr{ 0.51} & \fdr{ 0.44} & \fdr{ 0.45} & \fdr{ 0.47} & \fdr{ 0.53}\\
\hline
\multirow{3}{*}{$\mathrm{sodinn_{LSTM}}$} & 0.67 & \emph{u} & 0.83 & 0.75 & 0.33 & 0.00 & \emph{u} & 0.17 & 0.33 & 0.57 & 0.55 & 0.48 & 0.47\\
 & \tpr{0.50} & \tpr{ \emph{u}} & \tpr{ 0.50} & \tpr{ 0.50} & \tpr{ 0.19} & \tpr{ 0.16} & \tpr{ \emph{u}} & \tpr{ 0.17} & \tpr{ 0.50} & \tpr{ 0.50} & \tpr{ 0.50} & \tpr{ 0.50} & \tpr{ 0.39}\\
 & \fdr{0.47} & \fdr{ \emph{u}} & \fdr{ 0.43} & \fdr{ 0.45} & \fdr{ 0.43} & \fdr{ 0.34} & \fdr{ \emph{u}} & \fdr{ 0.38} & \fdr{ 0.47} & \fdr{ 0.47} & \fdr{ 0.48} & \fdr{ 0.48} & \fdr{ 0.44}\\
\hline
\multirow{3}{*}{$\mathrm{sodinn_{BiLSTM}}$} & 0.40 & \emph{u} & 0.71 & 0.56 & 0.00 & 0.50 & \emph{u} & 0.25 & 0.11 & 0.33 & 0.20 & 0.21 & 0.34\\
 & \tpr{0.50} & \tpr{ \emph{u}} & \tpr{ 0.50} & \tpr{ 0.50} & \tpr{ 0.08} & \tpr{ 0.17} & \tpr{ \emph{u}} & \tpr{ 0.13} & \tpr{ 0.50} & \tpr{ 0.50} & \tpr{ 0.50} & \tpr{ 0.50} & \tpr{ 0.38}\\
 & \fdr{0.48} & \fdr{ \emph{u}} & \fdr{ 0.45} & \fdr{ 0.47} & \fdr{ 0.45} & \fdr{ 0.22} & \fdr{ \emph{u}} & \fdr{ 0.34} & \fdr{ 0.49} & \fdr{ 0.49} & \fdr{ 0.49} & \fdr{ 0.49} & \fdr{ 0.43}\\
\hline
\multirow{3}{*}{$\mathrm{sodinn_{3D}}$} & 0.29 & \emph{u} & 0.83 & 0.56 & 0.25 & 0.29 & \emph{u} & 0.27 & 0.50 & 1.00 & 0.86 & 0.79 & 0.54\\
 & \tpr{0.84} & \tpr{ \emph{u}} & \tpr{ 0.50} & \tpr{ 0.67} & \tpr{ 0.29} & \tpr{ 0.35} & \tpr{ \emph{u}} & \tpr{ 0.32} & \tpr{ 0.47} & \tpr{ 0.50} & \tpr{ 0.50} & \tpr{ 0.49} & \tpr{ 0.50}\\
 & \fdr{0.65} & \fdr{ \emph{u}} & \fdr{ 0.40} & \fdr{ 0.53} & \fdr{ 0.70} & \fdr{ 0.68} & \fdr{ \emph{u}} & \fdr{ 0.69} & \fdr{ 0.48} & \fdr{ 0.46} & \fdr{ 0.48} & \fdr{ 0.47} & \fdr{ 0.56}\\
\hline
\end{tabular}
\end{center}
\end{table} 
     

\subsubsection{Ranking the ADI subchallenge results} 
Based on the three scores defined in Sect.~\ref{sec:metrics}, we compute the F1-score, the AUC of the TPR and AUC of the FDR for all data sets. The different rankings for these three metrics and the five families of methods described are shown in Fig.~\ref{fig:rnk_adi}. 
Based on the ADI subchallenge results, we can already draw a few conclusions. 

(1) The three rankings based on each of the three scores give consistent results: for a given method, a high F1-score indeed corresponds to a high AUC of the TPR and low AUC of the FDR. However, it is possible that the AUC of the FDR differs a lot from other scores. For instance when some algorithms are more robust to false alarms (high AUC of the FDR) but are not very sensitive to faint signals (low AUC of the TPR). 
Depending on the science case (such as deep search for a point source or homogeneous processing of a large sample of data) one may use the most relevant metric to choose some methods over the others. For further analysis, the results of the data challenge are available upon request to the first author, in agreement with the EIDC collaboration. 

(2) As expected, we observe that globally the speckle subtraction based techniques (in red in Fig.~\ref{fig:rnk_adi}) do not perform as good as more advanced techniques: the most recent techniques indeed show better detection capabilities than former techniques. 

(3) The exception goes to the supervised machine learning techniques (in green in Fig.~\ref{fig:rnk_adi}) for which we observe numerous false positives. Applying supervised machine learning techniques to the challenging field of high-contrast imaging is very recent and investigations are on-going to improve this type of methods. Another bias in the current ranking of these techniques stems from the type of detection map they deliver. The detection maps are a binary classifier, normalized to one (see detection maps Fig.~\ref{fig:img_sph3_sml}, top row) and the TPR is abruptly dropping to zero for thresholds greater than the submitted threshold. Therefore calculating the AUC of the TPF  for a range of threshold from 0 to twice the submitted threshold does not make sense and this specific score disfavor these methods. 

(4) Figure~\ref{fig:rnk_adi} also highlights that the performance are indeed dependent on the type of instrument. Indeed coronagraphic images (light and medium colors) do not show the same type of stellar light residuals as non-coronagraphic images (dark colors). In addition, the  wavelength and therefore the exposure time (number of frames) do play a role as some algorithms are working along the temporal axis. For instance, it is expected from Ref.~\cite{Samland2020} that TRAP will not perform as good on temporally binned data, which is the case for the Keck/NIRC2 data sets (medium color). 

(5) At first glance we can see that the latest algorithms, RSM and SLIMask for the speckle subtraction techniques, and FMMF, PACO and TRAP for the inverse problem approaches, are providing with the best performance along with the three scores used here. 
Note that in this phase of the EIDC we did not account for practical aspects such as the running time or the number of user-parameters to fine-tune. In future phases of the EIDC, we intend to take more factors into account to better guide users towards understanding better the performance of the various methods dedicated to exoplanet detection in high-contrast images.

\subsection{Subchallenge 2: ADI+mSDI}
For this challenge, we received $6$ valid submissions from $4$ participants. Two algorithms belong to the speckle subtraction family ($\mathrm{PCA_{Padova}}$ and $\mathrm{STIM_{ADI}}$). The other four methods belong to the inverse problem approach family ($\mathrm{ANDROMEDA_{ADI}}$,$\mathrm{ANDROMEDA_{ASDI}}$,$\mathrm{FMMF_{ASDI}}$, and $\mathrm{PACO_{ASDI}}$). 
In the following, we show the detailed plots for the fifth Gemini-S/GPI data set (gpi5), which contains 4 injected planetary signals (indicated with blue circles in Fig.~\ref{fig:img_sph3_baseline}, right). On this data set, the result of the baseline algorithm is shown in Fig.~\ref{fig:img_sph3_baseline} (detection map on the left and analysis in the middle). The 6 submitted detection maps obtained for the ten data sets of this subchallenge can be found in App.~\ref{app_asdi}.

\begin{figure}[H]
    \centering
    \resizebox{\hsize}{!}{\includegraphics{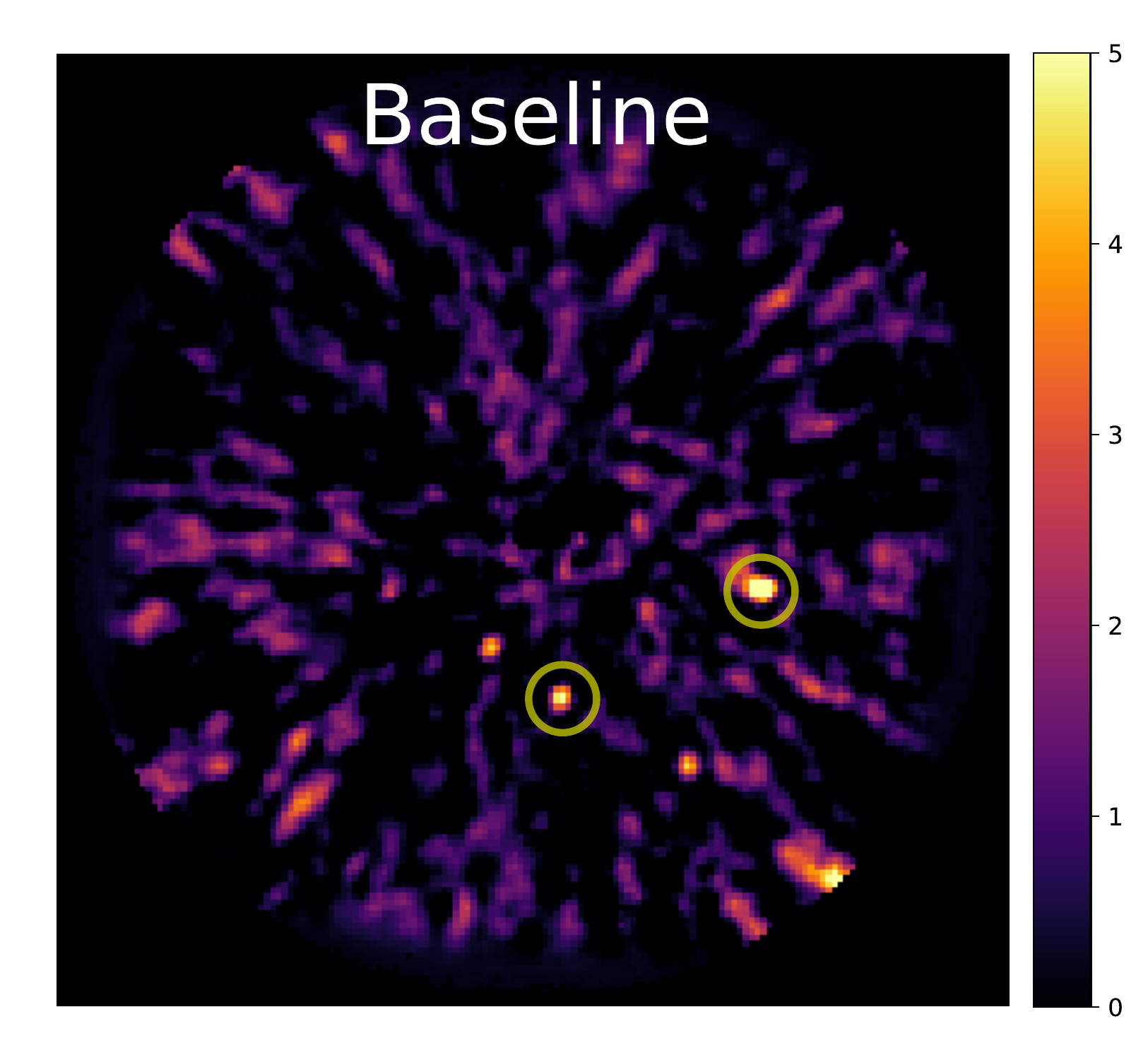}
                          \includegraphics{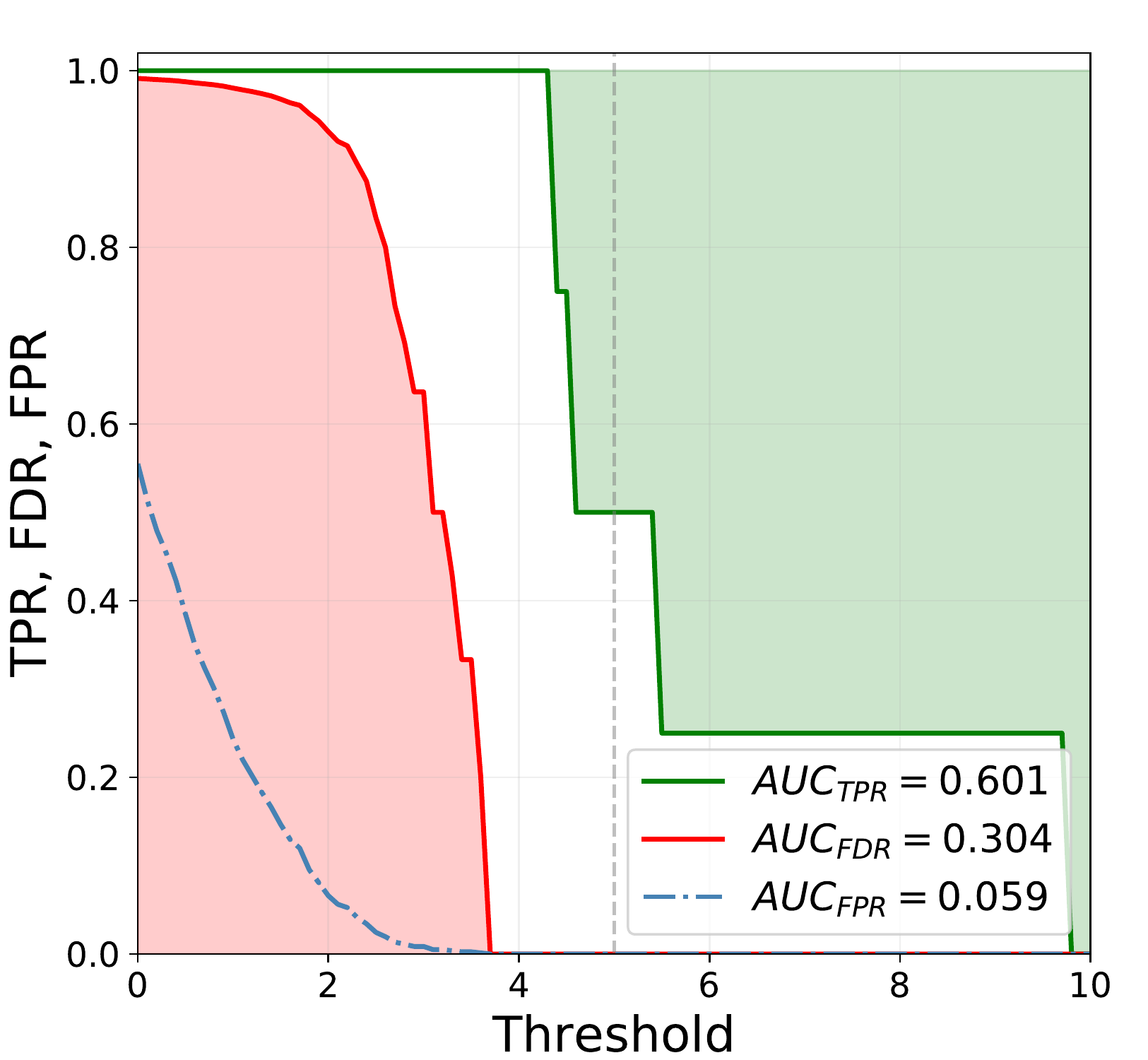}
                          \includegraphics{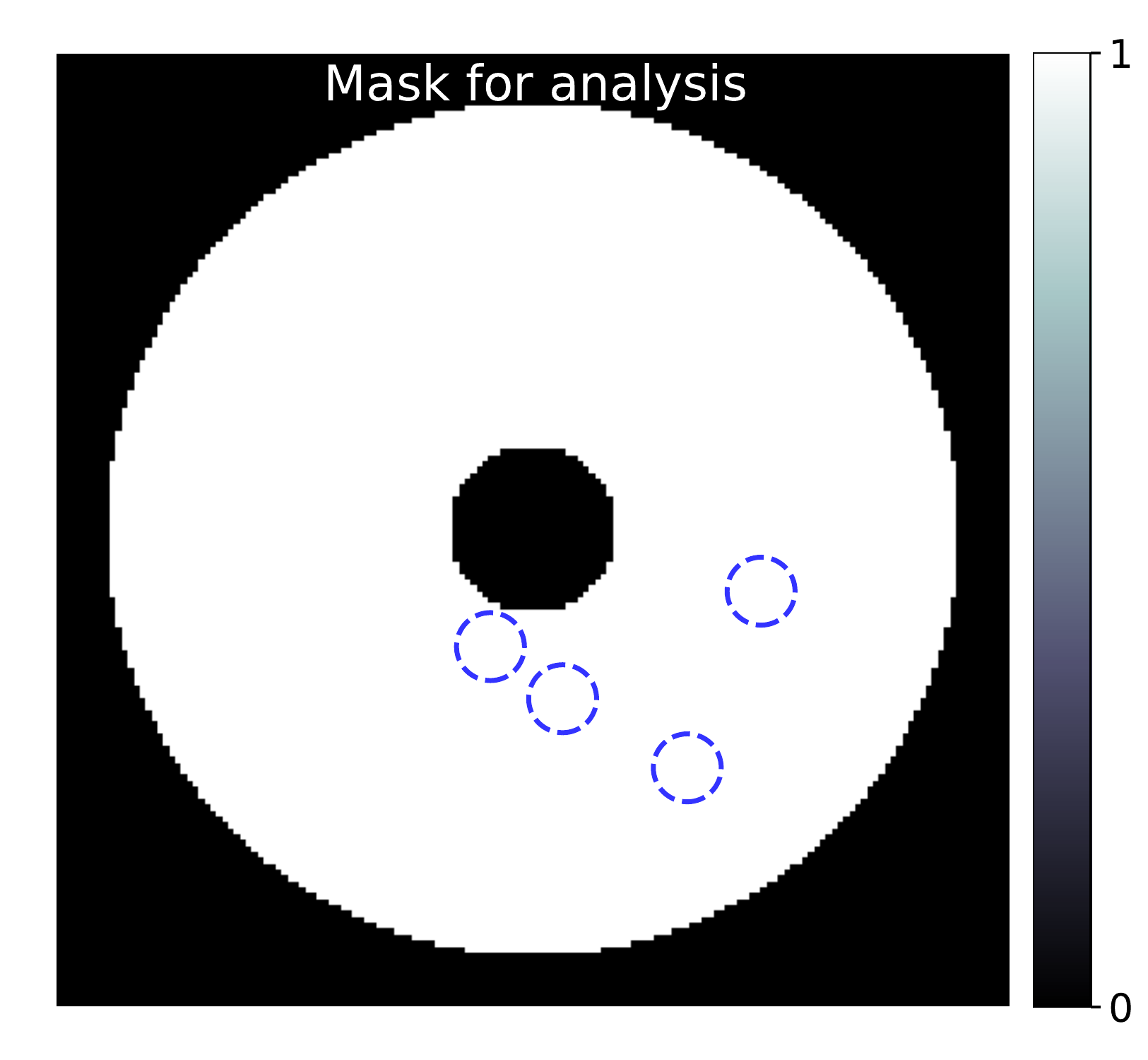}}
    \caption{Example of the results displayed for the chosen baseline algorithm (annular PCA) applied on the fifth Gemini-S/GPI data set (gpi5). 
    Left: Detection map with a color-bar ranging from 0 to the submitted threshold. True positives at the submitted threshold are indicated with yellow circles (the diameter of the circles correspond to the instrumental PSF FWHM). False positives at the submitted threshold are indicated with red squares. 
    Middle: True Positive Rate (TPR, in green), False Discovery Rate (FDR, in red) and False Positive Rate (FPR, dash-dotted blue line), for a range of threshold varying from 0 to twice the submitted threshold. 
    Right: Common mask applied to all the detection maps of this specific gpi5 data set. The dashed blue circles indicate the positions of the 4 synthetic planetary signals injected in this data set.}
    \label{fig:img_sph3_baseline}
\end{figure}

$\mathrm{PCA_{Padova}}$ builds the library of speckle field models by applying a principal component analysis on the whole 4D cube. This method is implemented in the SpeCal pipeline and used for analysing the SPHERE guaranteed time observations data. 
$\mathrm{STIM_{ADI}}$, consists in building a STIM detection map for each spectral channel (using the high-pass filtered version and a full frame PCA) and then combining the detection maps of each spectral channel by summing the squared images. 
$\mathrm{ANDROMEDA_{ADI}}$, consists in running ANDROMEDA on each spectral channel and then combining the signal-to-noise ratio maps of each channel by summing the squared images. 
$\mathrm{ANDROMEDA_{SADI}}$, performs a simple spectral differential imaging (SDI) of the data cube before running ANDROMEDA on each spectral channel and then combining the signal-to-noise ratio maps of each channel by summing the squared images. The reference channels to be rescaled and subtracted (used as a SDI starlight residuals model) are chosen according to typical expected exoplanet spectrum. 
$\mathrm{FMMF_{SADI}}$ consists in taking into account the spectral dimension included in FMMF and weights each channel with a typical substellar spectral template, as described in Ref.~\cite{Ruffio2017}. 
$\mathrm{PACO_{ASDI}}$\cite{flasseur2020pacoasdi} uses the PACO framework to model both the temporal and spectral variation of the stellar residuals and track the planetary signal.

\begin{figure}
    \centering
    \resizebox{\hsize}{!}{\includegraphics{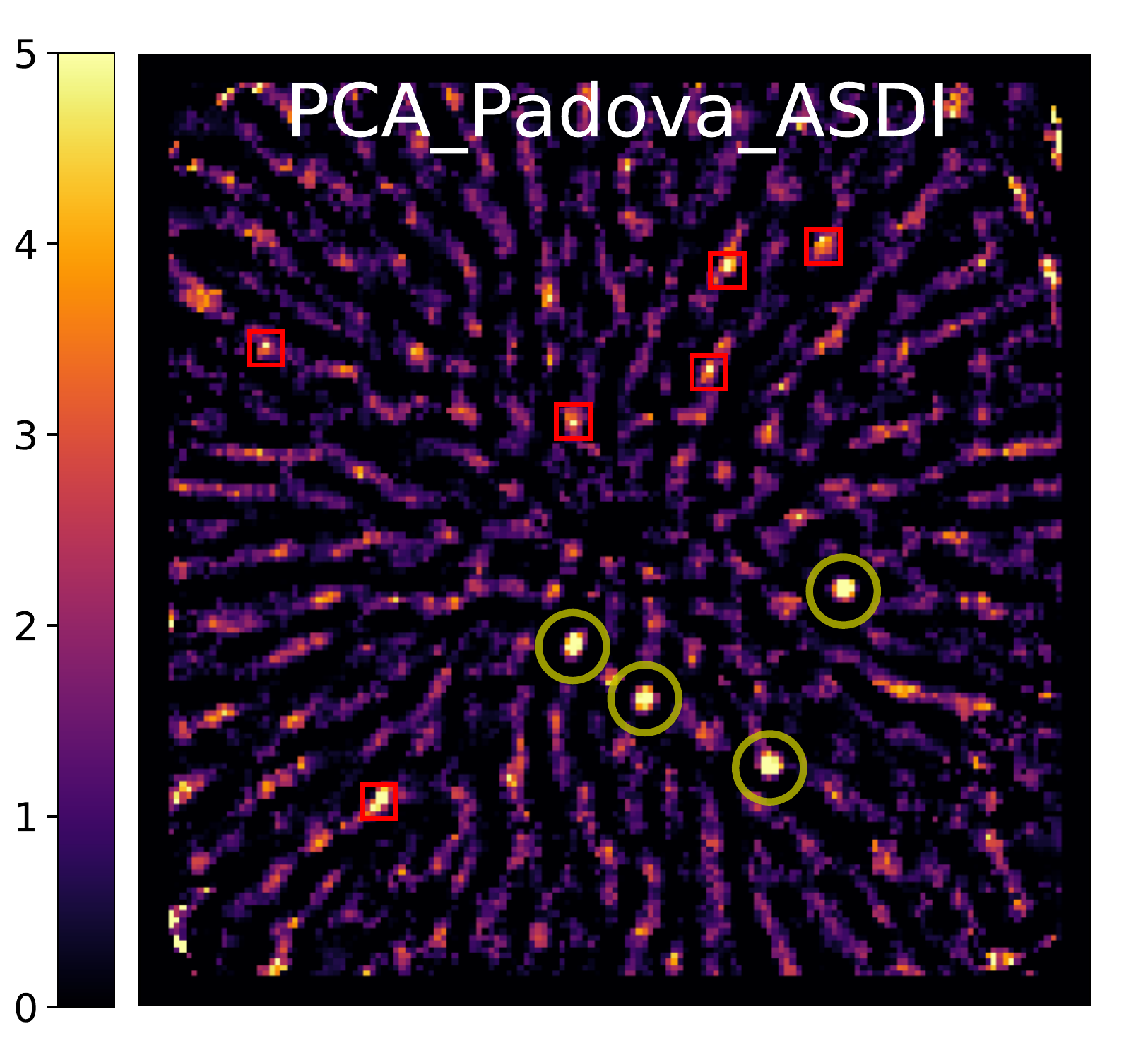}
                            \includegraphics{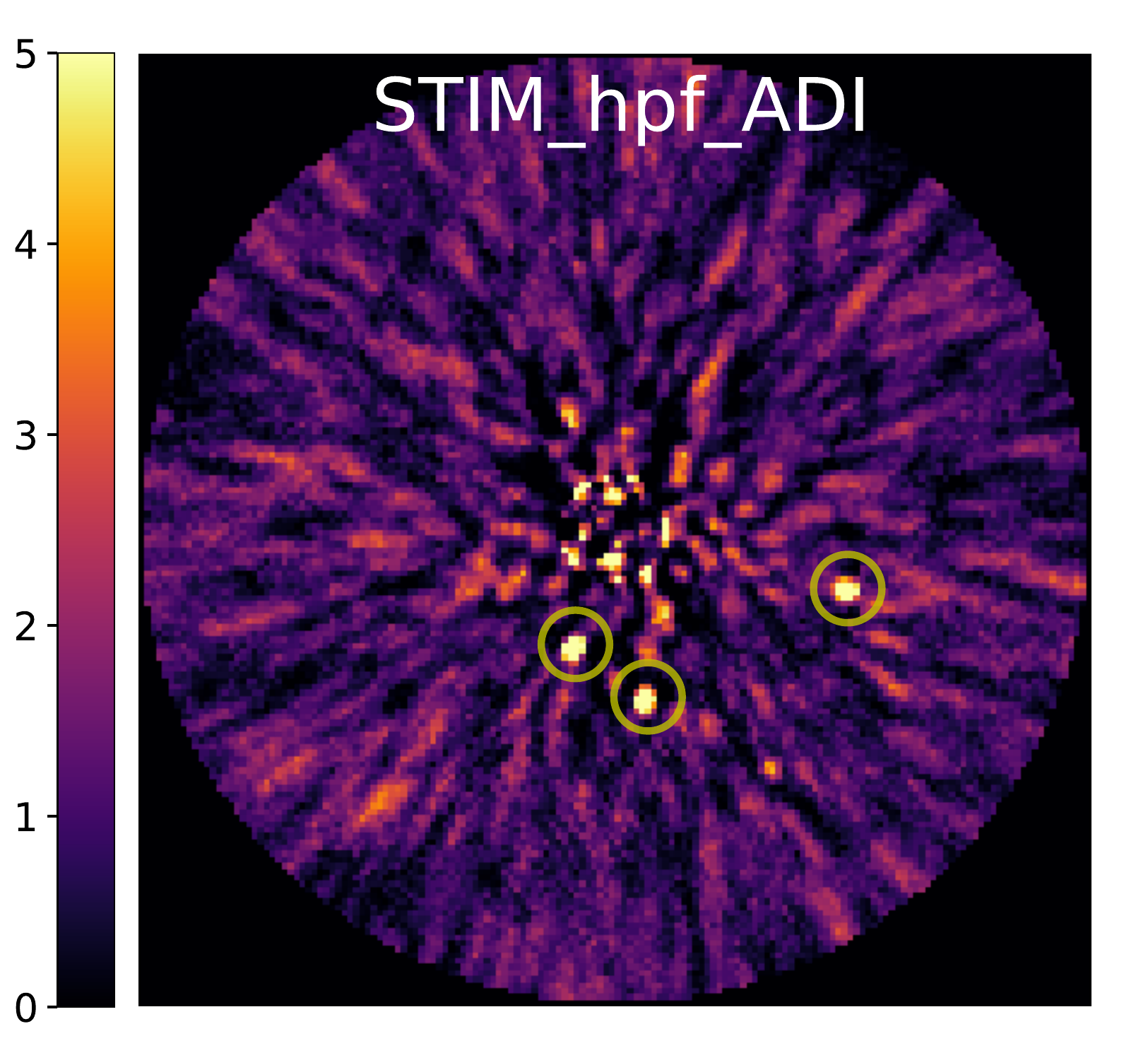}
                            \includegraphics{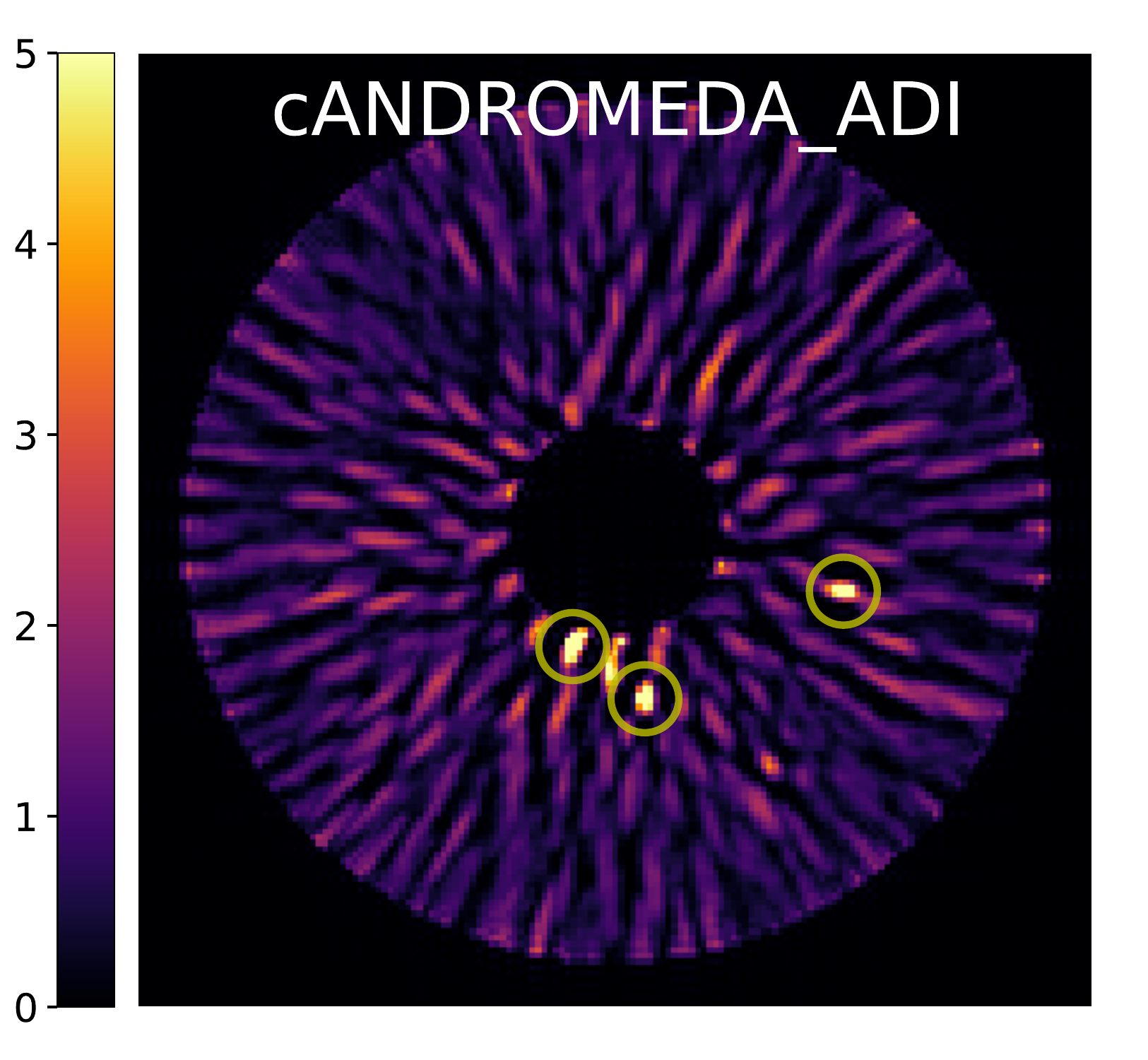}}
    \resizebox{\hsize}{!}{\includegraphics{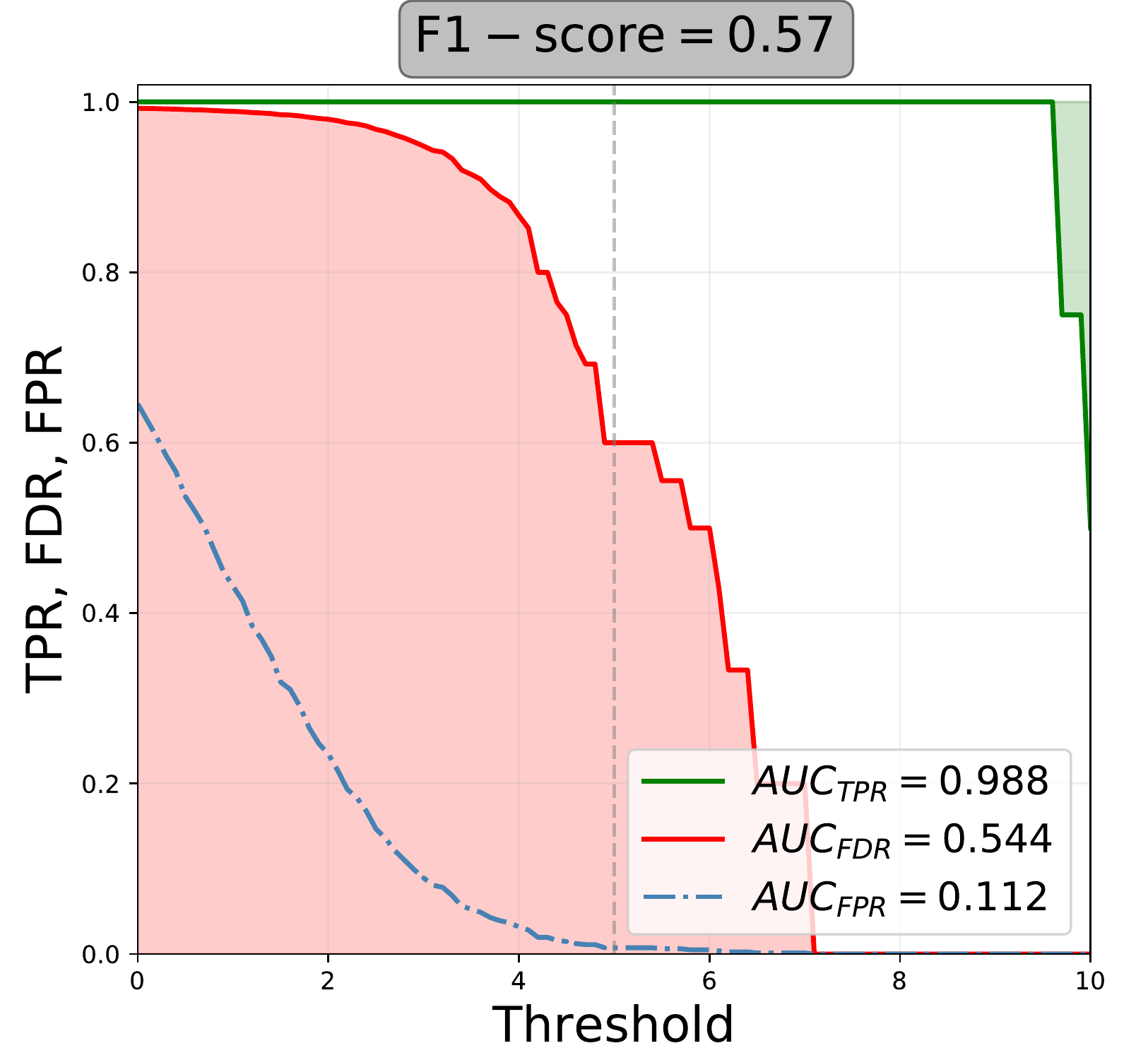}
                            \includegraphics{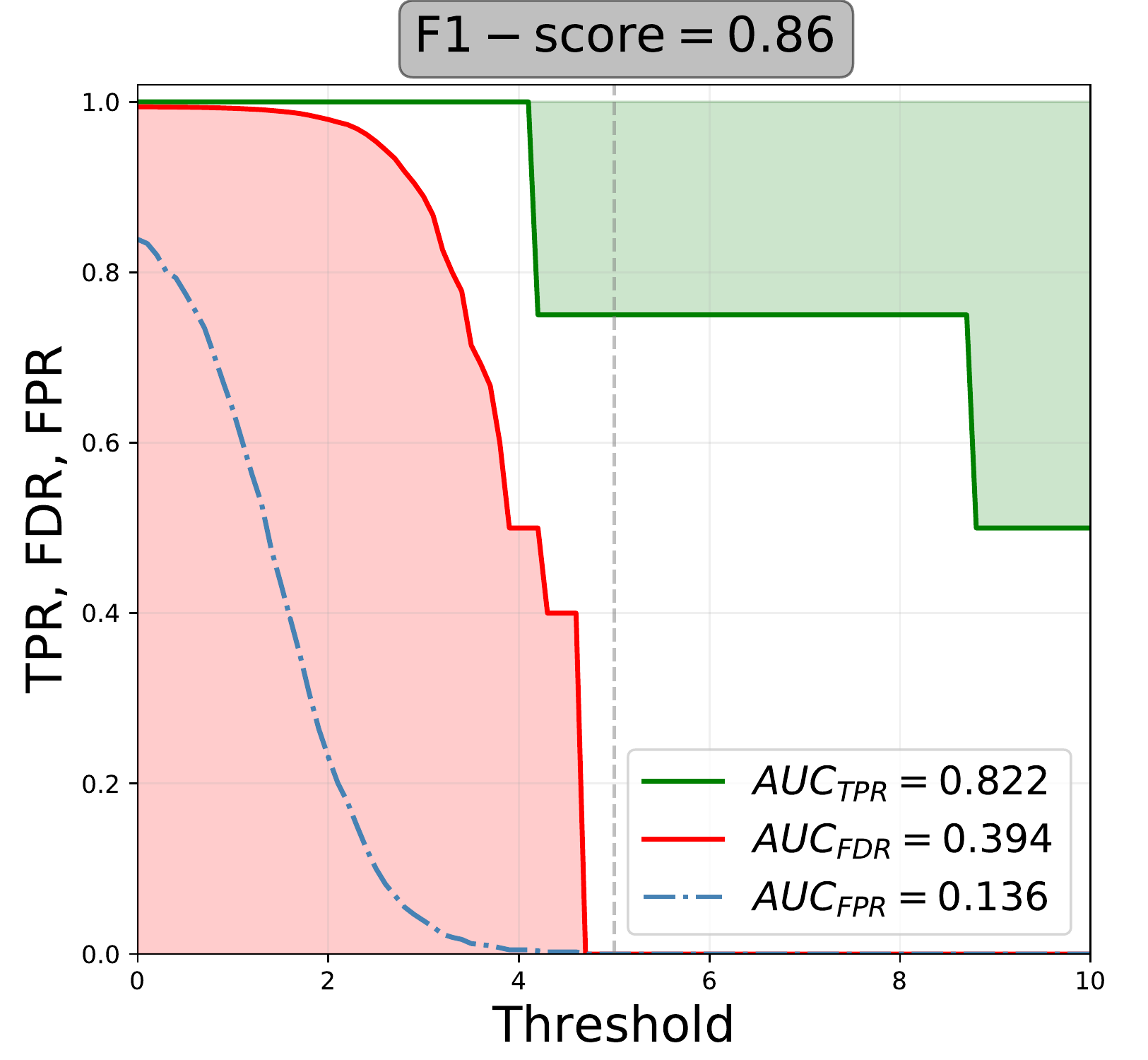}
                            \includegraphics{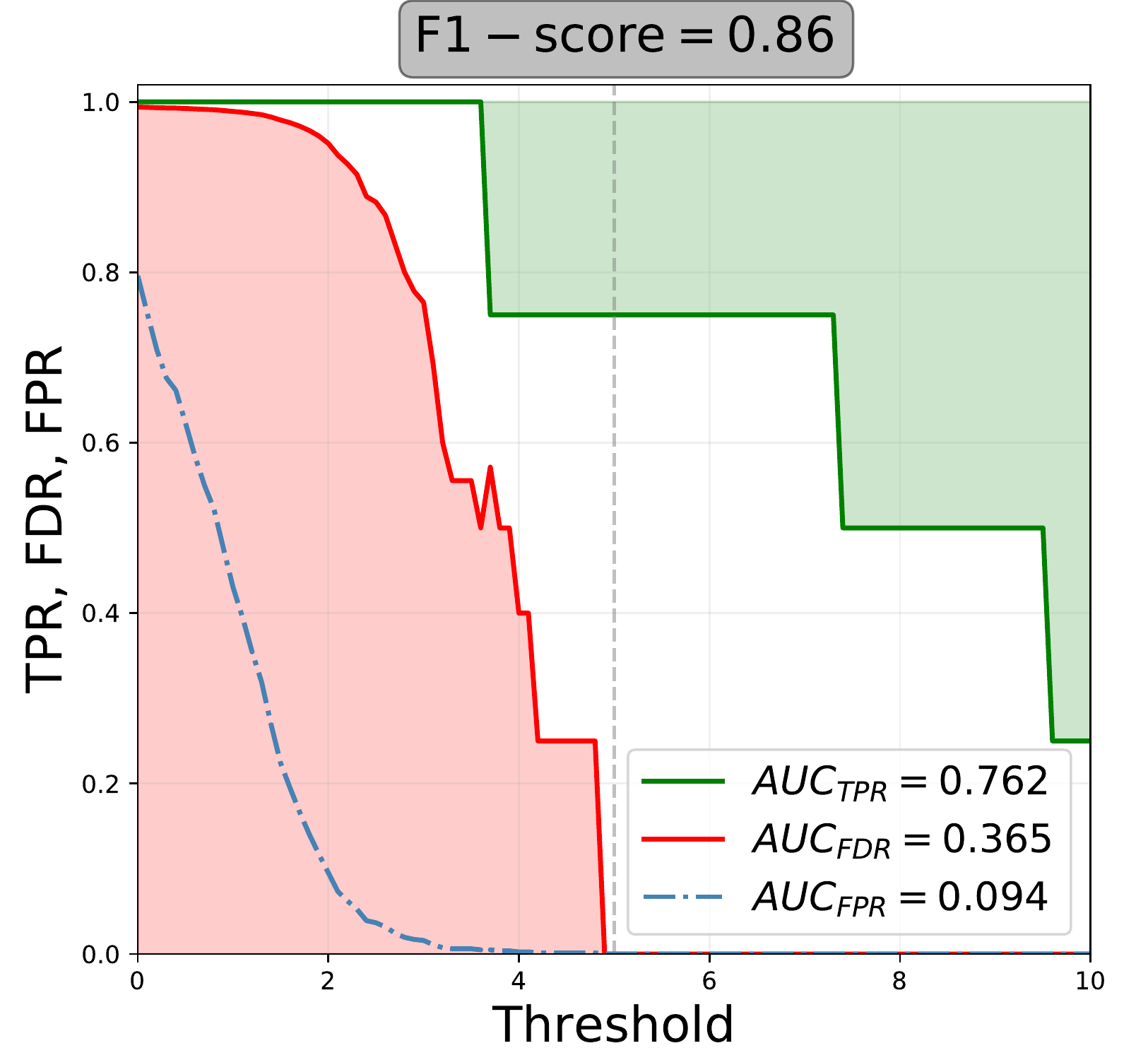}}
    \smallskip
    
    \resizebox{\hsize}{!}{\includegraphics{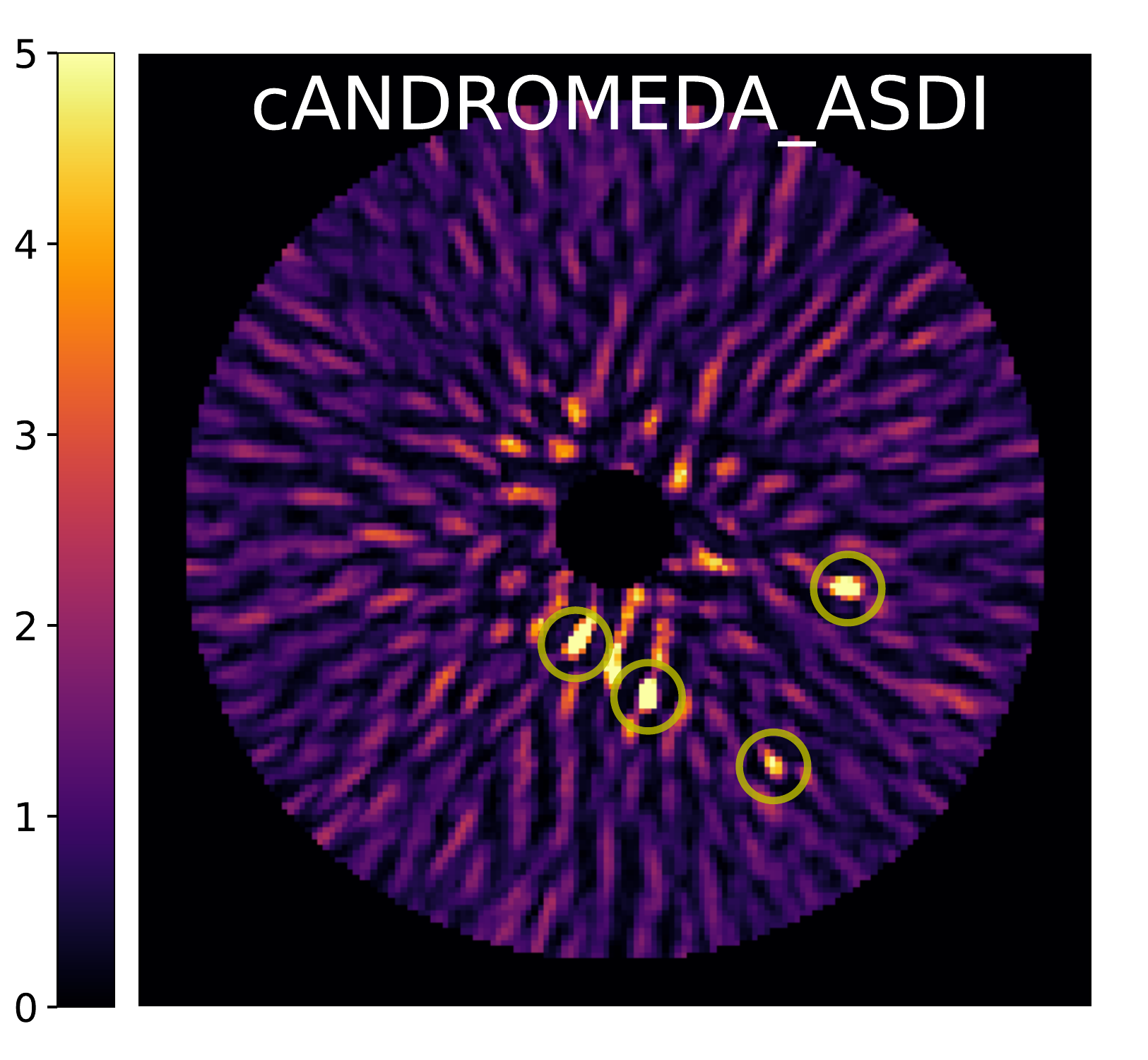}
                            \includegraphics{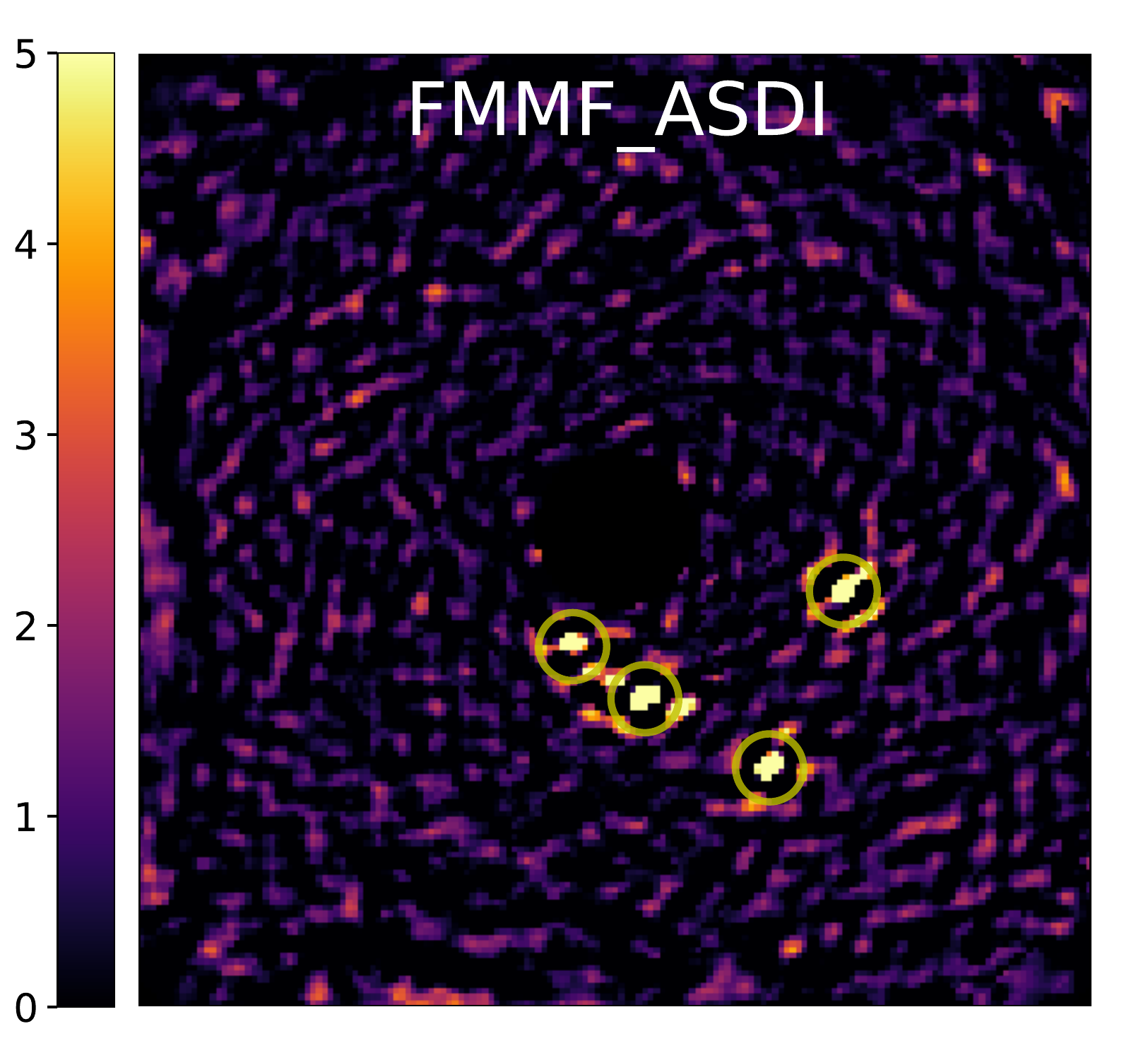}
                            \includegraphics{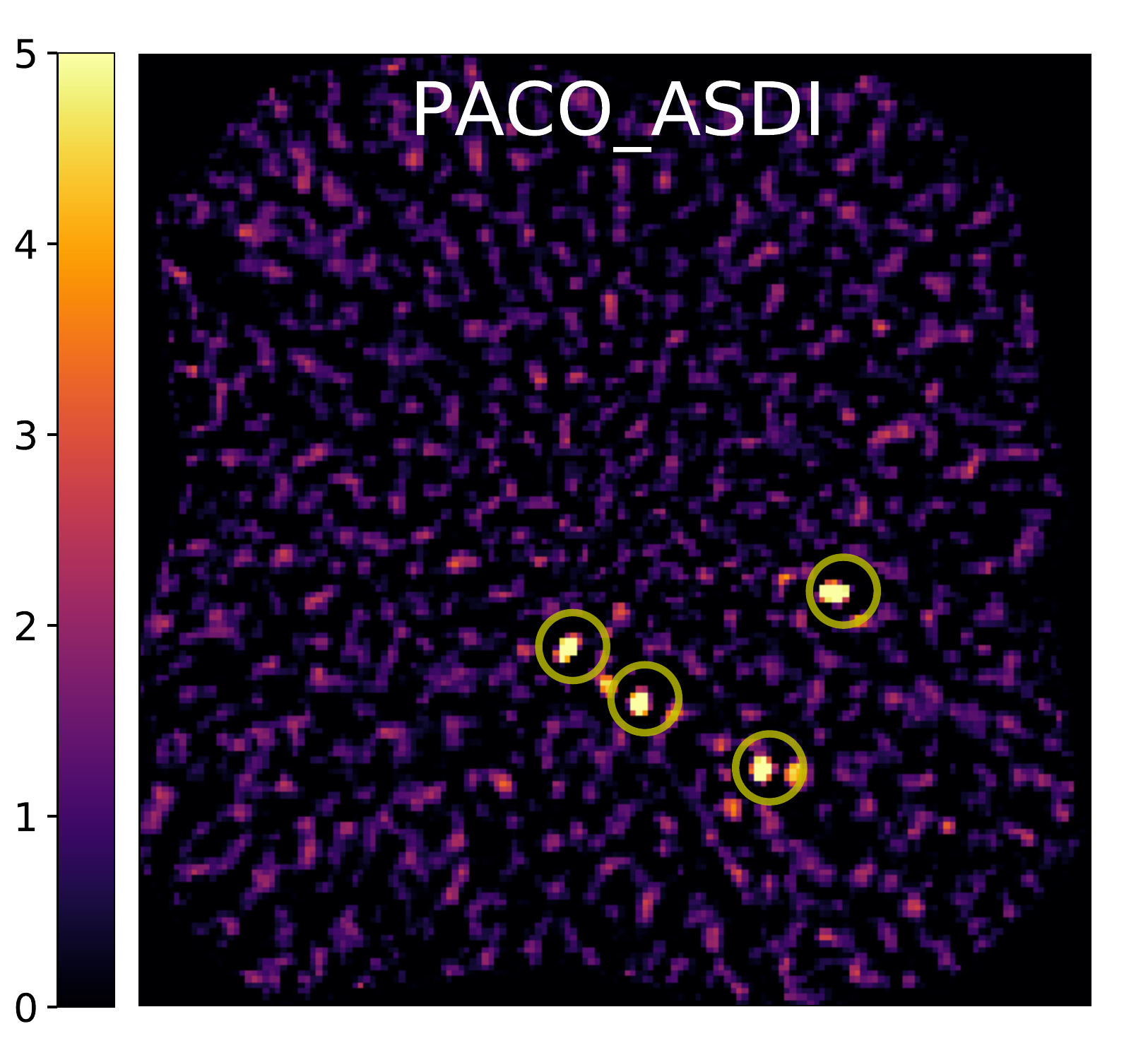}}
    \resizebox{\hsize}{!}{\includegraphics{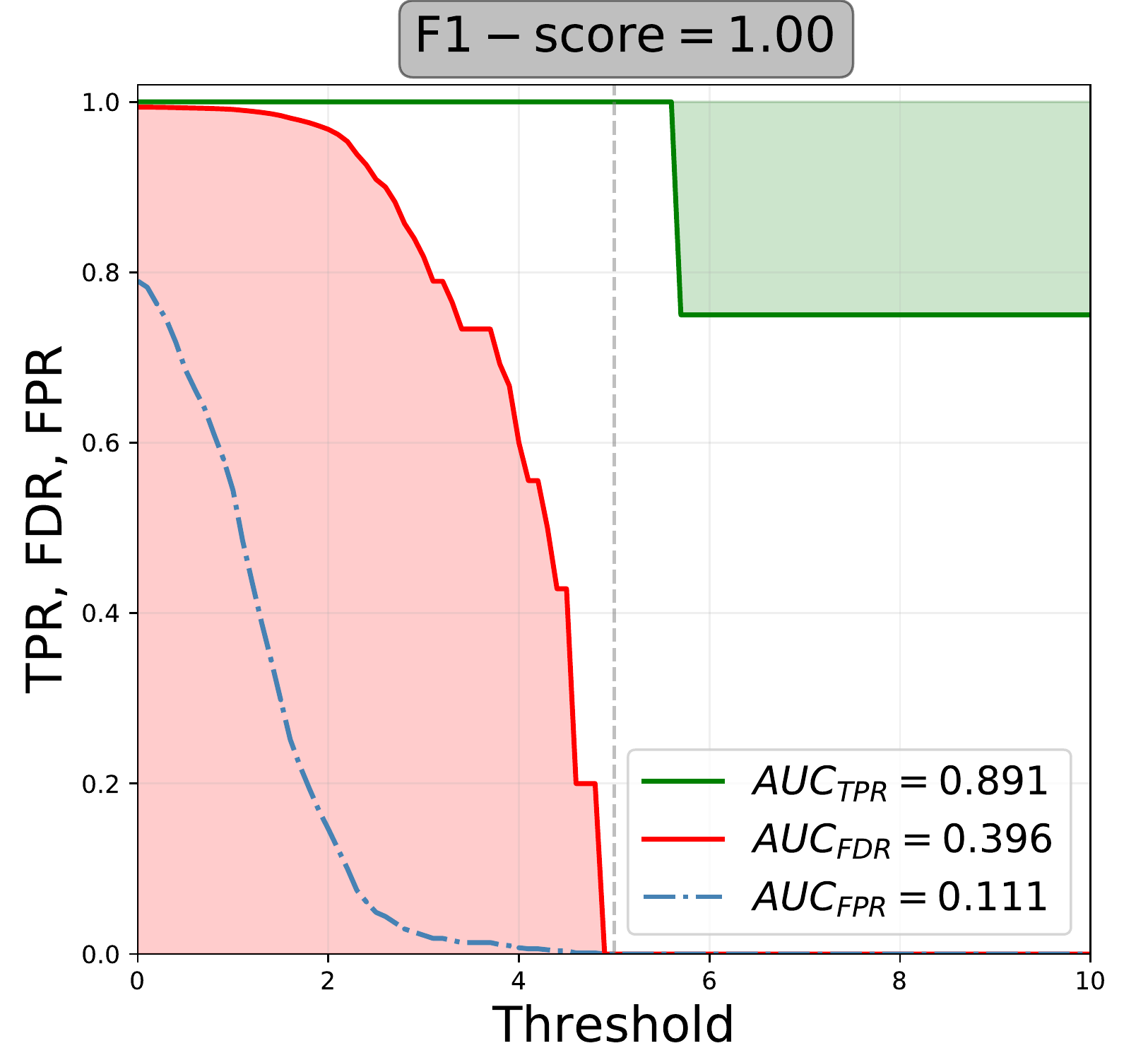}
                            \includegraphics{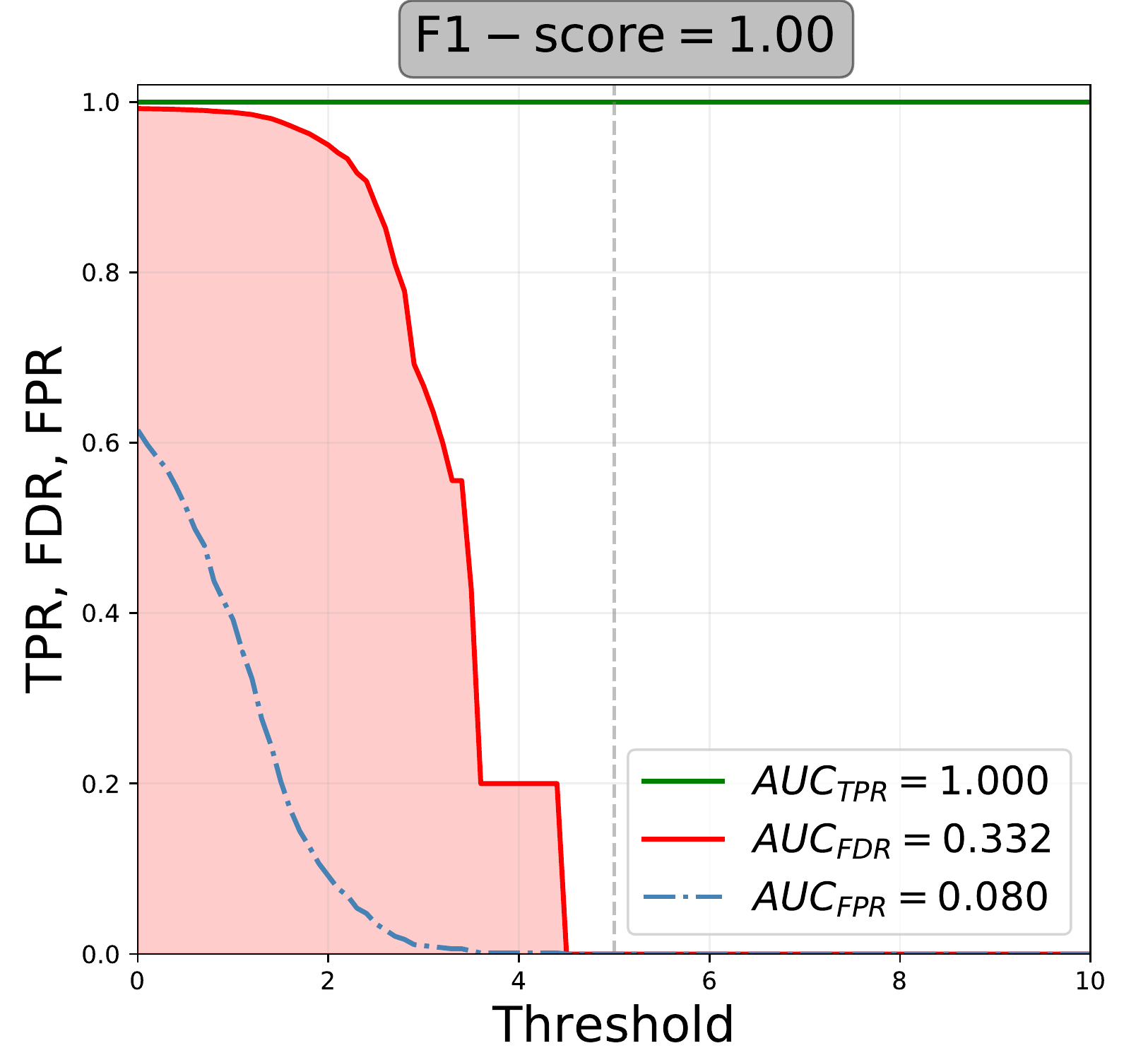}
                            \includegraphics{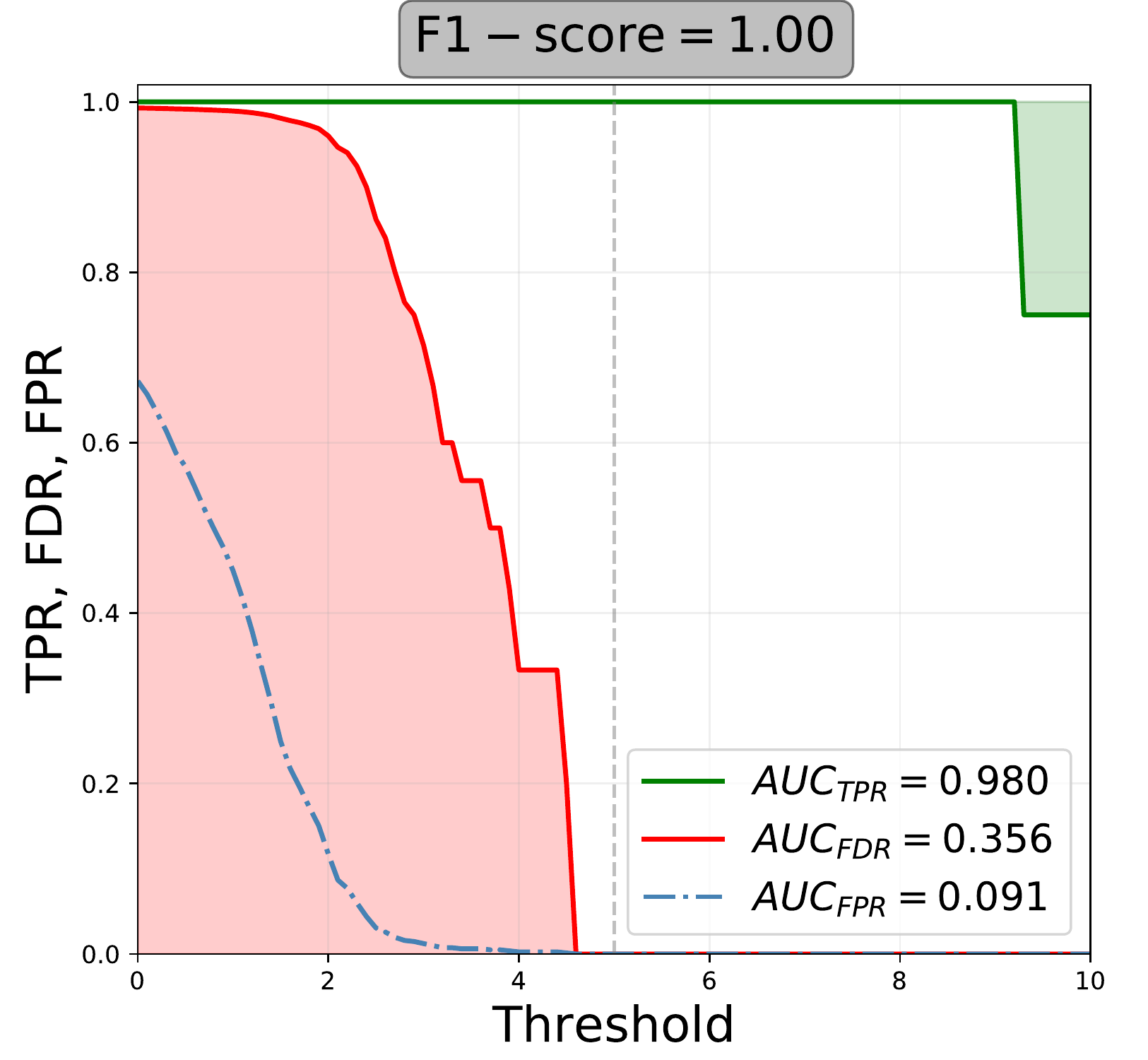}}                            
      \caption{Results of the ADI+mSDI subchallenge for the fifth Gemini-S/GPI dataset (gpi5).}
    \label{fig:img_gpi5}
\end{figure}

\begin{figure}
    \centering
    \resizebox{\hsize}{!}{\includegraphics{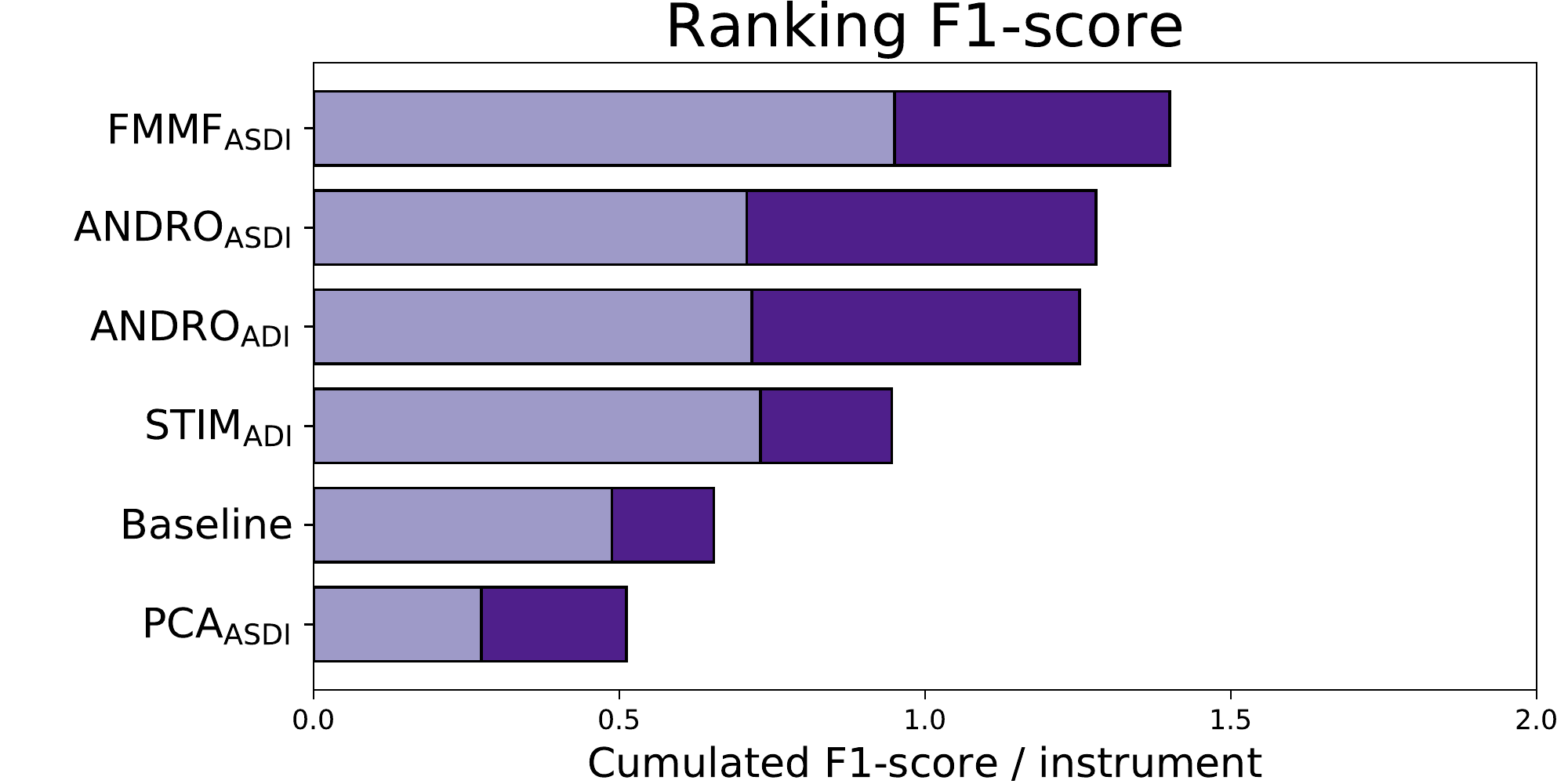}
    \includegraphics{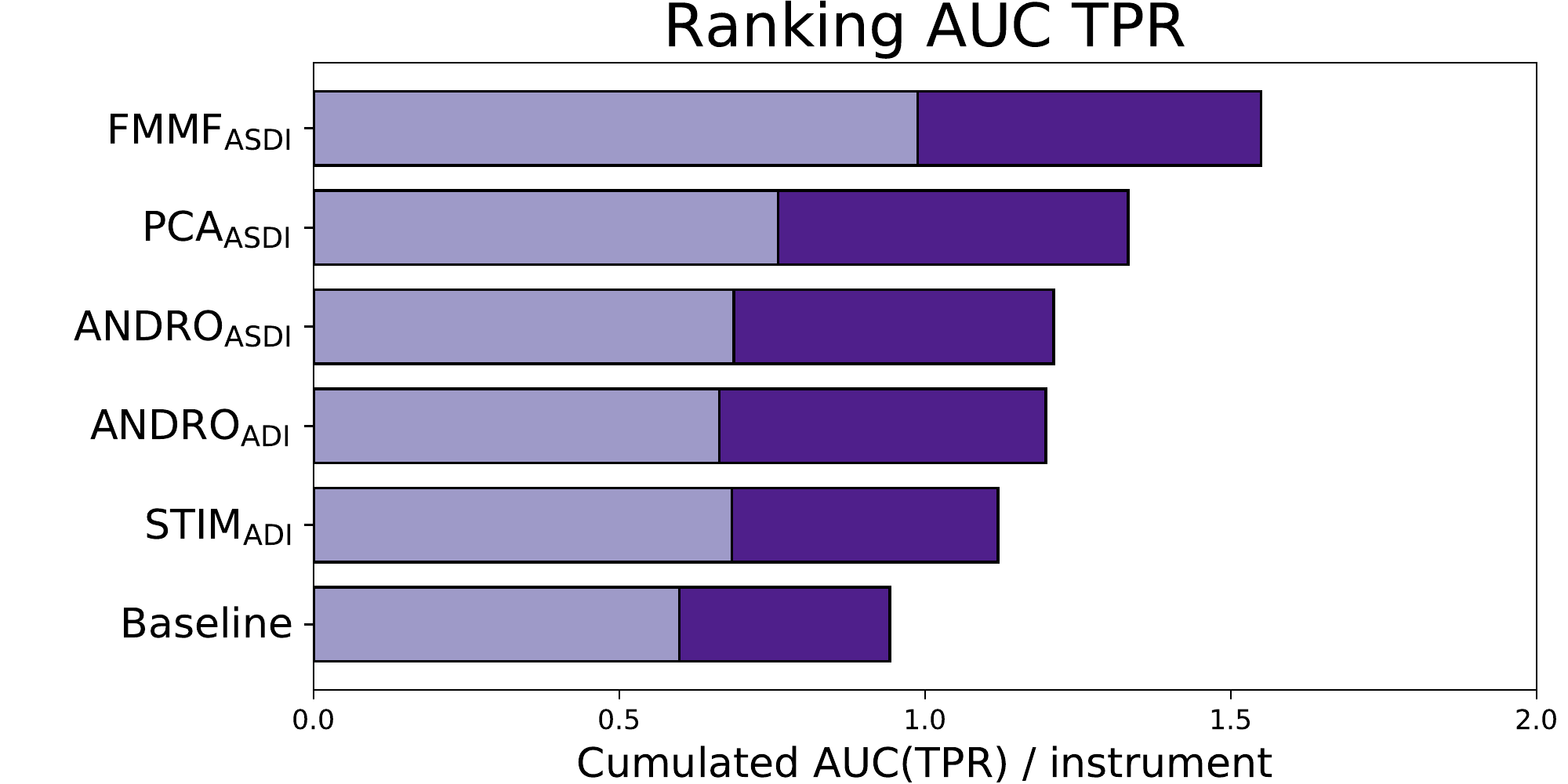}
    \includegraphics{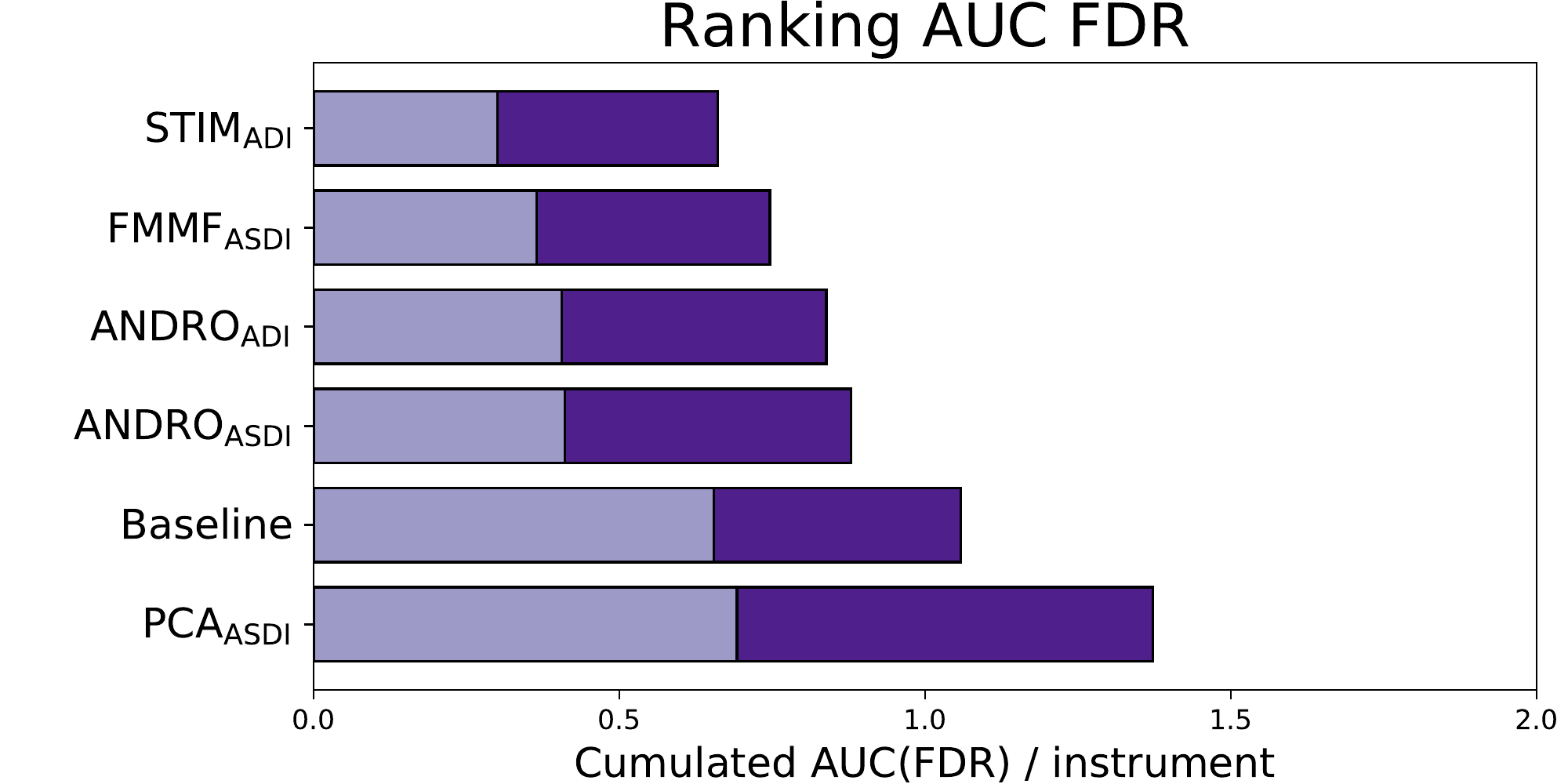}}                          
    \caption{Ranking of the 5 valid algorithms submitted for the EIDC ADI+mSDI subchallenge. 
    Left column: Ranking based on the F1-score. 
    Middle column: Ranking based on the area under curve of the TPR. 
    Right column: Ranking based on the area under curve of the FDR. 
    The light and dark colors correspond to the five VLT/SPHERE-IFS and five Gemini-S/GPI data sets respectively.}
    \label{fig:rnk_asdi}
\end{figure}

The images and corresponding counts of detections and non-detections as a function of the threshold are displayed in Fig.~\ref{fig:img_gpi5} for the case of the fifth Gemini-S/GPI data set (gpi5). The scores proposed in Sect.~\ref{sec:metrics} computed on the 10 data sets from this ADI+mSDI subchallenge are gathered in Tab.~\ref{tab:res_sadi}. The ranking of this subchallenge is shown in Fig.~\ref{fig:rnk_asdi}.

\begin{table}[h]
\caption[example] 
   {\label{tab:res_sadi} 
   Results from the second subchallenge (ADI+mSDI) of the EIDC. The 'X' symbol means that the detection map has not been submitted. The corresponding algorithm, PACO-ASDI, is not be taken into account in the ranking.}
\begin{center}
\begin{tabular}{|l|| c | c | c | c | c | c || c | c | c | c | c | c || c |} 
\hline
\multirow{2}{*}{\textbf{Algorithm}} & \multicolumn{6}{c||}{VLT/SPHERE-IFS} & \multicolumn{6}{c||}{Gemini-S/GPI} & \multirow{1}{*}{\textbf{All}} \\ 
\cline{2-13}
& (1) & (2) & (3) & (4) & (5) & Mean & (1) & (2) & (3) & (4) & (5) & Mean & \\
\hline\hline
\multirow{3}{*}{$\mathrm{Baseline}$} & 0.80 & 0.75 & 0.40 & 0.00 & \emph{u} & 0.49 & 0.00 & 0.00 & \emph{u} & 0.00 & 0.67 & 0.17 & 0.33\\
 & \tpr{0.78} & \tpr{ 0.77} & \tpr{ 0.63} & \tpr{ 0.21} & \tpr{ \emph{u}} & \tpr{ 0.60} & \tpr{ 0.34} & \tpr{ 0.24} & \tpr{ \emph{u}} & \tpr{ 0.19} & \tpr{ 0.60} & \tpr{ 0.34} & \tpr{ 0.47}\\
 & \fdr{0.57} & \fdr{ 0.36} & \fdr{ 0.71} & \fdr{ 0.98} & \fdr{ \emph{u}} & \fdr{ 0.65} & \fdr{ 0.46} & \fdr{ 0.39} & \fdr{ \emph{u}} & \fdr{ 0.46} & \fdr{ 0.30} & \fdr{ 0.40} & \fdr{ 0.53}\\
\hline
\multirow{3}{*}{$\mathrm{PCA_{Padova,ASDI}}$} & 0.50 & 0.25 & 0.18 & 0.17 & \emph{u} & 0.27 & 0.00 & 0.15 & \emph{u} & 0.22 & 0.57 & 0.24 & 0.26\\
 & \tpr{0.94} & \tpr{ 0.87} & \tpr{ 0.73} & \tpr{ 0.50} & \tpr{ \emph{u}} & \tpr{ 0.76} & \tpr{ 0.46} & \tpr{ 0.44} & \tpr{ \emph{u}} & \tpr{ 0.41} & \tpr{ 0.99} & \tpr{ 0.57} & \tpr{ 0.67}\\
 & \fdr{0.59} & \fdr{ 0.78} & \fdr{ 0.83} & \fdr{ 0.58} & \fdr{ \emph{u}} & \fdr{ 0.69} & \fdr{ 0.60} & \fdr{ 0.97} & \fdr{ \emph{u}} & \fdr{ 0.61} & \fdr{ 0.54} & \fdr{ 0.68} & \fdr{ 0.69}\\
\hline
\multirow{3}{*}{$\mathrm{STIM_{ADI}}$} & 1.00 & 0.86 & 0.67 & 0.40 & \emph{u} & 0.73 & 0.00 & 0.00 & \emph{u} & 0.00 & 0.86 & 0.21 & 0.47\\
 & \tpr{0.86} & \tpr{ 0.80} & \tpr{ 0.65} & \tpr{ 0.42} & \tpr{ \emph{u}} & \tpr{ 0.68} & \tpr{ 0.39} & \tpr{ 0.29} & \tpr{ \emph{u}} & \tpr{ 0.24} & \tpr{ 0.82} & \tpr{ 0.44} & \tpr{ 0.56}\\
 & \fdr{0.32} & \fdr{ 0.30} & \fdr{ 0.29} & \fdr{ 0.29} & \fdr{ \emph{u}} & \fdr{ 0.30} & \fdr{ 0.36} & \fdr{ 0.38} & \fdr{ \emph{u}} & \fdr{ 0.30} & \fdr{ 0.39} & \fdr{ 0.36} & \fdr{ 0.33}\\
\hline
\multirow{3}{*}{$\mathrm{ANDROMEDA_{ADI}}$} & 0.80 & 1.00 & 0.67 & 0.40 & \emph{u} & 0.72 & 1.00 & 0.00 & \emph{u} & 0.29 & 0.86 & 0.54 & 0.63\\
 & \tpr{0.75} & \tpr{ 0.81} & \tpr{ 0.66} & \tpr{ 0.44} & \tpr{ \emph{u}} & \tpr{ 0.66} & \tpr{ 0.73} & \tpr{ 0.26} & \tpr{ \emph{u}} & \tpr{ 0.38} & \tpr{ 0.76} & \tpr{ 0.53} & \tpr{ 0.60}\\
 & \fdr{0.46} & \fdr{ 0.36} & \fdr{ 0.39} & \fdr{ 0.41} & \fdr{ \emph{u}} & \fdr{ 0.41} & \fdr{ 0.38} & \fdr{ 0.44} & \fdr{ \emph{u}} & \fdr{ 0.55} & \fdr{ 0.36} & \fdr{ 0.43} & \fdr{ 0.42}\\
\hline
\multirow{3}{*}{$\mathrm{ANDROMEDA_{ASDI}}$} & 1.00 & 1.00 & 0.50 & 0.33 & \emph{u} & 0.71 & 1.00 & 0.00 & \emph{u} & 0.29 & 1.00 & 0.57 & 0.64\\
 & \tpr{0.81} & \tpr{ 0.76} & \tpr{ 0.65} & \tpr{ 0.53} & \tpr{ \emph{u}} & \tpr{ 0.69} & \tpr{ 0.51} & \tpr{ 0.35} & \tpr{ \emph{u}} & \tpr{ 0.34} & \tpr{ 0.89} & \tpr{ 0.52} & \tpr{ 0.60}\\
 & \fdr{0.44} & \fdr{ 0.32} & \fdr{ 0.44} & \fdr{ 0.44} & \fdr{ \emph{u}} & \fdr{ 0.41} & \fdr{ 0.37} & \fdr{ 0.53} & \fdr{ \emph{u}} & \fdr{ 0.58} & \fdr{ 0.40} & \fdr{ 0.47} & \fdr{ 0.44}\\
\hline
\multirow{3}{*}{$\mathrm{FMMF_{ASDI}}$} & 1.00 & 1.00 & 0.80 & 1.00 & \emph{u} & 0.95 & 0.00 & 0.80 & \emph{u} & 0.00 & 1.00 & 0.45 & 0.70\\
 & \tpr{1.00} & \tpr{ 0.97} & \tpr{ 1.00} & \tpr{ 0.98} & \tpr{ \emph{u}} & \tpr{ 0.99} & \tpr{ 0.37} & \tpr{ 0.55} & \tpr{ \emph{u}} & \tpr{ 0.33} & \tpr{ 1.00} & \tpr{ 0.56} & \tpr{ 0.77}\\
 & \fdr{0.39} & \fdr{ 0.34} & \fdr{ 0.40} & \fdr{ 0.33} & \fdr{ \emph{u}} & \fdr{ 0.36} & \fdr{ 0.43} & \fdr{ 0.38} & \fdr{ \emph{u}} & \fdr{ 0.39} & \fdr{ 0.33} & \fdr{ 0.38} & \fdr{ 0.37}\\
\hline
\multirow{3}{*}{$\mathrm{PACO_{ASDI}}$} & 1.00 & 0.86 & 0.67 & 0.86 & \emph{u} & 0.85 & x & x & \emph{u} & x & 1.00 & x & x\\
 & \tpr{1.00} & \tpr{ 0.86} & \tpr{ 0.57} & \tpr{ 0.78} & \tpr{ \emph{u}} & \tpr{ 0.81} & \tpr{x} & \tpr{ x} & \tpr{ \emph{u}} & \tpr{x} & \tpr{ 0.98} & \tpr{x} & \tpr{x}\\
 & \fdr{0.34} & \fdr{ 0.37} & \fdr{ 0.38} & \fdr{ 0.33} & \fdr{ \emph{u}} & \fdr{ 0.36} & \fdr{ x} & \fdr{ x} & \fdr{ \emph{u}} & \fdr{x} & \fdr{ 0.36} & \fdr{ x} & \fdr{x}\\
\hline
 \end{tabular}
\end{center}
\end{table}

\subsubsection{Ranking the ADI+mSDI subchallenge results}
Based on this ADI+mSDI subchallenge, we can already draw a few conclusions. 
(1) Exploiting properly the spectral information does help to detect fainter sources. As an example, stacking each spectral channel after processing each spectral channel separately using ADI-based techniques, does not allow one to detect the lower faint companion shown in Fig.~\ref{fig:img_gpi5} with $\mathrm{STIM_{ADI}}$ compared to $\mathrm{PCA_{Padova,ASDI}}$ for the speckle subtraction techniques, and with $\mathrm{ANDROMEDA_{ADI}}$ compared to $\mathrm{ANDROMEDA_{ASDI}}$ for the inverse problem approaches. 
(2) The spectral template used as a prior for the planetary signal in some methods plays an important role. For instance, in Fig.~\ref{fig:app_ifs2}, $\mathrm{PACO_{ASDI}}$ uses a uniform spectral template prior and does not detect one of the four planetary signal. However by using a non-flat spectrum as a prior, the signal of the fourth planet is above the submitted threshold.  
(3) As a general rule, based on the image galleries in App.~\ref{app_asdi}, the more advanced/recent techniques perform better at detecting planetary signals but the results are not homogeneous. This shows once again that applying various algorithms based on different concept is the most appropriate way of validating or invalidating a candidate based on a single data set. 
(4) As for the ADI subchallenge, inverse problem approaches are the only methods providing directly with the contrast of the planetary signal candidates. This allows to additionally help disentangling the candidate from a stellar residual, by comparing the extracted spectrum of the candidate with the extracted spectrum of a residual speckle.

\section{Conclusion \& Perspectives}
\label{sec:ccl}  
In this paper, we presented the Exoplanet Imaging Data Challenge and the
results from its first phase, consisting in assessing the detection performance of various post-processing techniques dedicated to exoplanet imaging. 
For this first phase, we gathered typical data sets from several instruments offering a high-contrast imaging mode, in which we injected synthetic planetary signals. The data sets are available permanently on a \emph{Zenodo} repository, and are described on the EIDC website\footnote{\url{https://exoplanet-imaging-challenge.github.io/datasets/}}. 
Participants were invited to submit their results, consisting in one detection map for each data set and one single threshold value for all the data sets, on our competition  platform hosted on \emph{CodaLab}. 
For the analysis of the submitted detection maps, we opted for counting the detections and non-detections at a given threshold and derive relevant scores to compare the different submissions. 
By organizing and maintaining this data challenge, we intend to provide the user-community with a global vision of all the existing post-processing methods and their different capabilities. Within the developer-community, we intend to trigger new collaborations and ideas, while providing with a platform for a common comparison  between the algorithms. 

The first phase of the EIDC presented in this paper is limited in several aspects: the data chosen (via, for instance, the planetary signal injection procedure) and the evaluation method proposed (by counting detections and non-detections) are not perfectly representative and might favor or disfavor some methods. Further work will make use of the data challenge results and apply evaluation methods more adapted to the planet detection problem. On the practical aspect, we encountered a few difficulties with the CodaLab platform, in particular because it is not designed for participants to submit more than one algorithm products (it displays only one result per participant in the leader-board). In addition, the position of the injections are revealed to the participants after their first submission, potentially biasing the following submissions (one participant can submit as many results as they wanted). At last, the leader-board is currently displaying incorrect results since the backend calculating the F1-score is erroneous and it was unfortunately not possible to modify the backend while the competition is open. 

In the future, we intend to increase the number of data sets available. For instance, we put a link towards the NEAR campaign\cite{Kasper2019near} data on the EIDC website\footnote{\url{https://exoplanet-imaging-challenge.github.io/near/}}, to allow anyone to test advanced algorithms on this challenging observation sequence. On top of using comparison metrics that are more adapted, we also plan to launch new phases of the data challenge, adding, for example, the characterisation of the detections (position and contrast with their uncertainties) or adding extended sources to be detected. 
We also plan to offer alternative data set enabling the direct detection of exoplanets, such as high spectral resolution images\cite{Hoeijmakers2018,Haffert2019}. 
We aim to regularly publish the results of the EIDC at every future SPIE astronomical telescopes+instrumentation conference.

\appendix  

\section{Gallery of detection maps: ADI subchallenge}
\label{app_adi}
In this appendix, we display the detection maps of the 22 submissions for the nine data sets of the ADI subchallenge. Each detection map is shown with a color-bar ranging from 0 to the submitted threshold. The yellow circles indicate true positives at the given threshold, while red squares indicate false positives. 
The first image (top left) shows the detection map for the chosen baseline algorithm (annular PCA). 
The last image (bottom right) shows the mask used to conduct the analysis of each map, with blue circles indicating the location of the injected synthetic planetary signals.

\begin{figure}
    \centering
    \includegraphics[scale=0.71]{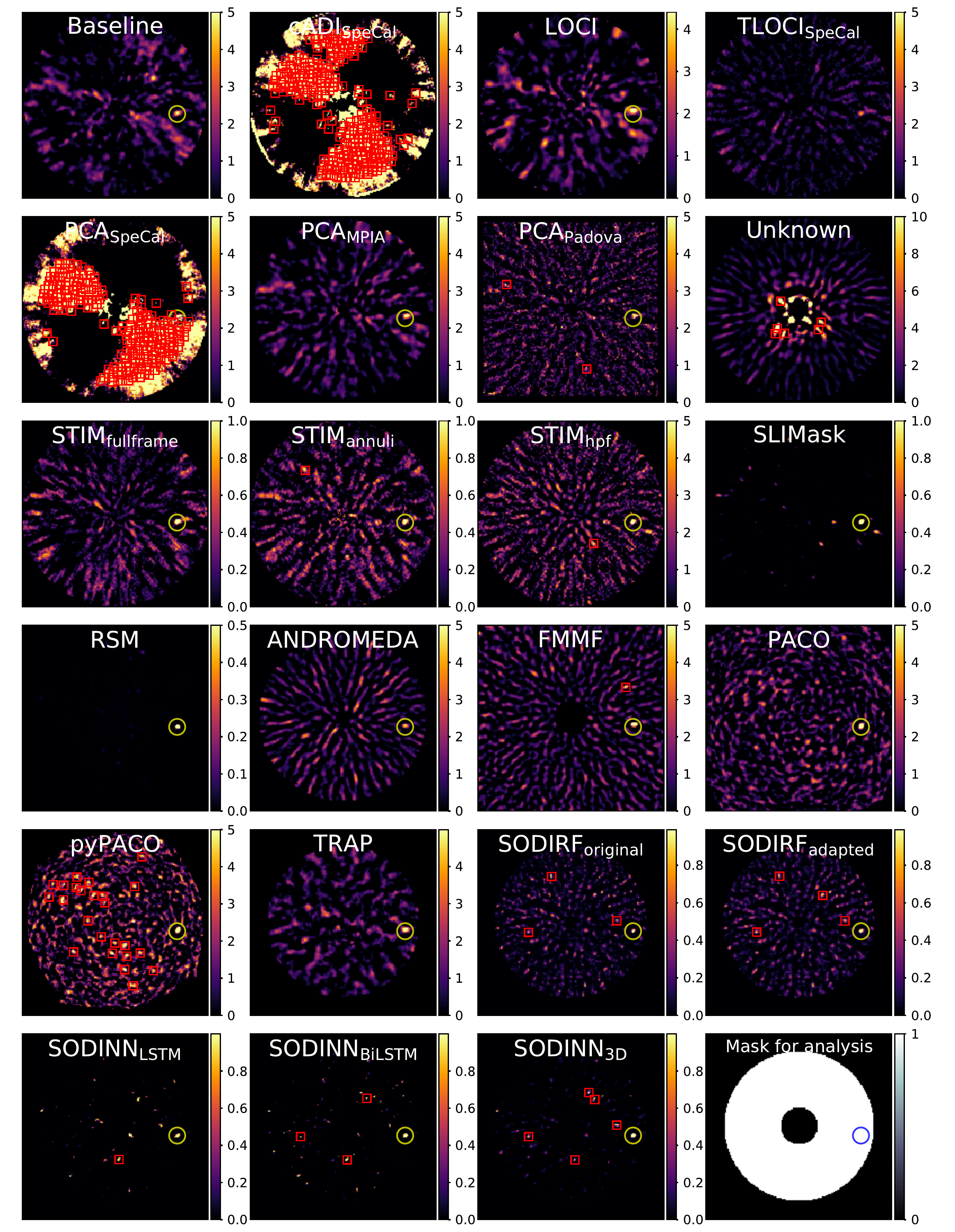}
    \caption{Results of the ADI subchallenge for the first VLT/SPHERE-IRDIS dataset (sph1).}
    \label{fig:app_sph1}
\end{figure}

\begin{figure}
    \centering
    \includegraphics[scale=0.71]{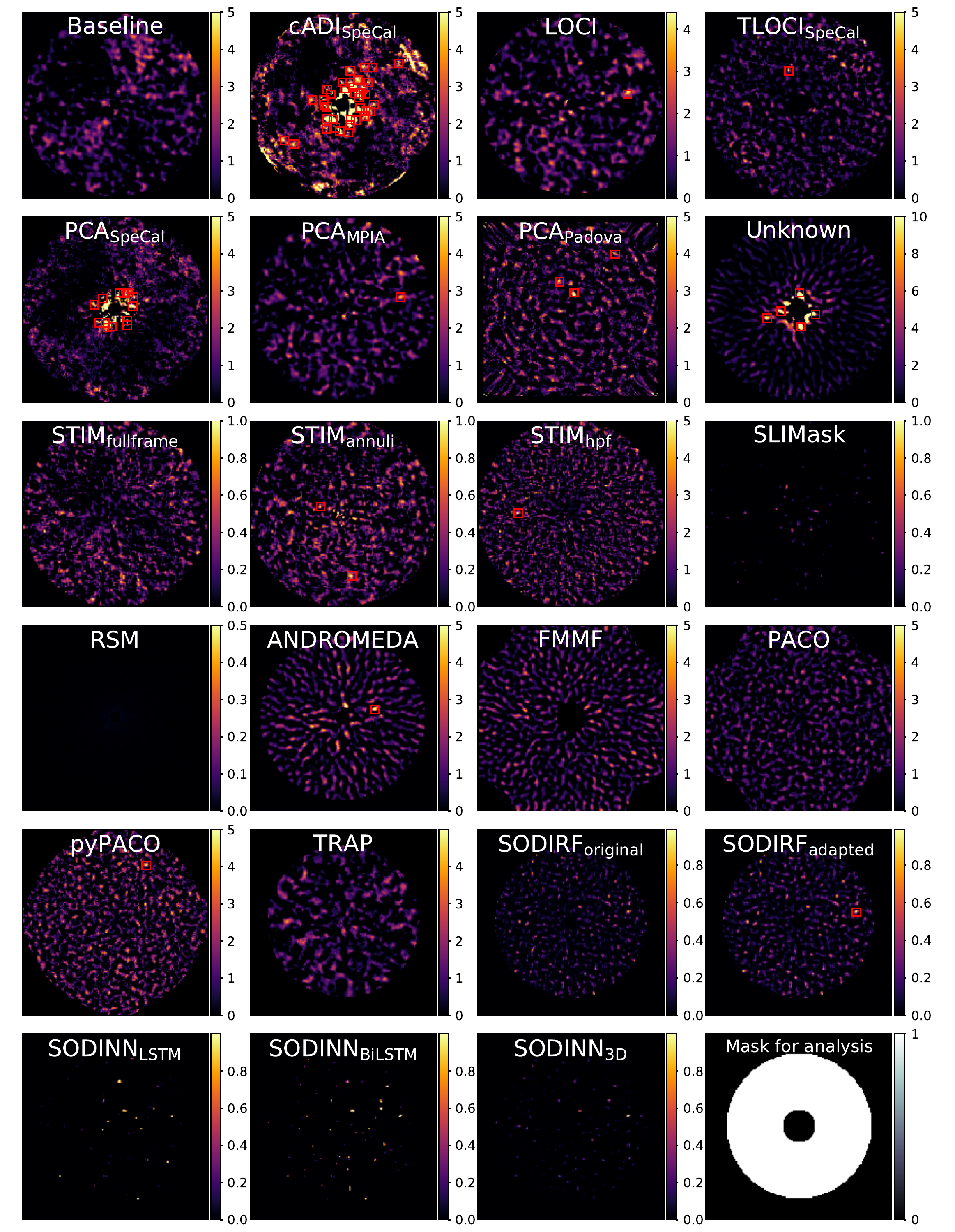}
    \caption{Results of the ADI subchallenge for the second VLT/SPHERE-IRDIS dataset (sph2).}
    \label{fig:app_sph2}
\end{figure}

\begin{figure}
    \centering
    \includegraphics[scale=0.71]{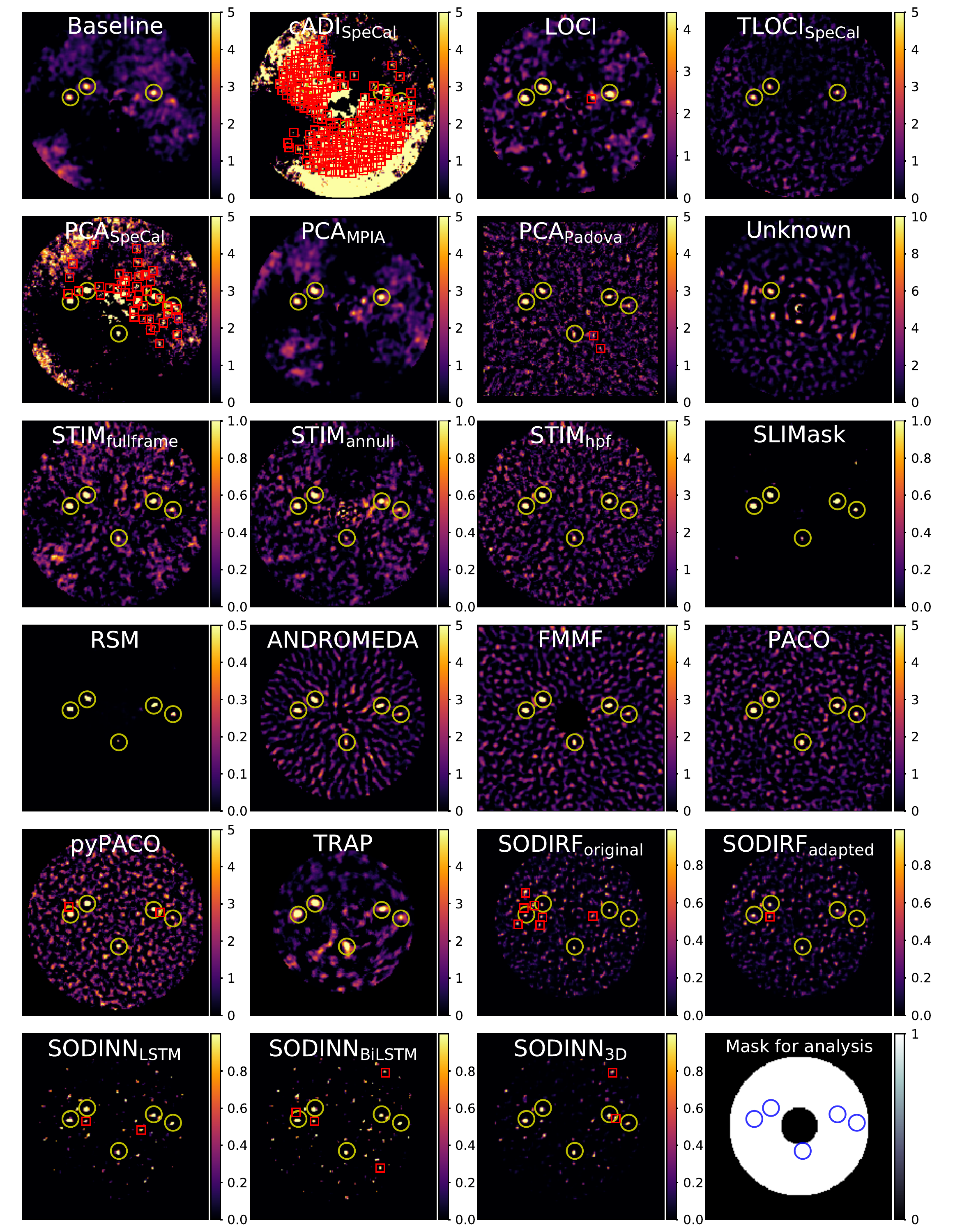}
    \caption{Results of the ADI subchallenge for the third VLT/SPHERE-IRDIS dataset (sph3).}
    \label{fig:app_sph3}
\end{figure}

\begin{figure}
    \centering
    \includegraphics[scale=0.71]{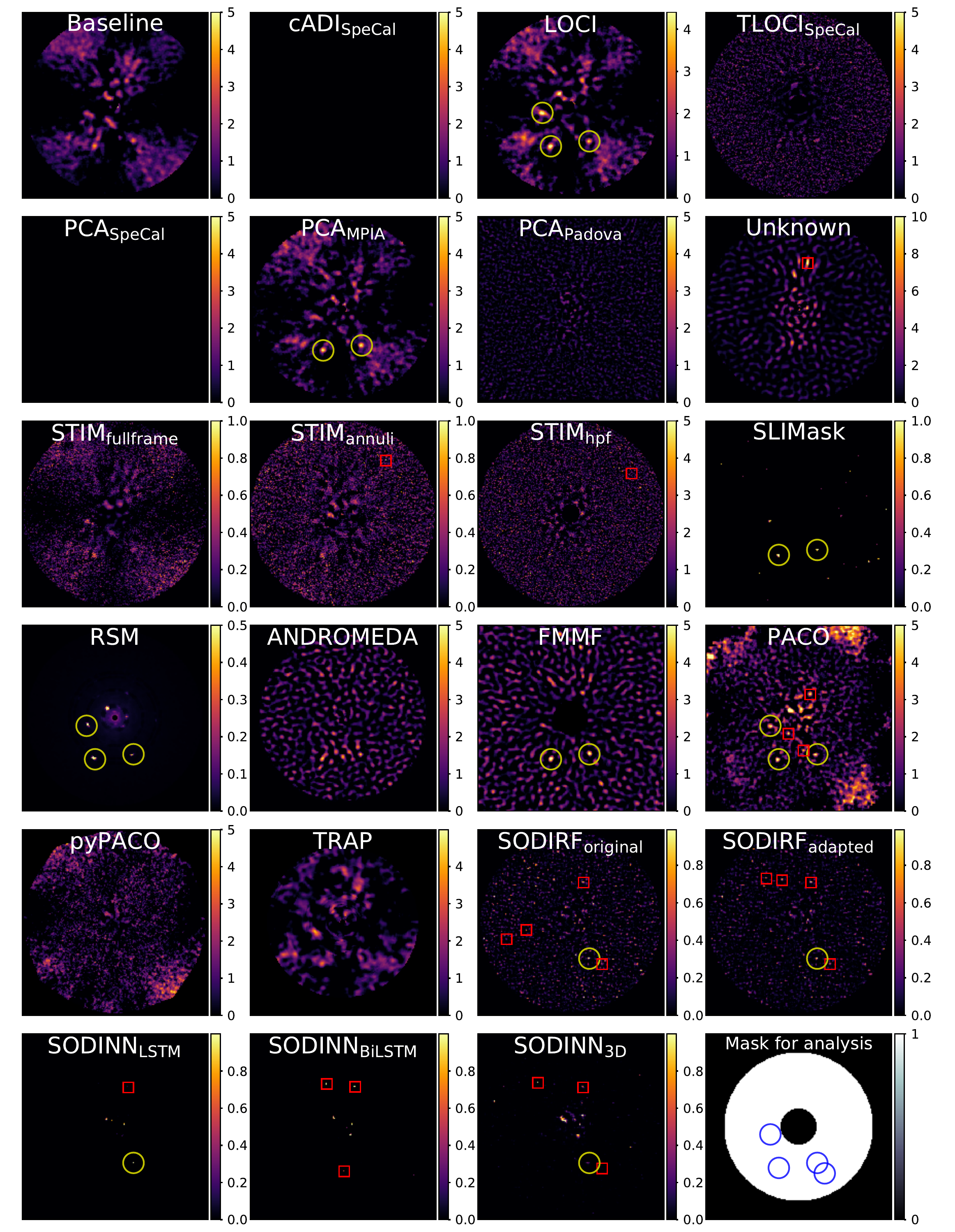}
     \caption{Results of the ADI subchallenge for the first Keck/NIRC2 dataset (nrc1).}
    \label{fig:app_nrc1}
\end{figure}

\begin{figure}
    \centering
    \includegraphics[scale=0.71]{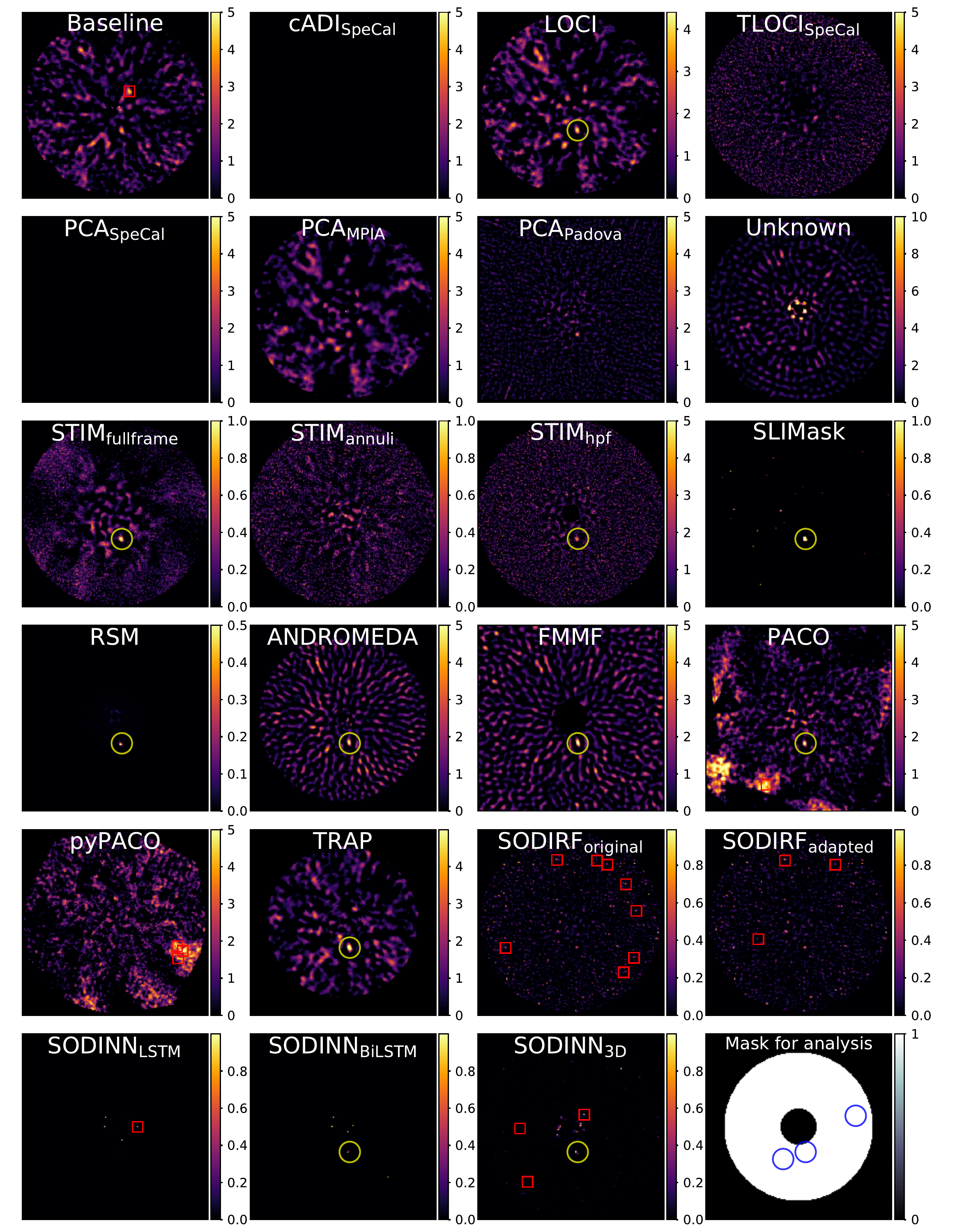}
    \caption{Results of the ADI subchallenge for the second Keck/NIRC2 dataset (nrc2).} 
    \label{fig:app_nrc2}
\end{figure}

\begin{figure}
    \centering
    \includegraphics[scale=0.71]{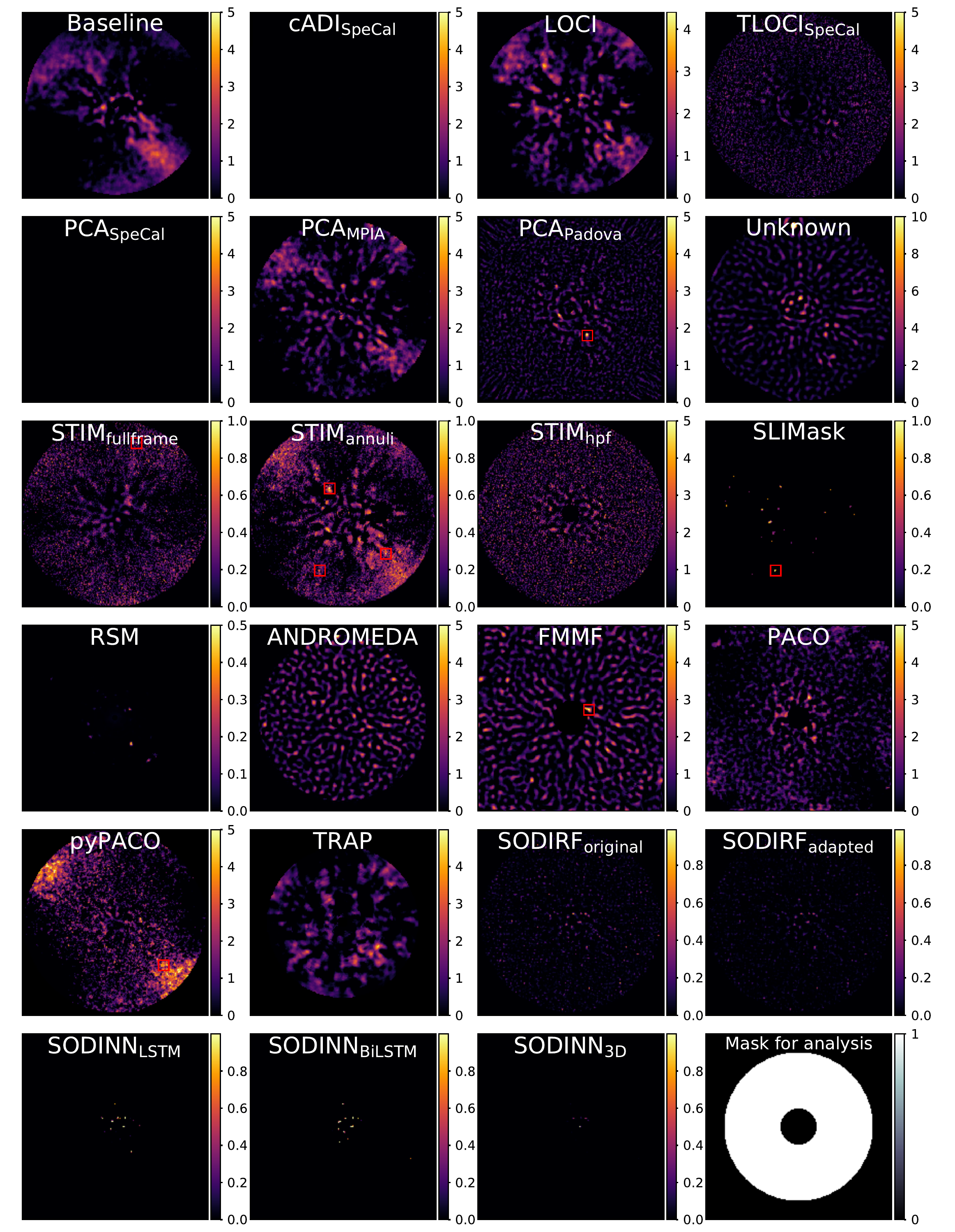}
    \caption{Results of the ADI subchallenge for the third Keck/NIRC2 dataset (nrc3).} 
    \label{fig:app_nrc3}
\end{figure}

\begin{figure}
    \centering
    \includegraphics[scale=0.71]{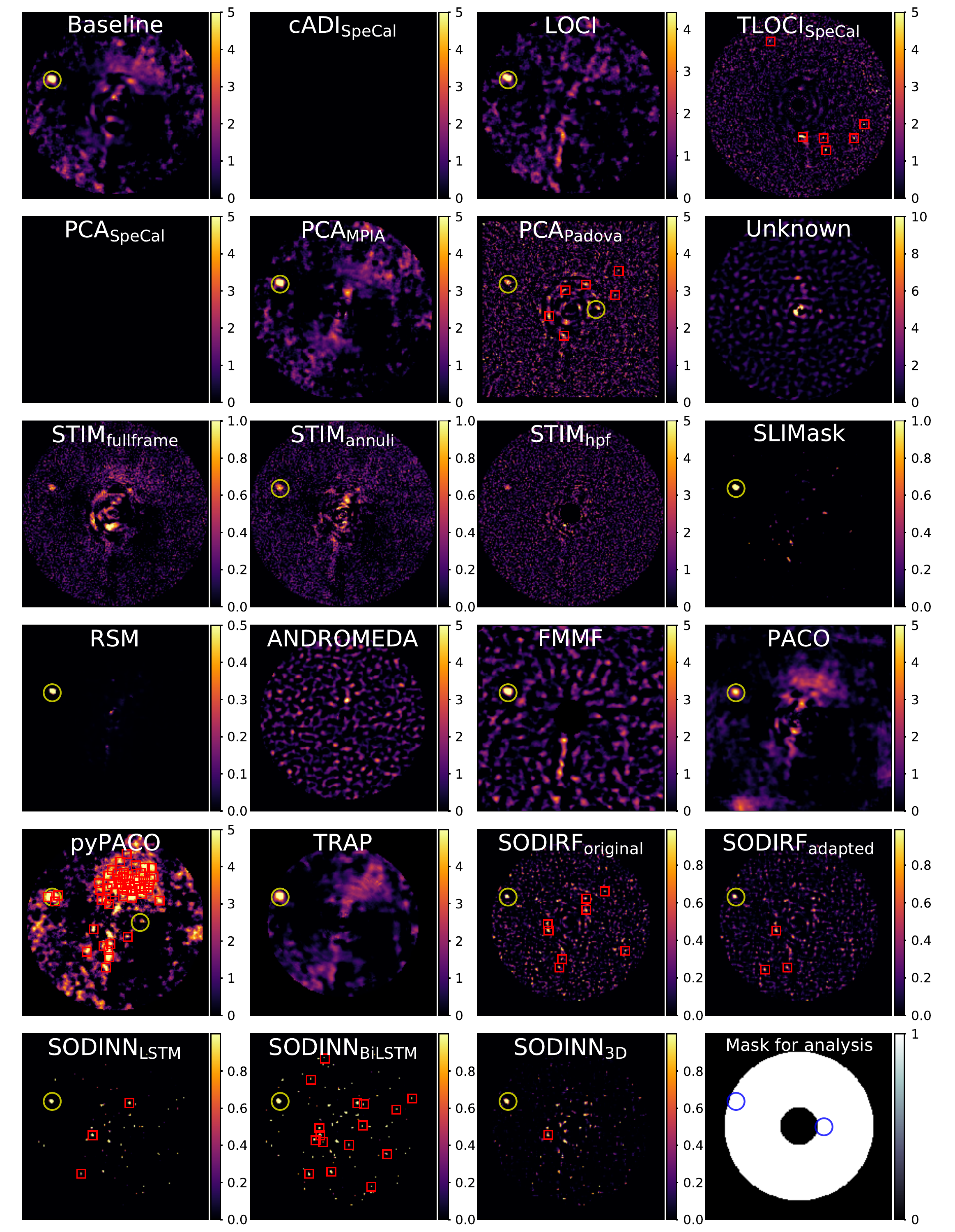}
    \caption{Results of the ADI subchallenge for the first LBT/LMIRCam dataset (lmr1).}
    \label{fig:app_lmr1}
\end{figure}

\begin{figure}
    \centering
    \includegraphics[scale=0.71]{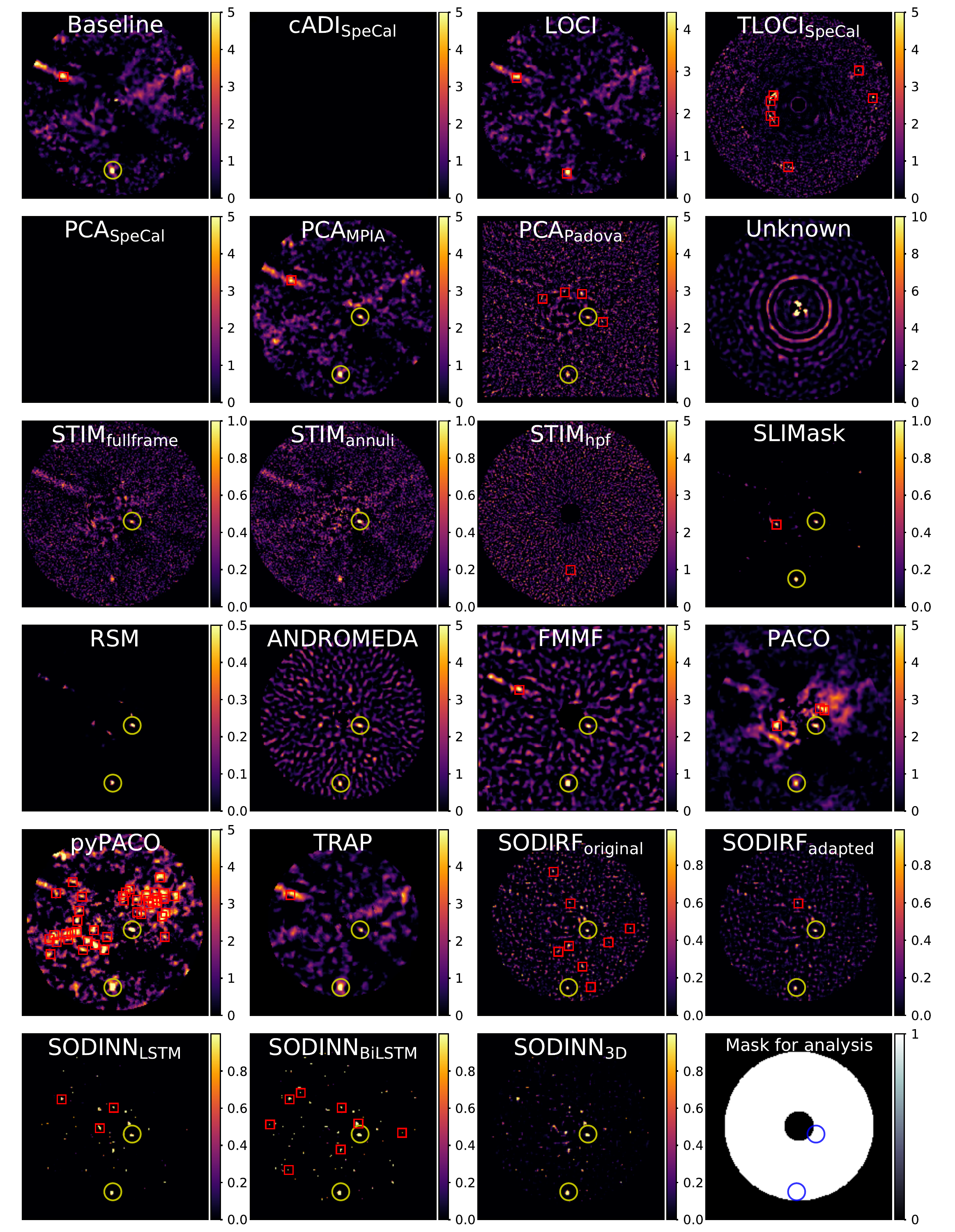}
  \caption{Results of the ADI subchallenge for the second LBT/LMIRCam dataset (lmr2).}
    \label{fig:app_lmr2}
\end{figure}

\begin{figure}
    \centering
    \includegraphics[scale=0.71]{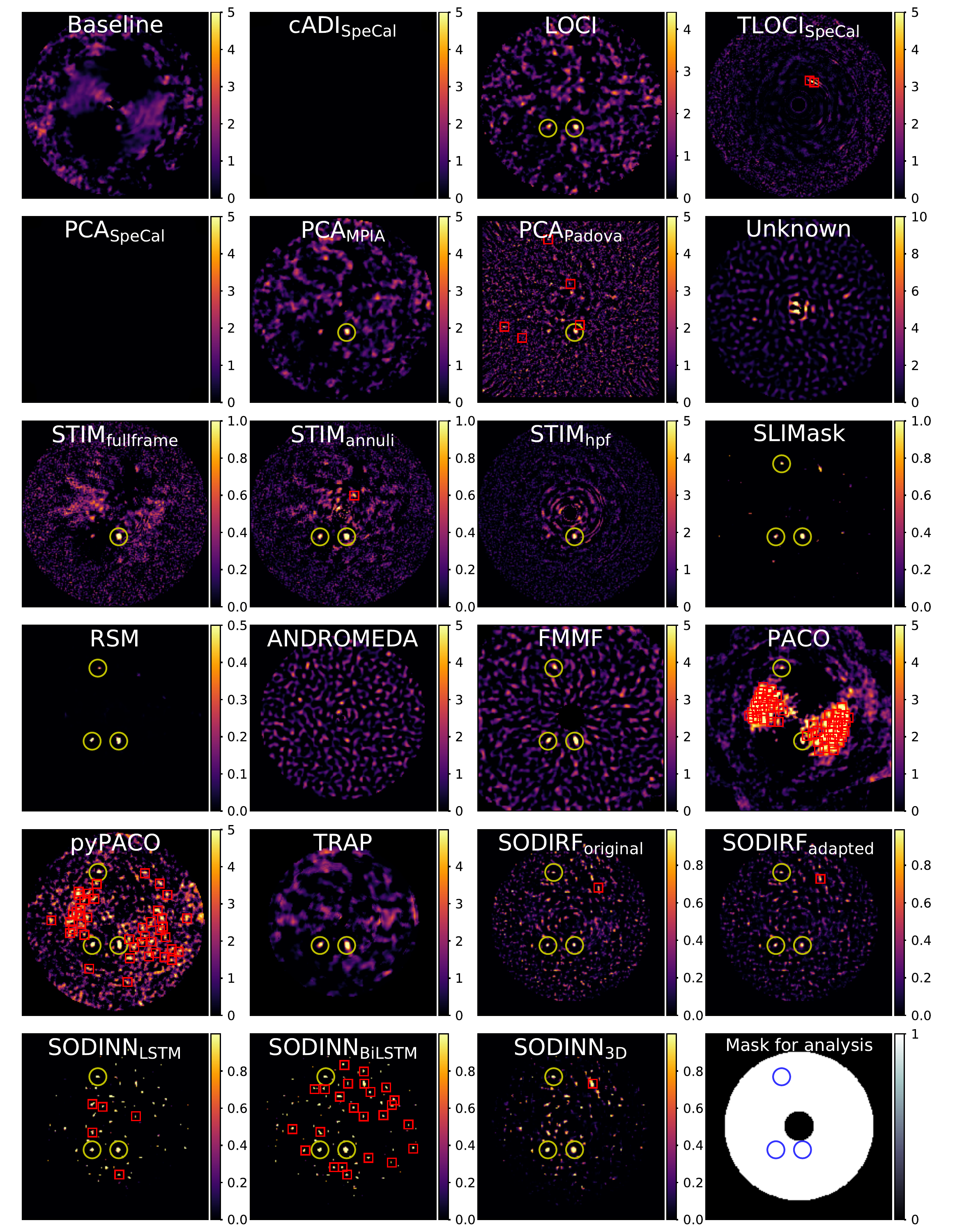}
    \caption{Results of the ADI subchallenge for the third LBT/LMIRCam dataset (lmr3).}
    \label{fig:app_lmr3}
\end{figure}

\section{Gallery of detection maps: ADI+mSDI subchallenge}
\label{app_asdi}
In this appendix, we display the detection maps of the 6 submissions for the ten data sets of the ADI+mSDI subchallenge. Each detection map is shown with a color-bar ranging from 0 to the submitted threshold. The yellow circles indicate true positives at the given threshold, while red squares indicate false positives. 
The first image (top left) shows the detection map for the chosen baseline algorithm (annular PCA). The last image (bottom right) shows the mask used to conduct the analysis of each map, with blue circles indicating the location of the injected synthetic planetary signals.

\begin{figure}
    \centering
    \resizebox{\hsize}{!}{\includegraphics{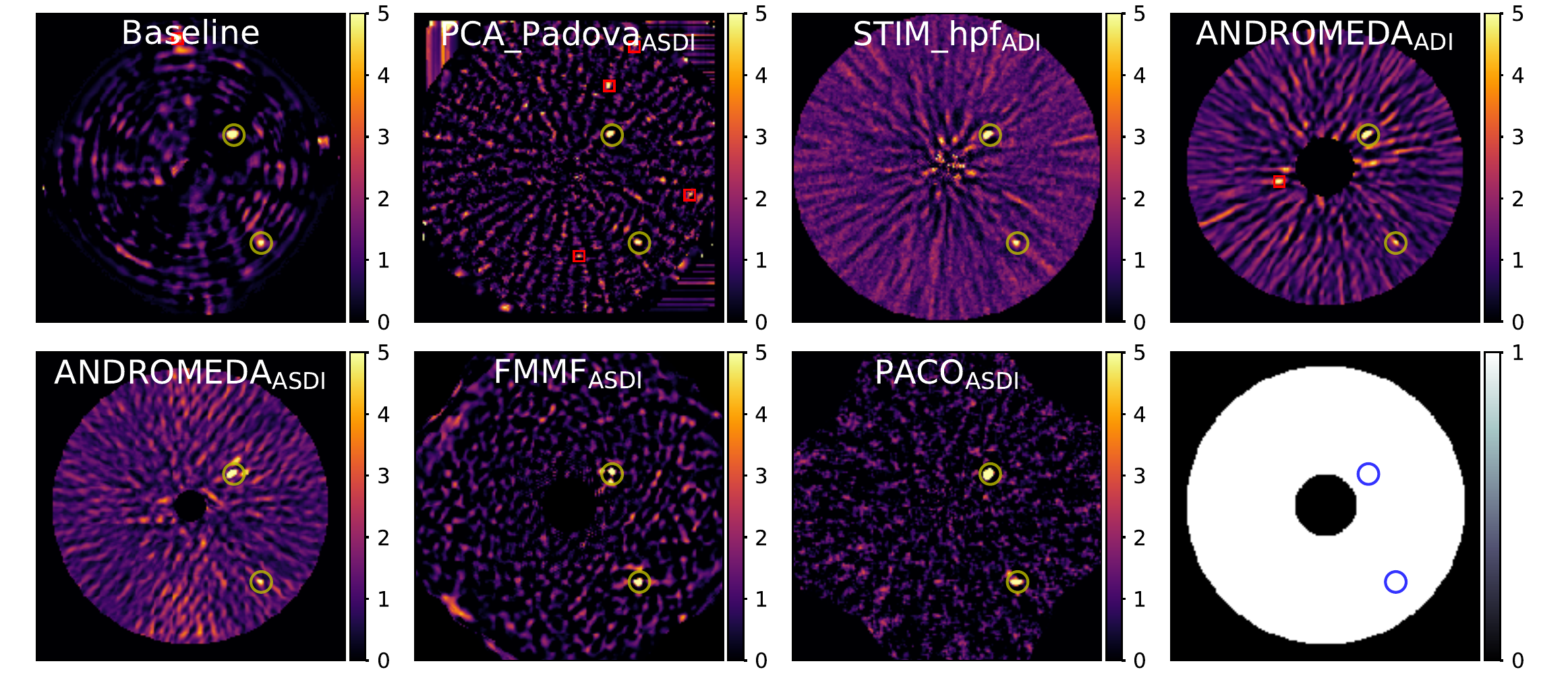}}
    \caption{Results of the ADI+mSDI subchallenge for the first VLT/SPHERE-IFS dataset (ifs1).}
    \label{fig:app_ifs1}
\end{figure}

\begin{figure}
    \centering
    \resizebox{\hsize}{!}{\includegraphics{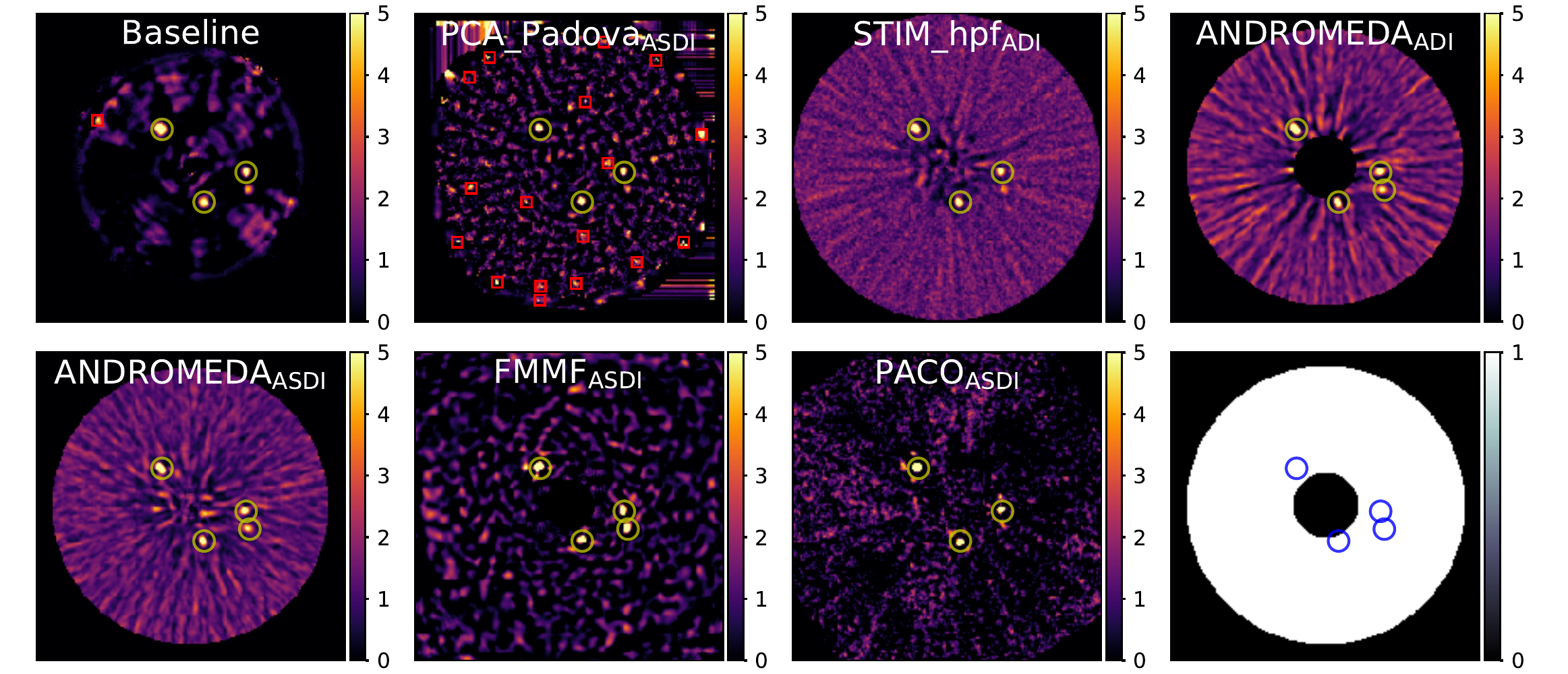}}
    \caption{Results of the ADI+mSDI subchallenge for the second VLT/SPHERE-IFS dataset (ifs2).}
    \label{fig:app_ifs2}
\end{figure}

\begin{figure}
    \centering
    \resizebox{\hsize}{!}{\includegraphics{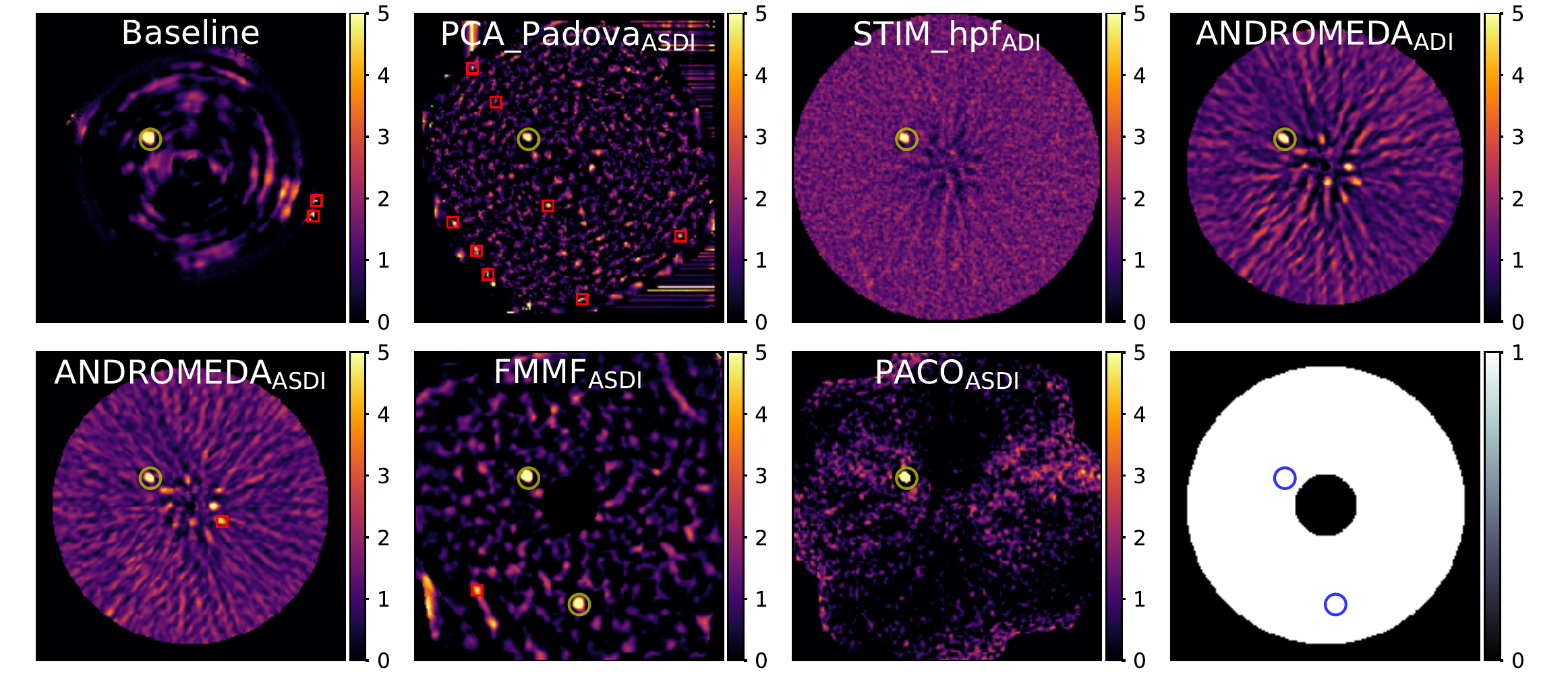}}
    \caption{Results of the ADI+mSDI subchallenge for the third VLT/SPHERE-IFS dataset (ifs3).}
    \label{fig:app_ifs3}
\end{figure}

\begin{figure}
    \centering
    \resizebox{\hsize}{!}{\includegraphics{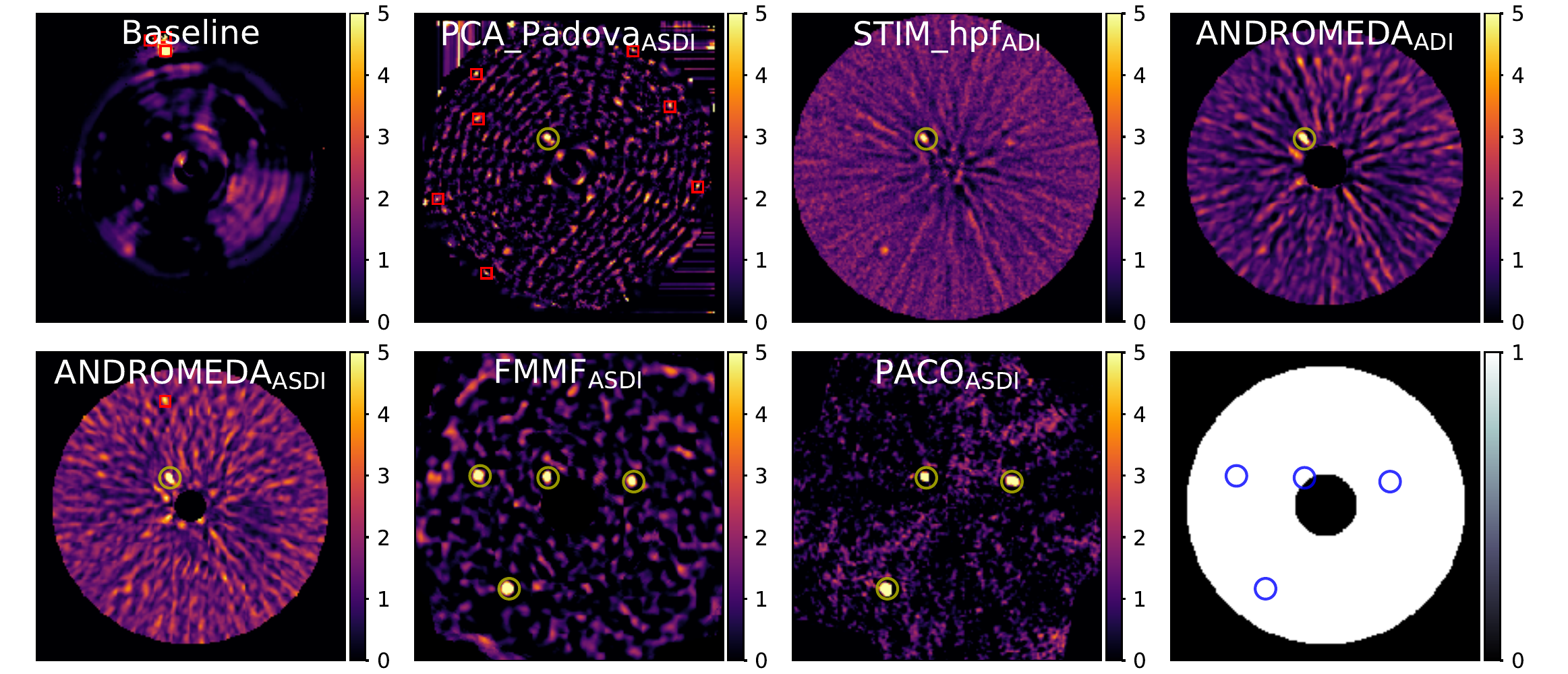}}
    \caption{Results of the ADI+mSDI subchallenge for the fourth VLT/SPHERE-IFS dataset (ifs4).}
    \label{fig:app_if4}
\end{figure}

\begin{figure}
    \centering
    \resizebox{\hsize}{!}{\includegraphics{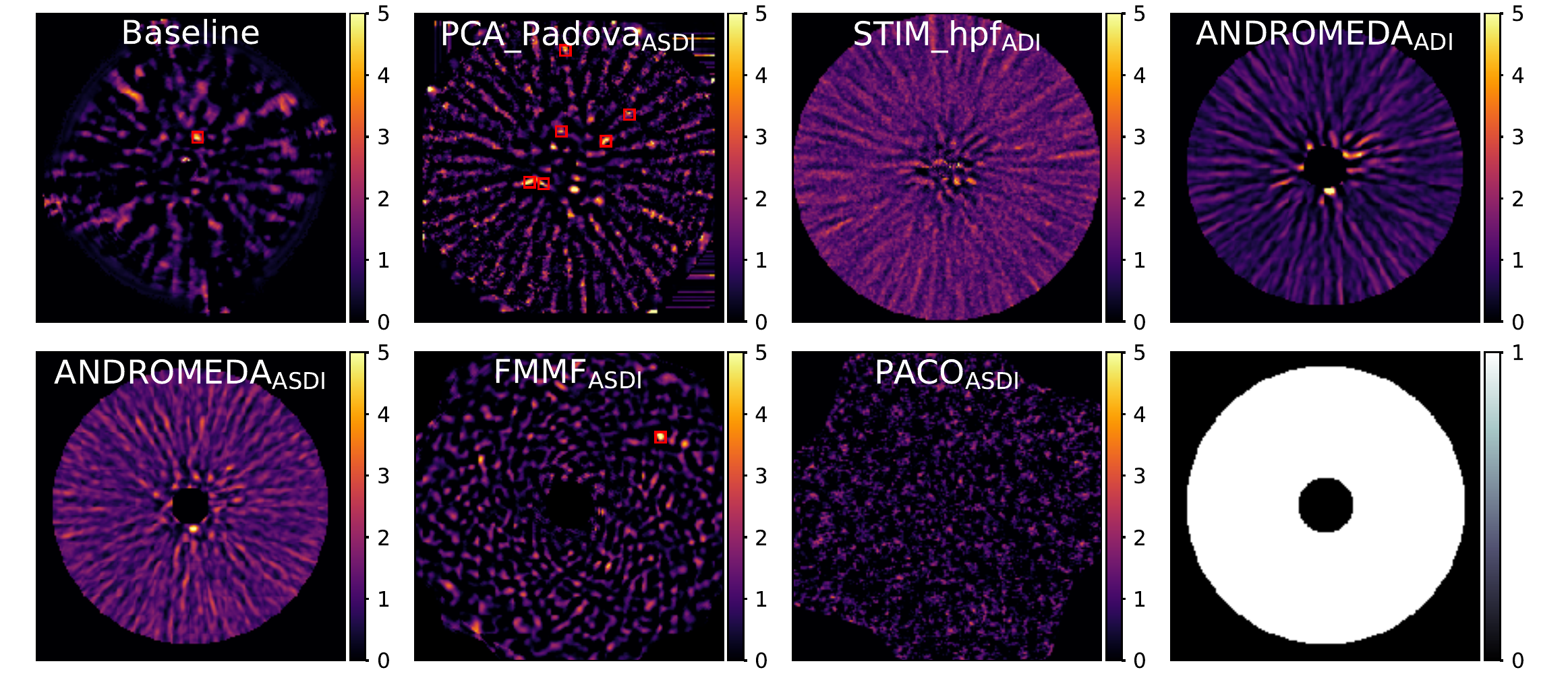}}
    \caption{Results of the ADI+mSDI subchallenge for the fifth VLT/SPHERE-IFS dataset (ifs5).}
    \label{fig:app_ifs5}
\end{figure}

\begin{figure}
    \centering
    \resizebox{\hsize}{!}{\includegraphics{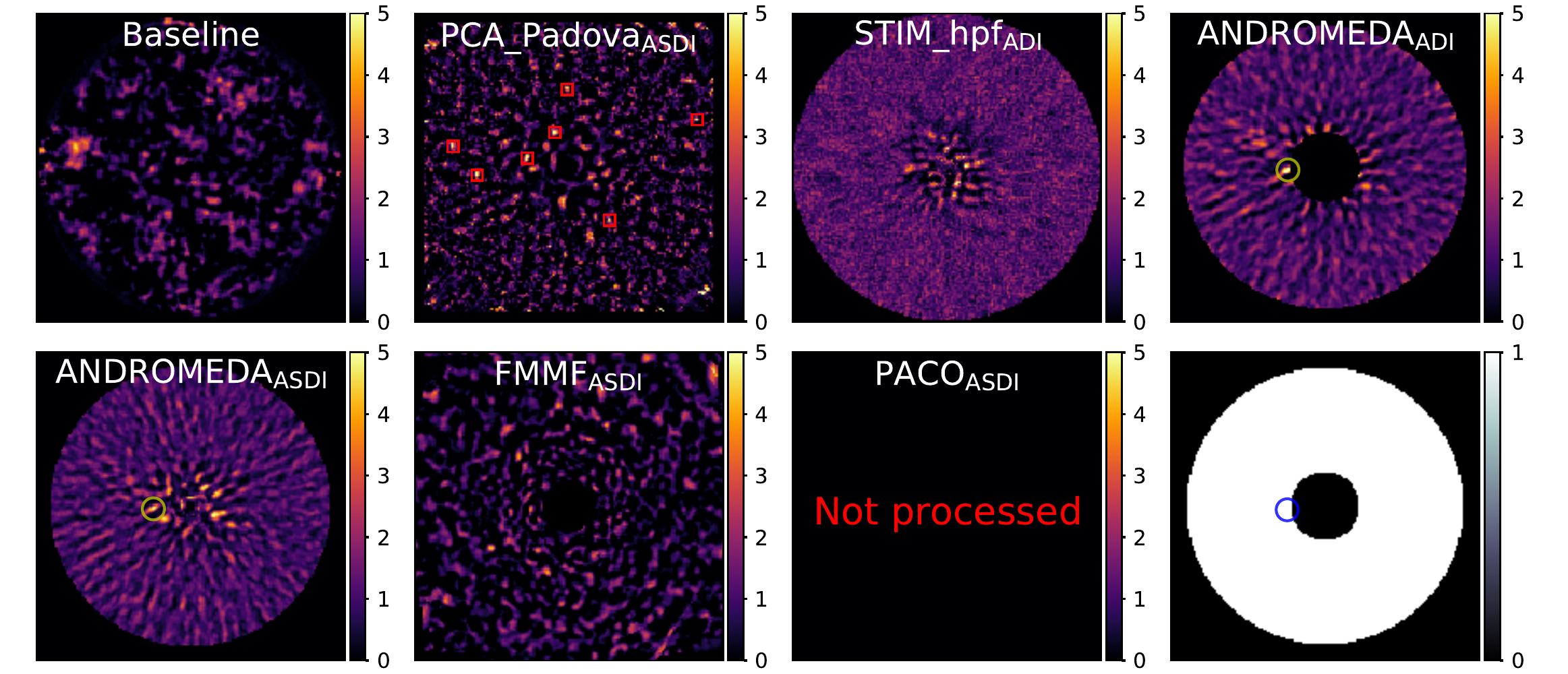}}
    \caption{Results of the ADI+mSDI subchallenge for the first Gemini-S/GPI dataset (gpi1).}
    \label{fig:app_gpi1}
\end{figure}

\begin{figure}
    \centering
    \resizebox{\hsize}{!}{\includegraphics{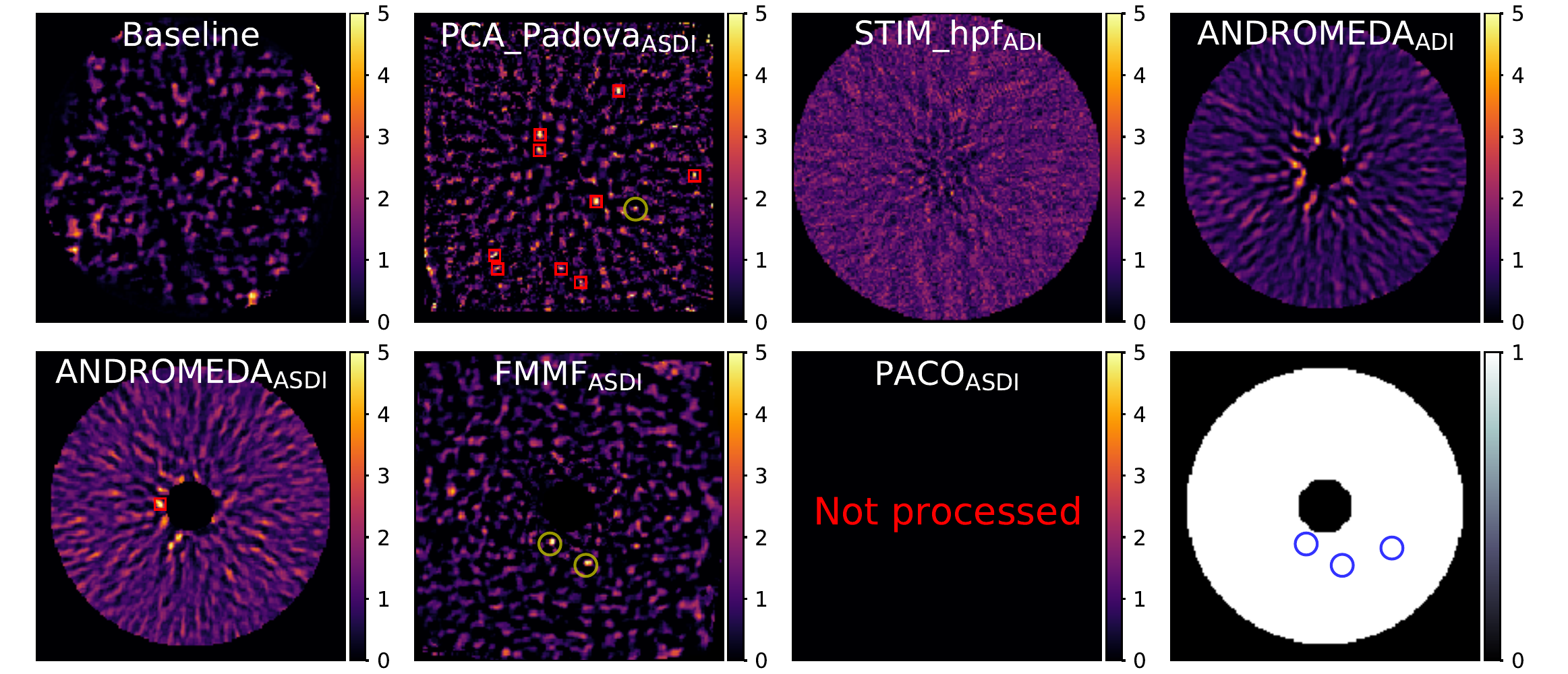}}
    \caption{Results of the ADI+mSDI subchallenge for the second Gemini-S/GPI dataset (gpi2).}
    \label{fig:app_gpi2}
\end{figure}

\begin{figure}
    \centering
    \resizebox{\hsize}{!}{\includegraphics{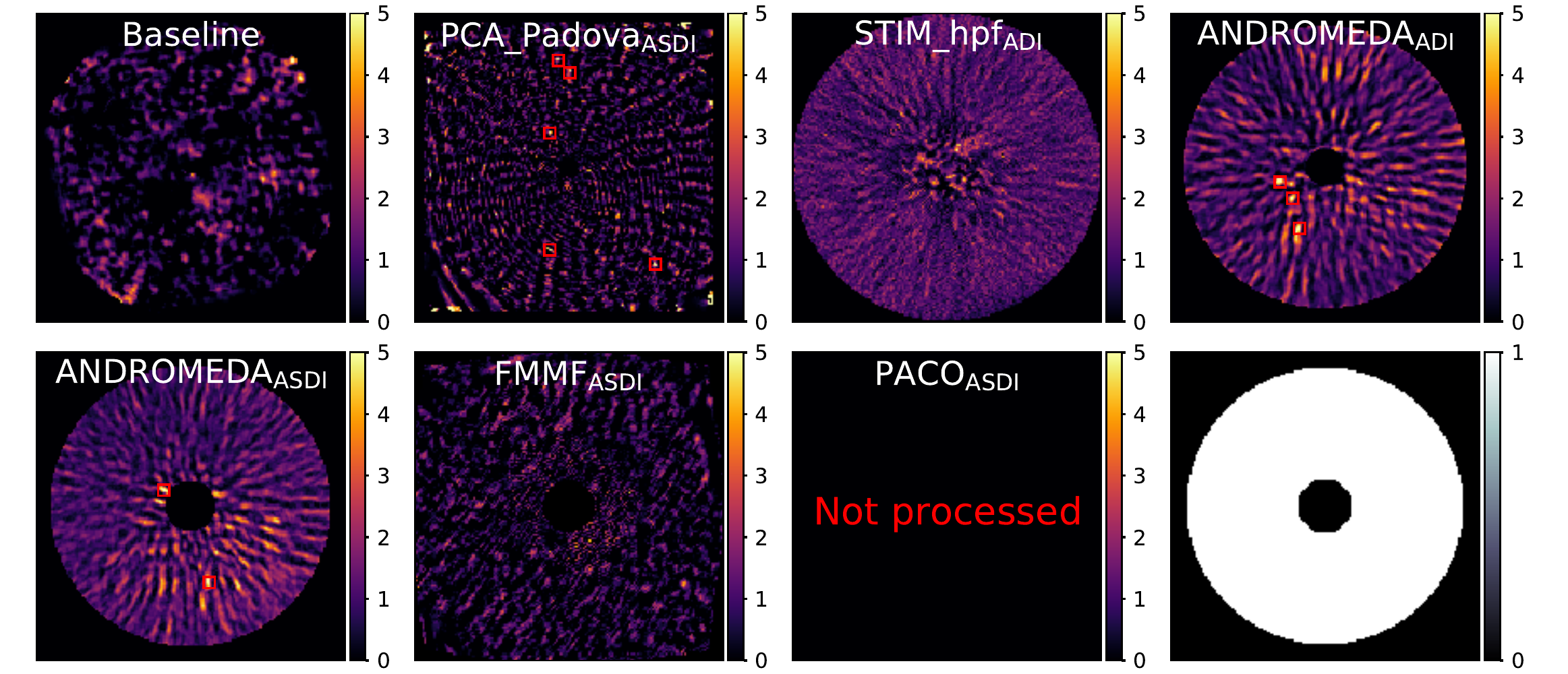}}
    \caption{Results of the ADI+mSDI subchallenge for the third Gemini-S/GPI dataset (gpi3).}
    \label{fig:app_gpi3}
\end{figure}

\begin{figure}
    \centering
    \resizebox{\hsize}{!}{\includegraphics{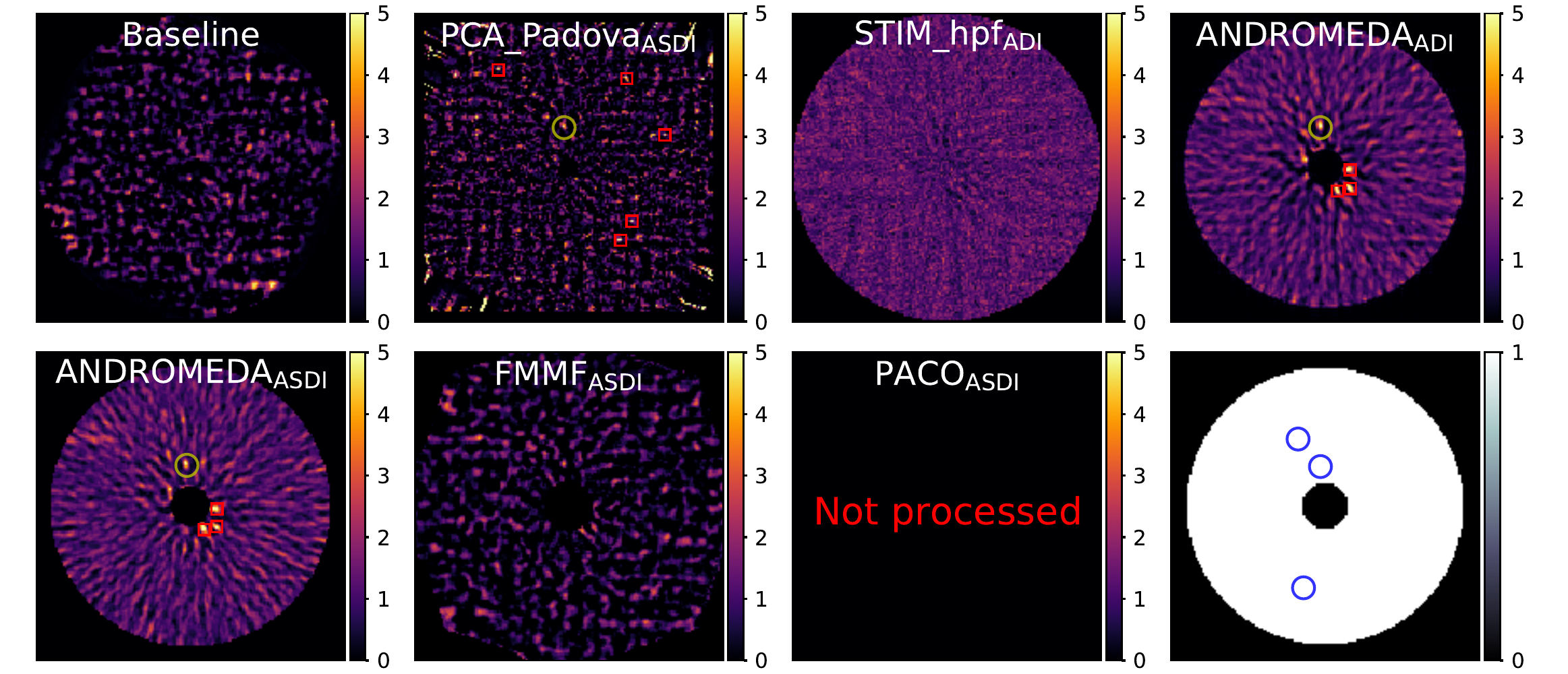}}
    \caption{Results of the ADI+mSDI subchallenge for the fourth Gemini-S/GPI dataset (gpi4).}
    \label{fig:app_gpi4}
\end{figure}

\begin{figure}
    \centering
    \resizebox{\hsize}{!}{\includegraphics{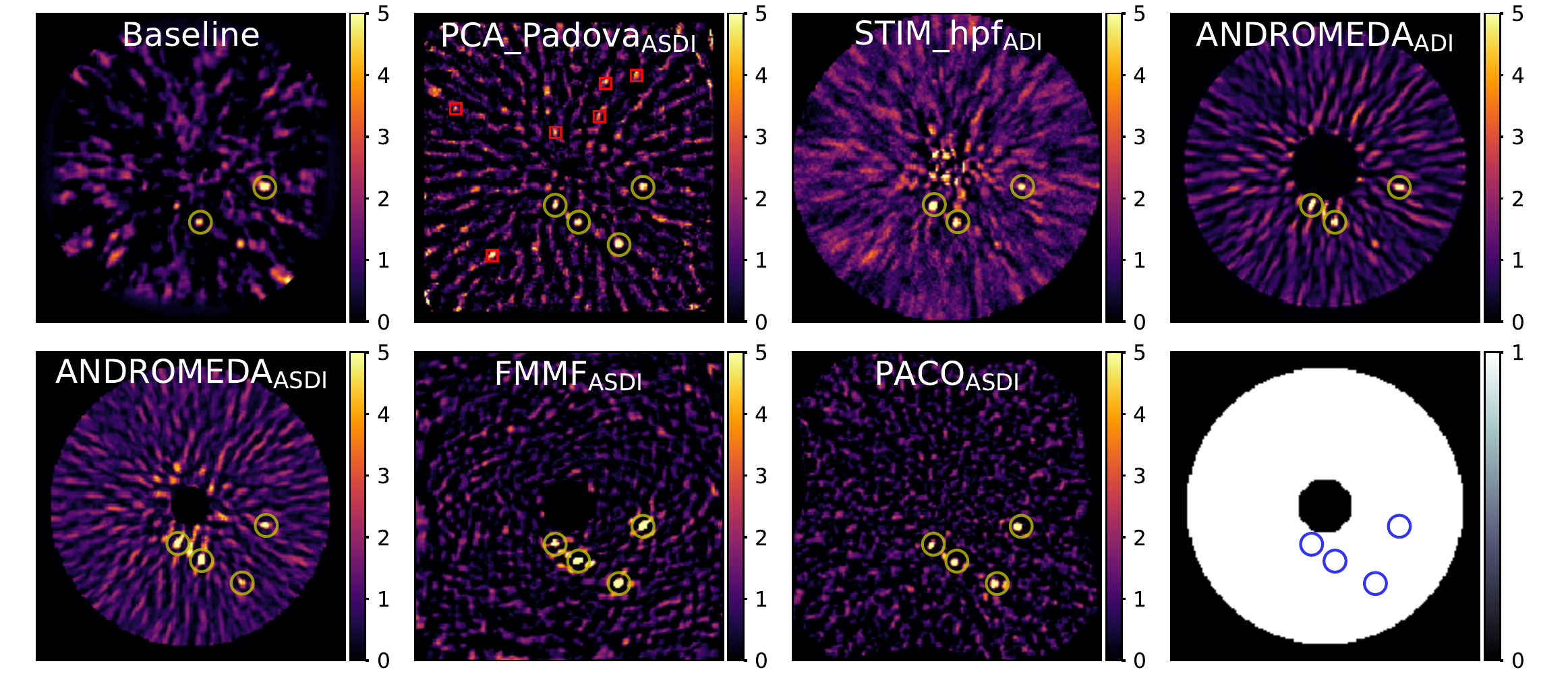}}
    \caption{Results of the ADI+mSDI subchallenge for the fifth Gemini-S/GPI dataset (gpi5).}
    \label{fig:app_gpi5}
\end{figure}

\acknowledgments
The EIDC collaboration would like to thank the GPIES collaboration and the SHINE collaboration, for providing us the pre-reduced data from Gemini/GPI and VLT/SPHERE respectively. 
This research has benefited from the SpeX Prism Spectral Libraries, maintained by Adam Burgasser at \url{http://pono.ucsd.edu/~adam/browndwarfs/spexprism}.
This work was supported by the Fonds de la Recherche Scientifique-FNRS under Grant nb F.4504.18 and by the European Research Council (ERC) under the European Union’s Horizon 2020 research and innovation program (grant agreement n 819155). 
T.H. acknowledges support from the European Research Council under the Horizon 2020 Framework Program via the ERC Advanced Grant Origins 83 24 28.

\bibliography{report} 
\bibliographystyle{spiebib} 




\end{document}